\documentclass[a4paper,final]{article}

\pdfoutput=1

\usepackage[T1]{fontenc} 
\usepackage[utf8]{inputenc} 
\usepackage{lmodern} 

\usepackage{subfiles}
\providecommand{\MainFolder}{.}
\providecommand{\MyPackagesFolder}{\MainFolder/LatexPackages}

\providecommand{\BibliographyFile}{\MainFolder/bibliography.bib}
\providecommand{\GraphicsFolder}{\MainFolder/Graphics}

\usepackage[
style=numeric,
giveninits=true,
backend=bibtex
]{biblatex}
\DeclareSourcemap{
 \maps[datatype=bibtex]{ 
 \map{\step[fieldsource=doi,final] 
      \step[fieldset=isbn,null]}   
 \map{\step[fieldsource=arxivId,final] 
      \step[fieldset=url,null]}        
 \map[overwrite=true]{ 
      \step[fieldsource=arxivId] 
      \step[fieldset=eprint, origfieldval] 
      \step[fieldsource=eprint, match={/}, final] 
      \step[fieldset=primaryClass, null]} 
 }}
\DefineBibliographyExtras{english}{} 
\usepackage{microtype} 

\usepackage{\MyPackagesFolder/my_math_external_packages}
\usepackage{\MyPackagesFolder/my_tools}
\usepackage{\MyPackagesFolder/my_math_1}
\usepackage{\MyPackagesFolder/my_amsthm}
\usepackage{\MyPackagesFolder/my_todo}
\usepackage{hyperref}

\newlist{RemarkList}{enumerate}{5}
\setlist[RemarkList]{
	leftmargin=0em,
	itemindent=*,
	label=(\roman*),
listparindent=\parindent 
}

\newlist{EqnList}{enumerate}{5}
\setlist[EqnList]{label=\Roman*.,listparindent=\parindent}

\newlist{PlainList}{enumerate}{5}
\setlist[PlainList]{label=(\arabic*),listparindent=\parindent}

\newlist{ExampleList}{enumerate}{5}
\setlist[ExampleList]{leftmargin=0em,itemindent=*,label=(\alph*),listparindent=\parindent}

\newlist{ClaimList}{enumerate}{5}
\setlist[ClaimList]{label=(\alph*),listparindent=\parindent}

\newlist{ProofList}{enumerate}{5}
\setlist[ProofList]{leftmargin=0em,itemindent=*,label=(\alph*),listparindent=\parindent}

\usepackage{authblk} 

\MHInternalSyntaxOn
\def\MT_mult_invisible_line: { 
  \crcr
  \global\MH_set_boolean_F:n {mult_firstline}
  \hbox to \l_MT_multwidth_dim{}\crcr
  \noalign{
    \vskip-\baselineskip 
    \vskip-\normallineskip
  }
}
\MHInternalSyntaxOff

\usepackage{ulem} 

\begin{document}

\ToDoList
\Modify[inline,caption={Stars}]{Remove * from the definitions of bar complexes and place them by hand where needed! But how to distinguish bar and dual bar complex then?}

\Modify[inline,caption={DONE Degree shift symbol}]{The formal symbol $\SuspU$ should be used as the symbol of degree $+2$ and not $-1$! Let use e.g. $\theta$ as the formal symbol of degree $-1$.}

\Modify[inline,caption={DONE Labels of lists}]{I changed 1), a), ... to (1), (a), ...!! Go through the text and check this!}

\title{Twisted $\IBLInfty$-algebra and string topology:\\ First look and examples} 
\author{Pavel H\'ajek\thanks{\textit{E-mail address:} \href{mailto:pavel.hajek@math.uni-augsburg.de}{pavel.hajek@math.uni-augsburg.de}}}
\affil{\normalsize Institut für Mathematik \\
Universit\"at Augsburg\\
Universit\"atsstrasse 14 \\
86159 Augsburg (Germany) \\
}
\maketitle
\begin{abstract}
\Modify[caption={DONE Vanishing results}]{Results about vanishing of $\PMC$.}
Cieliebak \& Fukaya \& Latschev proposed to twist the canonical $\IBL$-structure on cyclic cochains of $\HDR(M)$ for a closed oriented manifold~$M$ with a Maurer-Cartan element $\PMC$ built up from Chern-Simons like integrals associated to trivalent ribbon graphs. They conjectured that this construction gives a chain model for Chas-Sullivan string topology. In this text, we assume that the integrals converge and explicitly compute the case of~$\Sph{n}$, supporting the conjecture. We generalize this computation and show that the twist with $\PMC$ is often trivial.

\end{abstract}
\clearpage
\tableofcontents
%
%
\clearpage
\section{Introduction and summary}

An \emph{$\IBLInfty$-algebra} is essentially a collection of multilinear operations $\OPQ_{klg}$ with~$k$ inputs, $l$ outputs and ``genus'' $g$ satisfying certain relations; in particular,~$\OPQ_{110}$ is a boundary operator, and the pair $\OPQ_{210}, \OPQ_{120}$ induces the structure of an involutive Lie bialgebra on the homology of $\OPQ_{110}$. It was introduced in \cite{Cieliebak2015} and applications to string topology, symplectic field theory and higher genus Lagrangian Floer theory were proposed.

This text is an attempt to understand the application to \emph{string topology}. The idea was to carry out some explicit computations according to the plan sketched in~\cite[Section~13]{Cieliebak2015} and test the string topology conjecture (see below).


The following results from~\cite[Corollary~11.9]{Cieliebak2015} are our starting point (precise definitions of all the notions will be given in Section~\ref{Section:Review}; our $\IBLInfty$-algebras will be strict and filtered in the terminology of \cite{Cieliebak2015}): 
\begin{enumerate}[listparindent=\parindent,label=\textbf{(\Alph*)}]
\item For a finite-dimensional cyclic cochain complex $(V,\Pair,m_1)$ of degree $2-n$, 
there is a canonical $\dIBL$-structure $\OPP_{110}$, $\OPP_{210}$, $\OPP_{120}$ of bidegree $(n-3,2)$ on the degree shifted dual cyclic bar complex
$$ \CycC(V):= \DBCyc V[2-n] \simeq \Bigl(\bigoplus_{k\ge 1} \bigl(V[1]^{\otimes k} / \text{cyc}\bigr)' \Bigr)[2-n], $$
where $\mathrm{cyc}$ stands for cyclic permutations with the Koszul sign, $'$ denotes the graded dual and $[\cdot]$ the degree shift. This structure is denoted by $\dIBL(\CycC(V))$.

 \item Let $(\Harm,\Pair,m_1) \subset (V,\Pair,m_1)$ be a subcomplex such that the restriction of $\Pair$ to $\Harm[1]$ is non-degenerate. We apply (A) to $(\Harm,\Pair,m_1)$ to get the canonical $\dIBL$-algebra $\dIBL(\CycC(\Harm)) = (\CycC(\Harm),\OPQ_{110},\OPQ_{210},\OPQ_{120})$. 
Suppose that $\pi: V[1] \rightarrow V[1]$ is a projection to $\Harm[1]$ which satisfies
\begin{equation*}
 \begin{aligned}
 \pi \circ m_1 &= m_1 \circ \pi \quad\text{and}\\ \Pair(\pi(v_1),v_2) &= \Pair(v_1,\pi(v_2))
\end{aligned}
\end{equation*}
for all $v_1$, $v_2 \in V[1]$, and let $\iota: \Harm[1]\rightarrow V[1]$ be the inclusion. A linear map $\GOp: V[1]\rightarrow V[1]$ of degree $-1$ such that
\begin{equation}\label{Eq:ConditionOnG}
\begin{aligned}
m_1 \circ \GOp + \GOp \circ m_1 &= \iota\circ \pi - \Id_{V[1]}\quad\text{and} \\ 
\Pair(\GOp(v_1),v_2) &= (-1)^{\Abs{v_1}}\Pair(v_1,\GOp(v_2)) 
\end{aligned}
\end{equation}
for all $v_1$, $v_2 \in V[1]$ induces the $\IBLInfty$-homotopy equivalence 
$$\HTP=(\HTP_{klg}): \dIBL(\CycC(V)) \longrightarrow \dIBL(\CycC(\Harm)) $$
such that $\HTP_{110}: \CycC(V)[1] \rightarrow \CycC(\Harm)[1]$ is the map given by the precomposition with~$\iota$ in every component. We recall from \cite{Cieliebak2015} that $\HTP_{klg}: \Ext_k \CycC(V) \rightarrow \Ext_l \CycC(\Harm)$ is a linear map between exterior powers.
\end{enumerate}

The map $\HTP_{klg}$ is constructed as a sum of contributions coming from isomorphism classes of \emph{ribbon graphs} (=:\,multigraphs with a cyclic ordering of half-edges at every internal vertex) with~$k$ internal vertices, $l$ boundary components and genus $g$. To compute the contribution of a labeled ribbon graph~$\Gamma$ to the value 
$$ \HTP_{klg}(\Psi_1\otimes \dotsb \otimes \Psi_k)(\W_1 \otimes \dotsb\otimes \W_l)$$ 
for $\Psi_1$, $\dotsc$, $\Psi_k \in \DBCyc V[3-n]$ and  $\W_1$, $\dotsc$, $\W_l\in \BCyc \Harm [3-n]$, we decorate the~$i$-th internal vertex of $\Gamma$ with $\Psi_i$, external vertices lying on the $i$-th boundary component with components $v_{i1}$, $\dotsc$, $v_{i s_i}\in V[1]$ of $\W_i = \Susp (v_{i 1} \otimes \dotsb \otimes v_{i s_i} / \text{cyc})$, where $\Susp$ is a formal symbol of degree $n-3$, and internal edges with the Schwartz kernel $\GKer$ of $\GOp$ with respect to $\Pair$. Decorated ribbon graphs are then evaluated in a consistent way to obtain real numbers (see Appendix \ref{Section:Appendix} for an invariant formalism or \cite[Section 10]{Cieliebak2015} for a coordinate version of this construction).

We will also use the following results from \cite[Proposition~12.5 and Theorem~12.9]{Cieliebak2015} about deformations of $\IBLInfty$-algebras:
\begin{enumerate}[resume,listparindent=\parindent,label=\textbf{(\Alph*)}]
 \item If in addition to (A) there is the product $m_2 : V[1]\otimes V[1] \rightarrow V[1]$ making $(V,m_1,m_2)$ into a cyclic dga, then $(-1)^{n-2} m_2^+$ defines a canonical Maurer-Cartan element $\MC:=(\MC_{10})$ for $\dIBL(\CycC(V))$. The twisted $\IBLInfty$-algebra is again a $\dIBL$-algebra of bidegree $(n-3,2)$; it is denoted by $\dIBL^\MC(\CycC(V))$ and satisfies
\begin{equation} \label{Eq:CanonMCTwist}
\begin{aligned}
&\dIBL^\MC(\CycC(V)) \\ 
&\qquad = (\CycC(V),\OPP^\MC_{110}=\OPP_{110}+\OPP_{210}\circ_1\MC_{10},\ \OPP^\MC_{210}=\OPP_{210},\ \OPP^\MC_{120} = \OPP_{120}).
\end{aligned}
\end{equation}
\item The $\IBLInfty$-morphism $\HTP$ from (B) can be used to pushforward $\MC$ and obtain the Maurer-Cartan element $\PMC = (\PMC_{lg})$ for $\dIBL(\CycC(\Harm))$. The twist by $\PMC$ is an $\IBLInfty$-algebra of bidegree $(n-3,2)$; it is denoted by $\dIBL^\PMC(\CycC(\Harm))$ and satisfies
$$\begin{aligned}
& \dIBL^\PMC(\CycC(\Harm)) \\
& \quad = \begin{multlined}[t]
\bigl(\CycC(\Harm), \OPQ_{110}^\PMC = \OPQ_{110} + \OPQ_{210}\circ_1 \PMC_{10},\ \OPQ_{210}^\PMC=\OPQ_{210},\ \OPQ_{120}^\PMC = \OPQ_{120} \\+ \OPQ_{210} \circ_1 \PMC_{20},
\text{ plus the higher operations }\OPQ_{1lg}^\PMC = \OPQ_{210}\circ_1 \PMC_{lg} \bigr).
\end{multlined} \end{aligned} $$
This $\IBLInfty$-algebra is $\IBLInfty$-homotopy equivalent to $\dIBL^\MC(\CycC(V))$ via the twisted $\IBLInfty$-morphism 
$$\HTP^\MC=(\HTP^\MC_{klg}): \dIBL^\MC(\CycC(V)) \longrightarrow \dIBL^\PMC(\CycC(\Harm)). $$
\end{enumerate}

The pushforward Maurer-Cartan element $\PMC = \HTP_* \MC$ can be expressed as a sum of contributions of isomorphism classes of \emph{trivalent ribbon graphs} ($m_2^+$ has namely three inputs), where a labeled ribbon graph $\Gamma$ is decorated with $m_2^+$ at internal vertices, with the components of the $i$-th argument of $\PMC_{lg}$, i.e., elements of $\Harm(V)[1]$, at the $i$-th boundary component and with $\GKer$ at internal edges.
Note that whereas  (A) -- (C) can be formulated without completions, infinite sums appear in~$\PMC_{lg}$, and hence filtration and completions necessarily come into play.

The application to string topology of an oriented closed manifold $M$ of dimension $n$ comes from studying generalizations of (A) -- (D) to the infinite-dimensional cyclic dga $(\DR^*(M),\allowbreak \Pair,\allowbreak m_1,\allowbreak m_2)$. Here $\DR^*(M)$ is the de Rham complex of $M$ and the maps $\Pair: \DR(M)[1]^{\otimes 2} \rightarrow \R$, $m_1: \DR(M)[1]\rightarrow \DR(M)[1]$ and $m_2: \DR(M)[1]^{\otimes 2} \rightarrow \DR(M)[1]$ are defined for all $\eta$, $\eta_1$, $\eta_2 \in \DR(M)$ as follows:
\begin{equation} \label{Eq:DeRhamDGA}
 \qquad \mathllap{\text{de Rham cyc. dga}}\left\{ \begin{aligned}
 \Pair(\SuspU\eta_1,\SuspU\eta_2) &:= (-1)^{\eta_1} \int_M \eta_1\wedge \eta_2, \\ 
 m_1(\SuspU\eta) &:= \SuspU \Dd \eta,\\[\jot] 
 \quad m_2(\SuspU\eta_1,\SuspU\eta_2) &:= (-1)^{\eta_1} \SuspU(\eta_1\wedge\eta_2), \end{aligned}\right.
\end{equation}
where $\Dd$ is the de Rham differential, $\wedge$ the wedge product, $\SuspU$ a formal symbol of degree $-1$ and $\eta_1$ in the exponent denotes the form-degree of $\eta_1$. By picking a Riemannian metric on $M$, we obtain the subcomplex of harmonic forms 
$$ (\Harm^*(M), \Pair, m_1 \equiv 0 )$$ 
with the projection $\pi_{\Harm}: \DR(M) \rightarrow \Harm(M)$ coming from the Hodge decomposition. This cyclic cochain complex shall be taken as the subcomplex in (B).

From technical reasons stemming from the fact that the non-degenerate pairing $\Pair$ on $\DR(M)[1]$ is not perfect, one has to restrict the construction in (A) to the subspace $\DBCyc \DR(M)_\infty$ of elements with a smooth Schwartz kernel. Then (A) and (B) work in the setting of the so called Fr\'echet $\IBLInfty$-algebras introduced in \cite[Section 13]{Cieliebak2015}. However, the element $\MC_{10} \in \DBCyc \DR(M)[3-n]$, which translates into the \emph{Chern-Simons term}
\begin{equation*} \label{Eq:ChernSimons}
 m_2^+(\SuspU\eta_1,\SuspU\eta_2,\SuspU\eta_3) := (-1)^{\eta_2}  \int_M \eta_1\wedge \eta_2\wedge\eta_3\quad\text{for all }\eta_1, \eta_2, \eta_3 \in \DR(M),
\end{equation*}
does not define the canonical Maurer-Cartan element $\MC$ in (C) directly because $m_2^+ \not\in \CDBCyc \DR(M)_\infty$. This also means that one cannot use (D) to conclude the existence of the pushforward Maurer-Cartan element $\PMC$.

Nevertheless, it was proposed to define $\PMC$ formally using the summation over trivalent ribbon graphs as in the finite-dimensional case. We call such $\PMC$ a \emph{formal pushforward Maurer-Cartan element.} In order to compute the contribution of a labeled trivalent ribbon graph $\Gamma$ with $k$ internal vertices, $l$ boundary components and genus $g$ to the value
$$ \PMC_{lg}(\Omega_1\otimes \dotsb \otimes \Omega_l), $$
where $\Omega_i = \Susp \omega_i$ for $\omega_1$,~$\dotsc$, $\omega_l \in \BCyc\Harm(M)$, one starts by decorating internal vertices with integration variables $x_1$, $\dotsc$, $x_k$ on the $k$-fold product $M\times \dotsb \times M$, external vertices on the $i$-th boundary component with the components $\alpha_{i1}$,~$\dotsc$, $\alpha_{i s_i} \in\Harm(M)[1]$ of $\omega_i$ and internal edges with the Green kernel $\GKer$. In this setting, $\GKer$ becomes the Schwartz kernel of $\GOp$ in the sense of pseudo-differential operators; this $\GKer$ is necessarily singular at the diagonal $\Diag$, so that we have only $\GKer \in \DR^{n-1}(M\times M \backslash \Diag)$. One then takes the wedge product of all forms in the decorated graph in the order and with the sign deduced from the labeling of~$\Gamma$ and computes the integral over $x_1$, $\dotsc$, $x_k$. Similar integrals appear in \emph{perturbative Chern-Simons quantum field theory}.

Because of the singularity of $\GKer$ at $\Diag$, the integrand described above is smooth only on the $k$-th configuration space of $M$. It is not clear that all the integrals converge and that the resulting $\PMC_{lg}$ are well-defined and satisfy the Maurer-Cartan equation. The idea of work in progress~\cite{Cieliebak2018} of K. Cieliebak and E. Volkov is to use iterated spherical blow-ups of the diagonals to resolve the singularities and obtain integrals of smooth forms on compact manifolds with corners; this guarantees integrability. The Maurer-Cartan equation for $\PMC = (\PMC_{lg})$ is then proven with the help of Stokes' formula and by showing that the contributions of hidden codimension-$1$ faces cancel. This method is similar to the method from~\cite{Axelrod1991} and~\cite{Axelrod1993}, where Feynman integrals of perturbative Chern-Simons theory were considered.

Having $\PMC$, the twisted $\IBLInfty$-algebra $\dIBL^\PMC(\CycC(\Harm(M)))$, which can be equivalently written as $\dIBL^\PMC(\CycC(\H_{\mathrm{dR}}(M)))$ using the Hodge isomorphism $\Harm(M)\simeq \H_{\mathrm{dR}}(M)$, should satisfy the following conjecture:

\newtheorem*{Conject}{String topology conjecture}
\begin{Conject}[Conjecture~1.12 in~\cite{Cieliebak2015}]
Let $M$ be a closed oriented manifold of dimension $n$ and $\H_{\mathrm{dR}}(M)$ its de Rham cohomology. Then there exists an $\IBLInfty$-structure on (a suitable version of) $\DBCyc \H_{\mathrm{dR}}[2-n]$ whose homology equals the cyclic cohomology of the de Rham complex of $M$.
\end{Conject}

The idea is that the \emph{$\Sph{1}$-equivariant homology of the free loop space} $\StringH_*(\Loop M)$ is isomorphic to a version of \emph{Connes' cyclic cohomology of the de Rham algebra} $\CycCoH^*(\DR^*(M))$, at least for simply-connected $M$. The precise relation will be established in yet another work in progress~\cite{Cieliebak2018} of K. Cieliebak and E. Volkov using a chain-map coming from a cyclic version of Chen's iterated integrals. Now, a suitable degree shift of $\CycCoH^*(\DR^*(M))$ is isomorphic to the homology of the boundary operator~$\OPQ_{110}^\MC$ of the only formally defined $\dIBL$-algebra $\dIBL^\MC(\CycC(\DR(M)))$, which is according to (D) (formally) quasi-isomorphic to $\dIBL^\PMC(\CycC(\Harm(M)))$ via the twisted $\IBLInfty$-morphism $\HTP^\MC$.

The space $\StringH_*(\Loop M)$ is equipped with an $\IBL$-structure coming from the Chas-Sullivan string bracket $\StringOp_2$ and string cobracket $\StringCoOp_2$; \Add[caption={DONE Reduced loop space}]{Write that they are actually defined on the homology relative to constant loops. Actually, write here modulo problems with moding out constant loops or one constant loop.} these operations were defined geometrically on suitably transverse smooth chains in \cite{Sullivan1999} and \cite{Sullivan2002}, respectively.\footnote{In fact, $\StringCoOp_2$ is geometrically defined only on the homology relative to constant loops and~$\StringOp_2$ does not always restrict to it.} The natural question is: How is the $\IBL$-structure $\StringOp_2$, $\StringCoOp_2$ related to the $\IBL$-structure $\OPQ_{210}^\PMC$, $\OPQ_{120}^\PMC$ induced on $\StringH_*(\Loop M)$ via the isomorphism from the string topology conjecture? The \emph{extended string topology conjecture} asserts that these structures agree, and hence the operations $\OPQ_{210}^\PMC$, $\OPQ_{120}^\PMC$ defined on cyclic cochains provide a \emph{chain model} for~$\StringOp_2$, $\StringCoOp_2$. Based on our observations and explicit computations, we formulate an up-to-date version of the string topology conjecture for simply-connected manifolds (see Conjecture~\ref{Conj:StringTopology}).

A large part of this text consists of setting up the algebraic base for the work with $\dIBL^\PMC(\CycC(\Harm(M)))$. In addition to repeating the theory from~\cite{Cieliebak2015} in a slightly different formalism, we also include the following topics:

\begin{itemize}
\item A formula for the partial composition $\circ_s$ in terms of operations of the canonical associative bialgebra on the symmetric algebra (Definition~\ref{Def:CircS}); formulas for $\OPQ_{klg}^\PMC$ (Proposition~\ref{Prop:Formulafortwisted}).
\item Definition of the cyclic cohomology of $\AInfty$-algebras (Definition~\ref{Def:CycHom}) and its relation to the homology of $\OPQ_{110}^\PMC$ (Proposition~\ref{Prop:CyclicHom}); definitions of the reduced versions (Definitions~\ref{Def:ReducedDual}, \ref{Def:ReduceddIBL} and \ref{Def:StrictlyReduced}) and their relation to the unreduced versions (Propositions~\ref{Prop:Ones} and \ref{Prop:Reduced}).
\item An invariant formulation of the evaluation of labeled ribbon graphs (Definition~\ref{Def:EvalRibGraph} and Proposition~\ref{Prop:GraphPairing}); formal analogy of the finite-dimensional and the de Rham case which we use to obtain signs for the definition of~$\PMC$ (Proposition~\ref{Prop:FinDimAnalog}).
\item Definition of the Green kernel (Definition~\ref{Def:GreenKernel}) and of the formal pushforward Maurer-Cartan element $\PMC$ (Definition~\ref{Def:PushforwardMCdeRham}). 
\end{itemize}

Our first result is an explicit computation of $\dIBL^\PMC(\CycC(\HDR({\Sph{n}})))$ by finding a particular Green kernel and showing that all integrals which contribute to $\PMC$ vanish for $n\ge 3$; for $n=1$, there is a non-vanishing integral whose value we compute (see Section~\ref{Sec:GreenSphere}); for $n=2$, the existence of a non-vanishing integral remains open.

\begin{IntroThm}[Explicit computation for $\Sph{n}$]\label{IntroThm:A}
Consider the round sphere $\Sph{n}\subset \R^{n+1}$. Define $\NOne:= \SuspU 1$, $\NVol:= \SuspU \Vol \in \HDR(\Sph{n})[1]$, where $\Vol$ is the volume form, $1$ the constant one and $\SuspU$ a formal symbol of degree $-1$. The following holds for the homology of the twisted boundary operator $\OPQ_{110}^\PMC$:
\begin{equation*}
\begin{aligned}
& \HIBL^\PMC(\CycC(\HDR(\Sph{n})))[1] := \H(\CDBCyc \HDR(\Sph{n})[3-n],\OPQ_{110}^\PMC) \\
& \qquad =\begin{cases}
\langle \Susp \NVol^{i*}, \Susp \NOne^{2j-1*} \mid i, j \ge 1\rangle & \text{for }n\ge 3 \text{ odd}, \\
\langle \Susp \NVol^{2 i-1*}, \Susp \NOne^{2j-1*} \mid i, j \ge 1\rangle &\text{for }n\text{ even}, \\
 \bigl\langle \Susp\sum_{k=1}^\infty c_k \NVol^{k*}, \Susp \NOne^{2j-1*}\mid c_k\in \R, j \ge 1\bigr\rangle & \text{for }n=1. 
\end{cases}
\end{aligned}
\end{equation*}
Here $\langle \cdot \rangle$ denotes the linear span over $\R$, $*$ the dual and $\Susp$ is a formal symbol of degree $n-3$. The product $\OPQ_{210}^\PMC$ vanishes on $\HIBL^\PMC$ except for the following relations for $n\ge 3$ odd
$$ \OPQ_{210}^\PMC(\Susp \NOne^* \otimes \Susp \NVol^{k*}) = \OPQ_{210}^\PMC(\Susp \NVol^{k*} \otimes \Susp \NOne^*) = -(k-1) \NVol^{k-1*} $$
and the following relations for $n=1$:
$$ \OPQ_{210}^\PMC\Bigl(\Susp \NOne^* \otimes \Susp\sum_{k=1}^\infty c_k \NVol^{k*}\Bigr) = - \Susp \sum_{k=1}^\infty k c_{k+1} \NVol^{k*}.  $$
The coproduct $\OPQ_{120}^\PMC$ as well as all higher operations $\OPQ_{1lg}^\PMC$ vanish on $\HIBL^\PMC$ in every  dimension $n$. For $\Sph{1}$, we have $\OPQ_{120}^\PMC \neq \OPQ_{120}$ on the chain level; i.e., the twisting is non-trivial. For $n\neq 2$, all higher operations vanish on the chain level.

\end{IntroThm}

If we mod out $\Susp \NOne^{2j-1*}$, i.e., if we consider the point-reduced version\Modify[caption={DONE Point reduced}]{Write Point reduced version.}, then, after dropping $\Susp$, the results agree with the string topology of $M$ relative to one constant loop and with Chas-Sullivan operations. The only exception is $M=\Sph{1}$. This supports the string topology conjecture for simply-connected manifolds and provides a counterexample for non-simply connected manifolds.

\Correct[caption={DONE Degree shift}]{Write here that we have to drop $\Susp$!}

Our second result generalizes the previous explicit computation and shows that in many cases, the twists with $\PMC$ and $\MC$ coincide. Its proof is a combination of facts from Section~\ref{Sec:Vanishing}.


\begin{IntroThm}[Triviality of the twist with $\PMC$ on the chain level] \label{IntroThm:B}
Let $M$ be a closed oriented $n$-manifold. There exists a Green kernel $\GKer$ such that the following holds for the twisted $\IBLInfty$-structure $\dIBL^\PMC(\CycC(\HDR(M)))$:
\begin{enumerate}[label=(\arabic*)]
\item For the basic operations $\OPQ_{110}^\PMC = \OPQ_{210}\circ_1 \PMC_{10}$, $\OPQ_{210}^\PMC = \OPQ_{210}$, $\OPQ_{120}^\PMC = \OPQ_{120} + \OPQ_{210}\circ_1 \PMC_{20}$, we have:
\begin{enumerate}
\item If $\HDR^1(M)=0$, then $\PMC_{20}=0$, and hence $\OPQ_{120}^\PMC = \OPQ_{120}$.
\item If $M$ is geometrically formal, then $\PMC_{10} = \MC_{10}$, and hence $\OPQ_{110}^\PMC = \OPQ_{110}^\MC$. (In fact, if in addition $\HDR^1(M)=0$, then $\PMC = \MC$, at least for $n\ne 2$.)
\end{enumerate}
\item For the higher operations $\OPQ_{1lg}^\PMC=\OPQ_{210}\circ_1 \PMC_{20}$ with $(l,g)\neq (1,0)$, $(2,0)$, we have $\PMC_{lg}=0$, and hence $\OPQ_{1lg}^\PMC=0$ with the possible exception of surfaces and $3$-manifolds with $\HDR^1(M)\neq 0$.
\end{enumerate}
\end{IntroThm}

In our future work,  we plan to concentrate on the following:
\begin{enumerate}[label=(\arabic*)]
\item We would like to improve Theorem~\ref{IntroThm:B} by showing that the higher operations for $\Sph{2}$ vanish. If this is the case,  then the statement that all higher operations vanish for every manifold $M$ with $\HDR^1(M)=0$ is true.
\item For a formal simply-connected manifold $M$, we would like to investigate whether $\dIBL^\PMC(\CycC(\HDR(M)))$ and $\dIBL^\MC(\CycC(\HDR(M)))$ are homotopy equivalent as $\IBLInfty$-algebras. If not, we would like to understand the obstruction.
\item We would like to compute $\dIBL^\PMC(\CycC(\HDR(M)))$ for surfaces $\Sigma_g$ with $g\ge 1$ and formulate a string topology conjecture for non-simply connected manifolds. 
\item We would like to know whether the Schwartz kernel $\GKerStd$ of $\GOpStd=-\CoDd \Delta^{-1}$ (the so called standard Green kernel), where $\CoDd$ is the codifferential and~$\Delta$ the Hodge-de~Rham Laplacian, extends smoothly to a blow-up. If yes, then it is a canonical Green kernel for which the statement of Theorem~\ref{IntroThm:B} holds.
\item We would like to define a generalization of an $\IBLInfty$-algebra---a weak, non-reduced $\IBLInfty$-algebra with a gauge group---and understand its precise relation to perturbative Chern-Simons theory within the $\BV$-formalism.
\end{enumerate}

In the end, let us summarize some existing work on $\IBLInfty$-algebras which helped us to understand $\IBLInfty$-algebras in broader context: In \cite{Muenster2011}, they find an $\IBLInfty$-structure in open-closed string field theory. In~\cite{Doubek2017}, they view $\IBLInfty$-algebras as algebras over a certain Frobenius properad. In~\cite{Markl2015}, they consider $\IBLInfty$-algebras as a particular case of $\BVInfty$- or,  more generally, $\mathrm{MV}$-algebras.

\vspace{2ex}
\textbf{Acknowledgements:} 
I thank my Ph.D.~supervisor Prof.~Dr.~Kai Cieliebak for his continuous support, helpful discussions and encouragement during the research. I thank Dr.~Evgeny Volkov for providing me \cite{Cieliebak2018} and \cite{Cieliebak2018b}, for explaining me his work and for helpful discussions. I~thank Prof.~Robert Bryant for suggesting a better notation for the Green kernel for spheres in an online discussion. I thank my colleagues Dr.~Alexandru Doicu for checking a tedious sign computation, Alexei Kudryashov for discussing the introductory paragraphs, English and notation, and Thorsten Hertl for answering my questions about algebraic topology, which he understands astonishingly well. I~thank the University of Augsburg for financial support and for being a nice place to pursue my Ph.D..


\clearpage
\section{Algebraic structures} \label{Section:Review} 

\label{Sec:Alg0}

In Section~\ref{Sec:Alg1a}, we recall weight-grading (Definition \ref{Def:Grading}), Koszul sign (Definition~\ref{Def:Koszul}), degree shift (Definition~\ref{Def:DegreeShift}), filtrations (Definition~\ref{Def:Filtrations}) and completions (Definition~\ref{Def:Completion}). We prove the K\"unneth formula for completed symmetric cohomology (Proposition~\ref{Prop:Kuenneth}).

In Section~\ref{Sec:Alg1}, we review basics of $\IBLInfty$-algebras from \cite{Cieliebak2015}. We define the exterior algebra $\Ext C$ over a graded vector space $C$ as the symmetric algebra~$\Sym$ over~$C[1]$ (Definition~\ref{Def:ExtAlg}) and use the operations $\mu$ and $\Delta$ of the structure of an associative bialgebra on $\Sym(C[1])$ to give explicit formulas for the partial compositions $\circ_{h_1, \dotsc, h_k}$  (Definition~\ref{Def:CircS}). We use the compositions to define the notion of an $\IBLInfty$-algebra $(\OPQ_{klg})$ on $C$ (Definition~\ref{Def:IBLInfty}), a~Maurer-Cartan element~$(\PMC_{lg})$ (Definition~\ref{Def:MaurerCartan}) and twisted operations $(\OPQ_{klg}^\PMC)$ (Definition~\ref{Def:TwistedOperations}). We mention that an $\IBL$-algebra according to our definition is an odd degree shift of a classical $\IBL$-algebra (Proposition~\ref{Prop:ClasModIBL}). We define the induced $\IBL$-structure on homology (Definition~\ref{Def:HomIBL}), briefly discuss the $\mathrm{BV}$-formalism (Remark~\ref{Rem:BVForm}) and mention weak $\IBLInfty$-algebras (Remark~\ref{Rem:Weak}). Finally, we summarize the situation for twisted $\dIBL$-algebras (Proposition~\ref{Prop:dIBL}) and briefly discuss higher operations (Remark~\ref{Rem:Higher}).

In Section~\ref{Sec:Alg2}, we define the (weight-reduced) dual cyclic bar-complex $\DBCyc V$ of a graded vector space $V$ (Definition~\ref{Def:BarComplex}) and introduce some notation (Notation~\ref{Def:Notation}). We then summarize some facts about the completions $\CDBCyc V$ and $\hat{\Ext}_k \DBCyc V$ (Proposition~\ref{Prop:Compl}). We define the notion of a cyclic $\AInfty$-structure on~$V$ (Definition~\ref{Def:CyclicAinfty}) and its Hochschild and cyclic cohomology (Definition~\ref{Def:CycHom}). We recall strict units and strict augmentations (Definition~\ref{Def:AugUnit}), define the reduced dual cyclic bar complex $\RedDBCyc V$ (Definition~\ref{Def:ReducedDual}) and sketch a proof of the fact that the cyclic cohomology is a direct sum of the reduced cyclic cohomology and the cyclic cohomology of the ground field (Proposition~\ref{Prop:Reduced}). We relate our version of the cyclic cohomology for dga's to the classical version from~\cite{LodayCyclic} (Proposition~\ref{Prop:DGA}). We also show that the reduced spaces for a simply connected and connected $V$ are complete (Proposition~\ref{Prop:SimplCon}).

In Section~\ref{Sec:Alg3}, we review the construction of the canonical $\dIBL$-structure $\dIBL(\CycC(V))$ (Definition~\ref{Def:CanonicaldIBL}) and the canonical Maurer-Cartan element $\MC$ (Definition~\ref{Def:CanonMC}) starting from a cyclic dga $(V,\Pair,m_1,m_2)$. We give formulas for the operations $(\OPQ_{1lg}^\PMC)$ of the $\IBLInfty$-algebra $\dIBL^\PMC(\CycC(V))$ twisted by a Maurer-Cartan element $\PMC$ (Proposition~\ref{Prop:Formulafortwisted}). We consider the $\AInfty$-structure induced on $V$ by $\PMC_{10}$ (Definition~\ref{Def:MukDef}) and relate its cyclic cohomology to the homology of $\OPQ_{110}^\PMC$ (Proposition~\ref{Prop:CyclicHom}). We define the reduced canonical $\dIBL$-algebra $\dIBL(\RedCycC(V))$ (Definition~\ref{Def:ReduceddIBL}) and the notion of a strictly reduced Maurer-Cartan element (Definition~\ref{Def:StrictlyReduced}). The twisted $\IBLInfty$-structure then splits into the reduced part and the part generated by $\NOne^{i*}$, which we can explicitly compute (Proposition~\ref{Prop:Ones}).

\subsection{Gradings, degree shifts and completions
}
\label{Sec:Alg1a}

We will work with vector spaces over $\R$, possibly infinite-dimensional, graded by the \emph{degree} $d\in \Z$ and the \emph{weight} $k\in \N_0$.

\begin{Definition}[Weight-graded vector spaces]\label{Def:Grading}
A \emph{graded vector space} is a vector space $W$ together with a collection  of subspaces $W^d \subset W$ for all $d\in \Z$ such that
$$ W=\bigoplus_{d\in \Z} W^d. $$
Elements of $W^d$ are called \emph{homogenous} of degree $d$; given $w\in W^d$, we denote the degree of $w$ by $\Abs{w} := d$. 

A linear map of graded vector spaces $f: W_1 \rightarrow W_2$ is called homogenous of degree $\Abs{f} \in \Z$ if it holds 
\begin{equation}\label{Eq:StdGrading}
f(W_1^d)\subset W_2^{d+\Abs{f}}\quad\text{for all }d\in \Z.
\end{equation}

A \emph{weight-graded vector space} is a graded vector space $W$ together with a collection of subspaces $W_k^d \subset W$ for all $k\in \N_0$ and $d\in \Z$ such that
$$ W^d=\bigoplus_{k\in \N_0} W_k^d\quad\text{for all }d\in \Z. $$
We define the \emph{weight-$k$ component} by
$$ W_k := \bigoplus_{d\in \Z} W_k^d\quad\text{for all }k\in \N_0. $$
If $W_0^d = 0$ for all $d\in \Z$, we say that $W$ is \emph{weight-reduced.} We define the \emph{weight-reduced subspace} of a weight-graded vector space $W$ by
$$ \bar{W} := \bigoplus_{d\in \Z} \bigoplus_{k\in \N} W_k^d. $$ 

We consider the following versions of the dual space of $W$:
\begin{equation}\label{Eq:Duals}\begin{aligned}
W^* := \{\psi: W \rightarrow \R \text{ linear}\} &\;\,\dots&& \text{\emph{linear dual,}}\\[2\jot]
W'  := \bigoplus_{d\in  \Z} \prod_{k\in \N_0} W_{k}^{d*} &\;\,\dots&& \text{\emph{graded dual,}}\\
W'' := \bigoplus_{d\in \Z} \bigoplus_{k\in \N_0} W_k^{d*} &\;\,\dots&& \text{\emph{weight-graded dual.}}
\end{aligned}\end{equation}
We identify $W'$ with the subspace of $W^*$ generated by homogenous maps and~$W''$ with the subspace of $W^*$ generated by maps which are non-zero only on finitely many~$W_k^d$; hence, we have 
$$W'' \subset W' \subset W^*.$$
\end{Definition}

The grading convention for $W'$ is the \emph{cohomological grading convention}, which differs from the convention \eqref{Eq:StdGrading} for maps $f: W \rightarrow \R$ by the degree reversal (see Definition~\ref{Def:DegreeShift}).

\begin{Definition}[Koszul sign] \label{Def:Koszul}
Let $k\ge 1$, and let $\sigma\in \Perm_k$ be a permutation on~$k$ elements. For $i=1$, $\dotsc$, $k$,  let $a_i$ and $b_i$ be graded symbols of degrees~$\Abs{a_i}$ and~$\Abs{b_i}$, respectively. We denote by 
$$\varepsilon(\sigma,a) \quad\text{and}\quad\varepsilon(a,b)$$
the \emph{Koszul signs} of the transformations
 $$ a_1 \dots a_k \longmapsto a_{\sigma_1^{-1}} \dots a_{\sigma_k^{-1}} \quad \text{and} \quad a_1 \dots a_k b_1 \dots b_k \longmapsto a_1 b_1 \dots a_k b_k, $$
respectively. Here $\sigma_i^{-1} := \sigma^{-1}(i)$. The Koszul sign is computed by permuting the left-hand side to the right-hand side using transpositions of two adjacent elements such that whenever we transpose two graded symbols, e.g., $a_i \longleftrightarrow a_{j}$, we multiply with $(-1)^{\Abs{a_i}\Abs{a_{j}}}$.
\end{Definition}

We emphasize that the Koszul sign depends only on the initial and the final order of the graded symbols; not on the sequence of transpositions.


\begin{Definition}[Degree shift and grading reversal]\label{Def:DegreeShift}
Let $A\in \Z$. The \emph{degree shift by $A$} is a functor which associates to a graded vector space $W$ the graded vector space $W[A]$ with
$$ W[A]^d := W^{d+A}\quad \text{for all }d\in \Z.$$
There is the canonical degree shift morphism 
\begin{equation}\label{Eq:DegreeShift}
 W\longrightarrow W[A]
\end{equation}
of degree $-A$ mapping $W^d$ identically to $W[A]^{d-A}$. We view this morphism as multiplication from the left with a formal symbol $\Susp_A$ of degree $\Abs{\Susp_A}=-A$, so that~\eqref{Eq:DegreeShift} is given by $w\in W \longmapsto \Susp_A w\in W[A]$.

Given graded vector spaces $W_1$, $W_2$ and constants $A_1$, $A_2\in \Z$, we associate to a morphism $f: W_1 \rightarrow W_2$ its \emph{degree shift} $f: W_1[A_1] \rightarrow W_2[A_2]$ by defining
\begin{equation}\label{Eq:DegreeShiftConv}
f(\Susp_{A_1} w) := \Susp_{A_2} f(w)\quad\text{for all } w \in W_1.
\end{equation} 
Notice that if $f:W_1 \rightarrow W_2$ has degree $\Abs{f}$, then $f: W_1[A_1] \rightarrow W_2[A_2]$ has degree $\Abs{f} + A_1 - A_2$.

The \emph{grading reversal} $r$ is a functor which associates to a graded vector space~$W$ the graded vector space $r(W)$ with
$$ r(W)^d := W^{-d}\quad\text{for all }d\in \Z. $$
There is the canonical morphism $W\rightarrow r(W)$ mapping $W^d$ identically to $W^{-d}$ for every $d\in \Z$. The degree reversal of a morphism $f: W_1 \rightarrow W_2$ is the morphism $f: r(W_1) \rightarrow  r(W_2)$ defined by conjugating $f$ with the canonical morphism. If $\Abs{f}$ is the degree of $f: W_1 \rightarrow W_2$, then $-\Abs{f}$ is the degree of $f: r(W_1) \rightarrow  r(W_2)$.
\end{Definition}

In our main reference \cite{Cieliebak2015}, they view $W$ and $W[A]$ as one vector space with two different gradings $\Deg(\cdot)$ and $\Abs{\cdot}$, respectively; these are related by
\begin{equation*}
\Abs{w} = \Deg(w) - A \quad\text{for all homogenous }w\in W. 
\end{equation*}
On the other hand, we think of $W$ and $W[A]$ as of two different graded vector spaces and never use the same symbol for an element $w\in W$ and its degree shift $\Susp_A w \in W[A]$. 
It allows us to use just one notation~$\Abs{\cdot}$ for the gradings on both~$W$ and~$W[A]$. However, in order to preserve compatibility with~\cite{Cieliebak2015}, we will also sometimes use the notation $\Deg(w)$ (in the exponent just $(-1)^w$) for the degrees on $W$.

For graded vector spaces $W_1$, $\dotsc$, $W_k$ and constants $A_1$, $\dotsc$, $A_k\in \Z$, we identify 
$$ W_1[A_1]\otimes \dotsb \otimes W_k[A_k] \simeq (W_1\otimes \dotsb \otimes W_k)[A_1+\dotsb+A_k] $$ using the \emph{Koszul convention for the tensor product}; for homogenous elements $w_1 \in W_1$,~$\dotsc$, $w_k \in W_k$, it reads
\begin{equation} \label{Eq:KoszulTensor}
\Susp_{A_1} w_1 \otimes \dotsb \otimes \Susp_{A_k} w_k = \varepsilon(\Susp_A,w) \underbrace{\Susp_{A_1}\dots\Susp_{A_k}}_{\displaystyle \mathclap{=: \Susp_{A_1 + \dotsb + A_k}}\rule{0ex}{2ex}} w_1 \otimes \dotsb \otimes w_k.
\end{equation}
If $A_1 = \dotsb = A_k = :A$ is fixed in the context, which is our usual case, we omit the subscript~$A$ and write just $\Susp$.

In the case of the multilinear map $f: W_1\otimes \dotsb \otimes W_k \rightarrow V_1\otimes \dotsb \otimes V_l$, the combination of \eqref{Eq:DegreeShiftConv} and \eqref{Eq:KoszulTensor} gives for $f: W_1[A_1]\otimes \dotsb\otimes W_k[A_k] \rightarrow V_1[B_1]\otimes \dotsb\otimes V_l[B_l]$ the following:
\begin{equation}\label{Eq:DegreeShiftConvII}
 f(\Susp_{A_1 +\dotsb + A_k} w_1 \otimes \dotsb \otimes w_k) = \Susp_{B_1 + \dotsb + B_l} f(w_1 \otimes \dotsb \otimes w_k).
\end{equation}

\begin{Remark}[Why is this sign convention bad?]\label{Rem:BadConvention}
Let us illustrate that \eqref{Eq:DegreeShiftConvII} is not compatible with the following standard Koszul rule:
\begin{equation*}
(K):\qquad (f_1 \otimes f_2)(w_1 \otimes w_2) = (-1)^{\Abs{f_2}\Abs{w_1}} f_1(w_1) \otimes f_2(w_2).
\end{equation*}
On one hand, we get 
$$\begin{aligned}
(f_1 \otimes f_2)(\Susp^2 w_1 \otimes w_2) &\overset{\eqref{Eq:DegreeShiftConvII}}{=} \Susp^2 (f_1\otimes f_2)(w_1 \otimes w_2) \\
& \overset{(K)}{=} (-1)^{\Abs{f_2}\Abs{w_1}} \Susp^2 f_1(w_1) \otimes f_2(w_2)  \\
& \overset{\eqref{Eq:KoszulTensor}}{=} (-1)^{\Abs{f_2}\Abs{w_1} + A(\Abs{f_1} + \Abs{w_1})} \Susp f_1(w_1) \otimes \Susp f_2(w_2).
\end{aligned}$$
On the other hand, we get
$$\begin{aligned}
(f_1 \otimes f_2)(\Susp^2 w_1 \otimes w_2) &\overset{\eqref{Eq:KoszulTensor}}{=} (-1)^{A\Abs{w_1}}(f_1 \otimes f_2)(\Susp w_1 \otimes \Susp w_2)\\
&\overset{(K)}{=} (-1)^{A\Abs{w_1} + \Abs{f_2}(A+\Abs{w_1})} f_1(\Susp w_1) \otimes f_2(\Susp w_2)\\ &\overset{\eqref{Eq:DegreeShiftConvII}}{=}(-1)^{A\Abs{w_1}+ \Abs{f_2}(A+\Abs{w_1})} \Susp f_1(w_1) \otimes \Susp f_2(w_2).
\end{aligned}$$
The results differ by $(-1)^{A(\Abs{f_1}+\Abs{f_2})}$. Therefore, we can not use (K) to identify $\Hom(W_1,V_1)\otimes \Hom(W_2,V_2)$ with a subspace of $\Hom(W_1\otimes W_2, V_1\otimes V_2)$ in general. We will rather define an ad-hoc pairing in the case where we need it (see Definition~\ref{Def:Pairings}).

Another caveat is that in the case of the tensor product, the degree shift by $A_1$ followed by the degree shift by~$A_2$ is not the same as the degree shift by $A_1 + A_2$. Indeed, we compute
$$\begin{aligned}
(\Susp_{A_1 + A_2}w_1) \otimes (\Susp_{A_1 + A_2}w_2) &=
(\Susp_{A_2} \Susp_{A_1}w_1) \otimes (\Susp_{A_2}\Susp_{A_1} w_2) \\ &= (-1)^{A_2(A_1 + \Abs{w_1})}\Susp_{A_2}^2 (\Susp_{A_1} w_1) \otimes (\Susp_{A_1} w_2) \\
&= (-1)^{A_2 A_1 + (A_1 + A_2)\Abs{w_1}} \Susp_{A_2}^2 \Susp_{A_1}^2 (w_1 \otimes w_2) \\ &= (-1)^{A_2 A_1 + (A_1 + A_2)\Abs{w_1}} \Susp_{2(A_1 + A_2)} (w_1 \otimes w_2),
\end{aligned}$$
which differs by $(-1)^{A_1 A_2}$ from the direct degree shift by $A_1 + A_2$. Therefore, we have to always remember the vector spaces which we started with and the sequence of degree shifts. 

Note that we also have the unnatural $\Susp_{A_1} \Susp_{A_2} = \Susp_{A_2} \Susp_{A_1}$ due to \eqref{Eq:KoszulTensor}.
\end{Remark}

\begin{Remark}[Is there a better sign convention?]
The author originally respected the Koszul rule for the algebra with formal symbols and considered the following map $\Susp^{l}_*\overline{\Susp}^{k*} f: W[A]^{\otimes k} \rightarrow V[A]^{\otimes l}$ as the degree shift of $f: W^{\otimes k} \rightarrow V^{\otimes l}$:
\begin{equation}\label{Eq:AltConv}
(\Susp^{l}_*\overline{\Susp}^{k*} f)(\Susp^{k} w_1 \otimes \dotsb \otimes w_k ) = (-1)^{k \Abs{f}A + \frac{1}{2}k(k-1) A} \Susp^l f(w_1 \otimes \dotsb \otimes w_k).
\end{equation}
Here $\DeSusp$ denotes the ``inverse'' of $\Susp$ with $\Abs{\overline{\Susp}} = - \Abs{\Susp}$, $\Susp_*^l f = \Susp^l \circ f$ is the post-composition, $\DeSusp^{k*}f = (-1)^{k A \Abs{f}} f\circ \DeSusp^k$ the pre-composition, and the sign $ \varepsilon(\Susp, \DeSusp) = (-1)^{\frac{1}{2}k(k-1)A}$ comes from the ``collision'' $\DeSusp_1\dots \DeSusp_k \Susp_1 \dots \Susp_k \mapsto \DeSusp_1 \Susp_1 \dots \DeSusp_k \Susp_k$. 

However, the author did not manage to reprove the theory in~\cite{Cieliebak2015} using~\eqref{Eq:AltConv} (because of too many ``external'' signs appearing and a problem with disconnected graphs). A~motivation to try a different sign convention was to explain some artificial signs in~\cite{Cieliebak2015} and formulate their coordinate constructions invariantly in order to generalize them to the ``continuous'' de Rham case.

It might be possible to deduce a ``universal'' sign convention ``respecting'' the Koszul rules by considering the category of chain complexes and graded morphisms $\mathcal{C}$ as the category enriched in the closed monoidal category of chain complexes and chain maps of degree~$0$. One can then define the enriched degree shift functor $\Susp_{A}: \mathcal{C} \rightarrow \mathcal{C}$, embed $\mathcal{C}^{\otimes k} \subset \mathcal{C}$ using~$(K)$ and study enriched natural transformations in the algebra of functors consisting of tensor products and compositions of $\Susp_{A}$, $\Hom(\cdot, \cdot)$ and the dual $*$.
\end{Remark}



\begin{Definition}[Permutations]\label{Def:Permutations}
For $k\ge 1$ and $\sigma\in \Perm_k$ (:=\,the group of permutations on $k$ elements), we define the action of $\sigma$ on $W^{\otimes k}$ by 
\begin{equation}\label{Eq:Perm}
\sigma(w_1 \otimes \dotsb \otimes w_k) :=  \varepsilon(\sigma,w) w_{\sigma_1^{-1}}\otimes \dotsb \otimes w_{\sigma_k^{-1}}
\end{equation}
for all homogenous $w_1$, $\dotsc$, $w_k\in W$.
\end{Definition}

Notice that the $i$-th vector is permuted to the $\sigma_i$-th place --- this is the ``active'' convention for permutations.

\begin{Definition}[Symmetric algebra]\label{Def:SymAlgebra}
Let $\Ten(V):= \bigoplus_{k\ge 0} V^{\otimes k}$ be the tensor algebra over a graded vector space~$V$. The \emph{symmetric algebra} over $V$ is defined by $\Sym(V):= \bigoplus_{k\ge 0} \Sym_k(V)$, where
$$ \Sym_k(V) := V^{\otimes k} \bigl/ \sum_{\sigma\in \Perm_k} \Im(\Id-\sigma)\quad(=: \Perm_k\text{-coinvariants}). $$
It is a weight-graded vector space with components denoted by $(\Sym_k V)^d$ for all $d\in \Z$ and $k\in \N_0$. Note that $\Sym_0 V = \R$ has degree $0$ by definition. Consider the canonical projection
$$\begin{aligned}
\pi : \Ten(V) &\longrightarrow \Sym(V) \\
v_1\otimes \dotsb \otimes v_k &\longmapsto v_1\dotsb v_k.
\end{aligned}$$
The dot $\cdot$ indicates the symmetric product. If $v_i\in V$ are homogenous, we call $v_1 \dotsb v_k$ a \emph{generating word}; we have
$$ v_1 \dotsb  v_k = \varepsilon(\sigma,v) v_{\sigma_1^{-1}} \dotsb v_{\sigma_k^{-1}}\quad \text{for every }\sigma\in \Perm_k. $$
Let $\iota: \Sym(V) \rightarrow \Ten(V)$ be the section of $\pi$ defined by 
$$ \iota(v_1\dotsb v_k) := \frac{1}{k!}\sum_{\sigma\in\Perm_k} \varepsilon(\sigma,v) v_{\sigma_1^{-1}}\otimes \ldots \otimes v_{\sigma_k^{-1}}. $$
We use it to identify $\Sym(V)$ with the subspace of symmetric tensors
$$ \iota(\Sym_k(V)) = \bigcap_{\sigma\in \Perm_k} \Ker(\Id - \sigma) \subset \Ten_k(V)\quad(=:\Perm_k\text{-invariants}). $$
\end{Definition}

\begin{Definition}[Filtrations] \label{Def:Filtrations}
Let $W$ be a graded vector space. A filtration of~$W$ is a collection of linear subspaces $\mathcal{F}_\lambda W \subset W$ for $\lambda\in \R$ such that we have either
\begin{itemize}
\item $\mathcal{F}_{\lambda_1}W\subset \mathcal{F}_{\lambda_2}W$ for all $\lambda_1 \le \lambda_2$\quad$\Longleftrightarrow:$\quad\emph{increasing filtration}, or
\item  $\mathcal{F}_{\lambda_1}W\supset \mathcal{F}_{\lambda_2}W$ for all $\lambda_1 \le \lambda_2$\quad$\Longleftrightarrow:$\quad\emph{decreasing filtration.}
\end{itemize}
We will assume that our filtrations are \emph{graded} in the following sense:
$$ \forall \lambda\in \R:\quad \F_\lambda W = \bigoplus_{d\in \Z} \F_\lambda W^d,\quad \text{where}\quad\F_\lambda W^d := \F_\lambda W \cap W^d.$$
A filtration $\F_\lambda W$ is called:
\begin{itemize}
 \def\lenabcnd{3.5cm}
 \item\makebox[\lenabcnd][l]{\emph{exhaustive}}$:\Equiv$\quad$\bigcup_{\lambda\in \R} \F_\lambda W = W$;
 \item\makebox[\lenabcnd][l]{\emph{Hausdorff}}$:\Equiv$\quad$\bigcap_{\lambda\in \R} \F_\lambda W = 0$;
\item\makebox[\lenabcnd][l]{\emph{$\Z$-gapped}}$:\Equiv$\quad$\F_\lambda W = \F_{\lfloor\lambda\rfloor} W$ for all $\lambda \in \R$; 
 \item\makebox[\lenabcnd][l]{\parbox[t]{\lenabcnd-.5em}{\emph{bounded from below}}}$:\Equiv$\quad$\exists \lambda \in \R: \F_\lambda W =0$; 
 \item\makebox[\lenabcnd][l]{\parbox[t]{\lenabcnd-.5em}{\emph{bounded from above}}}$:\Equiv$\quad$\exists \lambda\in \R: \F_\lambda W =W$.
\end{itemize} 

Given a graded vector space $W$ filtered by a $\Z$-gapped filtration $\F_\lambda W$, we associate to it the bi-graded vector space 
$$ \Gr(W) = \bigoplus_{d\in\Z}\bigoplus_{\lambda\in \Z} \Gr(W)_\lambda^d $$
called the \emph{graded module} whose components are given as follows:
$$ \forall d, \lambda\in \Z: \quad \Gr(W)_\lambda^d := \begin{cases}
                   \F_\lambda W^d /\F_{\lambda-1} W^d & \text{for increasing }\F_\lambda W,\\
                   \F_{\lambda-1} W^d / \F_{\lambda} W^d & \text{for decreasing } \F_\lambda W.
                 \end{cases}$$
We naturally extend a filtration over degree shifts, graded duals, direct sums, tensor products and symmetric products as follows: \allowdisplaybreaks
\begin{align*}
\F_\lambda W[A]^d &:= \F_\lambda W^{d+A},\\[\jot]
\F_\lambda (W')^d &:= \{\psi\in W^{d*} \mid \Restr{\psi}{\F_\lambda W} = 0\}, \\
\F_\lambda\bigl(\bigoplus_{i\in I} W_i\bigr)^d &:= \bigoplus_{i\in I} \F_{\lambda} W_i^d, \\
\F_{\lambda}(W_1 \otimes \dotsb \otimes W_k)^d &:= \bigoplus_{\substack{d_1, \dotsc, d_k\in \Z \\ d_1 + \dotsb + d_k = d}}\ \sum_{\substack{\lambda_1, \dotsc, \lambda_k \in \R \\ \lambda_1 + \dotsb + \lambda_k = \lambda}} \F_{\lambda_1} W_1^{d_1} \otimes \dotsb \otimes \F_{\lambda_k} W_k^{d_k},\\
\F_\lambda (\Sym_k V)^d &:= \pi(\F_\lambda(V^{\otimes k})^d),
\end{align*}
where $\pi: \Ten(V) \rightarrow \Sym(V)$ is the canonical projection. If $(W,\Bdd)$ is a filtered chain complex, we filter the homology as follows:
$$ \forall \lambda \in \R, d\in \Z: \quad \F_\lambda \H_d(W,\Bdd) := \{\alpha \in \H_d(C,\Bdd) \mid \exists w\in \alpha: w\in \F_\lambda W^d \}. $$
\end{Definition}

\begin{Definition}[Completions]\label{Def:Completion}
Let $W$ be a graded vector space filtered by a decreasing filtration $\F_\lambda W$. The \emph{filtration degree} of $w\in W$ is defined by
$$ \Norm{w} := \sup\{ \lambda\in \R \mid w\in \mathcal{F}_\lambda W\}. $$
The filtration degree of a linear map $f: W_1 \rightarrow W_2$ is defined by
$$ \Norm{f} := \sup \{ \lambda\in \R \mid \Norm{f(w)} \ge \Norm{w} + \lambda\ \forall w\in W_1\}. $$
We say that the \emph{filtration degree is finite} if $\Norm{f}>-\infty$. Note that $\Norm{0} = \infty$.

The \emph{completion} of $W$ is the graded vector space
$$ \hat{W}:= \bigoplus_{d\in \Z} \hat{W}^d, $$
where for all $d\in \Z$  we define
$$ \hat{W}^d := \Bigl\{ \sum_{i=0}^\infty w_i\ \Bigl|\   \forall i\in \N_0: w_i\in W^d; \Norm{w_i} \to \infty \text{ as }i\to \infty \Bigl\}\Bigl/\sim.
$$
Here $\sum_{i=0}^\infty w_i \sim \sum_{i=0}^\infty w_i'$ if and only if $\Norm{\sum_{i=0}^n (w_i - w_i')}\to \infty$ as $n\to \infty$.\footnote{\label{Footnote:Compl}In fact, $\hat{W}$ is the inverse limit $\varprojlim_\lambda^{\mathrm{gr}}(W/\mathcal{F}_\lambda W)$ in the category of graded vector spaces and~$\hat{W}^d$ the inverse limit $\varprojlim_\lambda(W^d/\mathcal{F}_\lambda W^d)$ in the category of vector spaces. As a side-remark, if we forget the grading on $W$, we might also consider $\varprojlim_\lambda (W/\mathcal{F}_\lambda W)$, which would be a vector space containing~$\hat{W}$ as a subspace\vphantom{$W^d$}.}
The completion~$\hat{W}$ is canonically filtered by the filtration $\mathcal{F}_\lambda \hat{W}$ defined as follows:
$$ \forall \lambda\in \R, d\in \Z: \quad \mathcal{F}_\lambda\hat{W}^d:= \Bigl\{\sum_{i=0}^\infty w_i \in \hat{W}^d \ \Bigr|\ \forall i\in \N_0: w_i \in \mathcal{F}_\lambda W^d \Bigr\}. $$
We denote the completion of $W_1 \otimes \dotsb \otimes W_k$ by $W_1 \hat{\otimes} \dotsb \hat{\otimes} W_k$ and the completion of $\Sym_k V$ by $\hat{\Sym}_k V$.

A map $f: W_1 \rightarrow W_2$ of finite filtration degree
\emph{extends continuously} to a linear map $f: \hat{W}_1 \rightarrow \hat{W}_2$; this extension is defined by
$$ f\Bigl(\sum_{i=0}^\infty w_i\Bigr):= \sum_{i=0}^\infty f(w_i)\quad\text{for all }\sum_{i=0}^\infty w_i \in \hat{W}. $$
\end{Definition}

\begin{Remark}[Completed tensor product]\label{Rem:ComplTens}
Using Proposition~\ref{Prop:IsoCrit} below, one can show that the \emph{completed tensor product} $\hat{\otimes}$ is associative and that $W_1 \hat{\otimes} W_2 \simeq \hat{W}_1 \hat{\otimes} \hat{W}_2$. By refining this argument, one can show that $\hat{\Sym}_k V \simeq \hat{\Sym}_k\hat{V}$ for any $k\in \N$.
\end{Remark}

A weight-graded vector space $W$ is canonically \emph{filtered by weights:}
\begin{equation}\label{Eq:FiltrWeights}
\forall \lambda\in \R, d\in \Z:\quad\F_\lambda W^d := \bigoplus_{k\le \lambda} W_k^d.
\end{equation}
This filtration is $\Z$-gapped, exhaustive, Hausdorff, increasing and bounded from below. The induced filtration on the graded dual $W'$ is $\Z$-gapped, Hausdorff, decreasing and bounded from above (and thus automatically exhaustive). It holds $\Gr(W) \simeq W$, and it is easy to see from \eqref{Eq:Duals} that the canonical map $W'' \rightarrow W'$ induces the isomorphism
$$ \widehat{W''} \simeq W'. $$
We also see that the condition
$$ (WG0): \quad \forall d\in \Z\ \exists J\subset \N_0, \Abs{J}<\infty\ \forall k\in \N_0\backslash J: \quad W_k^d = 0 $$
is equivalent to $W''= W'$.

A useful tool to compare completions is the following proposition:

\begin{Proposition}[{\cite[Proposition 7.3.7]{Fresse}}, Isomorphism criterion]\label{Prop:IsoCrit}
Let $W_1$ and~$W_2$ be graded vector spaces filtered by $\Z$-gapped filtrations which are decreasing and bounded from above. Suppose that $f: W_1 \rightarrow W_2$ is a filtration preserving homogenous linear map. Then the continuous extension $f: \hat{W}_1 \rightarrow \hat{W}_2$ is an isomorphism if and only if the induced map $f: \Gr(W_1) \rightarrow \Gr(W_2)$ is an isomorphism.
\end{Proposition}
\begin{proof}
The implication from the right to the left is obtained from the diagram 
$$\begin{tikzcd}
0 \arrow{r} & \Gr(W_1)_\lambda \arrow[hook]{r} \arrow{d}{f} & W_1/\F_{\lambda}W_1 \arrow[two heads]{r} \arrow{d}{f} & W_1/\F_{\lambda-1} W_1 \arrow{r}   \arrow{d}{f} & 0 \\
0 \arrow{r} & \Gr(W_2)_\lambda \arrow[hook]{r} & W_2/\F_{\lambda}W_2 \arrow[two heads]{r} & \arrow{r}W_2/\F_{\lambda-1} W_2   & 0
\end{tikzcd}$$
by induction using the definition of $\hat{W}$ as the inverse limit of $W/\F_\lambda W$ (see Footnote \ref{Footnote:Compl} on page \pageref{Footnote:Compl}).
\end{proof}

For a graded vector space $W$ filtered by a $\Z$-gapped filtration, consider the following conditions:
\begin{equation}\label{Eq:WGs}
\begin{aligned}
(WG1):\quad& \forall \lambda\in \Z\ \exists I \subset \Z, \Abs{I}<\infty\ \forall d\in \Z\backslash I:& \Gr(W)_\lambda^d &= 0, \\
(WG2):\quad & \forall d, \lambda\in \Z:& \dim(\Gr(W)_\lambda^d) &< \infty.
\end{aligned}
\end{equation}

\begin{Lemma}[Completion of symmetric powers of the graded dual]\label{Lem:Terrible}
Let $W$ be a graded vector space filtered by an exhaustive $\Z$-gapped filtration~$\F_\lambda W$ which is increasing and bounded from below. If (WG1) \& (WG2) are satisfied, then the natural map $\Sym_k(W') \rightarrow (\Sym_k W)'$ induces the isomorphism 
$$ \hat{\Sym}_k(W') \simeq  (\Sym_k W)' \quad\text{for every }k\in \N.$$
Note that we filter graded duals by the induced filtration from Definition~\ref{Def:Filtrations}.
\end{Lemma}
\begin{proof}
The natural map $\Sym_k(W') \rightarrow (\Sym_k W)'$ is clearly filtration preserving, and hence it extends continuously to a map of completions. The target space $(\Sym_k W)'$ is already complete (the dual space $W'$ is complete, provided that the filtration of~$W$ is exhaustive), and thus we obtain the map $\hat{\Sym}_k(W') \rightarrow (\Sym_k W)'$. According to Proposition~\ref{Prop:IsoCrit}, this map is an isomorphism if and only if the induced map $\Gr(\Sym_k(W')) \rightarrow \Gr((\Sym_k W)')$ is. This is shown by the following computation (the maps involved are natural in at least one direction):
\allowdisplaybreaks
\begin{align*} 
\frac{\F_\lambda ({W^{\otimes k}}')^d}{\F_{\lambda+1} ({W^{\otimes k}}')^d} &\simeq \frac{\F_\lambda (W^{\otimes k})^{d*}}{\F_{\lambda+1} (W^{\otimes k})^{d*}} \simeq  \Bigl(\frac{\F_{\lambda+1}(W^{\otimes k})^d}{\F_{\lambda}(W^{\otimes k})^d}\Bigr)^* \\ 
 &\simeq \Biggl( \frac{\bigoplus_{\Abs{\vec{d}}= d} \sum_{\Abs{\vec{\lambda}}= \lambda + 1} \F_{\lambda_1} W^{d_1} \otimes \dotsb \otimes \F_{\lambda_k} W^{d_k}}{\bigoplus_{\Abs{\vec{d}} = d}\sum_{\Abs{\vec{\lambda}} = \lambda} \F_{\lambda_1} W^{d_1} \otimes \dotsb \otimes \F_{\lambda_k} W^{d_k}}\Biggr)^* \\
&\simeq \Biggl( \bigoplus_{\Abs{\vec{d}}= d} \frac{\sum_{\Abs{\vec{\lambda}} = \lambda + 1} \F_{\lambda_1} W^{d_1} \otimes \dotsb \otimes \F_{\lambda_k} W^{d_k}}{\sum_{\Abs{\vec{\lambda}} = \lambda} \F_{\lambda_1} W^{d_1} \otimes \dotsb \otimes \F_{\lambda_k} W^{d_k}}\Biggr)^* \\
&\simeq\Bigl( \bigoplus_{\Abs{\vec{d}} = d} \bigoplus_{\Abs{\vec{\lambda}}=\lambda} \frac{\F_{\lambda_1 + 1}W^{d_1}}{\F_{\lambda_1} W^{d_1}} \otimes \dotsb \otimes \frac{\F_{\lambda_k+1}W^{d_k}}{\F_{\lambda_k} W^{d_k}}\Bigr)^*\\
&\simeq\Bigl( \bigoplus_{\Abs{\vec{\lambda}}=\lambda} \bigoplus_{\Abs{\vec{d}} = d} \frac{\F_{\lambda_1 + 1} W^{d_1}}{\F_{\lambda_1} W^{d_1}} \otimes \dotsb \otimes \frac{\F_{\lambda_k + 1} W^{d_k}}{\F_{\lambda_k} W^{d_k}}\Bigr)^* \\
&\mathclap{\substack{\Z-\text{gapped}  \\[1ex] \&\ \text{bounded below}  \\[1ex] \&\ (WG1)}\rightarrow\rule{7.7em}{0pt}}\simeq  \bigoplus_{\Abs{\vec{\lambda}}= \lambda} \bigoplus_{\Abs{\vec{d}} = d} \Bigl( \frac{\F_{\lambda_1 + 1} W^{d_1}}{\F_{\lambda_1} W^{d_1}} \otimes \dotsb \otimes \frac{\F_{\lambda_k + 1} W^{d_k}}{\F_{\lambda_k} W^{d_k}}\Bigr)^* \\
&\mathclap{{\scriptstyle (WG2)}\rightarrow\rule{4em}{0pt}}\simeq  \bigoplus_{\Abs{\vec{\lambda}}= \lambda} \bigoplus_{\Abs{\vec{d}} = d}  \Bigl( \frac{\F_{\lambda_1 + 1} W^{d_1}}{\F_{\lambda_1} W^{d_1}}\Bigr)^* \otimes \dotsb \otimes \Bigl( \frac{\F_{\lambda_k + 1} W^{d_k}}{\F_{\lambda_k} W^{d_k}}\Bigr)^* \\
&\simeq \bigoplus_{\Abs{\vec{d}} = d} \bigoplus_{\Abs{\vec{\lambda}}=\lambda} \frac{\F_{\lambda_1}(W')^{d_1}}{\F_{\lambda_1 + 1}(W')^{d_1}}\otimes \dotsb \otimes \frac{\F_{\lambda_k}(W')^{d_k}}{\F_{\lambda_k + 1}(W')^{d_k}} \\
&\simeq \frac{\F_{\lambda} ({W'}^{\otimes k})^d}{\F_{\lambda + 1} ({W'}^{\otimes k})^d}.
\end{align*}
In fact, this computation shows that $\hat{\Ten}_k(W') \simeq (\Ten_k W)'$. The conclusion for $\Sym_k$ follows by checking that the maps above are $\Perm_k$-equivariant.
%
\end{proof}

Given a chain complex $(W,\Bdd)$, the boundary operator $\Bdd$ induces the boundary operator $\Bdd_k : W^{\otimes k} \rightarrow W^{\otimes k}$ for all $k\in \N$; for all $w_1$, $\dotsc$, $w_k\in W$, it is defined~by
\begin{equation}\label{Eq:BddExt}
\Bdd_k(w_1 \otimes \dotsb \otimes w_k):= \sum_{i=1}^k (-1)^{\Abs{w_1} + \dotsb + \Abs{w_{i-1}}} w_1 \otimes \dotsb \otimes \Bdd w_i \otimes \dotsb \otimes w_k.
\end{equation}
The map $\Bdd_k$ is clearly $\Perm_k$-equivariant, and thus induces the boundary operator $\Bdd_k : \Sym_k W \rightarrow \Sym_k W$.

\begin{Proposition}[K\"unneth formula for completed symmetric cohomology]\label{Prop:Kuenneth}
Let $(W,\Bdd)$ be a $\Z$-graded chain complex over $\R$ filtered by an exhaustive $\Z$-gapped filtration~$\F_\lambda W$ which is increasing and bounded from below. Consider the dual cochain complex $(W',\Dd:= \Bdd^*)$. Suppose that $\Dd$ has finite filtration degree, so that $\Dd_k: \Sym_k(W') \rightarrow \Sym_k(W')$ extends continuously to $\Dd_k : \hat{\Sym}_k(W') \rightarrow \hat{\Sym}_k(W')$ for every $k\in \N$. If (WG1) \& (WG2) are satisfied, then the natural map $\Sym_k \H(W',\Dd) \rightarrow \H(\hat{\Sym}_k(W'),\Dd_k)$ induces the isomorphism
\begin{equation*}
\hat{\Sym}_k \H(W',\Dd) \simeq \H(\hat{\Sym}_k(W'), \Dd_k)\quad \text{for all }k\in \N.
\end{equation*}
\end{Proposition}
\begin{proof}
The natural map $\Sym_k \H(W',\Dd) \rightarrow \H(\hat{\Sym}_k W',\Dd_k)$ is clearly filtration preserving, and hence it extends continuously to a map of completions. The target space $\H(\hat{\Sym}_k W',\Dd_k)$ is already complete (the homology of a complete space is complete), and hence we obtain the map $\hat{\Sym}_k \H(W',\Dd) \rightarrow \H(\hat{\Sym}_k W',\Dd_k)$. The following facts are easy to verify:
\begin{enumerate}[label=(\arabic*)]
 \item The isomorphism from Lemma~\ref{Lem:Terrible} is an isomorphism of cochain complexes 
 $$ (\hat{\Sym}_k W', \Dd_k) \simeq ((\Sym_k W)', \Bdd_k^*). $$
\item If the filtration on~$W$ satisfies (WG1) and (WG2), then the filtration on $\H(W)$ also satisfies (WG1) and (WG2), respectively. Consequently, Lemma~\ref{Lem:Terrible} holds for symmetric powers of $\H(W,\Bdd)'$ as well.
\item The Künneth formula $\H(W^{\otimes k}) \simeq \H(W)^{\otimes k}$ implies $\H(\Sym_k W) \simeq \Sym_k \H(W)$ for any $\Z$-graded chain complex $W$ over $\R$.
\item We have $(\H(W))' \simeq \H(W')$ over $\R$ by the universal coefficient theorem.
\end{enumerate}
Now, we compute
\begin{align*}
\H(\hat{\Sym}_k W',\Dd_k ) &
\underset{\substack{\uparrow\rule{0pt}{1.5ex} \\ (1)}}{\simeq}
\H((\Sym_k W)', \Bdd_k^*)
\underset{\substack{\uparrow\rule{0pt}{1.5ex} \\ (4)}}{\simeq}
\H(\Sym_k W, \Bdd_k)' 
\underset{\substack{\uparrow\rule{0pt}{1.5ex} \\ (3)}}{\simeq}
(\Sym_k \H(W,\Bdd))' \\ 
&\underset{\substack{\uparrow\rule{0pt}{1.5ex} \\ (2)}}{\simeq}
\hat{\Sym}_k (\H(W,\Bdd)') 
\underset{\substack{\uparrow\rule{0pt}{1.5ex} \\ (4)}}{\simeq}
\hat{\Sym}_k \H(W',\Dd).
\end{align*}
This proves the proposition.
\end{proof}

\subsection{Basics of \texorpdfstring{$\IBLInfty$-algebras}{IBL-infinity-algebras}
}
\label{Sec:Alg1}


\begin{Definition}[Exterior algebra]\label{Def:ExtAlg}
Given a graded vector space $C$ over $\R$, we define the \emph{exterior algebra} over $C$ by
$$ \Ext C := \Sym(C[1]). $$
The weight-$k$ component is denoted by $\Ext_k C$ and the weight-reduced part by~$\RExt C$. If $C$ is in addition filtered, then $\Ext_k C$ is filtered by the induced filtration and its completion is denoted by $\hat{\Ext}_k C$.
\end{Definition}

We have the product $\mu : \Ext C \otimes \Ext C \rightarrow \Ext C$ and coproduct $\Delta: \Ext C \rightarrow \Ext C\otimes \Ext C$ defined~by
$$ \begin{aligned}
&\mu(c_{11}\dots c_{1k} \otimes c_{21} \dots c_{2k'}) := c_{11} \dots c_{1k} c_{21} \dots c_{2k'}\quad\text{and}\\[\jot]
&\Delta(c_1 \dots c_k) := \sum_{\substack{k_1,\,k_2 \ge 0\\ k_1 + k_2 = k}} \sum_{\sigma\in \Perm_{k_1, k_2}} \varepsilon(\sigma,c) c_{\sigma^{-1}_1} \dots c_{\sigma_{k_1}^{-1}}\otimes c_{\sigma_{k_1+1}^{-1}}\dots c_{\sigma_{k_1 + k_2}^{-1}}
\end{aligned} $$
for all homogenous $c_{ij}$, $c_i \in C[1]$ and $k$, $k'\ge 0$, respectively, where $\Perm_{k_1, k_2}\subset \Perm_{k_1+k_2}$ denotes the set of shuffle permutations with blocks of lengths $k_1$ and $k_2$. These operations satisfy relations of an \emph{associative bialgebra} (see \cite{Loday2012}):
\begin{equation}\label{Eq:Bialgebra}
\text{Ass. bialg.}\quad \left\{
\begin{aligned}
\mu\circ (\Id\otimes \mu) &= \mu\circ (\mu \otimes \Id), \\ 
(\Id\otimes \Delta)\circ\Delta &= (\Delta\otimes \Id)\circ \Delta, \\
\Delta \circ \mu &= (\mu\otimes \mu) \circ (\Id \otimes \tau\otimes \Id) \circ (\Delta\otimes \Delta).
\end{aligned} \right.
\end{equation}
Here $\tau: C_1 \otimes C_2 \rightarrow C_2 \otimes C_1$, $c_1\otimes c_2 \mapsto (-1)^{\Abs{c_1}\Abs{c_2}}c_2\otimes c_1$ denotes the \emph{twist map}.

We will use the bialgebra calculus (:=\,relations \eqref{Eq:Bialgebra}) to write down explicit formulas for the operations $\circ_{h_1, \dotsc, h_r}$ which were briefly introduced in \cite{Cieliebak2015}; these operations take symmetric maps $f_1$, $\dotsc$, $f_r$ and connect $h_1$, $\dotsc$, $h_r$ of their outputs to the inputs of a symmetric map $f$ in all possible ways, so that the result, which we denote by $f\circ_{h_1, \dotsc, h_r}(f_1, \dotsc,f_r)$, becomes a symmetric map again.


\begin{Definition}[Partial compositions] \label{Def:CircS}
Let $C$ be a graded vector space. For $i$, $j\ge 0$, we denote by 
$$ \begin{aligned} \pi_i : \Ext C \longrightarrow \Ext_i C,& \quad \iota_i : \Ext_i C \longrightarrow \Ext C, \\
 \Id_i : \Ext_i C \longrightarrow \Ext_i C,& \quad \begin{aligned}[t]\tau_{i,j}: \Ext_i C\otimes \Ext_j C &\longrightarrow \Ext_j C \otimes \Ext_i C \end{aligned}
 \end{aligned} $$ 
the components of the canonical projection $\pi$, the canonical inclusion $\iota$, the identity $\Id$ and the twist map $\tau$, respectively. We also set 
$$ \Delta_{i,j} := (\pi_i \otimes \pi_j)\circ \Delta\circ \iota_{i+j} \quad\text{and}\quad \mu_{i,j}:= \pi_{i+j}\circ \mu\circ (\iota_i \otimes \iota_j). $$
For $k'$, $k_1$, $l'$, $l_1\ge 0$, let $f: \Ext_{k'}C \rightarrow \Ext_{l'} C$ and $f_1: \Ext_{k_1} C \rightarrow \Ext_{l_1} C$ be linear maps, and let $0 \le h \le \min(k', l_1)$. We set  
$$k := k' + k_1 - h \quad\text{and}\quad l := l' + l_1 - h $$
and define the \emph{composition of $f$ and $f_1$ at $h$ common outputs} to be the linear map $f \circ_h f_1: \Ext_k C \rightarrow \Ext_l C$ given by 
\begin{equation}\label{Eq:CompositionSimple}
f \circ_h f_1 := \begin{multlined}[t]\mu_{l', l_1 - h} \circ (f\otimes \Id_{l_1-h})\circ (\mu_{h, k'-h}\otimes \Id_{l_1-h})\circ (\Id_{h} \otimes \tau_{\rule{0pt}{7pt}l_1-h,k'-h}) \\[\jot] \circ (\Delta_{h,l_1-h}\otimes \Id_{k'-h}) \circ (f_1 \otimes \Id_{k'-h} ) \circ \Delta_{k_1, k'-h}. \end{multlined}
\end{equation}
More generally, we define the composition of $f: E_{k'}\rightarrow E_{l'}$ with $r\ge 1$ linear maps $f_{i}: E_{k_i} \rightarrow E_{l_i}$ with $k_i$, $l_i \ge 0$ for $i=1$,~$\dotsc$, $r$ at $0 \le h_i \le l_i$ common outputs such that $h:= h_1 + \dotsb + h_r \le k'$ as follows. We set 
$$ k:= k' + k_1 + \dotsb + k_r-h\quad\text{and}\quad l:= l' +  l_1 + \dotsb + l_r  - h $$
and define $f\circ_{h_1, \dotsc, h_r}(f_1, \dotsc, f_r): \Ext_k C \rightarrow \Ext_l C$ by
\begin{equation} \label{Eq:CompositionFormula}
\begin{aligned}
&f\circ_{h_1, \dotsc, h_r}(f_1, \dotsc, f_r)\\
&\qquad := \begin{multlined}[t]
\mu\circ (f\otimes \Id)\circ(\mu\otimes\Id)\circ(\Id\otimes \tau) \\[\jot] \circ \big(\big[(\mu^{(r)}\otimes \mu^{(r)})\circ (F_{h_1,\dotsc,h_r} \otimes \Id^{\otimes r})\circ \sigma_r\circ \Delta^{\otimes r}\big] \otimes \Id \big) \\[\jot] \circ (f_1\otimes \dotsb \otimes f_r\otimes \Id)\circ\Delta^{(r+1)},
\end{multlined}\end{aligned}
\end{equation}
where we have:
\begin{itemize}
\item The operation $\mu^{(r)}$ is the ``product with $r$ inputs'' and the operation $\Delta^{(r)}$ the ``coproduct with $r$ outputs''; they are defined~by
$$ \begin{aligned}
\mu^{(r)} &:= \mu(\Id \otimes \mu)\dotsb(\Id^{\otimes r-2} \otimes \mu), & \mu^{(1)}&:= \Id,\\
\Delta^{(r)} &:= (\Id^{\otimes r-2} \otimes \Delta)\dotsb(\Id \otimes \Delta)\Delta, &\Delta^{(1)}&:= \Id.
\end{aligned} $$

\item $F_{h_1,\dotsc, h_r} := (\iota_{h_1}\pi_{h_1}) \otimes \dotsb \otimes (\iota_{h_r}\pi_{h_r})$.
\item The permutation $\sigma_r \in \Perm_{2r}$ is given by  
$$\sigma_r: (1,2,\dotsc, 2r-1, 2r) \longmapsto (1,r+1, \dotsc, r, 2r).$$
\item The symbols $f$ and $f_i$ inside the formula denote the \emph{trivial extensions} of $f$ and $f_i$, respectively; we extend a linear map $f: E_{k'} C \rightarrow E_{l'} C$ trivially to $f: \Ext C \rightarrow \Ext C$ by defining $f(\Ext_i C)=0$ for $i\neq k'$.
\end{itemize}
\end{Definition}

\begin{Remark}[On partial compositions]\phantomsection\label{Rem:Compositions}
\begin{RemarkList}
\item Defining $f\circ_{h_1, \dotsc, h_r}(f_1, \dotsc, f_r): \Ext_k C \rightarrow \Ext_l C$ using~\eqref{Eq:CompositionFormula} makes sense because the right hand side is a trivial extension of its component $\Ext_k C \rightarrow \Ext_l C$. In fact, all $\mu$, $\Delta$, $\pi$, $\iota$ in \eqref{Eq:CompositionFormula} can be replaced with $\mu_{i,j}$, $\Delta_{i,j}$, $\pi_i$, $\iota_i$ for unique $i$, $j$, so that trivial extensions do not have to be used at all. In this way, it can be seen that \eqref{Eq:CompositionSimple} is indeed a special case of~\eqref{Eq:CompositionFormula}. 
\item If $h = k' = l_1$, then $f \circ_{h} f_1 = f\circ f_1$.
\item It holds $f \circ_0 f_1 = (-1)^{\Abs{f}\Abs{f_1}} f_1 \circ_0 f$ and 
$$ f\circ_{h_1,\dotsc,h_r}(f_1,\dotsc,f_r) = \varepsilon(\sigma,f) f\circ_{h_{\sigma_1^{-1}},\dotsc,h_{\sigma_r^{-1}}}(f_{\sigma_1^{-1}},\dotsc,f_{\sigma_r^{-1}}). $$
\item Consider the (``non-trivial'') extension $\hat{f}:= \mu(f\otimes \Id)\Delta: \Ext C \rightarrow \Ext C$ and the symmetric product $f_1 \odot \dotsb \odot f_r := \mu^{(r)}(f_1 \otimes \dotsb \otimes f_r)\Delta^{(r)}: \Ext C \rightarrow \Ext C$.
 The proof of the following formulas appearing in \cite{Cieliebak2015} is now an exercise on the bialgebra calculus:
\begin{equation} \label{Eq:Mix}
\begin{aligned}
f\circ_{h_1,\dotsc,h_{r-1},0}(f_1,\dotsc, f_r) &= f\circ_{h_1,\dotsc, h_{r-1}}(f_1,\dotsc, f_{r-1}) \odot f_r, \\
\hat{f} \circ \hat{f}_1 &= \sum_{h = 0}^{\min(k',l_1)} \widehat{f\circ_h f_1}, \\ 
   \hat{f} \circ (f_1 \odot \dotsb \odot f_r) &= \sum_{\substack{h_1, \dotsc, h_r \ge 0 \\ h_1 + \dotsb + h_r = k'}} f\circ_{h_1,\dotsc, h_r}(f_1,\dotsc, f_r).
\end{aligned}
\end{equation}
We also have the ``weak associativity''
\begin{equation}\label{Eq:WeakAssoc}
\qquad\qquad \mathclap{\sum_{\substack{0 \le h_2 \le \min(f_3^-, f_2^+) \\
0 \le h_1 \le \min(f_1^+,f_2^- + f_3^- - h_2) \\
h_1 + h_2 = h}}}\quad\qquad f_1 \circ_{h_1} (f_2 \circ_{h_2} f_3) = \qquad\quad \mathclap{\sum_{\substack{0 \le h_1 \le \min(f_1^+,f_2^-) \\ 0 \le h_2 \le \min(f_1^+ + f_2^+ - h_1, f_3^-) \\ h_1 + h_2 = h}}}\qquad\quad (f_1 \circ_{h_1} f_2) \circ_{h_2} f_3
\end{equation}
for every $0\le h \le \min(k_1 + k_2 + k_3, l_1 + l_2 + l_3)$, where $f^+$ denotes the number of inputs and $f^-$ the number of outputs of $f$. The weak associativity of $\circ_h$ can be proven using the associativity of $\,\hat{\cdot}$ and the second relation of \eqref{Eq:Mix}.\qedhere
\end{RemarkList}
\end{Remark}

If $C$ is filtered by a decreasing filtration, then the bialgebra operations extend continuously to 
$$\begin{aligned}
\mu: \hat{\Ext}_{k_1} C \hat{\otimes} \hat{\Ext}_{k_2} C &\longrightarrow \hat{\Ext}_{k_1+k_2} C \quad \text{and}\\ 
\Delta: \hat{\Ext}_k C &\longrightarrow \bigoplus_{\substack{l_1, l_2 \ge 0 \\ l_1 + l_2 = k}} \hat{\Ext}_{l_1}C\hat{\otimes}\hat{\Ext}_{l_2}C
\end{aligned}$$ 
for all $k_1$, $k_2$, $k\in \N_0$ because they preserve the filtration degree (see~\cite{Fresse} for a similar construction). Next, if $f_1: \hat{\Ext}_{k_1}C \rightarrow \hat{\Ext}_{l_1}C$ and $f_2: \hat{\Ext}_{k_2}C\rightarrow \hat{\Ext}_{l_2}C$ have finite filtration degrees, then $f_1 \otimes f_2: \hat{\Ext}_{k_1} C \otimes \hat{\Ext}_{k_2} C \rightarrow \hat{\Ext}_{l_1}C\otimes\hat{\Ext}_{l_2}C$ has finite filtration degree too, and hence it extends continuously to $f_1 \otimes f_2: \hat{\Ext}_{k_1} C\hat{\otimes} \hat{\Ext}_{k_2} C \rightarrow \hat{\Ext}_{l_1}C\hat{\otimes}\hat{\Ext}_{l_2}C$.
Using these facts, we can canonically extend Definition~\ref{Def:CircS} to maps $f: \hat{\Ext}_{k'} C \rightarrow \hat{\Ext}_{l'} C$ and $f_i: \hat{\Ext}_{k_i}C \rightarrow \hat{\Ext}_{l_i}C$ of finite filtration degrees. The resulting map $f\circ_{h_1,\dotsc,h_r}(f_1,\dotsc,f_r): \hat{\Ext}_k C \rightarrow \hat{\Ext}_l C$ will have finite filtration degree too. Moreover, the formulas in Remark~\ref{Rem:Compositions} will still hold.

We will now rephrase the definitions of an $\IBLInfty$-algebra, a Maurer-Cartan element and twisted operations from~\cite{Cieliebak2015} in terms of~$\circ_{h_1, \dotsc, h_r}$.



\begin{Def}[$\IBLInfty$-algebra] \label{Def:IBLInfty} Let $C$ be a graded vector space equipped with a decreasing filtration, and let $d\in \Z$ and $\gamma\ge 0$ be fixed constants. An \emph{$\IBLInfty$-algebra of bidegree $(d,\gamma)$} on~$C$ is a collection of linear maps $\OPQ_{klg}: \hat{\Ext}_k C \rightarrow \hat{\Ext}_l C$ for all $k,l\ge 1$, $g\ge 0$ which are homogenous, of finite filtration degree and satisfy the following conditions: 
\begin{enumerate}[label=\arabic*)]
\item $\Abs{\OPQ_{klg}} = - 2d(k+g-1) - 1$.
\item $\Norm{\OPQ_{klg}} \ge \gamma \chi_{klg}$,
where $\chi_{klg}:=2-2g-k-l$. 
\item The \emph{$\IBLInfty$-relations} hold: for all $k,l\ge 1$, $g\ge 0$, we have
\begin{equation} \label{Eq:IBLInfRel}
\sum_{h=1}^{g+1} \sum_{\substack{k_1, k_2, l_1, l_2 \ge 1 \\ g_1, g_2 \ge 0 \\k_1 + k_2 = k + h \\ l_1 + l_2 = l+ h\\ g_1 + g_2 = g+ 1 -h }} \OPQ_{k_2 l_2 g_2} \circ_h \OPQ_{k_1 l_1 g_1} = 0.
\end{equation}
\end{enumerate}
We denote a given $\IBLInfty$-algebra structure on $C$ by $\IBLInfty(C)$; i.e., we write $\IBLInfty(C)=(C,(\OPQ_{klg}))$.

If $\OPQ_{klg} \equiv 0$ for all $(k,l,g)\neq (1,1,0)$, $(2,1,0)$, $(1,2,0)$, then we call $\IBLInfty(C)$ a \emph{$\dIBL$-algebra} and denote it by $\dIBL(C)$. If in addition $\OPQ_{110} \equiv 0$, then we have an \emph{$\IBL$-algebra} $\IBL(C)$. If the operations on the completed exterior powers~$\hat{\Ext}_k C$ arise as continuous extensions of operations $\OPQ_{klg}: \Ext_k C \rightarrow \Ext_l C$, then we call the $\IBLInfty$-algebra \emph{completion-free} and denote $C$ together with the operations $\OPQ_{klg}: \Ext_k C \rightarrow \Ext_l C$ by $\ShortIBLInfty(C)$.
\end{Def}

The acronym $\IBL$ stands for an \emph{involutive Lie bialgebra.} It follows namely from the $\IBLInfty$-relations \eqref{Eq:IBLInfRel} that for $\IBL(C) = (C,\OPQ_{210},\OPQ_{120})$ the following holds: 
\begin{equation*}
\raisebox{2ex}{$\text{Lie bialg.}\;\left\{\rule{0pt}{5ex}\right.$}\;
\begin{aligned}   
   0&= \OPQ_{210}\circ_1 \OPQ_{210} &&\leftarrow\text{Jacobi id.}\\
   0 &= \OPQ_{120} \circ_1 \OPQ_{120} &&\leftarrow\text{co-Jacobi id.}\\
   0 &= \OPQ_{120}\circ_1 \OPQ_{210} + \OPQ_{210}\circ_1 \OPQ_{120}&&\leftarrow\text{Drinfeld id.} \\
   0 &= \OPQ_{210} \circ_2 \OPQ_{120} &&\leftarrow\text{Involutivity}
\end{aligned}
\end{equation*}

\begin{Proposition}[Odd degree shift of an $\IBL$-algebra]\label{Prop:ClasModIBL}
Let $(C,\OPQ_{210}, \OPQ_{120})$ be an $\IBL$-algebra of degree $d$ from Definition~\ref{Def:IBLInfty}, and let $\tilde{\OPQ}_{210} : C^{\otimes 2} \rightarrow C$ and $\tilde{\OPQ}_{120}: C \rightarrow C^{\otimes 2}$ be the linear maps defined by
\begin{equation}\label{Eq:ClasModIBL}
\begin{aligned}
\SuspU \tilde{\OPQ}_{210}(x_1 \otimes x_2) &:= \OPQ_{210}(\pi(\SuspU^2 x_1 \otimes x_2)) \quad\text{and} \\
\SuspU^2 \tilde{\OPQ}_{120}(x) &:= \iota(\OPQ_{120}(\SuspU x))
\end{aligned}
\end{equation}
for all $x_1$, $x_2$, $x\in C$, where $\iota: \Sym_2(C[1]) \rightarrow C[1]^{\otimes 2}$ is the section of $\pi: C[1]^{\otimes 2} \rightarrow \Sym_2(C[1])$ from Definition~\ref{Def:SymAlgebra} and~$\SuspU$ is a formal symbol of degree $\Abs{\SuspU} = -1$. Then the degrees satisfy
$$ \Abs{\tilde{\OPQ}_{210}} = \Abs{\OPQ_{210}} - 1 = -2d - 2\quad\text{and}\quad\Abs{\tilde{\OPQ}_{120}} = \Abs{\OPQ_{120}} + 1 = 0, $$
the operations $\tilde{\OPQ}_{210}$ and $\tilde{\OPQ}_{120}$ are graded antisymmetric, i.e., we have
$$ \tilde{\OPQ}_{210} \circ\tau = - \tilde{\OPQ}_{210}\quad\text{and}\quad\tau \circ \tilde{\OPQ}_{120} = - \tilde{\OPQ}_{120} $$
for the twist map $\tau$, and the relations
\begin{equation*}
\begin{aligned}
0&=\tilde{\OPQ}_{210}\circ (\tilde{\OPQ}_{210}\otimes \Id)\circ (\Id^{\otimes 3}+ t_3 + t_3^2), \\
0&=(\Id^{\otimes 3}+t_3 + t_3^2)\circ (\tilde{\OPQ}_{120}\otimes\Id)\circ \tilde{\OPQ}_{120}, \\
0&= x_1 \cdot \tilde{\OPQ}_{120}(x_2) - (-1)^{ x_1 x_2} x_2 \cdot \tilde{\OPQ}_{120}(x_1) - \tilde{\OPQ}_{120}(\tilde{\OPQ}_{210}(x_1,x_2)), \\
0& = \tilde{\OPQ}_{210} \circ \tilde{\OPQ}_{120},
\end{aligned}
\end{equation*}
hold for all $x_1$, $x_2\in C$. Hhere $t_3 \in \Perm_3$ denotes the cyclic permutation with $t_3(1) = 2$ acting on $C^{\otimes 3}$ and we define
$$ x\cdot (y_1 \otimes y_2) := \tilde{\OPQ}_{210}(x,y_1)\otimes y_2 + (-1)^{ x y_1} y_1 \otimes \tilde{\OPQ}_{210}(x,y_2) $$
for all $x$, $y_1$, $y_2 \in C$.
\end{Proposition}

\begin{proof}
The proof is a lengthy but straightforward computation.
\end{proof}
\begin{figure}[t]
\centering
\begin{subfigure}{\textwidth}
\centering
\begin{tikzpicture}
\tikzset{decorate sep/.style 2 args=
{decorate,decoration={shape backgrounds,shape=circle,shape size=#1,shape sep=#2}}}

 \def\ecc{0.1} 
    \def\gencanc{0.05} 
  \def\genecc{20} 
  \def\distI{0.25}
  \def\radI{0.5}
  \def\radIII{0.2}
  \def\eccI{0.1}  
  \def\gencancI{0.05} 
  \def\geneccI{20} 
  \def\genradI{0.45} 
  \def\mezI{0.4} 
  \def\mezII{3.8} 
  \def\mezIII{0.3} 
  \def\mezIV{0.8} 
  \def\mezV{0.1} 
  \def\triI{0.2mm}
  \def\triII{0.6mm}
  \def\vysI{1} 
  \def\vysII{0.2} 
  \def\vysIII{1} 
  \def\distIII{3*\radI} 
  
  
  \coordinate (A1) at (0,0);
  \coordinate (A2) at ($(A1)+(2*\radI,0)$);
  \coordinate (A3) at ($(A2)+(\mezI,0)$);
  \coordinate (A4) at ($(A3)+(2*\radI,0)$);
  \coordinate (A5) at ($(A4)+(\mezIII,0)$);
  \coordinate (A6) at ($(A5)+(2*\radI,0)$);
  \coordinate (A7) at ($(A6)+(\mezI,0)$);
  \coordinate (A8) at ($(A7)+(2*\radI,0)$);
  \coordinate (A9) at ($(A8)+(\mezI,0)$);

  \coordinate (B1) at ($(A1)+(0,-\vysI)$);
  \coordinate (B2) at ($(B1)+(2*\radI,0)$);
  \coordinate (B3) at ($(B2)+(\mezI,0)$);
  \coordinate (B4) at ($(B3)+(2*\radI,0)$);
  \coordinate (B5) at ($(B4)+(\mezIII,0)$);
  \coordinate (B6) at ($(B5)+(2*\radI,0)$);
  \coordinate (B7) at ($(B6)+(\mezI,0)$);
  \coordinate (B8) at ($(B7)+(2*\radI,0)$);
  \coordinate (B9) at ($(B8)+(\mezI,0)$);
  
  \coordinate (C1) at ($(B5)+(0,-\distI)$);
  \coordinate (C2) at ($(C1)+(2*\radI,0)$);
  \coordinate (C3) at ($(C2)+(\mezI,0)$);
  \coordinate (C4) at ($(C3)+(2*\radI,0)$);
  \coordinate (C5) at ($(C4)+(\mezIII,0)$);
  \coordinate (C6) at ($(C5)+(2*\radI,0)$);
  \coordinate (C7) at ($(C6)+(\mezI,0)$);
  \coordinate (C8) at ($(C7)+(2*\radI,0)$);
  \coordinate (C9) at ($(C8)+(\mezI,0)$);

  \coordinate (D1) at ($(C1)+(0,-\vysI)$);
  \coordinate (D2) at ($(D1)+(2*\radI,0)$);
  \coordinate (D3) at ($(D2)+(\mezI,0)$);
  \coordinate (D4) at ($(D3)+(2*\radI,0)$);
  \coordinate (D5) at ($(D4)+(\mezIII,0)$);
  \coordinate (D6) at ($(D5)+(2*\radI,0)$);
  \coordinate (D7) at ($(D6)+(\mezI,0)$);
  \coordinate (D8) at ($(D7)+(2*\radI,0)$);
  \coordinate (D9) at ($(D8)+(\mezI,0)$);
  
\draw (A1) arc (180:360:{\radI} and {\ecc});
\draw (A1) arc (180:0:{\radI} and {\ecc});
\draw (A3) arc (180:360:{\radI} and {\ecc});
\draw (A3) arc (180:0:{\radI} and {\ecc});
\draw (A5) arc (180:360:{\radI} and {\ecc});
\draw (A5) arc (180:0:{\radI} and {\ecc});
\draw (A7) arc (180:360:{\radI} and {\ecc});
\draw (A7) arc (180:0:{\radI} and {\ecc});

\draw (C1) arc (180:360:{\radI} and {\ecc});
\draw (C1) arc (180:0:{\radI} and {\ecc});
\draw (C3) arc (180:360:{\radI} and {\ecc});
\draw (C3) arc (180:0:{\radI} and {\ecc});
\draw (C5) arc (180:360:{\radI} and {\ecc});
\draw (C5) arc (180:0:{\radI} and {\ecc});
\draw (C7) arc (180:360:{\radI} and {\ecc});
\draw (C7) arc (180:0:{\radI} and {\ecc});

\draw (D1) arc (180:360:{\radI} and {\ecc});
\draw[dashed] (D1) arc (180:0:{\radI} and {\ecc});
\draw (D3) arc (180:360:{\radI} and {\ecc});
\draw[dashed] (D3) arc (180:0:{\radI} and {\ecc});
\draw (D5) arc (180:360:{\radI} and {\ecc});
\draw[dashed] (D5) arc (180:0:{\radI} and {\ecc});
\draw (D7) arc (180:360:{\radI} and {\ecc});
\draw[dashed] (D7) arc (180:0:{\radI} and {\ecc});

\draw (B1) arc (180:360:{\radI} and {\ecc});
\draw[dashed] (B1) arc (180:0:{\radI} and {\ecc});
\draw (B3) arc (180:360:{\radI} and {\ecc});
\draw[dashed] (B3) arc (180:0:{\radI} and {\ecc});
\draw (B5) arc (180:360:{\radI} and {\ecc});
\draw[dashed] (B5) arc (180:0:{\radI} and {\ecc});
\draw (B7) arc (180:360:{\radI} and {\ecc});
\draw[dashed] (B7) arc (180:0:{\radI} and {\ecc});

\draw (A1)--(B1);
\draw (A8)--(B8);
\draw (C1)--(D1);
\draw (C8)--(D8);

\draw (A4) to[out=-90,in=-90] (A5);
\draw (C4) to[out=-90,in=-90] (C5);
\draw (B4) to[out=90,in=90] (B5);
\draw (D4) to[out=90,in=90] (D5);

\draw[decorate sep={\triI}{\triII},fill] ($(A2)+(\mezV,0)$) to ($(A3)+(-\mezV,0)$);

\draw[decorate sep={\triI}{\triII},fill] ($(A6)+(\mezV,0)$) to ($(A7)+(-\mezV,0)$);

\draw[decorate sep={\triI}{\triII},fill] ($(B2)+(\mezV,0)$) to ($(B3)+(-\mezV,0)$);

\draw[decorate sep={\triI}{\triII},fill] ($(B6)+(\mezV,0)$) to ($(B7)+(-\mezV,0)$); 

\draw[decorate sep={\triI}{\triII},fill] ($(C2)+(\mezV,0)$) to ($(C3)+(-\mezV,0)$);

\draw[decorate sep={\triI}{\triII},fill] ($(C6)+(\mezV,0)$) to ($(C7)+(-\mezV,0)$);

\draw[decorate sep={\triI}{\triII},fill] ($(D2)+(\mezV,0)$) to ($(D3)+(-\mezV,0)$);

\draw[decorate sep={\triI}{\triII},fill] ($(D6)+(\mezV,0)$) to ($(D7)+(-\mezV,0)$);

\coordinate (U1) at ($(B1)-(0,\distI)$);
\coordinate (U2) at ($(U1)+(2*\radI,0)$);
\coordinate (U3) at ($(U2)+(\mezI,0)$);
\coordinate (U4) at ($(U3)+(2*\radI,0)$);
\coordinate (U5) at ($(U1)-(0,\vysI)$);
\coordinate (U6) at ($(U5)+(2*\radI,0)$);
\coordinate (U7) at ($(U6)+(\mezI,0)$);
\coordinate (U8) at ($(U7)+(2*\radI,0)$);

\coordinate (V1) at ($(A8)+(\mezIII,0)$);
\coordinate (V2) at ($(V1)+(2*\radI,0)$);
\coordinate (V3) at ($(V2)+(\mezI,0)$);
\coordinate (V4) at ($(V3)+(2*\radI,0)$);
\coordinate (V5) at ($(V1)-(0,\vysI)$);
\coordinate (V6) at ($(V5)+(2*\radI,0)$);
\coordinate (V7) at ($(V6)+(\mezI,0)$);
\coordinate (V8) at ($(V7)+(2*\radI,0)$);

\draw (U5) arc (180:360:{\radI} and {\ecc});
\draw[dashed] (U5) arc (180:0:{\radI} and {\ecc});
\draw (U7) arc (180:360:{\radI} and {\ecc});
\draw[dashed] (U7) arc (180:0:{\radI} and {\ecc});
\draw (V5) arc (180:360:{\radI} and {\ecc});
\draw[dashed] (V5) arc (180:0:{\radI} and {\ecc});
\draw (V7) arc (180:360:{\radI} and {\ecc});
\draw[dashed] (V7) arc (180:0:{\radI} and {\ecc});

\draw (U1) arc (180:360:{\radI} and {\ecc});
\draw (U1) arc (180:0:{\radI} and {\ecc});
\draw (U3) arc (180:360:{\radI} and {\ecc});
\draw (U3) arc (180:0:{\radI} and {\ecc});
\draw (V1) arc (180:360:{\radI} and {\ecc});
\draw (V1) arc (180:0:{\radI} and {\ecc});
\draw (V3) arc (180:360:{\radI} and {\ecc});
\draw (V3) arc (180:0:{\radI} and {\ecc});

\draw (U1)--(U5);
\draw (U4)--(U8);
\draw (U2)--(U6);
\draw (U3)--(U7);

\draw (V1)--(V5);
\draw (V4)--(V8);
\draw (V2)--(V6);
\draw (V3)--(V7);

\draw[decorate sep={\triI}{\triII},fill] ($(U2)+(\mezV,0)$) to ($(U3)+(-\mezV,0)$);
\draw[decorate sep={\triI}{\triII},fill] ($(U6)+(\mezV,0)$) to ($(U7)+(-\mezV,0)$);
\draw[decorate sep={\triI}{\triII},fill] ($(V2)+(\mezV,0)$) to ($(V3)+(-\mezV,0)$);
\draw[decorate sep={\triI}{\triII},fill] ($(V6)+(\mezV,0)$) to ($(V7)+(-\mezV,0)$);

\coordinate (G1) at ($(B1)+(\radI,0.5*\vysI)$);
  \draw (G1) to[out=-\geneccI,in=180+\geneccI] coordinate[pos=\gencancI] (G11) coordinate[pos=1-\gencancI] (G12) ($(G1) + (2*\genradI,0)$) ;
\draw (G11) to[out=\genecc,in=180-\genecc] (G12);

\coordinate (G2) at ($(G11)+(\distIII,0)$);
  \draw (G2) to[out=-\geneccI,in=180+\geneccI] coordinate[pos=\gencancI] (G21) coordinate[pos=1-\gencancI] (G22) ($(G2) + (2*\genradI,0)$) ;
\draw (G21) to[out=\genecc,in=180-\genecc] (G22);

\coordinate (GG2) at ($(D8)+(-2*\genradI-\radI,0.5*\vysI)$);
  \draw (GG2) to[out=-\geneccI,in=180+\geneccI] coordinate[pos=\gencancI] (GG21) coordinate[pos=1-\gencancI] (GG22) ($(GG2) + (2*\genradI,0)$) ;
\draw (GG21) to[out=\genecc,in=180-\genecc] (GG22);

\coordinate (GG1) at ($(GG2)+(-\distIII,0)$);
  \draw (GG1) to[out=-\geneccI,in=180+\geneccI] coordinate[pos=\gencancI] (GG11) coordinate[pos=1-\gencancI] (GG12) ($(GG1) + (2*\genradI,0)$) ;
\draw (GG11) to[out=\genecc,in=180-\genecc] (GG12);

\draw[decorate sep={\triI}{\triII},fill] ($(G12)+(\mezV+\gencancI,0)$) to ($(G21)+(-\mezV-\gencancI,0)$);
\draw[decorate sep={\triI}{\triII},fill] ($(GG12)+(\mezV+\gencancI,0)$) to ($(GG21)+(-\mezV-\gencancI,0)$);

\node at ($(G22)+(2*\radI,0)$) {$\mathfrak{q}_{k_1 l_1 g_1}$};
\node at ($(GG11)+(-2*\radI,0)$) {$\mathfrak{q}_{k_2 l_2 g_2}$};
\end{tikzpicture}
\caption{The term $\OPQ_{k_2 l_2 g_2} \circ_h \OPQ_{k_1 l_1 g_1}$ in the $\IBLInfty$-equation \eqref{Eq:IBLInfRel}.}
\end{subfigure}
\par\bigskip
\begin{subfigure}{\textwidth}
\centering
\input{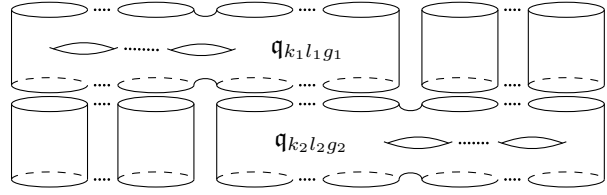}
\caption{The term $\OPQ_{k' l' g'}\circ_{h_1,\dotsc,h_r} (\PMC_{l_1 g_1},\dotsc,\PMC_{l_r g_r})$ in the Maurer-Cartan equation \eqref{Eq:MaurerCartanEquation}. We remark that the contour of the surface corresponding to $\OPQ_{k'l'g'}$ starts on the left and continues to the right along the dotted line behind the two trivial cylinders.}
\end{subfigure}
\par\bigskip
\begin{subfigure}{\textwidth}
\centering
\input{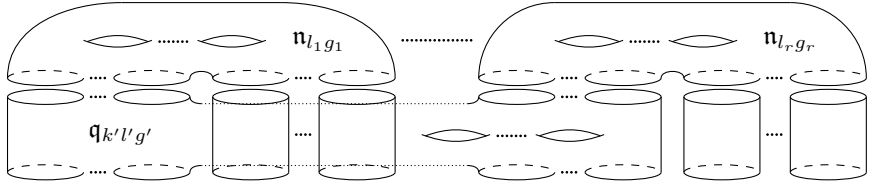}
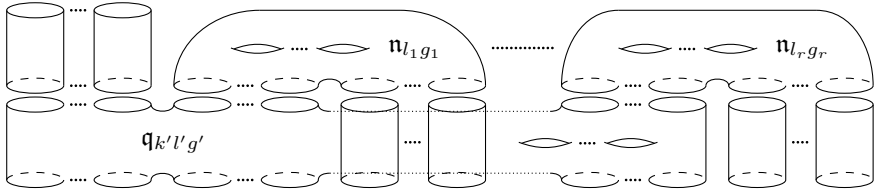
\caption{The term $\OPQ_{k' l' g'}\circ_{l_1,\dotsc, l_r} (\PMC_{l_1 g_1},\dotsc,\PMC_{l_r g_r})$ in the twisted operation \eqref{Eq:TwistedOperations}. The remark to Figure (b) applies too.}
\end{subfigure}
\caption{Graphical representation of compositions appearing in Definitions \ref{Def:IBLInfty}, \ref{Def:MaurerCartan} and \ref{Def:TwistedOperations} as gluing of connected Riemannian surfaces. The figure is to be read from the top  to the bottom, the empty cylinder represents the identity, and the resulting surface must be connected. We emphasize that the gluing is not associative (c.f., weak associativity \eqref{Eq:WeakAssoc}).}
\label{Fig:Surfaces}
\end{figure}
%
%
\begin{Def}[Maurer-Cartan element] \label{Def:MaurerCartan}
A \emph{Maurer-Cartan element} for an $\IBLInfty$-algebra $\IBLInfty(C)$ from Definition~\ref{Def:IBLInfty} is a collection $\PMC := (\PMC_{lg})_{l\ge 1, g\ge 0}$ of elements $\PMC_{lg}\in \hat{\Ext}_l C$ which are homogenous, of finite filtration degree and satisfy the following conditions:
\begin{enumerate}[label=\arabic*)]
\item $\Abs{\PMC_{lg}} = - 2d(g-1)$.
\item $\Norm{\PMC_{lg}}\ge\gamma \chi_{0lg}$ with $>$ for $(l,g)=(1,0)$, $(2,0)$ (see Definition~\ref{Def:IBLInfty} for~$\chi_{klg}$).
\item The \emph{Maurer-Cartan equation} holds: for all $l\ge 1$, $g\ge 0$, we have
\begin{equation} \label{Eq:MaurerCartanEquation}
\begin{aligned}
\sum_{r\ge 1}\frac{1}{r!}  \sum_{\substack{l', k', l_1, \dotsc, l_r\ge 1 \\ g', g_1, \dotsc, g_r \ge 0 \\ h_1, \dotsc, h_r \ge 1 \\ 
l_1 + \dotsb + l_r + l' - k'= l \\ g_1 + \dotsb + g_r + g' +  k' = g + r \\ h_1 + \dotsb + h_r - k' =0 } } \OPQ_{k' l' g'}\circ_{h_1,\dotsc,h_r} (\PMC_{l_1 g_1},\dotsc,\PMC_{l_r g_r}) = 0,
\end{aligned} 
\end{equation}
where we view $\PMC_{lg}$ as a linear map $\PMC_{lg}: \hat{\Ext}_0 C = \R \rightarrow \hat{\Ext}_l C$ with $\PMC_{lg}(1) = \PMC_{lg}$.

\end{enumerate}
\end{Def}


\begin{Def}[Twisted operations] \label{Def:TwistedOperations}
In the setting of Definition~\ref{Def:MaurerCartan}, the \emph{twisted operations} $\OPQ_{klg}^\PMC: \hat{\Ext}_k C\rightarrow \hat{\Ext}_l C$ for $k,l\ge 1$, $g\ge 0$ are defined by
\begin{equation}\label{Eq:TwistedOperations}
 \OPQ_{klg}^\PMC =\sum_{r\ge 0} \frac{1}{r!} \sum_{\substack{k', l', l_1, \dotsc, l_r \ge 1 \\ g', g_1, \dotsc, g_r \ge 0\\ h_1, \dotsc, h_r \ge 1 \\
l_1 + \dotsb + l_r + l' - k' = l-k \\ g_1 + \dotsb +g_r + g' + k' = g + r + k \\ h_1 + \dotsb + h_r - k' = -k}} \OPQ_{k' l' g'}\circ_{h_1,\dotsc, h_r} (\PMC_{l_1 g_1},\dotsc,\PMC_{l_r g_r}).
\end{equation}
In \cite[Proposition~9.3]{Cieliebak2015}, they prove that $(\OPQ_{klg}^\PMC)_{k,l\ge 1, g\ge 0}$ is again an $\IBLInfty$-algebra of bidegree $(d,\gamma)$ on $C$ --- \emph{the twisted $\IBLInfty$-algebra}.  We denote it by $\IBLInfty^\PMC(C)$.
\end{Def}

Let $(\OPQ_{klg})$ be an $\IBLInfty$-algebra on $C$. The boundary operator $\OPQ_{110}: C[1] \rightarrow C[1]$ induces the boundary operator $\Bdd_k : \Ext_k C \rightarrow \Ext_k C$ for every $k\in \N$ (see \eqref{Eq:BddExt}). Because of the finite filtration degree, $\Bdd_k$ continuously extends to $\Bdd_k: \hat{\Ext}_k C \rightarrow \hat{\Ext}_k C$.
The following is easy to see using \eqref{Eq:CompositionSimple}:
$$\begin{aligned}
 \OPQ_{klg} \circ_1 \OPQ_{110} &= \OPQ_{klg} \circ \Bdd_k, \\
 \OPQ_{110} \circ_1 \OPQ_{klg} &= \Bdd_l \circ \OPQ_{klg}.
\end{aligned}$$
Because $\OPQ_{klg}$ are odd (:=\,have odd degree), we have
$$ \begin{aligned}
  [\Bdd,\OPQ_{klg}] &:= \Bdd_l \circ \OPQ_{klg} - (-1)^{\Abs{\Bdd}\Abs{\OPQ_{klg}}} \OPQ_{klg}\circ \Bdd_k \\
   &= \Bdd_l \circ \OPQ_{klg} + \OPQ_{klg}\circ \Bdd_k \\
   &= \OPQ_{110}\circ_1 \OPQ_{klg} + \OPQ_{klg}\circ_1 \OPQ_{110}.
  \end{aligned}$$
With this notation, the $\IBLInfty$-relations \eqref{Eq:IBLInfRel} for $(k,l,g) = (2,1,0)$ and $(1,2,0)$ become $[\Bdd,\OPQ_{210}] = 0$ and $[\Bdd,\OPQ_{120}] = 0$, respectively. Therefore, $\OPQ_{210}$ and $\OPQ_{120}$ descend to the homology.

\begin{Definition}[Homology and the induced $\IBL$-algebra]\label{Def:HomIBL}
We define the \emph{homology} of an $\IBLInfty$-algebra $\IBLInfty(C)$ by
$$\HIBL(C)[1] := \H(\hat{C}[1],\OPQ_{110}). $$
It is a graded vector space with the induced filtration. 
If the canonical map $\Ext_k \HIBL(C) \rightarrow \H(\hat{\Ext}_k C, \Bdd_k)$ induces the isomorphism $\hat{\Ext}_k \HIBL(C) \simeq \H(\hat{\Ext}_k C, \Bdd_k)$, then the induced maps
$$ \OPQ_{210}: \hat{\Ext}_2\HIBL(C) \rightarrow \hat{\Ext}_1\HIBL(C)\quad \text{and}\quad \OPQ_{120}: \hat{\Ext}_1\HIBL(C) \rightarrow \hat{\Ext}_2\HIBL(C) $$
define an $\IBL$-structure on $\HIBL(C)$ --- the \emph{induced $\IBL$-algebra on homology}.

If $\PMC$ is a Maurer-Cartan element for $\IBLInfty(C)$, we denote by $\HIBL^\PMC(C)$ the homology of~$\IBLInfty^\PMC(C)$. 
\end{Definition}



\begin{Remark}[$\mathrm{BV}$-formalism]\label{Rem:BVForm}
Consider the weight-reduced exterior algebra $\RExt C$. Let $\CRExt C[[\hbar]]$ and $\CRExt C((\hbar))$ be the spaces of power and Laurent series in a formal variable $\hbar$ of degree $\Abs{\hbar} = 2d$ with coefficients in $\CRExt C$, respectively, where $\CRExt C$ is a suitable completion of $\RExt C$.
Operations of an $\IBLInfty$-algebra on $C$ can be encoded in a degree~$-1$ operator $\Delta: \CRExt C[[\hbar]] \rightarrow \CRExt C[[\hbar]]$ called the \emph{$\mathrm{BV}_\infty$-operator,} while the data of a Maurer-Cartan element $(\PMC_{lg})$ give rise to an operator $e^{\PMC}: \CRExt C[[\hbar]] \rightarrow \CRExt C((\hbar))$ called the exponential of $\PMC$. These operators are given by
$$ \Delta := \sum_{i\ge 0}\Delta_{i+1} \hbar^{i}\quad\text{and}\quad e^{\PMC} := \sum_{j\in \Z} (e^{\PMC})_j \hbar^{j}, $$
where the maps $\Delta_i$, $(e^\PMC)_j: \CRExt C\rightarrow \CRExt C$ for $i\ge 1$, $j\in \Z$ are defined by 
$$\begin{aligned}
\Delta_i & := \sum_{\substack{k\ge 1, g\ge 0 \\k+g=i}} \sum_{l\ge 1} \hat{\OPQ}_{klg}\quad\text{and} \\
(e^{\PMC})_j &:= \sum_{r=0}^\infty \frac{1}{r!} \sum_{\substack{g_1, \dotsc, g_r \ge 0 \\ g_1 + \dotsb +g_r - r= j }} \sum_{l_1, \dotsc, l_r\ge 1} \PMC_{l_1 g_1} \odot \dotsb \odot \PMC_{l_r g_r}.
\end{aligned}$$
It can be shown that the $\IBLInfty$-relations~\eqref{Eq:IBLInfRel} and the Maurer-Cartan equation~\eqref{Eq:MaurerCartanEquation} are equivalent to  
\begin{equation}\label{Eq:BVEquat}
 \Delta\circ \Delta = 0\quad\text{and}\quad \Delta \circ e^\PMC = 0,
\end{equation}
respectively, and that the $\BVInfty$-operator $\Delta^\PMC$ for the twisted $\IBLInfty$-structure $(\OPQ_{klg}^\PMC)$ satisfies
\begin{equation} \label{Eq:TwistBV}
\Delta^\PMC = e^{-\PMC}\circ \Delta\circ (e^\PMC \cdot). 
\end{equation}
The notation $(e^\PMC \cdot)$ means that we take the input $\cdot$ and multiply it, using the extension of $\mu$ to $\CRExt C[[\hbar]]$, with $e^\PMC$ evaluated at $1\in \Ext_0 C = \R$. These facts were shown in~\cite{Cieliebak2015} using \eqref{Eq:Mix}.\footnote{One has to check that the compositions \eqref{Eq:BVEquat} and \eqref{Eq:TwistBV} are well-defined and pick a suitable completion $\CRExt C$ so that all the constructions work. The details will be discussed in~\cite{MyPhD}.}
\end{Remark}

\begin{Remark}[Weak $\IBLInfty$-algebras]\label{Rem:Weak}
A possible generalization of the $\IBLInfty$-theory above is to allow $k=0$ and $l=0$, so that $\Ext C$ must be used instead of $\RExt C$ in Remark~\ref{Rem:BVForm}. Such structures would be called \emph{weak $\IBLInfty$-algebras}.
\end{Remark}

In our application in string topology, a canonical $\dIBL$-algebra $\dIBL(C)$ with a natural Maurer-Cartan element $\PMC$ are given, and we want to study $\dIBL^\PMC(C)$; in particular, we are interested in $\HIBL^\PMC(C)$, $\IBL(\HIBL^\PMC(C))$ and possible higher operations on $\HIBL^\PMC(C)$ induced by $\OPQ_{klg}^\PMC$ (these are not chain maps in general). The following proposition summarizes some observations in this situation:

\begin{Proposition}[Twist of a $\dIBL$-algebra]\label{Prop:dIBL}
Let $\dIBL(C) = (C,\OPQ_{110},\OPQ_{210},\OPQ_{120})$ be a $\dIBL$-algebra, and let $\PMC = (\PMC_{lg})$ be a Maurer-Cartan element. The Maurer-Cartan equation~\eqref{Eq:MaurerCartanEquation} reduces to the following:
$$ 
\begin{multlined}[b] 0 = \OPQ_{110}\circ_1 \PMC_{lg} + \OPQ_{120} \circ_1 \PMC_{l-1,g} +  \OPQ_{210}\circ_2 \PMC_{l+1,g-1} \\[\jot]+  \frac{1}{2}\sum_{\substack{l_1, l_2\ge 1 \\ g_1, g_2 \ge 0 \\ l_1 + l_2 = l + 1 \\ g_1 + g_2 = g}} \OPQ_{210}\circ_{1,1}(\PMC_{l_1 g_1}, \PMC_{l_2 g_2}) \end{multlined}\quad \forall l\ge1 , g\ge 0.
$$
In particular, the ``lowest'' equation is given by\footnote{In \cite[Definition 2.4.]{Cieliebak2015}, they define a partial ordering on the signatures $(k,l,g)$.}
\begin{equation} \label{Eq:MCEq}
(l,g) = (1,0): \qquad \OPQ_{110}(\PMC_{10}) + \frac{1}{2}\OPQ_{210}(\PMC_{10}, \PMC_{10}) = 0.
\end{equation}
This can be visualized as
{\begingroup \def\dist{0.25} 
  \def\rad{0.5} 
  \def\ecc{0.1} 
  \def\hght{1} 
  \def\dif{1.5} 
  \def\radO{\rad} 
  \def\eccO{\ecc} 
  \def\hghtO{2*\hght+\dist} 
  \def\difO{\dif} 
  \def\gencanc{0.05} 
  \def\genecc{20} 
  \def\genrad{0.45} 
$$0 =\quad \vcenterline{
\begin{tikzpicture}
  \coordinate (P8) at (0,0);
  \coordinate (P9) at ($(P8)+(0,\hghtO)$);
  
  \coordinate (P10) at ($(P8)+(0,\hght)$);
  \coordinate (P11) at ($(P10)+(0,\dist)$);
  \draw (P10) arc (180:0:{\radO} and {\eccO});

  \draw (P10) arc (180:360:{\radO} and {\eccO});

  \draw (P8) arc (180:360:{\radO} and {\eccO});
  \draw[dashed] (P8) arc (180:0:{\radO} and {\eccO});
  \draw (P8) -- (P10);
  \draw ($(P8)+(2*\radO,0)$) -- ($(P10)+(2*\radO,0)$);
  
  \node at ($(P8)+(\radO,0.21*\hghtO-.7*\dist)$) {$\OPQ_{110}$};
  
   \node at ($(P11)+(\rad,0.5*\hght)$) {$\PMC_{10}$};
 

 \draw (P11) arc (180:360:{\rad} and {\ecc});
 \draw[dashed] (P11) arc (180:0:{\rad} and {\ecc});
 
 \draw (P11) to[out=90,in=180] ($(P11)+(\rad,\hght)$) to[out=0,in=90] ($(P11)+(2*\rad,0)$);
  
\end{tikzpicture}}\; + \frac{1}{2} \quad \vcenterline{
\begin{tikzpicture}
  \coordinate (P1) at (0,0);
  \coordinate (P2) at (-0.5*\dif,\hght);
  \coordinate (P3) at (0.5*\dif,\hght);
  \coordinate (P4) at ($(P2)+(0,\dist)$);
  \coordinate (P8) at ($(P3)+(0,\dist)$);
  \coordinate (P9) at ($(P8)+(0,\hght)$);

  
  \draw (P1) arc (180:360:{\rad} and {\ecc});
  \draw[dashed] (P1) arc (180:0:{\rad} and {\ecc});
  
   \draw (P3) arc (180:360:{\rad} and {\ecc});
  \draw (P3) arc (180:0:{\rad} and {\ecc});
  
  \draw (P2) arc (180:360:{\rad} and {\ecc});
  \draw (P2) arc (180:0:{\rad} and {\ecc});
  
 \draw (P2) to[out=270,in=90] (P1);
 \draw ($(P3)+(2*\rad,0)$) to[out=270,in=90] ($(P1)+(2*\rad,0)$);
 \draw ($(P2)+(2*\rad,0)$) to[out=270,in=270] (P3); 
 

 \draw (P4) arc (180:360:{\rad} and {\ecc});
 \draw[dashed] (P4) arc (180:0:{\rad} and {\ecc});
 
 \draw (P4) to[out=90,in=180] ($(P4)+(\rad,\hght)$) to[out=0,in=90] ($(P4)+(2*\rad,0)$);

 
 \node at ($(P1)+(\rad,0.5*\hght)$) {$\OPQ_{210}$};
 \node at ($(P4)+(\rad,0.5*\hght)$) {$\PMC_{10}$};
 \node at ($(P8)+(\rad,0.5*\hght)$) {$\PMC_{10}$};
 

 \draw (P8) arc (180:360:{\rad} and {\ecc});
 \draw[dashed] (P8) arc (180:0:{\rad} and {\ecc});
 
 \draw (P8) to[out=90,in=180] ($(P8)+(\rad,\hght)$) to[out=0,in=90] ($(P8)+(2*\rad,0)$);
  
\end{tikzpicture}}. $$
\endgroup}

The twisted $\IBLInfty$-algebra $\dIBL^\PMC(C)$ consists of the operations $\OPQ_{110}^\PMC$, $\OPQ_{210}^\PMC$ and~$\OPQ_{120}^\PMC$, which we call the \emph{basic operations}, and of the operations $\OPQ_{1lg}^\PMC$ for $(l,g)\in \N \times \N_0 \backslash \{(1,0),(2,0)\}$, which we call the \emph{higher operations}. These operations are given by 
$$ \begin{aligned}
\OPQ_{110}^\PMC &= \OPQ_{110} + \OPQ_{210}\circ_1 \PMC_{10},\\
\OPQ_{210}^\PMC &= \OPQ_{210}, \\
\OPQ_{120}^\PMC & = \OPQ_{120} + \OPQ_{210}\circ_1 \PMC_{20},\\
\OPQ_{1lg}^\PMC & = \OPQ_{210}\circ_1 \PMC_{lg}.
\end{aligned}$$
This can be visualized as
{ \begingroup \allowdisplaybreaks
\def\dist{0.25} 
  \def\rad{0.5} 
  \def\ecc{0.1} 
  \def\hght{1} 
  \def\dif{1.5} 
  \def\radO{\rad} 
  \def\eccO{\ecc} 
  \def\hghtO{2*\hght+\dist} 
  \def\difO{\dif} 
  \def\gencanc{0.05} 
  \def\genecc{20} 
  \def\genrad{0.45} 
\begin{align*}
\OPQ_{110}^\PMC & =\quad\vcenterline{
\begin{tikzpicture}
  \coordinate (P8) at (0,0);
  \coordinate (P9) at ($(P8)+(0,\hghtO)$);
  
  \coordinate (P10) at ($(P8)+(0,\hght)$);
  \coordinate (P11) at ($(P10)+(0,\dist)$);
  \draw[dashed] (P11) arc (180:0:{\radO} and {\eccO});
  \draw (P10) arc (180:0:{\radO} and {\eccO});

  \draw (P10) arc (180:360:{\radO} and {\eccO});
  \draw (P11) arc (180:360:{\radO} and {\eccO});

  \draw (P9) arc (180:360:{\radO} and {\eccO});
  \draw (P9) arc (180:0:{\radO} and {\eccO});  
  \draw (P8) arc (180:360:{\radO} and {\eccO});
  \draw[dashed] (P8) arc (180:0:{\radO} and {\eccO});
  \draw (P8) -- (P10);
  \draw (P9) -- (P11);
  \draw ($(P8)+(2*\radO,0)$) -- ($(P10)+(2*\radO,0)$);
  \draw ($(P9)+(2*\radO,0)$) -- ($(P11)+(2*\radO,0)$);
  
  \node at ($(P8)+(\radO,0.21*\hghtO-.7*\dist)$) {$\OPQ_{110}$};
\end{tikzpicture}}\quad +\quad \vcenterline{
\begin{tikzpicture}
  \coordinate (P1) at (0,0);
  \coordinate (P2) at (-0.5*\dif,\hght);
  \coordinate (P3) at (0.5*\dif,\hght);
  \coordinate (P4) at ($(P2)+(0,\dist)$);
  \coordinate (P8) at ($(P3)+(0,\dist)$);
  \coordinate (P9) at ($(P8)+(0,\hght)$);

  
  \draw (P1) arc (180:360:{\rad} and {\ecc});
  \draw[dashed] (P1) arc (180:0:{\rad} and {\ecc});
  
   \draw (P3) arc (180:360:{\rad} and {\ecc});
  \draw (P3) arc (180:0:{\rad} and {\ecc});
  
  \draw (P2) arc (180:360:{\rad} and {\ecc});
  \draw (P2) arc (180:0:{\rad} and {\ecc});
  
 \draw (P2) to[out=270,in=90] (P1);
 \draw ($(P3)+(2*\rad,0)$) to[out=270,in=90] ($(P1)+(2*\rad,0)$);
 \draw ($(P2)+(2*\rad,0)$) to[out=270,in=270] (P3); 
 

 \draw (P4) arc (180:360:{\rad} and {\ecc});
 \draw[dashed] (P4) arc (180:0:{\rad} and {\ecc});
 
 \draw (P4) to[out=90,in=180] ($(P4)+(\rad,\hght)$) to[out=0,in=90] ($(P4)+(2*\rad,0)$);

 
 \node at ($(P1)+(\rad,0.5*\hght)$) {$\OPQ_{210}$};
 \node at ($(P4)+(\rad,0.5*\hght)$) {$\PMC_{10}$};


 \draw (P9) arc (180:360:{\rad} and {\ecc});
 \draw (P9) arc (180:0:{\rad} and {\ecc});  
 \draw (P8) arc (180:360:{\rad} and {\ecc});
 \draw[dashed] (P8) arc (180:0:{\rad} and {\ecc});
 \draw (P8) -- (P9);
 \draw ($(P8)+(2*\rad,0)$) -- ($(P9)+(2*\rad,0)$);
  
\end{tikzpicture}}, \\[1ex] 
\OPQ_{210}^\PMC &=\quad \vcenterline{
\begin{tikzpicture}
  \coordinate (P1) at (0,0);
  \coordinate (P2) at (-0.5*\dif,\hght);
  \coordinate (P3) at (0.5*\dif,\hght);
  \coordinate (P4) at ($(P2)+(0,\dist)$);
  \coordinate (P8) at ($(P3)+(0,\dist)$);
  \coordinate (P9) at ($(P8)+(0,\hght)$);

  
  \draw (P1) arc (180:360:{\rad} and {\ecc});
  \draw[dashed] (P1) arc (180:0:{\rad} and {\ecc});
  
   \draw (P3) arc (180:360:{\rad} and {\ecc});
  \draw (P3) arc (180:0:{\rad} and {\ecc});
  
  \draw (P2) arc (180:360:{\rad} and {\ecc});
  \draw (P2) arc (180:0:{\rad} and {\ecc});
  
 \draw (P2) to[out=270,in=90] (P1);
 \draw ($(P3)+(2*\rad,0)$) to[out=270,in=90] ($(P1)+(2*\rad,0)$);
 \draw ($(P2)+(2*\rad,0)$) to[out=270,in=270] (P3); 
 

 
 \node at ($(P1)+(\rad,0.5*\hght)$) {$\OPQ_{210}$};


 \draw (P9) arc (180:360:{\rad} and {\ecc});
 \draw (P9) arc (180:0:{\rad} and {\ecc});  
 \draw (P8) arc (180:360:{\rad} and {\ecc});
 \draw[dashed] (P8) arc (180:0:{\rad} and {\ecc});
 \draw (P8) -- (P9);
 \draw ($(P8)+(2*\rad,0)$) -- ($(P9)+(2*\rad,0)$);

  \draw (P4) arc (180:360:{\rad} and {\ecc});
 \draw[dashed] (P4) arc (180:0:{\rad} and {\ecc});  
 \coordinate (P12) at ($(P4)+(0,\hght)$); 
 \draw (P12) arc (180:360:{\rad} and {\ecc});
 \draw (P12) arc (180:0:{\rad} and {\ecc});
 \draw (P4) -- (P12);
 \draw ($(P4)+(2*\rad,0)$) -- ($(P12)+(2*\rad,0)$);
\end{tikzpicture}}, \\[1ex] 
\OPQ_{120}^\PMC &=\quad \vcenterline{
\begin{tikzpicture}

  \coordinate (P1) at (0,\hght);
  \coordinate (P2) at (-0.5*\difO,0);
  \coordinate (P3) at (0.5*\difO,0);
  \coordinate (P4) at ($(P1)+(0,\dist)$);
  \coordinate (P5) at ($(P4)+(0,\hght)$);
 
  \draw (P5) arc (180:360:{\radO} and {\eccO});
  \draw (P5) arc (180:0:{\radO} and {\eccO});
  \draw (P4) arc (180:360:{\radO} and {\eccO});
  \draw[dashed] (P4) arc (180:0:{\radO} and {\eccO});
  \draw (P4) -- (P5);
  \draw ($(P4)+(2*\rad0,0)$) -- ($(P5)+(2*\rad0,0)$);
  
  \draw (P1) arc (180:360:{\radO} and {\eccO});
  \draw (P1) arc (180:0:{\radO} and {\eccO});
  
   \draw (P3) arc (180:360:{\radO} and {\eccO});
  \draw[dashed] (P3) arc (180:0:{\radO} and {\eccO});
  
  \draw (P2) arc (180:360:{\radO} and {\eccO});
  \draw[dashed] (P2) arc (180:0:{\radO} and {\eccO});
  
 \draw (P2) to[out=90,in=270] (P1);
 \draw ($(P3)+(2*\radO,0)$) to[out=90,in=270] ($(P1)+(2*\radO,0)$);
 \draw ($(P2)+(2*\radO,0)$) to[out=90,in=90] (P3); 
 
 
 \node at ($0.5*(P2)+(\radO,0)+0.5*(P3)+(0,0.21*\hghtO-.7*\dist)$) {$\OPQ_{120}$};

\end{tikzpicture}}+\quad\vcenterline{
\begin{tikzpicture}
  \coordinate (P1) at (0,0);
  \coordinate (P2) at (-0.5*\dif,\hght);
  \coordinate (P3) at (0.5*\dif,\hght);
  \coordinate (P4) at ($(P2)+(0,\dist)$);
  \coordinate (P5) at ($(P4)+(-\dif,0)$);
  \coordinate (P6) at ($(P5)+(0,-\dist)$);
  \coordinate (P7) at ($(P6)+(0,-\hght)$);
  \coordinate (P8) at ($(P3)+(0,\dist)$);
  \coordinate (P9) at ($(P8)+(0,\hght)$);

  
  \draw (P1) arc (180:360:{\rad} and {\ecc});
  \draw[dashed] (P1) arc (180:0:{\rad} and {\ecc});
  
   \draw (P3) arc (180:360:{\rad} and {\ecc});
  \draw (P3) arc (180:0:{\rad} and {\ecc});
  
  \draw (P2) arc (180:360:{\rad} and {\ecc});
  \draw (P2) arc (180:0:{\rad} and {\ecc});
  
 \draw (P2) to[out=270,in=90] (P1);
 \draw ($(P3)+(2*\rad,0)$) to[out=270,in=90] ($(P1)+(2*\rad,0)$);
 \draw ($(P2)+(2*\rad,0)$) to[out=270,in=270] (P3); 
 

 \draw (P4) arc (180:360:{\rad} and {\ecc});
 \draw[dashed] (P4) arc (180:0:{\rad} and {\ecc});
 
 \draw (P5) arc (180:360:{\rad} and {\ecc});
 \draw[dashed] (P5) arc (180:0:{\rad} and {\ecc});
 
 \draw ($(P5)+(2*\rad,0)$) to[out=90,in=90] (P4);
 
 \draw (P5) to[out=90,in=180] ($0.5*(P4)+(\rad,0)+0.5*(P5)+(0,\hght)$) to[out=0,in=90] ($(P4)+(2*\rad,0)$);

 
 \node at ($(P1)+(\rad,0.5*\hght)$) {$\OPQ_{210}$};
 \node at ($0.5*(P4)+(\rad,0)+0.5*(P5)+(0,0.5*\hght)$) {$\PMC_{20}$};


 \draw (P6) arc (180:360:{\rad} and {\ecc});
 \draw (P6) arc (180:0:{\rad} and {\ecc});  
 \draw (P7) arc (180:360:{\rad} and {\ecc});
 \draw[dashed] (P7) arc (180:0:{\rad} and {\ecc});
 \draw (P6) -- (P7);
 \draw ($(P6)+(2*\rad,0)$) -- ($(P7)+(2*\rad,0)$);

 \draw (P9) arc (180:360:{\rad} and {\ecc});
 \draw (P9) arc (180:0:{\rad} and {\ecc});  
 \draw (P8) arc (180:360:{\rad} and {\ecc});
 \draw[dashed] (P8) arc (180:0:{\rad} and {\ecc});
 \draw (P8) -- (P9);
 \draw ($(P8)+(2*\rad,0)$) -- ($(P9)+(2*\rad,0)$);
  
\end{tikzpicture}}, \\[1ex]
\OPQ^{\PMC}_{1lg} & =\quad \vcenterline{
\begin{tikzpicture}
  \coordinate (P1) at (0,0);
  \coordinate (P2) at (-0.5*\dif,\hght);
  \coordinate (P3) at (0.5*\dif,\hght);
  \coordinate (P4) at ($(P2)+(0,\dist)$);
  \coordinate (P5) at ($(P4)+(-\dif,0)$);
  \coordinate (P6) at ($(P5)+(0,-\dist)$);
  \coordinate (P7) at ($(P6)+(0,-\hght)$);
  \coordinate (P8) at ($(P3)+(0,\dist)$);
  \coordinate (P9) at ($(P8)+(0,\hght)$);
  \coordinate (P51) at ($(P5)+(-\dif,0)$);
  \coordinate (P52) at ($(P51)+(-\dif,0)$);
  \coordinate (P53) at ($(P52)+(-\dif,0)$);
  \coordinate (P62) at ($(P53)+(0,-\dist)$);
  \coordinate (P72) at ($(P62)+(0,-\hght)$);
  \coordinate (PG1) at ($(P51)+(1.6*\rad,0.5*\hght)$);
  \coordinate (PG2) at ($(P53)+(2*\rad,0.5*\hght)$);

  
  \draw (P1) arc (180:360:{\rad} and {\ecc});
  \draw[dashed] (P1) arc (180:0:{\rad} and {\ecc});
  
   \draw (P3) arc (180:360:{\rad} and {\ecc});
  \draw (P3) arc (180:0:{\rad} and {\ecc});
  
  \draw (P2) arc (180:360:{\rad} and {\ecc});
  \draw (P2) arc (180:0:{\rad} and {\ecc});
  
 \draw (P2) to[out=270,in=90] (P1);
 \draw ($(P3)+(2*\rad,0)$) to[out=270,in=90] ($(P1)+(2*\rad,0)$);
 \draw ($(P2)+(2*\rad,0)$) to[out=270,in=270] (P3); 
 

 \draw (P4) arc (180:360:{\rad} and {\ecc});
 \draw[dashed] (P4) arc (180:0:{\rad} and {\ecc});
 
 \draw (P5) arc (180:360:{\rad} and {\ecc});
 \draw[dashed] (P5) arc (180:0:{\rad} and {\ecc});
 
 \draw (P53) arc (180:360:{\rad} and {\ecc});
 \draw[dashed] (P53) arc (180:0:{\rad} and {\ecc});
 
 \draw ($(P51)+(2*\rad,0)$) to[out=90,in=90] (P5);
 \draw ($(P53)+(2*\rad,0)$) to[out=90,in=90] (P52);  
 \draw ($(P5)+(2*\rad,0)$) to[out=90,in=90] (P4);
 
 \coordinate (P5m) at ($0.8*(P4)+(\rad,0)+0.2*(P53)+(0,1*\hght)$);
 \coordinate (P5mm) at ($0.2*(P4)+(\rad,0)+0.8*(P53)+(0,1*\hght)$); 
 
 \draw (P53) to[out=90,in=180] (P5mm);
 \draw (P5m) to[out=0,in=90] ($(P4)+(2*\rad,0)$);
 \draw (P5mm) to[out=0,in=180] (P5m);
 
\tikzset{decorate sep/.style 2 args=
{decorate,decoration={shape backgrounds,shape=circle,shape size=#1,shape sep=#2}}} 
 
 \draw[decorate sep={0.3mm}{2mm},fill] ($0.5*(P62)+0.5*(P72) + (\dif,0)$) to ($0.5*(P6)+0.5*(P7)+(-\dif+2*\rad,0)$);

 
 \node at ($(P1)+(\rad,0.5*\hght)$) {$\OPQ_{210}$};
 \node at ($0.5*(P4)+(\rad,0)+0.5*(P5)+(0,0.5*\hght)$) {$\PMC_{lg}$};


  \draw (P62) arc (180:360:{\rad} and {\ecc});
 \draw (P62) arc (180:0:{\rad} and {\ecc});  
 \draw (P72) arc (180:360:{\rad} and {\ecc});
 \draw[dashed] (P72) arc (180:0:{\rad} and {\ecc});
 \draw (P62) -- (P72);
 \draw ($(P62)+(2*\rad,0)$) -- ($(P72)+(2*\rad,0)$);

 \draw (P6) arc (180:360:{\rad} and {\ecc});
 \draw (P6) arc (180:0:{\rad} and {\ecc});  
 \draw (P7) arc (180:360:{\rad} and {\ecc});
 \draw[dashed] (P7) arc (180:0:{\rad} and {\ecc});
 \draw (P6) -- (P7);
 \draw ($(P6)+(2*\rad,0)$) -- ($(P7)+(2*\rad,0)$);

 \draw (P9) arc (180:360:{\rad} and {\ecc});
 \draw (P9) arc (180:0:{\rad} and {\ecc});  
 \draw (P8) arc (180:360:{\rad} and {\ecc});
 \draw[dashed] (P8) arc (180:0:{\rad} and {\ecc});
 \draw (P8) -- (P9);
 \draw ($(P8)+(2*\rad,0)$) -- ($(P9)+(2*\rad,0)$);
 

 \draw (PG1) to[out=-\genecc,in=180+\genecc] coordinate[pos=\gencanc] (PG11) coordinate[pos=1-\gencanc] (PG12) ($(PG1) + (2*\genrad,0)$) ;
 \draw (PG11) to[out=\genecc,in=180-\genecc] (PG12);
 
\draw (PG2) to[out=-\genecc,in=180+\genecc] coordinate[pos=\gencanc] (PG21) coordinate[pos=1-\gencanc] (PG22) ($(PG2) + (2*\genrad,0)$) ;
 \draw (PG21) to[out=\genecc,in=180-\genecc] (PG22);
 
 \draw[decorate sep={0.3mm}{2mm},fill] ($(PG2) + (3*\genrad,0)$) to ($(PG1)-(\genrad,0)$);

%
 
\end{tikzpicture}}.
\end{align*}
\endgroup}
The $\IBLInfty$-relations satisfied by $(\OPQ_{klg}^\PMC)$ read for all $l\ge 1$, $g\ge 0$ as follows:
\begin{equation}\label{Eq:IBLInftydIBL}
\begin{aligned}
(3,1,0):\quad 0& = \OPQ_{210}^\PMC \circ_1 \OPQ_{210}^\PMC, \\[\jot]
(2,l,g):\quad 0&=\OPQ^\PMC_{1lg}\circ_1 \OPQ^\PMC_{210} + \OPQ^\PMC_{210}\circ_1\OPQ^\PMC_{1lg}, \\[\jot]
(1,l,g):\quad 0&= \begin{multlined}[t] \sum_{\substack{l_1, l_2 \ge 1 \\ g_1, g_2 \ge 0 \\ l_1 + l_2 = l+1 \\ g_1 + g_2 = g}} \OPQ^\PMC_{1l_1 g_1}\circ_1 \OPQ^\PMC_{1 l_2 g_2}+\OPQ^\PMC_{210} \circ_2 \OPQ^\PMC_{1, l+1, g-1}.\end{multlined}
\end{aligned}
\end{equation}
We call the relations for $(k,l,g) = (1,1,0)$, $(2,1,0)$, $(1,2,0)$, $(3,1,0)$, $(1,3,0)$, $(2,2,0)$, $(1,1,1)$ \emph{basic relations} because they contain all compositions of basic operations. In the order above, they read:
$$\begin{aligned} 
 0 & =\OPQ^\PMC_{110} \circ_1 \OPQ^\PMC_{110}, && \\
0 &=\OPQ^\PMC_{110}\circ_1\OPQ^\PMC_{210} + \OPQ^\PMC_{210}\circ_1 \OPQ^\PMC_{110}, && \\
0 &= \OPQ^\PMC_{110}\circ_1 \OPQ^\PMC_{120} + \OPQ^\PMC_{120}\circ_1 \OPQ^\PMC_{110}, && \\
0 &= \OPQ^\PMC_{210} \circ_1 \OPQ^\PMC_{210}, && \leftarrow\text{Jacobi identity} \\
0 &=\OPQ^\PMC_{120} \circ_1 \OPQ^\PMC_{120} + \OPQ^\PMC_{110}\circ_1 \OPQ^\PMC_{130} + \OPQ^\PMC_{130}\circ_1 \OPQ^\PMC_{110}, && \leftarrow\text{co-Jacobi id. up to htpy.} \\
0 & = \OPQ^\PMC_{120}\circ_1 \OPQ^\PMC_{210} + \OPQ^\PMC_{210}\circ_1 \OPQ^\PMC_{120}, && \leftarrow\text{Drinfeld identity} \\
0 & = \OPQ^\PMC_{210}\circ_2 \OPQ^\PMC_{120} + \OPQ^\PMC_{111}\circ_1 \OPQ^\PMC_{110} + \OPQ^\PMC_{110}\circ_1 \OPQ^\PMC_{111}. && \leftarrow\text{Involutivity up to htpy.}
\end{aligned}$$
The last four equations can be visualized as
{ \begingroup \allowdisplaybreaks
\def\dist{0.25} 
  \def\rad{0.4} 
  \def\ecc{0.1} 
  \def\hght{1} 
  \def\dif{1.1} 
  \def\difbig{1.5*\dif} 
  \def\radO{\rad} 
  \def\eccO{\ecc} 
  \def\hghtO{2*\hght+\dist} 
  \def\difO{\dif} 
  \def\gencanc{0.05} 
  \def\genecc{20} 
  \def\genrad{0.3} 
\begin{align*}
0 & =\quad\vcenterline{
\begin{tikzpicture}
  
  \coordinate (P1) at (0,0);
  \coordinate (P2) at (-0.5*\difbig,\hght);
  \coordinate (P3) at ($(0.5*\difbig,\hght)$);

  \coordinate (P4) at ($(P2)+(0,\dist)$);
  \coordinate (P5) at ($(P4)+(-0.5*\dif,\hght)$);
  \coordinate (P6) at ($(P4)+(0.5*\dif,\hght)$);

  \coordinate (P7) at ($(P6)+(0,-\hght)$);
  \coordinate (P8) at ($(P3)+(0,\dist)$);
  \coordinate (P9) at ($(P8)+(0,\hght)$);

  
  \draw (P1) arc (180:360:{\rad} and {\ecc});
  \draw[dashed] (P1) arc (180:0:{\rad} and {\ecc});
  
   \draw (P3) arc (180:360:{\rad} and {\ecc});
  \draw (P3) arc (180:0:{\rad} and {\ecc});
  
  \draw (P2) arc (180:360:{\rad} and {\ecc});
  \draw (P2) arc (180:0:{\rad} and {\ecc});
  
 \draw (P2) to[out=270,in=90] (P1);
 \draw ($(P3)+(2*\rad,0)$) to[out=270,in=90] ($(P1)+(2*\rad,0)$);
 \draw ($(P2)+(2*\rad,0)$) to[out=270,in=270] (P3); 
 

  \draw (P4) arc (180:360:{\rad} and {\ecc});
  \draw[dashed] (P4) arc (180:0:{\rad} and {\ecc});
  
   \draw (P6) arc (180:360:{\rad} and {\ecc});
  \draw (P6) arc (180:0:{\rad} and {\ecc});
  
  \draw (P5) arc (180:360:{\rad} and {\ecc});
  \draw (P5) arc (180:0:{\rad} and {\ecc});
  
 \draw (P5) to[out=270,in=90] (P4);
 \draw ($(P6)+(2*\rad,0)$) to[out=270,in=90] ($(P4)+(2*\rad,0)$);
 \draw ($(P5)+(2*\rad,0)$) to[out=270,in=270] (P6); 

%
%
%

 
 \node at ($(P1)+(\rad,0.42*\hght)$) {$\OPQ_{210}^\PMC$};
 \node at ($(P4)+(\rad,0.5*\hght)$) {$\OPQ_{210}^\PMC$};


%
 
 \draw (P9) arc (180:360:{\rad} and {\ecc});
 \draw (P9) arc (180:0:{\rad} and {\ecc});  
 \draw (P8) arc (180:360:{\rad} and {\ecc});
 \draw[dashed] (P8) arc (180:0:{\rad} and {\ecc});
 \draw (P8) -- (P9);
 \draw ($(P8)+(2*\rad,0)$) -- ($(P9)+(2*\rad,0)$);
  
\end{tikzpicture}}, \\[1ex] 
0&=\quad \vcenterline{
\begin{tikzpicture}
  \coordinate (P1) at (0,\hght);
  \coordinate (P2) at (-0.5*\difO,0);
  \coordinate (P3) at (0.5*\difO,0);
  \coordinate (P4) at ($(P1) + (-0.5*\difbig,\dist+\hght)$);
  \coordinate (P5) at ($(P4)+(-0.5*\difbig,-\hght)$);
  \coordinate (P6) at ($(P4) + (0.5*\difbig,-\hght)$);
 
  
  \draw (P1) arc (180:360:{\radO} and {\eccO});
  \draw (P1) arc (180:0:{\radO} and {\eccO});
  
   \draw (P3) arc (180:360:{\radO} and {\eccO});
  \draw[dashed] (P3) arc (180:0:{\radO} and {\eccO});
  
  \draw (P2) arc (180:360:{\radO} and {\eccO});
  \draw[dashed] (P2) arc (180:0:{\radO} and {\eccO});
  
 \draw (P2) to[out=90,in=270] (P1);
 \draw ($(P3)+(2*\radO,0)$) to[out=90,in=270] ($(P1)+(2*\radO,0)$);
 \draw ($(P2)+(2*\radO,0)$) to[out=90,in=90] (P3);

  \draw (P4) arc (180:360:{\radO} and {\eccO});
  \draw (P4) arc (180:0:{\radO} and {\eccO});
  
   \draw (P6) arc (180:360:{\radO} and {\eccO});
  \draw[dashed] (P6) arc (180:0:{\radO} and {\eccO});
  
  \draw (P5) arc (180:360:{\radO} and {\eccO});
  \draw[dashed] (P5) arc (180:0:{\radO} and {\eccO});
  
 \draw (P5) to[out=90,in=270] (P4);
 \draw ($(P6)+(2*\radO,0)$) to[out=90,in=270] ($(P4)+(2*\radO,0)$);
 \draw ($(P5)+(2*\radO,0)$) to[out=90,in=90] (P6); 
 
 
 \node at ($0.5*(P2)+(\radO,0)+0.5*(P3)+(0,0.21*\hghtO-.7*\dist)$) {$\OPQ_{120}^\PMC$};
 
  \node at ($0.5*(P5)+(\radO,0)+0.5*(P6)+(0,0.21*\hghtO-.7*\dist)$) {$\OPQ_{120}^\PMC$};

\coordinate (P7) at ($(P5) + (0,-\dist)$);
\coordinate (P8) at ($(P7) + (0,-\hght)$);
\draw (P7)--(P8);
\draw ($(P7)+(2*\rad,0)$)--($(P8)+(2*\rad,0)$);

  \draw (P7) arc (180:360:{\radO} and {\eccO});
  \draw (P7) arc (180:0:{\radO} and {\eccO});
  
   \draw (P8) arc (180:360:{\radO} and {\eccO});
  \draw[dashed] (P8) arc (180:0:{\radO} and {\eccO});

\end{tikzpicture}}+
\vcenterline{
\begin{tikzpicture}
  \coordinate (P1) at (0,0);
  \coordinate (P2) at ($(P1)+(\dif,0)$);
  \coordinate (P3) at ($(P1)+(2*\dif,0)$);
  \coordinate (P4) at ($(P1) + (\dif,\hght)$);

  
  \draw (P1) arc (180:360:{\rad} and {\ecc});
  \draw[dashed] (P1) arc (180:0:{\rad} and {\ecc});
  
   \draw (P2) arc (180:360:{\rad} and {\ecc});
  \draw[dashed] (P2) arc (180:0:{\rad} and {\ecc});
  
  \draw (P3) arc (180:360:{\rad} and {\ecc});
  \draw[dashed] (P3) arc (180:0:{\rad} and {\ecc});
  
  \draw (P4) arc (180:360:{\rad} and {\ecc});
  \draw (P4) arc (180:0:{\rad} and {\ecc});
  
 \draw ($(P1) + (2*\rad,0)$) to[out=90,in=90] (P2);
 \draw ($(P2) + (2*\rad,0)$) to[out=90,in=90] (P3); 
 \draw (P1) to[out=90,in=-90] (P4);
 \draw ($(P3)+(2*\rad,0)$) to[out=90,in=-90] ($(P4)+(2*\rad,0)$);

 \node at ($(P1)+(\dif + \rad,0.5*\hght)$) {$\OPQ_{130}^\PMC$};

\coordinate (P5) at ($(P4)+(0,\dist)$);
\coordinate (P6) at ($(P5)+(0,\hght)$);
\draw (P5) arc (180:360:{\rad} and {\ecc});
\draw[dashed] (P5) arc (180:0:{\rad} and {\ecc});
\draw (P6) arc (180:360:{\rad} and {\ecc});
\draw (P6) arc (180:0:{\rad} and {\ecc}); 
\draw (P5) -- (P6);
\draw ($(P5) + (2*\rad,0)$) -- ($(P6) + (2*\rad,0)$);

 \node at ($(P5)+(\rad,0.5*\hght)$) {$\OPQ_{110}^\PMC$};
  
\end{tikzpicture}}
+\quad\vcenterline{
\begin{tikzpicture}
  \coordinate (P1) at ($(0,\hght+\dist)$);
  \coordinate (P2) at ($(P1)+(\dif,0)$);
  \coordinate (P3) at ($(P1)+(2*\dif,0)$);
  \coordinate (P4) at ($(P1) + (\dif,\hght)$);

  
  \draw (P1) arc (180:360:{\rad} and {\ecc});
  \draw[dashed] (P1) arc (180:0:{\rad} and {\ecc});
  
   \draw (P2) arc (180:360:{\rad} and {\ecc});
  \draw[dashed] (P2) arc (180:0:{\rad} and {\ecc});
  
  \draw (P3) arc (180:360:{\rad} and {\ecc});
  \draw[dashed] (P3) arc (180:0:{\rad} and {\ecc});
  
  \draw (P4) arc (180:360:{\rad} and {\ecc});
  \draw (P4) arc (180:0:{\rad} and {\ecc});
  
 \draw ($(P1) + (2*\rad,0)$) to[out=90,in=90] (P2);
 \draw ($(P2) + (2*\rad,0)$) to[out=90,in=90] (P3); 
 \draw (P1) to[out=90,in=-90] (P4);
 \draw ($(P3)+(2*\rad,0)$) to[out=90,in=-90] ($(P4)+(2*\rad,0)$);

 \node at ($(P1)+(\dif + \rad,0.5*\hght)$) {$\OPQ_{130}^\PMC$};

\coordinate (P5) at (\dif,0);
\coordinate (P6) at ($(P5)+(0,\hght)$);
\draw (P5) arc (180:360:{\rad} and {\ecc});
\draw[dashed] (P5) arc (180:0:{\rad} and {\ecc});
\draw (P6) arc (180:360:{\rad} and {\ecc});
\draw (P6) arc (180:0:{\rad} and {\ecc}); 
\draw (P5) -- (P6);
\draw ($(P5) + (2*\rad,0)$) -- ($(P6) + (2*\rad,0)$);

\coordinate (P6) at ($(P1) + (0,-\dist)$);
\coordinate (P7) at ($(P6) + (0,-\hght)$);
\draw (P6) arc (180:360:{\rad} and {\ecc});
\draw (P6) arc (180:0:{\rad} and {\ecc});
\draw (P7) arc (180:360:{\rad} and {\ecc});
\draw[dashed] (P7) arc (180:0:{\rad} and {\ecc}); 
\draw (P6) -- (P7);
\draw ($(P6) + (2*\rad,0)$) -- ($(P7) + (2*\rad,0)$);

\coordinate (P8) at ($(P3) + (0,-\dist)$);
\coordinate (P9) at ($(P8) + (0,-\hght)$);
\draw (P8) arc (180:360:{\rad} and {\ecc});
\draw (P8) arc (180:0:{\rad} and {\ecc});
\draw (P9) arc (180:360:{\rad} and {\ecc});
\draw[dashed] (P9) arc (180:0:{\rad} and {\ecc}); 
\draw (P8) -- (P9);
\draw ($(P8) + (2*\rad,0)$) -- ($(P9) + (2*\rad,0)$);

\node at ($(P7)+(\rad,0.5*\hght)$) {$\OPQ_{110}^\PMC$};

\end{tikzpicture}}, \\[1ex] 
0&=\quad \vcenterline{
\begin{tikzpicture}
  
  \coordinate (P1) at (0,0);
  \coordinate (P2) at ($(P1) + (-0.5*\dif,\hght)$);  
  \coordinate (P3) at ($(P1) + (0.5*\dif,\hght)$);
  
  \coordinate (P4) at (0,-\dist);
  \coordinate (P5) at ($(P4) + (-0.5*\dif,-\hght)$);  
  \coordinate (P6) at ($(P4) + (0.5*\dif,-\hght)$);
  
  \draw (P1) to[out=90, in=-90] (P2);
  \draw ($(P1)+(2*\rad,0)$) to[out=90, in=-90] ($(P3)+(2*\rad,0)$);  
  \draw ($(P2)+(2*\rad,0)$) to[out=-90,in=-90] (P3);
 
   \draw (P4) to[out=-90, in=90] (P5);
  \draw ($(P4)+(2*\rad,0)$) to[out=-90, in=90] ($(P6)+(2*\rad,0)$);  
  \draw ($(P5)+(2*\rad,0)$) to[out=90,in=90] (P6); 
  
  \draw (P1) arc (180:360:{\rad} and {\ecc});
  \draw[dashed] (P1) arc (180:0:{\rad} and {\ecc});
  \draw (P5) arc (180:360:{\rad} and {\ecc});
  \draw[dashed] (P5) arc (180:0:{\rad} and {\ecc});
  \draw (P6) arc (180:360:{\rad} and {\ecc});
  \draw[dashed] (P6) arc (180:0:{\rad} and {\ecc});
 
  \draw (P2) arc (180:360:{\rad} and {\ecc});
  \draw (P2) arc (180:0:{\rad} and {\ecc});
  \draw (P3) arc (180:360:{\rad} and {\ecc});
  \draw (P3) arc (180:0:{\rad} and {\ecc});
  \draw (P4) arc (180:360:{\rad} and {\ecc});
  \draw (P4) arc (180:0:{\rad} and {\ecc}); 
 
 \node at ($(P1)+(\rad,0.5*\hght)$) {$\OPQ_{210}^\PMC$};
 \node at ($0.5*(P5)+(\rad,0)+0.5*(P6)+(0,0.5*\hght)$) {$\OPQ_{120}^\PMC$};

\end{tikzpicture}}+\quad\vcenterline{
\begin{tikzpicture}

 \coordinate (P1) at (0,0);
 \coordinate (P2) at ($(P1)+(0.5*\dif,\hght)$);
 \coordinate (P3) at ($(P1) + (\dif,0)$);
 \coordinate (P4) at ($(P3) + (0,-\dist)$);
 \coordinate (P5) at ($(P4) + (0.5*\dif,-\hght)$);
 \coordinate (P6) at ($(P4) + (\dif,0)$);
 
 \draw (P1) to[out=90,in=-90] (P2);
 \draw ($(P1)+(2*\rad,0)$) to[out=90,in=90] (P3);
 \draw ($(P2)+(2*\rad,0)$) to[out=-90,in=90] ($(P3)+(2*\rad,0)$); 
 \draw (P4) to[out=-90,in=90] (P5);
 \draw ($(P4)+(2*\rad,0)$) to[out=-90,in=-90] (P6);
 \draw ($(P5)+(2*\rad,0)$) to[out=90,in=-90] ($(P6)+(2*\rad,0)$);

  \draw (P1) arc (180:360:{\rad} and {\ecc});
  \draw[dashed] (P1) arc (180:0:{\rad} and {\ecc});
  \draw (P3) arc (180:360:{\rad} and {\ecc});
  \draw[dashed] (P3) arc (180:0:{\rad} and {\ecc});
  \draw (P5) arc (180:360:{\rad} and {\ecc});
  \draw[dashed] (P5) arc (180:0:{\rad} and {\ecc});

  \draw (P2) arc (180:360:{\rad} and {\ecc});
  \draw (P2) arc (180:0:{\rad} and {\ecc});
  \draw (P4) arc (180:360:{\rad} and {\ecc});
  \draw (P4) arc (180:0:{\rad} and {\ecc});
  \draw (P6) arc (180:360:{\rad} and {\ecc});
  \draw (P6) arc (180:0:{\rad} and {\ecc});
  
    \node at ($(P2)+(\rad,-0.5*\hght)$) {$\OPQ_{120}^\PMC$};
 \node at ($(P5)+(\rad,0.5*\hght)$) {$\OPQ_{210}^\PMC$};
 
 \coordinate (P7) at ($(P1)+(0,-\dist)$);
 \coordinate (P8) at ($(P7)+(0,-\hght)$);
 \coordinate (P9) at ($(P6)+(0,\dist)$);
 \coordinate (P10) at ($(P9)+(0,\hght)$);
 
 \draw (P7)--(P8);
 \draw ($(P7)+(2*\rad,0)$)--($(P8)+(2*\rad,0)$);
 \draw (P9)--(P10);
 \draw ($(P9)+(2*\rad,0)$)--($(P10)+(2*\rad,0)$);

\draw (P8) arc (180:360:{\rad} and {\ecc});
\draw[dashed] (P8) arc (180:0:{\rad} and {\ecc});
\draw (P9) arc (180:360:{\rad} and {\ecc});
\draw[dashed] (P9) arc (180:0:{\rad} and {\ecc});
\draw (P7) arc (180:360:{\rad} and {\ecc});
\draw (P7) arc (180:0:{\rad} and {\ecc});
\draw (P10) arc (180:360:{\rad} and {\ecc});
\draw (P10) arc (180:0:{\rad} and {\ecc}); 
 
\end{tikzpicture}},\\[1ex]
0& =\quad \vcenterline{
\begin{tikzpicture}

\coordinate (P1) at (0,0);
\coordinate (P2) at (-0.5*\dif,-\hght);
\coordinate (P3) at (0.5*\dif,-\hght);

\coordinate (P4) at ($(P2)+(0,-\dist)$);
\coordinate (P5) at ($(P3)+(0,-\dist)$);
\coordinate (P6) at ($(P4)+(0.5*\dif,-\hght)$);

\draw (P1) to[out=-90,in=90] (P2);
\draw ($(P1)+(2*\rad,0)$) to[out=-90,in=90] ($(P3)+(2*\rad,0)$);
\draw ($(P2)+(2*\rad,0)$) to[out=90,in=90] (P3);

\draw ($(P4)+(2*\rad,0)$) to[out=-90,in=-90] (P5);
\draw (P4) to[out=-90,in=90] (P6);
\draw ($(P5)+(2*\rad,0)$) to[out=-90,in=90] ($(P6)+(2*\rad,0)$);

\draw (P1) arc (180:360:{\rad} and {\ecc});
\draw (P1) arc (180:0:{\rad} and {\ecc});
\draw (P2) arc (180:360:{\rad} and {\ecc});
\draw[dashed] (P2) arc (180:0:{\rad} and {\ecc});
\draw (P3) arc (180:360:{\rad} and {\ecc});
\draw[dashed] (P3) arc (180:0:{\rad} and {\ecc});
\draw (P4) arc (180:360:{\rad} and {\ecc});
\draw (P4) arc (180:0:{\rad} and {\ecc}); 
\draw (P5) arc (180:360:{\rad} and {\ecc});
\draw (P5) arc (180:0:{\rad} and {\ecc}); 
\draw (P6) arc (180:360:{\rad} and {\ecc});
\draw[dashed] (P6) arc (180:0:{\rad} and {\ecc}); 

\node at ($(P1)+(\rad,-0.5*\hght)$) {$\OPQ_{120}^\PMC$};
\node at ($(P6)+(\rad,0.5*\hght)$) {$\OPQ_{210}^\PMC$};

\end{tikzpicture}}\quad+\quad\vcenterline{
\begin{tikzpicture}
\coordinate (P1) at (0,0);
\coordinate (P2) at ($(P1)+(0,-\hght)$);
\coordinate (P3) at ($(P2)+(0,-\dist)$);
\coordinate (P4) at ($(P3)+(0,-\hght)$);

\draw (P2) arc (180:360:{\rad} and {\ecc});
\draw[dashed] (P2) arc (180:0:{\rad} and {\ecc});
\draw (P4) arc (180:360:{\rad} and {\ecc});
\draw[dashed] (P4) arc (180:0:{\rad} and {\ecc});
\draw (P1) arc (180:360:{\rad} and {\ecc});
\draw (P1) arc (180:0:{\rad} and {\ecc});
\draw (P3) arc (180:360:{\rad} and {\ecc});
\draw (P3) arc (180:0:{\rad} and {\ecc}); 

\draw (P1)--(P2);
\draw (P3)--(P4);

\draw ($(P1)+(2*\rad,0)$)--($(P2)+(2*\rad,0)$);
\draw ($(P3)+(2*\rad,0)$)--($(P4)+(2*\rad,0)$);

\node at ($(P1)+(\rad,-0.5*\hght)$) {$\OPQ_{110}^\PMC$};
\node at ($(P3)+(\rad,-0.37*\hght)$) {$\OPQ_{111}^\PMC$};

\coordinate (PG1) at ($(P3)+(\rad-\genrad,-0.72*\hght)$);
 \draw (PG1) to[out=-\genecc,in=180+\genecc] coordinate[pos=\gencanc] (PG11) coordinate[pos=1-\gencanc] (PG12) ($(PG1) + (2*\genrad,0)$) ;
 \draw (PG11) to[out=\genecc,in=180-\genecc] (PG12);
\end{tikzpicture}}\quad+\quad\vcenterline{
\begin{tikzpicture}
\coordinate (P3) at (0,0);
\coordinate (P2) at ($(P3)+(0,-\hght)$);
\coordinate (P1) at ($(P2)+(0,-\dist)$);
\coordinate (P4) at ($(P1)+(0,-\hght)$);

\draw (P2) arc (180:360:{\rad} and {\ecc});
\draw[dashed] (P2) arc (180:0:{\rad} and {\ecc});
\draw (P4) arc (180:360:{\rad} and {\ecc});
\draw[dashed] (P4) arc (180:0:{\rad} and {\ecc});
\draw (P1) arc (180:360:{\rad} and {\ecc});
\draw (P1) arc (180:0:{\rad} and {\ecc});
\draw (P3) arc (180:360:{\rad} and {\ecc});
\draw (P3) arc (180:0:{\rad} and {\ecc}); 

\draw (P1)--(P2);
\draw (P3)--(P4);

\draw ($(P1)+(2*\rad,0)$)--($(P2)+(2*\rad,0)$);
\draw ($(P3)+(2*\rad,0)$)--($(P4)+(2*\rad,0)$);

\node at ($(P1)+(\rad,-0.5*\hght)$) {$\OPQ_{110}^\PMC$};
\node at ($(P3)+(\rad,-0.37*\hght)$) {$\OPQ_{111}^\PMC$};

\coordinate (PG1) at ($(P3)+(\rad-\genrad,-0.72*\hght)$);
 \draw (PG1) to[out=-\genecc,in=180+\genecc] coordinate[pos=\gencanc] (PG11) coordinate[pos=1-\gencanc] (PG12) ($(PG1) + (2*\genrad,0)$) ;
 \draw (PG11) to[out=\genecc,in=180-\genecc] (PG12);
\end{tikzpicture}}.
\end{align*}
\endgroup}
\end{Proposition}
\begin{proof}
The proof is clear by specializing \eqref{Eq:IBLInfRel}, \eqref{Eq:MaurerCartanEquation} and \eqref{Eq:TwistedOperations}.
\end{proof}

\begin{Remark}[Higher operations]\label{Rem:Higher}
We see from Proposition~\ref{Prop:dIBL} that if $\OPQ_{120}^\PMC \circ_1 \OPQ_{120}^\PMC = 0$ and $\OPQ_{210}^\PMC \circ_{2} \OPQ_{120}^\PMC = 0$, then $[\Bdd^\PMC, \OPQ_{130}^\PMC] = 0$ and $[\Bdd^\PMC, \OPQ_{111}^\PMC] = 0$, respectively, and hence the operations $\OPQ_{130}^\PMC: \hat{\Ext}_1\HIBL^\PMC\rightarrow \hat{\Ext}_3\HIBL^\PMC$ and $\OPQ_{111}^\PMC: \hat{\Ext}_1\HIBL^\PMC\rightarrow \hat{\Ext}_1\HIBL^\PMC$ are well-defined (provided that the assumption of Definition \ref{Def:HomIBL} holds). Likewise, the higher operation $\OPQ_{1lg}^\PMC$ defines a map $\hat{\Ext}_1\HIBL^\PMC \rightarrow \hat{\Ext}_l\HIBL^\PMC$, provided that the following equation holds:
$$ \OPQ^\PMC_{210}\circ_2 \OPQ_{1,l+1,g-1}^\PMC + \sum_{\substack{l_1, l_2 \ge 1 \\ g_1, g_2 \ge 0 \\ l_1 + l_2 = l+1 \\ g_1 + g_2 = g \\ (l_i,g_i)\neq (1,0)}} \OPQ^\PMC_{1l_1 g_1}\circ_1 \OPQ^\PMC_{1 l_2 g_2} = 0. $$
This expression is just the left-over after subtracting the commutator $[\OPQ_{1lg}^\PMC,\OPQ_{110}^\PMC] = \OPQ_{110}^\PMC \circ_1 \OPQ_{1lg}^\PMC + \OPQ_{1lg}^\PMC \circ_1 \OPQ_{110}^\PMC$ from \eqref{Eq:IBLInftydIBL}.
\end{Remark}

%
%
%

\subsection{Dual cyclic bar complex and cyclic cohomology
}
\label{Sec:Alg2}

\begin{Definition}[Bar complexes] \label{Def:BarComplex}
Let $V$ be a graded vector space. The \emph{bar- and dual bar-complex of $V$} are the weight-graded vector spaces defined by 
$$ \B V:= \RTen(V[1])\quad\text{and}\quad\DB V := (\B V)'', $$
respectively, where $\bar{T}V := \bigoplus_{k=1}^\infty V^{\otimes k}$ is the weight-reduced tensor algebra. For every $k\in \N$, let $t_k \in \Perm_k$ be the cyclic permutation $t_k : (1,\dotsc,k) \mapsto (2,\dotsc,k,1)$,
so that for all $v_1$, $\dotsc$, $v_k \in V[1]$ we have
\begin{equation*}
t_k(v_1 \otimes \dotsb \otimes v_k) = (-1)^{\Abs{v_k}(\Abs{v_1} + \dotsb + \Abs{v_{k-1}})} v_k \otimes v_1 \otimes \dotsb \otimes v_{k-1}.
\end{equation*}
We set
$$ t:= \sum_{k=1}^\infty t_k : \B V \longrightarrow \B V. $$
The \emph{cyclic bar-complex} is defined by 
$$ \BCyc V := \B V / \Im(1-t). $$
We denote the image of $v_1 \otimes \dotsb \otimes v_k \in \B V$ under the canonical projection $\pi: \B V \rightarrow \BCyc V$ by $v_1\dots v_k$. If $v_i\in V[1]$ are homogenous, then $v_1\dots v_k$ is called a \emph{generating word}; we have
\begin{equation*}
v_1 \dots v_k = (-1)^{\Abs{v_k}(\Abs{v_1}+\dotsb + \Abs{v_{k-1}})} v_k v_1 \dots v_{k-1}.
\end{equation*}
We define the section $\iota: \BCyc V \rightarrow \B V$ of $\pi$ by
$$ \iota(v_1\dots v_k) := \frac{1}{k} \sum_{i=0}^{k-1} \underbrace{t_k^i}_{\mathrlap{\displaystyle =: t_k \circ \dotsb \circ t_k\ i\text{-times}}}(v_1\otimes\dotsb \otimes v_k) $$
and use it to identify $\BCyc V$ with the subspace $\Im \iota = \Ker(1-t) \subset \B V$ consisting of cyclic symmetric tensors.

We define the \emph{dual cyclic bar-complex} by 
$$ \DBCyc V := \{ \psi\in \DB V \mid \psi \circ t = \psi \}. $$
\end{Definition}

\begin{Remark}[Non-weight-reduced bar complex]\label{Rem:NWG}
In fact, our $\DBCyc V$ is weight-reduced. The non-weight-reduced version would be $\DBCyc V \oplus \R$ with $\R$ of degree~$0$. This might play a role in the theory of  weak $\AInfty$-algebras (:=\,operation $\mu_0$ added; c.f., Definition~\ref{Def:CyclicAinfty}), and it might also be possible to consider $\IBLInfty$-algebras on non-weight-reduced cyclic cochains (c.f., Section~\ref{Sec:Alg3}). This may be discussed more in \cite{MyPhD}.
\end{Remark}

Notice that $\psi \in \DB V$ is homogenous of degree $\Abs{\psi}\in \Z$ if and only if for all homogenous $v_1$,~$\dotsc$, $v_k \in V[1]$ the following implication holds:
\begin{equation*}
\Abs{v_1} + \dotsb + \Abs{v_k} \neq \Abs{\psi}\quad\Implies\quad \psi(v_1\otimes \dotsb \otimes v_k) = 0.
\end{equation*}
This is the cohomological grading convention.

\begin{Notation}[Degree shifts of bar complexes] \label{Def:Notation}
Let $A\in \Z$. In the following, we write $\DBCyc V$, but the convention applies to all complexes from Definition\,\ref{Def:BarComplex}. We denote by $\Susp_A$ and $\SuspU$ the formal symbols of degrees 
$$ \Abs{\Susp_A} = -A \quad \text{and}\quad\Abs{\SuspU} = -1, $$
respectively. The degree shift $V \mapsto V[1]$ will be realized as  multiplication with~$\SuspU$ and the degree shift $\DBCyc V\mapsto \DBCyc V[A]$ as multiplication with~$\Susp_A$. In addition, the following notation will be used consistently:
\begin{itemize}
 \item $\tilde{v}\in V \longleftrightarrow v = \SuspU \tilde{v} \in V[1]$
 
  To clarify this, given $\tilde{v} \in V$, then~$v$ automatically means $v = \SuspU \tilde{v} \in V[1]$, and the other way round. Recall that the degree of $\tilde{v}\in V$ is denoted by~$\Deg(\tilde{v})$ or simply by $\tilde{v}$ in the exponent, e.g., $(-1)^{\tilde{v}}$.
 \item  $\psi\in\DBCyc V\longleftrightarrow \Psi = \Susp_A \psi\in \DBCyc V[A]$.
 \item A generating word of $\BCyc V$ of weight $k$ will be denoted by the symbol $w$ and written as $w= v_1 \dots v_k$, where $v_i = \SuspU \tilde{v}_i \in V[1]$. A generating word of $\Ext_k \BCyc V$ is an element $w_1 \dotsb w_k \in \Ext_k \BCyc V$ such that each~$w_i$ is a generating word of $\BCyc V$.
 \item $w\in \BCyc V\longleftrightarrow \text{\footnotesize W} = \Susp_A w \in \BCyc V[A]$.
\end{itemize}
We abbreviate
$$ \DBCyc V[A]:= (\DBCyc V)[A]. $$
In contrast to this, we would write $\DBCyc(V[A])$ for the dual cyclic bar-complex of~$V[A]$. We also identify $(\DBCyc V[A])[1] = \DBCyc V[A+1]$ in $\Ext \DBCyc V[A]$.
\end{Notation}

\begin{Definition}[Pairing of tensor powers of bar complexes]\label{Def:Pairings}
For every $A\in \Z$ and  $k\in \N$, we define the pairing as follows:
\begin{equation} \label{Eq:Pairing}
\begin{aligned}
(\DB V[A])^{\otimes k} \otimes (\B V[A])^{\otimes k} & \longrightarrow \R \\ 
(\Psi_1 \otimes \dotsb \otimes \Psi_k, \W_1 \otimes \dotsb \otimes \W_k) & \longmapsto \underbrace{\psi_1(w_1) \dots \psi(w_k)}_{\mathllap{\textstyle{(\Psi_1 \otimes \dotsb \otimes \Psi_k)(\W_1\otimes \dotsb\otimes \W_k):=}}}.
\end{aligned}
\end{equation}
This means that we evaluate elements from the left-hand side on the elements from the right-hand side in this way without any signs (see the discussion in Remark~\ref{Rem:BadConvention}). We extend the pairing by $0$ if the number of $\Psi_i$'s and the number of $\W_i$'s differ.
\end{Definition}

\begin{Remark}[Dual bar complex and dual of the bar complex] \label{Rem:Identifications}
Because the pairing~\eqref{Eq:Pairing} is non-degenerate, we can embed the space on the left into the the linear dual of the space on the right.
From Definition~\ref{Def:BarComplex} we have $\DBCyc V \subset \DB V$, and $\BCyc V$ is identified with $\Im \iota \subset \B V$. Therefore, we can restrict \eqref{Eq:Pairing} to obtain the pairing of $\DBCyc V$ and $\BCyc V$. It is easy to see that for any $\psi\in \DBCyc V$ and any generating word $v_1\dots v_k \in \BCyc V$, we have
\begin{equation*}
\psi(v_1\dots v_k) = \psi(v_1 \otimes \dotsb \otimes v_k).
\end{equation*}
The subspace of $(\BCyc V)^*$ corresponding to $\DBCyc V$ is then precisely $(\BCyc V)''$.

More generally, for every $k\in \N$, the spaces $\Ext_k \DBCyc V$ and $\Ext_k \BCyc V$ are embedded into $(\DBCyc V[1])^{\otimes k}$ and $(\BCyc V[1])^{\otimes k}$, respectively, using $\iota$ and $\pi$ from Definition~\ref{Def:SymAlgebra}. Therefore, the restriction of~\eqref{Eq:Pairing} gives the pairing of $\Ext_k \DBCyc V$ and $\Ext_k \BCyc V$. It is easy to see that for any generating word $w_1\dotsb w_k \in \Ext_k \BCyc V$ and any $\psi_1\dotsb \psi_k\in \Ext_k \DBCyc V$, we have
$$ (\psi_1\dotsb \psi_k)(w_1\dotsb w_k) = \frac{1}{k!}\sum_{\sigma\in \Perm_k} \varepsilon(\sigma,w) \psi_1(w_{\sigma_1^{-1}})\dotsc \psi_k(w_{\sigma_k^{-1}}). $$
The subspace of $(\Ext_k \BCyc V)^*$ corresponding to $\Ext_k \DBCyc V$  lies in $(\Ext_k \BCyc V)''$; it is equal to $(\Ext_k \BCyc V)''$, provided that $V$ is finite-dimensional.\footnote{The problem is that if $\dim(V) = \infty$, then $(V\otimes V)^* \neq V^* \otimes V^*$.}
\end{Remark}

The weight-graded vector spaces $\B V$ and $\BCyc V$ are canonically filtered by the filtration by weights \eqref{Eq:FiltrWeights}. Their weight-graded duals $\DB V$ and $\DBCyc V$ are filtered by the dual filtrations and the exterior powers $\Ext_k \DB V$ and $\Ext_k \DBCyc V$ by the induced filtration from Definition~\ref{Def:Filtrations}. 

\begin{Proposition}[Completed dual cyclic bar complex] \label{Prop:Compl}
Let $V$ be a graded vector space and $A\in \Z$. The filtration of $\DBCyc V$ dual to the weight-filtration of $\BCyc V$ is $\Z$-gapped, Hausdorff,  decreasing and bounded from above. Moreover, the following holds:
$$ \dim(V)<\infty\quad\Implies\quad (WG1)\ \&\ (WG2)\text{ are satisfied.} $$
The same holds for the induced filtration of $\Ext_k \DBCyc V[A]$.

In the sense of Remark~\ref{Rem:Identifications}, we have
$$ \nCDBCyc V \simeq (\BCyc V)'\quad\text{and}\quad \hat{\Ext}_k \DBCyc V[A] \subset (\Ext_k \BCyc V[A+1])', $$
where ``='' holds if $V$ is finite-dimensional.

The \emph{filtration degree} of $\Psi\in \hat{\Ext}_m \DBCyc V[A]$ satisfies
$$ \Norm{\Psi} = \min\{k\in \N_0 \mid \exists \W\in (\Ext_m \BCyc V[A])_k: \Psi(\!\W)\neq 0 \}.$$
\end{Proposition}

\begin{proof}
The proof is clear.
\end{proof}

\begin{Def}[Cyclic $\AInfty$-algebra] \label{Def:CyclicAinfty}
A graded vector space $V$ together with a pairing 
$$ \Pair: V[1]\otimes V[1] \rightarrow \R $$
of degree $d\in \Z$ and a collection of homogenous linear maps 
$$\mu_k: V[1]^{\otimes k} \rightarrow V[1]\quad\text{for }k\ge 1$$
is called a \emph{cyclic $\AInfty$-algebra of degree~$d$} if the following conditions are satisfied:
\begin{PlainList}
 \item The pairing $\Pair$ is non-degenerate and graded antisymmetric; i.e., we have
  $$ \Pair(v_1,v_2) = (-1)^{1+\Abs{v_1}\Abs{v_2}} \Pair(v_2,v_1) \quad\text{for all }v_1, v_2 \in V[1]. $$
 \item The degrees satisfy $\Abs{\mu_k}=1$ for all $k\ge 1$.
 \item The \emph{$\AInfty$-relations} are satisfied: for all $k\ge 1$, we have
\begin{equation} \label{Eq:AInftyDef}
 \sum_{\substack{k_1, k_2 \ge 1 \\ k_1+k_2 = k+1}} \sum_{p=1}^{k_1} \mu_{k_1} \circ_1^p \mu_{k_2} = 0,
 \end{equation}
where for all $p=1$, $\dotsc$, $k$ and $v_1$, $\dotsc$, $v_{k}\in V[1]$ we define
$$(\mu_{k_1} \circ_1^p \mu_{k_2})(v_1, \dotsc, v_{k}) := \begin{multlined}[t] (-1)^{\Abs{v_1} + \dotsb + \Abs{v_{p-1}}} \mu_{k_1}(v_1, \dotsc, v_{p-1},\\ \mu_{k_2}(v_p,\dotsc,v_{p+k_2-1}),v_{p+k_2}\dotsc,v_{k}). \end{multlined}$$

 \item The operations $\mu_k^+: V[1]^{\otimes k+1} \rightarrow \R$ defined by 
 $$ \mu_k^+:= \Pair\circ (\mu_k \otimes \Id) $$
 for all $k\ge 1$ are cyclic symmetric; i.e., we have
 $$ \mu_k^+ \circ t_{k+1} = \mu_k^+. $$
\end{PlainList}
We denote by $\tilde{\Pair}: V\otimes V \rightarrow \R$ and $\tilde{\mu}_k: V^{\otimes k} \rightarrow \R$ the operations before the degree shift; i.e., for all $k\ge 1$ and $\tilde{v}_1$, $\dotsc$, $\tilde{v}_k \in V$ with $v_i = \SuspU \tilde{v}_i$, we have
$$\begin{aligned}
\tilde{\Pair}(\tilde{v}_1, \tilde{v}_2) &:= (-1)^{\tilde{v}_1} \Pair(v_1, v_2)\quad\text{and} \\[\jot]
\tilde{\mu}_k(\tilde{v}_1, \dotsc, \tilde{v}_k) &:= \varepsilon(\SuspU,\tilde{v}) \mu_k(v_1,\dotsc, v_k).
\end{aligned} $$
We define $\tilde{\mu}_k^+: V^{\otimes k+1}\rightarrow \R$ similarly.

If $\mu_k \equiv 0$ for all $k\ge 2$, then $(V,\Pair, \mu_1)$ is called a \emph{cyclic cochain complex}. If $\mu_k \equiv 0$ for all $k\ge 3$, then $(V,\Pair,\mu_1,\mu_2)$ is called a \emph{cyclic dga}. We use the same terminology but omit ``cyclic'' if there is no pairing $\Pair$ and 1) and 4) are thus irrelevant.
\end{Def}

\begin{Remark}[A difference in sign conventions]\label{Rem:mukplus}
Our definition of $\mu_k^+$ differs from the definition of $\mathrm{m}_k^+$ in~\cite[Definition 12.1]{Cieliebak2015} by a sign. To compensate this, we have to add this artificial sign in the definitions of Maurer-Cartan elements later; e.g., in Definition~\ref{Def:CanonMC} or in the formula~\eqref{Eq:PushforwardMC}.
\end{Remark}

\begin{Definition}[Cyclic (co)homology of $\AInfty$-algebras]\label{Def:CycHom}
Let $\mathcal{A}=(V,(\mu_k))$ be an $\AInfty$-algebra. For every $k\ge 1$, we consider the maps ${\Hd'}^k$, $R^k: V[1]^{\otimes k} \rightarrow \B V$ given by 
\begin{equation}\label{Eq:bRH} \begin{aligned}{\Hd'}^k & := \sum_{j=1}^k \sum_{i=0}^{k-j} t^i_{k-j+1}\circ(\mu_j \otimes \Id^{k-j})\circ t_k^{-i}\quad\text{and}\\
R^k &:= \sum_{j=1}^k \sum_{i=1}^{j-1} (\mu_j\otimes \Id^{k-j})\circ t_k^{i}, \end{aligned}
\end{equation}
respectively, and define the following maps $\B V \rightarrow \B V$:
\begin{equation*}
\Hd':= \sum_{k=1}^{\infty} {\Hd'}^k, \quad R:= \sum_{k=2}^\infty R^k\quad\text{and}\quad \Hd:= \Hd' + R.
\end{equation*}
We denote by $\Hd^*: \CDB V = (\B V)' \rightarrow \CDB V$ the dual map to $\Hd: \B V \rightarrow \B V$.  The following holds:\footnote{The facts \eqref{Eq:HH} are generally known in some form (see \cite{Mescher2016} or \cite{Lazarev2003}). We also show them in \cite{MyPhD} using a graphical formalism which simplifies computations.}
\begin{equation} \label{Eq:HH}
\Abs{\Hd} = 1\ (\Abs{\Hd^*}=-1), \quad \Hd\circ \Hd = 0 \quad\text{and}\quad \Hd(1-t) = (1-t)\Hd'.
\end{equation} 
From the last equation we see that $\Hd$ restricts to $\BCyc V = \B V / \Im(1-t)$.
We define the following graded vector spaces:
$$\begin{aligned}
D_*(V) &:= r(\B V)[1], & D^*(V) &:= r(\CDB V)[1], \\ D^\lambda_*(V) &:= r(\BCyc V)[1], & D_\lambda^*(V) &:= r(\CDBCyc V)[1]. \end{aligned}$$
For instance, we have
$$ D_\lambda^q(V) = r(\CDBCyc V)^{q+1} = (\CDBCyc V)^{-q-1}\quad \text{for all } q\in \Z.$$
Then $(D_*(V),\Hd)$ and $(D^\lambda_*(V),\Hd)$ are chain complexes and $(D^*(V),\Hd^*)$ and $(D_\lambda^*(V),\Hd^*)$ the dual cochain complexes, respectively. We define the following (co)homologies:
$$ \begin{aligned}
\H\H_*(\mathcal{A};\R)& := \H(D_*(V), \Hd), & \H\H^*(\mathcal{A};\R) &:= \H(D^*(V),\Hd^*),\\  
\H^\lambda_*(\mathcal{A};\R)& := \H(D^\lambda_*(V), \Hd), & \H^*_\lambda(\mathcal{A};\R), &:= \H(D^*_\lambda(V),\Hd^*).
\end{aligned} $$
We call $\H\H_*$ the \emph{Hochschild homology} and $\H^\lambda_*$ the \emph{cyclic homology} of the $\AInfty$-algebra $\mathcal{A}$. We call $\H\H^*$ the \emph{Hochschild cohomology} and $\H_\lambda^*$ the \emph{cyclic cohomology} of~$\mathcal{A}$.
\end{Definition}

For a dga $\mathcal{A} = (V,\mu_1,\mu_2)$, we have for all $v_1$, $\dotsc$, $v_k\in V[1]$ the formula
\begin{align*}
 \Hd(v_1 \dots v_k) &= \sum_{i=1}^k (-1)^{\Abs{v_1} + \dotsb + \Abs{v_{i-1}}} v_1 \dots \mu_1(v_i) \dots v_k  \\ 
   &+ \sum_{i=1}^{k-1} (-1)^{\Abs{v_1} + \dotsb + \Abs{v_{i-1}}} v_1 \dots \mu_2(v_i,v_{i+1}) \dots v_k \\
   &+ (-1)^{\Abs{v_k}(\Abs{v_1} + \dotsb + \Abs{v_{k-1}})} \mu_2(v_k,v_1)v_2\dots v_{k-1}.
\end{align*}

\begin{Definition}[Strict units and strict augmentations]\label{Def:AugUnit}
Let $\mathcal{A}= (V, (\mu_k))$ be an $\AInfty$-algebra. A non-zero homogenous element $\NOne \in V[1]$ with $\Abs{\NOne} = -1$ is called a \emph{strict unit} for $\mathcal{A}$ if the following holds:
$$\begin{aligned} \mu_2(\NOne, v) = (-1)^{\Abs{v} + 1}\mu_2(v,\NOne) &= v\qquad\forall v\in V[1], \\[\jot]
\mu_k(v_1, \dotsc, v_{i-1}, \NOne, v_{i+1}, \dotsc, v_k) &= 0\qquad\forall\ k\neq 2,\ 1\le i \le k,\ v_j \in V[1]. \end{aligned}$$
The pair $(\mathcal{A},\NOne)$ is called a \emph{strictly unital  $\AInfty$-algebra.}

A strictly unital $\AInfty$-algebra $(\mathcal{A},\NOne)$ is called \emph{strictly augmented} if it is equipped with a linear map $\varepsilon: V[1] \rightarrow \R[1]$ which satisfies
$$ \varepsilon(\NOne_V) = \NOne_\R, \quad \varepsilon \circ \mu_1 = 0\quad\text{and}\quad \varepsilon \circ \mu_2 = \mu_2\circ(\varepsilon \otimes \varepsilon), $$
where $\NOne_\R$ is the strict unit for~$\R$ endowed with the standard multiplication. The map $\varepsilon$ is called a \emph{strict augmentation.}.

If the \emph{homological dga} $\H(\mathcal{A}):= (\H(V,\tilde{\mu}_1), \mu_1 \equiv 0, \mu_2)$ of $\mathcal{A}$ is strictly unital and strictly augmented, then~$\mathcal{A}$ is called \emph{homologically unital} and \emph{homologically augmented}, respectively. A strictly unital and strictly augmented cochain complex $(V,\mu_1,\NOne,\varepsilon)$ is called just augmented. 
\end{Definition}

We denote by $u: \R[1] \rightarrow V[1]$ the injective linear map defined by $u(\NOne_\R):= \NOne_V$, and by $u^*: \DBCyc V \rightarrow \DBCyc \R$ and $\varepsilon^*: \DBCyc \R \rightarrow \DBCyc V$ the precompositions with $u^{\otimes k}$ and $\varepsilon^{\otimes k}$ in every weight-$k$ component, respectively. 

\begin{Remark}[On units and augmentations]\phantomsection
\begin{RemarkList}
\item A strict unit $\NOne_V$ for $\mathcal{A}$ induces an $\AInfty$-morphism $(u_k): \R \rightarrow V$ given by $u_1(\NOne_\R):= \NOne_V$ and $u_k \equiv 0$ for all $k\ge 2$. A (general) augmentation of $(\mathcal{A},\NOne_V)$ is by definition any $\AInfty$-morphism $(\varepsilon_k): V \rightarrow \R$ such that $(\varepsilon_k) \circ (u_k) = \Id$ as $\AInfty$-morphisms (see~\cite{Keller1999}). Strict augmentations are precisely the maps $\varepsilon_1$ coming from augmentations $(\varepsilon_k)$ with $\varepsilon_k \equiv 0$ for all~$k\ge 2$.

\item As for $(V,\mu_1,\NOne,\varepsilon)$, we need the chain map $\varepsilon$ to provide the splitting of the short exact sequence of chain complexes
$$\begin{tikzcd}
0 \arrow{r} & \R[1] \arrow[hook]{r}{u} & \arrow[bend left=50]{l}{\varepsilon} V[1] \arrow[two heads]{r} & \coker(u) \arrow{r} & 0,
\end{tikzcd}$$
so that we get $\H(V) \simeq \H_{\mathrm{red}}(V)\oplus \R$, where $\H_{\mathrm{red}}(V):= \H(\coker(u))$. If $(V,\mu_1)$ is non-negatively graded and we are given an injective chain map $u: \R[1] \rightarrow V[1]$ (=:\,the classical augmentation), then one can show that such $\varepsilon$ always exists. \qedhere
\end{RemarkList}
\end{Remark}

\begin{Definition}[Reduced dual cyclic bar complex]\label{Def:ReducedDual}
Let $(\mathcal{A}, \NOne)$ be a strictly unital $\AInfty$-algebra. Consider the injection $\iota_{\NOne}: \B V \rightarrow \B V$, $v_1 \otimes \dotsb \otimes v_k \mapsto \NOne \otimes v_1 \otimes \dotsb \otimes v_k$. We define the \emph{reduced dual cyclic bar-complex} by
$$ \RedDBCyc V := \{\psi \in \DBCyc V \mid \psi\circ \iota_{\NOne} = 0\}. $$
Under the assumption of strict unitality, $\Hd^*$ preserves $\RedDBCyc V$, and hence we can consider the reduced cyclic cochain complex 
$$ D_{\lambda,\mathrm{red}}^*(V) := r(\CDBCyc V)[1]$$
and define the \emph{reduced cyclic cohomology of $\mathcal{A}$} by
$$ \H_{\lambda, \mathrm{red}}^*(\mathcal{A};\R):= \H(D_{\lambda, \mathrm{red}}^*(V), \Hd^*). $$ 
\end{Definition}

\begin{Proposition}[Reduction to the reduced cyclic cohomology]\label{Prop:Reduced}
Let $\mathcal{A}= (V,(\mu_k))$ be an $\AInfty$-algebra with a strict unit $\NOne$ and a strict augmentation $\varepsilon$. Then the inclusions $\RedDBCyc V$, $\varepsilon^*(\DBCyc \R) \subset \DBCyc V$ induce the decomposition
$$\begin{aligned}
\H_\lambda^*(\mathcal{A};\R) &\simeq \H_{\lambda, \mathrm{red}}^*(\mathcal{A};\R) \oplus \H_\lambda^*(\R;\R).
\end{aligned} $$
Here we have
\begin{equation*}
 \H_\lambda^{q}(\R; \R) = \begin{cases} \langle \NOne^{q+1*} \rangle & \text{for }q\ge 0 \text{ even}, \\
0 & \text{for }q> 0 \text{ odd and }q<0, \\
\end{cases}
\end{equation*}
where $\NOne^{i*}: \R[1]^{\otimes i} \rightarrow \R$ is defined by $\NOne^{i*}(\NOne^{i}) := 1$.
\end{Proposition}
\begin{proof}[Sketch of the proof]
The maps $\varepsilon^*: D_\lambda(\R) \rightarrow D_\lambda(V)$ and $u^*: D_\lambda(V) \rightarrow D_\lambda(\R)$ are chain maps with $u^*\circ \varepsilon^* = \Id$. Therefore, we have the sequence of cochain complexes
$$\begin{tikzcd}
 0 \arrow{r} &D_{\lambda, \mathrm{red}}(V) \arrow[hook]{r} & D_\lambda(V) \arrow[two heads]{r}{u^*} & \arrow[bend left=50]{l}{\varepsilon^*} D_\lambda(\R) \arrow{r} & 0, 
\end{tikzcd}$$
which is exact everywhere except for the middle, and where $\varepsilon^*$ is a splitting map. The idea of \cite{LodayCyclic} is to replace these cochain complexes with quasi-isomorphic bicomplexes consisting of normalized Hochschild cochains $\bar{D}(V)$ such that the sequence becomes exact. The work then reduces to proving that $\bar{D}(V)$ computes $\H\H(\mathcal{A};\R)$; a variant of this result for $\AInfty$-algebras was proven in~\cite{Lazarev2003}. A detailed proof in our formalism will be provided in~\cite{MyPhD}.
\end{proof}

We will now compare our version of the cyclic cohomology of a dga $(V,\mu_1, \mu_2)$ to a version based on~\cite[Section 5.3.2]{LodayCyclic}. Let $\tilde{\Hd}$, $\tilde{\delta}: \bar{T}V \rightarrow \bar{T}V$ be the linear maps defined for all $\tilde{v}_1$, $\dotsc$, $\tilde{v}_k \in V$ by
\allowdisplaybreaks
\begin{align*}
   \tilde{\Hd}(\tilde{v}_1\otimes \dotsb \otimes \tilde{v}_k) & : = \begin{multlined}[t] \sum_{i=1}^{k-1} (-1)^{i-1} \tilde{v}_1 \otimes \dotsb \otimes \tilde{\mu}_2(\tilde{v}_i, \tilde{v}_{i+1}) \otimes \dotsb \otimes \tilde{v}_k  \\ {}+ (-1)^{k-1+ \tilde{v}_k(\tilde{v}_1 + \dotsb + \tilde{v}_{k-1})}\tilde{\mu}_2(\tilde{v}_k, \tilde{v}_1)\otimes\tilde{v}_2\otimes\dotsb\otimes\tilde{v}_{k-1}, 
\end{multlined} \\ 
\tilde{\delta}(\tilde{v}_1\otimes \dotsb \otimes \tilde{v}_k) & :=  \sum_{i=1}^k (-1)^{\tilde{v}_1 + \dotsb + \tilde{v}_{i-1}} \tilde{v}_1\otimes\dotsb \otimes \tilde{\mu}_1(\tilde{v}_i)\otimes \dotsb \otimes \tilde{v}_k.
\end{align*}
For all $q\ge 0$, we define
$$ \tilde{D}_q(V) := \bigoplus_{\substack{k\ge 1 \\ d\in \Z \\k-d= q + 1}} (V^{\otimes k})^d $$
and $\tilde{\Bdd}: \tilde{D}_{q+1}(V) \rightarrow \tilde{D}_{q}(V)$ by   
$$ \tilde{\Bdd}(\tilde{v}_1\dotsb \tilde{v}_k) = \tilde{b}(\tilde{v}_1\dotsb \tilde{v}_k) + (-1)^{k+1} \tilde{\delta}(\tilde{v}_1\dotsb \tilde{v}_k). $$
It can be checked that $\tilde{\Bdd}\circ\tilde{\Bdd}=0$ and $\tilde{\Bdd}(\Im(1-\tilde{t}))\subset \Im(1-\tilde{t})$, so that $\tilde{\Bdd}$ induces a boundary operator on
the chain complexes \Correct[caption={DONE Wrong cyclic permutation}]{Here the $t$ is modified i.e. $\tilde{t}(v_1\dotsc v_k) = (-1)^{k-1} t(v_1 \dotsc v_k)$}
$$ \tilde{D}_*(V):= \bigoplus_{q\in \Z} \tilde{D}_q(V)\quad\text{and}\quad\tilde{D}^\lambda_*(V) := \tilde{D}_*(V)/\Im(1-\tilde{t}). $$
Here, we have $\tilde{t}(\tilde{v}_1 \dotsb \tilde{v}_k) := (-1)^{k + \Abs{\tilde{v}_k}(\Abs{\tilde{v}_1} + \dotsb + \Abs{\tilde{v}_{k-1}})} \tilde{v}_k \tilde{v}_1 \dotsb \tilde{v}_{k-1}$. We call $(\tilde{D}_*(V),\tilde{\Bdd})$ the \emph{classical Hochschild complex} and $(\tilde{D}^\lambda_*(V), \tilde{\Bdd})$ the \emph{classical cyclic complex} of the dga $(V,\mu_1,\mu_2)$. The chain complex $(\tilde{D}_*(V),\tilde{\Bdd})$ is the total complex of the bicomplex
$$\begin{tikzcd}
{} & \arrow{d} &\arrow{d} & \arrow{d} & {} & {} \\
{} &\arrow{l} \arrow{d}{\tilde{\Hd}} (V^{\otimes 3})^1 & \arrow{l}{\tilde{\delta}} \arrow{d}{\tilde{\Hd}} (V^{\otimes 3})^0 & \arrow{l}{\tilde{\delta}} \arrow{d}{\tilde{\Hd}} (V^{\otimes 3})^{-1} & \arrow{l} & {} \\
{} &\arrow{l} \arrow{d}{\tilde{\Hd}} (V^{\otimes 2})^1 & \arrow{l}{-\tilde{\delta}} \arrow{d}{\tilde{\Hd}} (V^{\otimes 2})^0 & \arrow{l}{-\tilde{\delta}} \arrow{d}{\tilde{\Hd}} (V^{\otimes 2})^{-1} & \arrow{l} & {} \\
{} &\arrow{l} V^1 & \arrow{l}{\tilde{\delta}} V^0 & \arrow{l}{\tilde{\delta}} V^{-1} & \arrow{l}, & {} 
\end{tikzcd}$$
which differs from the bicomplex \cite[Equation (5.3.2.1)]{LodayCyclic} by the reversed grading and by the fact that it lies in the whole upper half-plane and not just in the first quadrant. Their convention for a dga is namely $\Abs{\tilde{\mu}_1} = -1$, whereas ours is $\Abs{\tilde{\mu}_1}=1$, and they consider $\N_0$-grading, whereas we have $\Z$-grading. 

\begin{Proposition}[The classical case] \label{Prop:DGA}
Let $\mathcal{A} = (V,\mu_1,\mu_2)$ be a dga. Then the degree shift map
$$ \begin{aligned} 
 U: \tilde{D}_q (V) & \longrightarrow D_q(V), \\
        \tilde{v}_1 \otimes \dotsb \otimes \tilde{v}_k & \longmapsto \varepsilon(\SuspU, \tilde{v}) v_1 \otimes \dotsb \otimes v_k,  \end{aligned}$$
where we denote $v_i = \SuspU \tilde{v}_i$, is an isomorphism of the chain complexes $(\tilde{D}_*(V),\tilde{\Bdd}) \simeq (D_*(V), \Hd)$ and $(\tilde{D}_*^\lambda(V),\tilde{\Bdd})\simeq (D_*^\lambda(V),\Hd)$, respectively.
\end{Proposition}
   
\begin{proof} 
First of all, for the degrees holds $\Abs{\tilde{\mu}_j} = 2 - j$ for every $j\ge 1$. For every $j$, $k$, $l\ge 1$ such that $j+l \le k+1$ and for every $\tilde{v}_1$, $\dotsc$, $\tilde{v}_k \in V$, we compute
$$ \begin{aligned}
&\bigl[U^{-1}(\Id^{l-1}\otimes \mu_j \otimes \Id^{k-j-l+1})U\bigr](\tilde{v}_1\dotsb \tilde{v}_k) \\[\jot] &\quad = (-1)^{l-1 + (j-2)(\tilde{v}_1 + \dotsb + \tilde{v}_{l-1} + k - l - j +1)} \tilde{v}_1\dotsb\tilde{v}_{l-1}\tilde{\mu}_j(\tilde{v}_l\dotsb \tilde{v}_{l+j-1})\tilde{v}_{l+j}\dotsb \tilde{v}_k, \\[\jot]
& [U^{-1} t_k U](\tilde{v}_1\dotsb \tilde{v}_k) = (-1)^{k-1} \tilde{v}_1 \dotsb \tilde{v}_k,
\end{aligned}$$
where we use the Koszul convention $(f_1\otimes f_2)(v_1\otimes v_2) = (-1)^{\Abs{f_2}\Abs{v_1}} f_1(v_1)\otimes f_2(v_2)$. Using this, we obtain
$$\begin{aligned}
U^{-1} {\Hd'}^k U &= \sum_{j=1}^k \sum_{i=0}^{k-1} (-1)^{i+j(i+k+1)} t^i_{k-j+1}(\tilde{\mu}_j \otimes \Id^{k-j})t_k^{-i}\quad\text{and} \\
U^{-1} R^k U &= \sum_{j=1}^k \sum_{i=1}^{j-1} (-1)^{(i+j)(k+1)} (\tilde{\mu}_j\otimes \Id^{k-j})t_k^i.
\end{aligned}$$
It is now easy to check that $U^{-1}\circ \Hd\circ U = \tilde{\Bdd}$.

If $k\in \N$ is a weight and $d\in\Z$ a degree such that $k-d-1 = q$ for some $q\in \Z$, we have schematically $U: (k,d)\mapsto (k, d - k) = (k,-q-1)$. Therefore, $U$ preserves the grading of chain complexes. This finishes the proof.
\end{proof}

\begin{Proposition}[Reduced cochains are complete in $0$,\,$1$-connected case]\label{Prop:SimplCon}
Suppose that $V = \bigoplus_{d\ge 0} V^d$ is a non-negatively graded vector space with $V^0=\langle 1 \rangle$ for some $1\in V$ ($=:$\,$V$ is \emph{connected}) and $V^1 = 0$ ($=:$\,$V$ is \emph{simply-connected}). Then for all $m\ge 1$, we have
$$ \hat{\Ext}_m \RedDBCyc V = \Ext_m \RedDBCyc V. $$
\end{Proposition}
\begin{proof}
Let $\bar{V}:= \bigoplus_{d\ge 2} V^d$. We clearly have $\RedDBCyc V \simeq \DBCyc \bar{V}$. Since $\bar{V}[1]$ is positively graded, we have $(\B \bar{V})_{k}^d = 0$ whenever $k>d$. Therefore, a map $\Psi\in \hat{\Ext}_m \bar{V}$, which is non-zero only on finitely many homogenous components of $\BCyc V[1]^{\otimes m}$, will be non-zero only on finitely many weights. This implies that $\Psi\in \Ext_m \bar{V}$.
\end{proof}

\begin{Remark}[Universal coefficient theorem]\label{Rem:UCT}
Because $(D^*_\lambda(V), \Hd^*)$ is dual to $(D_*^\lambda(V), \Hd)$ as a chain complex and because we work over $\R$, the universal coefficient theorem gives
\begin{equation*}
 \H^q_\lambda(\mathcal{A},\Hd^*) \simeq [\H_q^\lambda(\mathcal{A},\Hd)]^*\quad\text{for all } q\in \Z. 
\end{equation*}
Suppose that we have found closed homogenous elements $(w_i)_{i\in I}\subset D^\lambda_*(V)$ for some index set~$I$ which induce a basis of $\H^\lambda_*(\mathcal{A}; \R)$. For every $i\in I$, we define the linear map $w_i^*: D^\lambda_*(V) \rightarrow \R$ by prescribing
$$ w_i^*(w_j) = \delta_{ij}\qquad\text{for all }j\in I $$
and $w_i^* \equiv 0$ on $\Im \Hd$ and on a complement \Correct[caption={DONE Universal coefficient theorem}]{Here is enough an arbitrary complement of $\Ker)(b)$. That means that for every $i$, we can have a different complement $Z_i$} of $\Ker(\Hd)$ in $D^\lambda_*(V)$. Then $(w_i^*)_{i\in I} \subset D_\lambda^*(V)$ are closed homogenous elements which generate linearly independent cohomology classes in $\H_\lambda^*(\mathcal{A}; \R)$; if we denote $I_q := \{i\in I \mid w_i \in C^\lambda_q(V)\}$, then we can write
\begin{equation*}
\H_\lambda^q(\mathcal{A}; \R) = \Bigl\{ \sum_{i\in I_q} \alpha_i w_i^* \bigMid \alpha_i\in \R \Bigr\}\quad\text{for all }q\in\Z.\qedhere
\end{equation*}
\end{Remark}

\subsection{Canonical \texorpdfstring{$\dIBL$-structure}{dIBL-structure} on cyclic cochains
}

\label{Sec:Alg3}

In this section, we will consider a \underline{finite-dimensional} cyclic dga $(V,\Pair,m_1,m_2)$ of degree $2-n$ for some $n\in \N$.
This means that for all $v_1$, $v_2$, $v_3 \in V[1]$, the following relations holds:
\begin{equation}\label{Eq:CycDGA}{\hbadness=10000 \text{cyc. dga}\left\{ \begin{aligned} \Pair(v_1,v_2) &= (-1)^{1+ \Abs{v_1} \Abs{v_2}} \Pair(v_2, v_1), &
\mathclap{
\smash{
\raisebox{-0.58cm}{$
\hspace{-0.8cm}
\left.\begin{aligned}\mathstrut\\ \mathstrut\\ \mathstrut \end{aligned}\right\}
\text{\parbox{5em}{cyc. cochain complex}}
$}}}
\\
m_1(m_1(v_1)) &= 0, &\\
m_1^+(v_1, v_2) &= (-1)^{\Abs{v_1}\Abs{v_2}} m_1^+(v_2, v_1), &\\
 m_1(m_2(v_1,v_2)) &= \begin{multlined}[t]- m_2(m_1(v_1),v_2)  \\ - (-1)^{\Abs{v_1}} m_2(v_1,m_1(v_2)), \end{multlined} &\\
m_2(m_2(v_1, v_2), v_3) &= (-1)^{\Abs{v_1} + 1} m_2(v_1,m_2(v_2,v_3)), &\\
m_2^+(v_1, v_2, v_3) &= (-1)^{\Abs{v_3}(\Abs{v_1}+\Abs{v_2})}m_2^+(v_3, v_1, v_2).
\end{aligned}\right.}
\end{equation}
The facts (A) and (C) from the Introduction apply, and we get the canonical $\dIBL$-algebra $\dIBL(\DBCyc V[2-n])$ of bidegree $(n-3,2)$ and the canonical Maurer-Cartan element $\MC = (\MC_{10})$. We will denote
$$ \CycC(V):= \DBCyc V[2-n] $$
and call it the space of \emph{cyclic cochains on $V$}. If $V$ is fixed, we will write just $\CycC$.

\begin{Def}[The canonical $\dIBL$-algebra] \label{Def:CanonicaldIBL}
Let $(V,\Pair,m_1)$ be a cyclic cochain complex of degree $2-n$ which is finite-dimensional. Let $(e_0, \dotsc, e_m)\subset V[1]$ be a basis of $V[1]$, and let $(e^0,\dotsc, e^m)$ be the dual basis with respect to $\Pair$; this means that
$$ \Pair(e_i,e^j) = \delta_{ij}\quad\text{for all }i, j =0, \dotsc,  m. $$
We define the tensor $T = \sum_{i,j=0}^m T^{ij} e_i \otimes e_j \in V[1]^{\otimes 2}$ by\footnote{See Appendix~\ref{Section:Appendix} for the invariant meaning of $T$ as the Schwartz kernel of $\pm \Id$.}
\begin{equation} \label{Eq:PropagatorT}
 T^{ij} = (-1)^{\Abs{e_i}} \Pair(e^i,e^j) \quad\text{for all }i,j = 0,\dotsc,m.
\end{equation}
The \emph{canonical $\dIBL$-algebra} on $\CycC(V)$ is the quadruple
$$ \dIBL(\CycC(V)) := (\CycC(V),\OPQ_{110}, \OPQ_{210}, \OPQ_{120}), $$
where the operations $\OPQ_{110}$, $\OPQ_{210}$, $\OPQ_{120}$ are defined for all $\psi$, $\psi_1$, $\psi_2 \in \CDBCyc V$ and generating words $w = v_1 \dots v_k$, $w_1 = v_{11}\dots v_{1k_1}$, $w_2 = v_{21}\dots v_{2k_2}\in \BCyc V$ with $k$, $k_1$, $k_2\ge 1$ as follows:
\begin{itemize}
\item The \emph{$\dIBL$-boundary operator} $\OPQ_{110}: \hat{\Ext}_1 \CycC \rightarrow \hat{\Ext}_1 \CycC$ of degree $\Abs{\OPQ_{110}} = -1$ is defined by
\begin{equation}\label{Eq:Diff}
\OPQ_{110}(\Susp \psi)(\Susp w) := \Susp\sum_{i=1}^k (-1)^{\Abs{v_1}+ \dotsb + \Abs{v_{i-1}}} \psi(v_1 \dots v_{i-1}m_1(v_i)v_{i+1} \dots v_k).
\end{equation}
\item The \emph{product} $\OPQ_{210}: \hat{\Ext}_2 \CycC \longrightarrow \hat{\Ext}_1 \CycC$ of degree $\Abs{\OPQ_{210}}= -2(n-3)-1$ is written schematically as
$$ \OPQ_{210}(\Susp^2 \psi_1 \otimes \psi_2)(\Susp w) := \sum \varepsilon(w\mapsto w^1 w^2)(-1)^{\Abs{e_j}\Abs{w^1}}T^{ij}\psi_{1}(e_i w^1) \psi_2(e_j w^2) $$
and defined ``algorithmically'' as follows:

For every cyclic permutation $\sigma\in\Perm_k$, consider the tensor 
$$\sigma(w) := \varepsilon(\sigma,w) v_{\sigma_1^{-1}}\otimes \dotsb \otimes v_{\sigma_k^{-1}}, $$
and split it into two parts $w^1$ and $w^2$ of possibly zero length such that $v_{\sigma_1^{-1}}\otimes \dotsb \otimes v_{\sigma_k^{-1}} = w^1 \otimes w^2$. Feed $w^1$ and $w^2$ into $\psi_1$ and $\psi_2$ preceded by~$e_i$ and~$e_j$, respectively, and multiply the result with the sign $(-1)^{\Abs{e_j}\Abs{w^1}}$, which is the Koszul sign to order 
$$ e_i e_j w^1 w^2 \longmapsto e_i w^1 e_j w^2. $$
Finally, sum over all $\sigma \in \Perm_k$, all splittings of $\sigma(w)$ and all indices $i,j = 0$,~$\dotsc$, $m$. The sign $\varepsilon(\sigma,w)$ is denoted by $\varepsilon(w\mapsto w^1w^2)$ to indicate the splitting.

\item The \emph{coproduct} $\OPQ_{120}: \hat{\Ext}_1 \CycC \longrightarrow \hat{\Ext}_2 \CycC$ of degree $\Abs{\OPQ_{120}} = -1$ is written schematically~as
$$ \begin{aligned} &\OPQ_{120}(\Susp \psi)(\Susp^2 w_1 \otimes w_2) \\ & \qquad = \frac{1}{2} \sum \varepsilon(w_1\mapsto w_1^1)\varepsilon(w_2\mapsto w_2^1) (-1)^{\Abs{e_j}\Abs{w_1^1}} T^{ij} \psi(e_i w_1^1 e_j w_2^1) \end{aligned}$$
and defined ``algorithmically'' as follows:

For all cyclic permutations $\sigma\in \Perm_{k_1}$ and $\mu\in \Perm_{k_2}$, denote $w_1^1:= \sigma(w_1)$ and $w_2^1:= \mu(w_2)$ and let $\varepsilon(w_1\mapsto w_1^1)$ and $\varepsilon(w_2\mapsto w_2^1)$ be the corresponding Koszul signs, respectively. Feed~$w_1^1$ and~$w_2^1$ into~$\psi$ in the indicated order interleaved by~$e_i$ and~$e_j$ and multiply the result with the sign $(-1)^{\Abs{e_j}\Abs{w_1^1}}$, which is the Koszul sign to order
$$ e_i e_j w_1^1 w_2^1 \mapsto e_i w_1^1 e_j w_2^1. $$
Finally, sum over all $\sigma\in \Perm_{k_1}$, $\mu\in\Perm_{k_2}$ and all indices $i$, $j = 0$, $\dotsc$, $m$.
\end{itemize}
The operations are extended continuously to the completion.
\end{Def}

\begin{Definition}[The canonical Maurer-Cartan element] \label{Def:CanonMC}
Let $(V,\Pair,m_1,m_2)$ be a finite-dimensional cyclic dga of degree $2-n$. The \emph{canonical Maurer-Cartan element} $\MC$ for $\dIBL(\CycC(V))$  consists of only one element $\MC_{10}\in \hat{\Ext}_1 \CycC$ of degree $\Abs{\MC_{10}} = 2(n-3)$ which is defined by
\begin{equation*}
\MC_{10}(\Susp v_1 v_2 v_3) := (-1)^{n-2}  \mu_2^+(v_1,v_2,v_3)\quad\text{for all } v_1, v_2, v_3\in V[1]
\end{equation*}
on the weight-three component of $\BCyc V[3-n]$ and extended by $0$ to other weight-$k$ components.
\end{Definition}

\begin{Remark}[On canonical $\dIBL$-structure]\phantomsection
\begin{RemarkList}
\item Elements of the completion $\hat{\CycC}(V)$ which are not in $\CycC(V)$ will be called \emph{long cyclic cochains}. Because there are no infinite sums in Definition~\ref{Def:CanonicaldIBL}, $\dIBL(\CycC)$ is completion-free. Clearly, the twist $\dIBL^{\PMC}(\CycC)$ remains completion-free as long as $\PMC_{lg}\in \Ext_l \CycC$ for all $l$, $g$.
\item The constructions of $\OPQ_{210}$ and $\OPQ_{120}$ do not depend on the choice of a basis and can be rephrased in terms of summation over ribbon graphs (see Example~\ref{Ex:Canon}).
\item According to Proposition~\ref{Prop:Compl}, the filtration on $\CycC(V)$ satisfies (WG1) \& (WG2), and hence the $\IBL$-structures $\IBL(\HIBL(\CycC))$ and $\IBL(\HIBL^{\MC}(\CycC))$ are well-defined (see Definition~\ref{Def:HomIBL}).\qedhere
\end{RemarkList}
\end{Remark}

\begin{Proposition}[Formulas for twisted operations]\label{Prop:Formulafortwisted}
Let $\dIBL(\CycC(V))$ be the canonical $\dIBL$-algebra for a finite-dimensional cyclic cochain complex $(V,\Pair,m_1)$ of degree $2-n$, and let $\PMC=(\PMC_{lg})$ be a Maurer-Cartan element. Then for all $l\ge 1$, $g\ge 0$, $\Psi\in \CDBCyc V[3-n]$ and generating words $\W_1$, $\dotsc$, $\W_l \in \BCyc V [3-n]$, we have 
\begin{equation}\label{Eq:TwisteddIBL}
\begin{aligned}
& [(\OPQ_{210}\circ_1 \PMC_{lg})(\Psi)](\W_1 \otimes \dotsb \otimes \W_l) \\ 
& \quad = \begin{multlined}[t]\sum_{j=1}^l \sum \varepsilon' \varepsilon(w_j \mapsto w_j^1 w_j^2) T^{ab} \Psi(\Susp e_a w_j^1) \PMC_{lg}(\W_1 \otimes \dotsb \W_{j-1} \otimes (\Susp e_b w_j^2)  \\ \otimes \W_{j+1} \otimes \dotsb \otimes \W_l), \end{multlined}
\end{aligned}
\end{equation}
where the sum without limits is the sum in Definition \ref{Def:CanonicaldIBL} for $\OPQ_{210}$ and $\varepsilon'$ is the Koszul sign of the following operation:
$$\begin{multlined}(\Susp e_a e_b) \W_1 \dots \W_{j-1}(\Susp w_j^1 w_j^2) \W_{j+1} \dots \W_l \\ 
\longmapsto (\Susp e_a w_j^1) \W_1 \dots \W_{j-1} (\Susp e_b w_j^2)\W_{j+1} \dots \W_l.\end{multlined}$$
In particular, for $l=1$, $g\ge 0$ and $\W\in \BCyc V[3-n]$, we have 
\begin{equation}\label{Eq:TwistDif}
(\OPQ_{210} \circ_1 \PMC_{1g})(\W)= (-1)^{n-3} \sum T^{ab} \varepsilon(w \mapsto w^1 w^2) \PMC_{1g}(\Susp e_a w^1)\psi(e_b w^2),
\end{equation}
and for $l=2$, $g\ge 0$ and $\W_1$, $\W_2\in \BCyc V[3-n]$, we have
\begin{equation}\label{Eq:Twistn2}
\begin{aligned}
 & [(\OPQ_{210}\circ_1 \PMC_{2g})(\Psi)](\W_1 \otimes \W_2) \\
 &\quad = \begin{multlined}[t](-1)^{(n-3)(\Abs{\Psi} + 1)} \Bigl[ \sum T^{ab} \varepsilon(w_1 \mapsto w_1^1w_{1}^{2}) (-1)^{\Abs{e_b}\Abs{w_1^1}}\Psi(\Susp e_a w_1^1)  \\ \PMC_{20}(\Susp e_b w_1^2 \otimes \W_2) + (-1)^{\Abs{\W_1}\Abs{\W_2}}\sum T^{ab}  \varepsilon(w_2\mapsto w_2^1w_2^2) \\ (-1)^{\Abs{e_b}\Abs{w_2^1}} \Psi(\Susp e_a w_2^1)\PMC_{20}(\Susp e_b w_2^2 \otimes \W_1)\Bigr].\end{multlined} \end{aligned}
\end{equation}
\end{Proposition}

\begin{proof}
Let us first discuss the completions. Given $\PMC_{lg}\in \hat{\Ext}_l \CycC$, we can write it as $\PMC_{lg} = \sum_{i=1}^\infty \Phi_1^i \dotsb \Phi_l^i$ with generating words $\Phi_1^i\dotsb \Phi_l^i\in \Ext_l \CycC$ of weights approaching $\infty$. The canonical extension of $\circ_h$ to maps with finite filtration degree commutes with convergent infinite sums, and hence we have $\OPQ_{klg} \circ_h \PMC_{lg} = \sum_{i=1}^\infty \OPQ_{klg}\circ_h (\Phi_1^i \dotsb \Phi_l^i)$. Therefore, it suffices to prove the formulas for generating words $\Phi_1^i \dotsb \Phi_l^i \in \Ext_l \CycC$.

From \eqref{Eq:CompositionSimple}, we get for every $\Psi$, $\Phi_1$, $\dotsc$, $\Phi_l \in \CycC$ the equation
$$ [\OPQ_{210}\circ_1(\Phi_1\dotsb \Phi_l)](\Psi) = \sum_{i=1}^l (-1)^{\Abs{\Phi_i}(\Abs{\Phi_1}+\dotsb+\Abs{\Phi_{i-1}})}\OPQ_{210}(\Psi,\Phi_i) \Phi_1 \dotsb \hat{\Phi}_i \dotsb \Phi_l, $$
where $\Phi_1\dotsb \Phi_l$ on the left-hand-side is considered as a map $\Ext_0 \CycC = \R \rightarrow \Ext_l \CycC$ mapping $1$ to $\Phi_1\dotsb \Phi_l$.
For $\W_1$, $\dotsc$, $\W_l\in \BCyc V[3-n]$ and $\sigma\in \Perm_l$, we use
$$ [\sigma(\Phi_1\otimes \dotsb \otimes \Phi_l)](\W_1 \otimes \dotsb \otimes \W_l) = (\Phi_1 \otimes \dotsb \otimes \Phi_l)[\sigma^{-1}(\W_1\otimes \dotsb \otimes \W_l)] $$
and Definition~\ref{Def:CanonicaldIBL} to get
\allowdisplaybreaks
\begin{align*}
&\bigl([\OPQ_{210}\circ_1(\Phi_1\dotsb \Phi_l)](\Psi)\bigr)(\W_1 \otimes \dotsb \otimes \W_l) = \\
&\quad= \begin{multlined}[t] \sum_{i=1}^l (-1)^{\Abs{\Phi_i}(\Abs{\Phi_1}+\dotsb+\Abs{\Phi_{i-1}})} \frac{1}{l!}\sum_{\sigma\in \Perm_l} \varepsilon(\sigma^{-1},\W) [\OPQ_{210}(\Psi,\Phi_i)](\W_{\sigma_1}) \\ \Phi_1(\W_{\sigma_2}) \dotsb \hat{\Phi}_i(\emptyset) \dotsb \Phi_l(\W_{\sigma_l})\end{multlined} \\
&\quad=\begin{multlined}[t] \sum_{i=1}^l (-1)^{\Abs{\Phi_i}(\Abs{\Phi_1}+\dotsb+\Abs{\Phi_{i-1}})} \frac{1}{l!}\sum_{\sigma\in \Perm_l} \varepsilon(\sigma^{-1},\W) (-1)^{\Abs{\Susp}\Psi}\\ \sum \varepsilon(w_{\sigma_1}\mapsto w_{\sigma_1}^1 w_{\sigma_1}^2) 
(-1)^{\Abs{e_b}\Abs{w_{\sigma_1}^1}}T^{ab} \Psi(e_a w_{\sigma_1}^1) \Phi_i(e_b w_{\sigma_1}^2) \\ \Phi_1(\W_{\sigma_2}) \dots \hat{\Phi}_i(\emptyset) \dots \Phi_l(\W_{\sigma_l})\end{multlined} \\
&\; =: (*),
\end{align*}
where $\hat{\Phi}_i(\emptyset)$ means omission of the corresponding term. Consider the bijection 
$$\begin{aligned}
I: \{1,\dotsc, l\} \times \Perm_l &\longrightarrow \{1,\dotsc, l\} \times \Perm_l \\
(i,\sigma) &\longmapsto \Biggl(j:= \sigma_1,\mu:= \begin{pmatrix} 1 & \dots & i-1 & i & i+1 & \dots & l \\ \sigma_2 & \dots & \sigma_{i} & \sigma_1 & \sigma_{i+1} & \dots & \sigma_l \end{pmatrix}\Biggr).
\end{aligned}$$
Given $(i,\sigma) \in \{1,\dotsc,l\}\times \Perm_l$ and $b\in \{1,\dotsc,m\}$, let $(j,\mu):=I(i,\sigma)$ and
$$ \W':= \W_1 \otimes \dotsb \otimes \W_{j-1} \otimes (\Susp e_b w_{j}^2)\otimes \W_{j+1}  \otimes \dotsb \otimes \W_l. $$
Suppose that $(\Phi_1 \otimes \dotsb \otimes \Phi_l)(\W')\neq 0$. We compute the Koszul sign $\varepsilon(\mu^{-1},\W')$ in the following way:
$$ \begin{aligned}
\W' &\mapsto (-1)^{(\Abs{w_{j}^1} + \Abs{e_b} + \Abs{\W_j})(\Abs{\W_1} + \dotsb + \Abs{\W_{j-1}})} (\Susp e_b w_j^2)\W_1 \dots \hat{\W}_j \dots \W_l \\
 & \mapsto \underbrace{(-1)^{(\Abs{w_{j}^1} + \Abs{e_b})(\Abs{\W_1} + \dotsb + \Abs{\W_{j-1}})}}_{=:\varepsilon_1}\varepsilon(\sigma^{-1},\W) (\Susp e_b w_j^2) \W_{\sigma_2} \dots \W_{\sigma_l} \\
& \mapsto \underbrace{\varepsilon_1 \varepsilon(\sigma^{-1},\W) (-1)^{\Abs{\Phi_i}(\Abs{\Phi_1} + \dotsb + \Abs{\Phi_{i-1}})}}_{=\varepsilon(\mu^{-1},\W')}  \underbrace{\W_{\sigma_2} \dots \W_{\sigma_{i}} (\Susp e_b w_j^2) \W_{\sigma_{i+1}}\dots\W_{\sigma_l}}_{=\W'_{\mu_1}\dots \W'_{\mu_l}}.
\end{aligned}$$
Using this, we can rewrite $(*)$ as
$$\begin{aligned}
(*) &= \begin{multlined}[t](-1)^{\Abs{s}\Abs{\Psi}}\sum_{j=1}^l \sum\varepsilon(w_j \mapsto w_j^1 w_j^2)(-1)^{\Abs{e_b}\Abs{w_j^1}}T^{ab} \Psi(e_a w_j^1)  \\ 
\varepsilon_1 \frac{1}{l!}\sum_{\mu \in \Perm_l} \varepsilon(\mu^{-1}, \W')\Phi_1(\W'_{\mu_1}) \dots \Phi_l(\W'_{\mu_l}) \end{multlined} \\
 & = \begin{multlined}[t]\sum_{j=1}^l \sum \varepsilon(w_j \mapsto w_j^1 w_j^2) (-1)^{\Abs{s}\Abs{\Psi} + \Abs{e_b}\Abs{w_j^1} + (\Abs{w_j^1} + \Abs{e_b})(\Abs{\W_1}+ \dotsb + \Abs{\W_{j-1}}) }T^{ab} \\ 
 \Psi(\Susp e_a w_j^1) (\Phi_1 \dotsb \Phi_l)(\W_1 \otimes \dotsb \otimes \W_{j-1} \otimes (\Susp e_b w_j^2) \otimes \W_{j+1} \otimes \dotsb \otimes \W_l). \end{multlined}
\end{aligned}$$
Finally, we use
$$ T^{ab} \neq 0\; \Implies\; \Abs{e_a} + \Abs{e_b} = n-2 $$
to write
$$\begin{aligned}
\Abs{s}\Abs{\Psi} &= \Abs{s}(\Abs{w_{j}^1} + \Abs{e_a}) = (n-3)(\Abs{w_{j}^1} + n-2 - \Abs{e_b}) \\ 
&= \Abs{s}(\Abs{w_{j}^1} + \Abs{e_b}) \mod 2, \end{aligned}$$
and the formula \eqref{Eq:TwisteddIBL} follows.

As for \eqref{Eq:TwistDif}, we first compute $\varepsilon'$ for $l=1$ as follows:
$$ \begin{aligned}\ln_{-1}(\varepsilon')  &= \Abs{w_1} \Abs{e_b} + (\Abs{e_b} + \Abs{w_1})\Abs{s}  \underset{\mathclap{\substack{\uparrow\rule{0pt}{1.5ex} \\2(n-3) = \Abs{\PMC_{10}} = \Abs{s} + \Abs{e_b} + \Abs{w^2}}}}{=} \Abs{w^1}\Abs{w^2} + \Abs{s}\Abs{e_b} \\ &\underset{\mathclap{\substack{\uparrow\rule{0pt}{1.5ex} \\ \Abs{e_a} + \Abs{e_b} = \Abs{s} + 1}}}{=} \Abs{w^1}\Abs{w^2} + \Abs{e_a}\Abs{e_b} \mod 2.  \end{aligned}$$
Using this, we obtain
$$ \begin{aligned}
[(\OPQ_{210}\circ_1 \PMC_{1g})(\Psi)](W) &= \sum \varepsilon' \varepsilon(w\mapsto w^1 w^2) T^{ab} \Psi(\Susp e_a w^1) \PMC_{1g}(\Susp e_b w^2) \\
&\underset{\mathclap{\substack{\uparrow\rule{0pt}{1.5ex} \\ T^{ab} = (-1)^{\Abs{s} + \Abs{e_a}\Abs{e_b}} T^{ba} \\
\varepsilon(w\mapsto w^1 w^2) =  (-1)^{\Abs{w^1} \Abs{w^2}} \varepsilon(w\mapsto w^2 w^1)}}}{=} (-1)^{\Abs{s}} \sum \varepsilon(w\mapsto w^2 w^1) T^{ba} \PMC_{1g}(\Susp e_b w^2) \Psi(\Susp e_a w^1),
\end{aligned}$$
which implies \eqref{Eq:TwistDif}.

The proof of \eqref{Eq:Twistn2} is a combination of the same arguments.
\end{proof}

We will now relate homology of the twisted boundary operator $\OPQ_{110}^\PMC$ to cohomology of an $\AInfty$-algebra on $V$ induced by $\PMC_{10}$. 

\begin{Definition}[$\AInfty$-operations and compatible Maurer-Cartan element]\label{Def:MukDef}
Let $(V,\Pair,m_1)$ be a finite-dimensional cyclic cochain complex of degree $2-n$, and let $\PMC = (\PMC_{lg})$ be a Maurer-Cartan element for $\dIBL(\CycC(V))$. We define the operations $\mu_k: V[1]^{\otimes k} \rightarrow V[1]$ for all $k\ge 1$ by
\begin{equation*}
\mu_k(v_1,\dotsc, v_k) := (-1)^{n-3}\sum_{i, j} T^{ij} \PMC_{10}(\Susp e_i v_1 \dots v_k) e_j
\end{equation*}
for all $v_1$,~$\dotsc$, $v_k \in V[1]$, where $T^{ij}$ is the matrix from Definition~\ref{Def:CanonicaldIBL}.

If $(V,\Pair,m_1,m_2)$ is in addition a cyclic dga and $\MC$ the canonical Maurer-Cartan element for $\dIBL(\CycC(V))$, then we say that $\PMC$ is \emph{compatible with $\MC$} if
$$ \PMC_{10}(\Susp v_1 v_2 v_3) = \MC_{10}(\Susp v_1 v_2 v_3)\quad\text{for all }v_1, v_2, v_3 \in V[1]. $$
\end{Definition}

\begin{Proposition}[Twisted boundary operator $\OPQ^\PMC_{110}$ and $\AInfty$-cyclic cohomology]\label{Prop:CyclicHom}
In the setting of Definition~\ref{Def:MukDef}, the triple $\mathcal{A}_\PMC(V) := (V,\Pair, (\mu_k))$ is a cyclic $\AInfty$-algebra. We always have $\mu_1 = m_1$, and if $\PMC$ is compatible with $\MC$ for a cyclic dga $(V,\Pair,m_1,m_2)$, then also $\mu_2 = m_2$. 

The following holds for the homologies:
\begin{equation*}
 \HIBL^\PMC_*(\CycC(V))=  r(\H^*_\lambda(\mathcal{A}_\PMC(V);\R))[3-n].
\end{equation*}
\end{Proposition}

\begin{proof}
First of all, according to Definition~\ref{Def:MaurerCartan} we must have $\Norm{\PMC_{10}} > 2$, and hence 
$$ \PMC_{10}(\Susp v_1 v_2) = \PMC_{10}(\Susp v_1) = 0 \quad\text{for all }v_1, v_2 \in V[1].  $$
This implies $\mu_1 = m_1$.

Now, let $e_0$, $\dotsc$, $e_m$ be a basis of $V[1]$ and let $e^0$, $\dotsc$, $e^m$ be the dual basis with respect to $\Pair$. For all $k\ge 2$ and $v_1$, $\dotsc$, $v_k \in V[1]$, we compute the following:
\allowdisplaybreaks
\begin{align*}
& \Pair(\mu_k(v_1,\dotsc,v_k),v_{k+1}) \\
&\qquad = (-1)^{n-3}\sum_{ij} (-1)^{\Abs{e_i}}\Pair(e^i, e^j) \PMC_{10}(\Susp e_i v_1\dots v_k) \Pair(e_j, v_{k+1}) \\
&\qquad  \underset{\mathclap{\substack{\uparrow\rule{0pt}{1.5ex}\\\forall v\in V[1]: \ \sum_j \Pair(v,e^j)e_j = v}}}{=} (-1)^{n-3}\sum_i (-1)^{\Abs{e_i}} \PMC_{10}(\Susp e_i v_1\dotsc v_k) \Pair(e^i,v_{k+1}) \\
&\qquad \underset{\mathclap{\substack{\uparrow\rule{0pt}{1.5ex}\\(\Abs{v}_1 + \dotsb + \Abs{v_k}) \Abs{e_i} = \\ (\Abs{\PMC_{10}} + \Abs{s}+\Abs{e_i})\Abs{e_i} = (\Abs{s}+1)\Abs{e_i}}}}{=} (-1)^{n-3}\sum_i (-1)^{\Abs{e_i}(n-3)} \PMC_{10}(\Susp v_1 \dotsc v_k e_i) \Pair(e^i,v_{k+1}) \\
&\qquad \underset{\mathclap{\substack{\big\uparrow\rule{0pt}{2.5ex}\\ 1+\Abs{v_{k+1}}\Abs{e^i} = 1+(\Abs{e}^i + 2 - n)\Abs{e^i} \\= 1+(3-n)\Abs{e^i} = 1+(3-n)\Abs{e_i}}}}{=}  (-1)^{n-2}\sum_i \PMC_{10}(\Susp v_1 \dotsc v_k e_i) \Pair(v_{k+1},e^i) \\
&\qquad = (-1)^{n-2}\PMC_{10}(\Susp v_1 \dotsc v_{k+1}).
\end{align*}
Therefore, we have
\begin{equation*}
 \PMC_{10} = (-1)^{n-2}\sum_{k\ge 2} \mu_k^+.
\end{equation*}
In this case,~\cite[Proposition 12.3]{Cieliebak2015} asserts that the $\AInfty$-relations~\eqref{Eq:AInftyDef} for $(\mu_k)_{k\ge 1}$ are equivalent to the ``lowest'' Maurer-Cartan equation~\eqref{Eq:MCEq} for $\PMC_{10}$. 
The degree condition $\Abs{\mu_k}=1$ and the cyclic symmetry of $\mu_k^+$ are easy to check. Therefore, $\mathcal{A}_\PMC(V)$ is a cyclic $\AInfty$-algebra.

As for the compatibility with $\MC$, we have for all $v_1$, $v_2\in V[1]$ the following:
\begin{equation*}
 \begin{aligned}
  m_2(v_1, v_2) & = \sum_i \Pair(e_i, m_2(v_1,v_2)) e^i \\
 &\underset{\mathclap{\substack{\big\uparrow\rule{0pt}{2.5ex} \\ T^{ij} = (-1)^{\Abs{e_i}} \Pair(e^i,e^j)}}}{=}  \sum_{i,j} (-1)^{\Abs{e_i}} T^{ij} \Pair(e_i,m_2(v_1,v_2))e_j \\
 &\underset{\mathclap{\substack{\big\uparrow\rule{0pt}{2.5ex}\\ \Pair(v_1,v_2) = (-1)^{1+(n-3)\Abs{v_1}} \Pair(v_2,v_1)}}}{=} \sum_{i,j} (-1)^{1 + (n-2)\Abs{e_i}}T^{ij} \underbrace{\Pair(m_2(v_1,v_2),e_i)}_{\displaystyle{\mathclap{=(-1)^{n-2}\MC_{10}(\Susp v_1 v_2 e_i)}}} e_j\\
 & \underset{\mathclap{\substack{\big\uparrow\rule{0pt}{2.5ex} \\  (\Abs{v_1} + \Abs{v_2})\Abs{e_i} =(\Abs{\MC_{10}} - \Abs{s} - \Abs{e_i})\Abs{e_i}\\ = (n-2)\Abs{e_i} }}}{=} (-1)^{n-3} \sum_{i,j} T^{ij} \MC_{10}(\Susp e_i v_1 v_2)e_j \\ 
 &= \mu_2(v_1,v_2).
\end{aligned}
\end{equation*}

We will now clarify the relation to the cyclic cohomology of $\mathcal{A}_\PMC(V)$. Recall from Proposition~\ref{Prop:dIBL} that $\OPQ_{110}^\PMC(\Psi) = \OPQ_{110}(\Psi) + \OPQ_{210}(\PMC_{10}, \Psi)$ for $\Psi\in\CDBCyc V[3-n]$, where the first term is given by~\eqref{Eq:Diff} and the second by~\eqref{Eq:TwistDif}. Consider now ${\Hd'}^k$ and $R^k$ from~\eqref{Eq:bRH}, whose sum gives the Hochschild boundary operator $\Hd$. Using the cyclic symmetry, we can rewrite a summand of ${\Hd'}^k$ for $j=1$, $\dotsc$, $k$ and $i=0$, $\dotsc$, $k-j$ applied to a generating word $v_1\dots v_k \in \BCyc V$ as follows: 
\begin{equation}\label{Eq:Tmp1}
\begin{aligned}
& [t^i_{k-j+1} \circ (\mu_j \otimes \Id^{k-j}) \circ t_k^{-i}](\underbrace{v_1 \dots v_k}_{=:w}) = \\
&\quad =(-1)^{\Abs{v_1} + \dotsb + \Abs{v_{i}}} v_1 \dots v_i \mu_j(v_{i+1} \dots v_{i+j})v_{i+j+1} \dots v_k \\[\jot]
&\quad = \varepsilon(w\mapsto w^1 w^2) \mu_j(\underbrace{v_{i+1}\dots v_{i+j}}_{=:w^1}) \underbrace{v_{i+j+1} \dots v_k v_1 \dots v_i}_{=:w^2}
\end{aligned}
\end{equation}
Clearly, summing \eqref{Eq:Tmp1} over $j=1$ and $i=0$, $\dotsc$, $k-1$ gives the dual to $\OPQ_{110}$. For $j=2$, $\dotsc$, $k$, we can write \eqref{Eq:Tmp1} as
$$ 
(-1)^{n-3} \sum_{i,j} \varepsilon(w\mapsto w^1 w^2) T^{ij}\PMC_{10}(\Susp e_i w^1) e_j w^2.
$$
Therefore, the sum over $j=2$, $\dotsc$, $k$ and $i=0$, $\dotsc$, $k-j$ gives the part of the dual to $\OPQ_{210}(\PMC_{10},\Psi)$ corresponding to the cyclic permutations $\sigma\in \Perm_{k}$ with $\sigma_1 = 1$, $j+1$, $\dotsc$, $k$. The rest, i.e., the cyclic permutations with $\sigma_1 = 2$, $\dotsc$, $j$, is obtained analogously from the summands $(\mu_j \otimes \Id^{k-j})\circ t_k^i$ of $R^k$ for $j=2$,~$\dotsc$,\,$k$ and $i=1$, $\dotsc$, $j-1$. We conclude that $\OPQ_{110}^\PMC: \CDBCyc V[3-n] \rightarrow \CDBCyc V[3-n]$ is a degree shift of $\Hd^*: \CDBCyc V \rightarrow \CDBCyc V$. As for the gradings, we have:
$$ \begin{aligned}
r(D_\lambda(V))[3-n]^i &= r(D_\lambda(V))^{i+3-n} = (D_\lambda(V))^{-i-3+n} = (\CDBCyc V)^{i+3-n-1} \\
 &= \CDBCyc V[2-n]^i.
\end{aligned}$$
This finishes the proof.
\end{proof}

%

We will now turn to units and augmentations.

\begin{Definition}[Reduced canonical $\dIBL$-algebra]\label{Def:ReduceddIBL}
Let $(V, \Pair, m_1, \NOne, \varepsilon)$ be an augmented cyclic cochain complex of degree $2-n$ from Definition~\ref{Def:AugUnit}. We define the space of \emph{reduced cyclic cochains} on $V$ by
$$ \CycC_{\mathrm{red}}(V):= \RedDBCyc V[2-n]. $$
We define the \emph{reduced canonical $\dIBL$-algebra} by
$$ \dIBL(\RedCycC(V)):= (\RedCycC(V), \OPQ_{110}, \OPQ_{210}, \OPQ_{120}), $$
where $\OPQ_{110}$, $\OPQ_{210}$, $\OPQ_{120}$ are restrictions of the operations of $\dIBL(\CycC(V))$.
\end{Definition}

\begin{Definition}[Strictly reduced Maurer-Cartan element] \label{Def:StrictlyReduced}
In the setting of Definition~\ref{Def:ReduceddIBL}, we call a Maurer-Cartan element $\PMC=(\PMC_{lg})$ for $\dIBL(\CycC(V))$ \emph{strictly reduced} if $\PMC_{lg}\in \hat{\Ext}_l \RedCycC(V)$ for all $(l,g)\neq (1,0)$ and if the $\AInfty$-algebra $(\mathcal{A}_\PMC(V),\NOne,\varepsilon)$ induced by $\PMC_{10}$ is strictly unital and strictly augmented.
Given a strictly reduced Maurer-Cartan element $\PMC$, we can define the twisted $\IBLInfty$-algebra
$$ \dIBL^\PMC(\CycC_{\mathrm{red}}(V)) := (\CycC_{\mathrm{red}}(V), (\OPQ_{klg}^\PMC)), $$
where $\OPQ_{klg}^\PMC$ are the restrictions of the operations of $\dIBL^\PMC(\CycC(V))$. We denote the homology of $\dIBL^\PMC(\RedCycC)$ by $\HIBL^\PMC(\RedCycC)$ or $\HIBL^{\PMC,\mathrm{red}}(\CycC)$.\footnote{The latter option suggests that it might be possible to define the reduced homology with the induced $\IBL$-algebra even if $\PMC$ is not strictly reducible, e.g., if $(\mathcal{A}_\PMC(V),\NOne,\varepsilon)$ is only homologically unital and augmented.}
\end{Definition}

\Modify[caption={DONE Put remark to text}]{Include this remark to text because it is in fact a definition.}

\begin{Remark}[On strictly reduced Maurer-Cartan element]\phantomsection
\begin{RemarkList}
\item We see that the $\IBLInfty$-algebra $\dIBL^\PMC(\CycC_{\mathrm{red}})$ is a \emph{subalgebra} of $\dIBL^\PMC(\CycC)$, which means that the inclusion $\CycC_{\mathrm{red}}\xhookrightarrow{}\CycC$ induces the following commutative diagram for all $k,l\ge 1$, $g\ge 0$:
$$\begin{tikzcd}
\hat{\Ext}_k \CycC \arrow{r}{\OPQ_{klg}^\PMC} & \hat{\Ext}_l \CycC \\
\hat{\Ext}_k \CycC_{\mathrm{red}} \arrow{r}{\OPQ_{klg}^\PMC}\arrow[hook]{u} & \hat{\Ext}_l \CycC_{\mathrm{red}}.\arrow[hook]{u}
\end{tikzcd}$$
We denote this fact by $\dIBL^\PMC(\CycC_{\mathrm{red}})\subset \dIBL^\PMC(\CycC)$.

\item The canonical Maurer-Cartan element $\MC$ of a strictly augmented strictly unital dga $(V,m_1,m_2,\NOne,\varepsilon)$ is strictly reduced (this follows from Proposition~\ref{Prop:CyclicHom}).

\item In the situation of Definition~\ref{Def:StrictlyReduced}, we denote 
$$ \bar{V}[1]:= \Ker(\varepsilon),$$
so that $V=\bar{V}\oplus \langle 1 \rangle$. We use  the canonical projection $\pi: V \rightarrow \bar{V}$ to identify $\CDBCyc \bar{V} \xrightarrow{\simeq} \CRedDBCyc V$ via the componentwise pullback $\pi^*$. In this way, we obtain the $\IBLInfty$-algebras $\dIBL(\CycC(\bar{V}))$ and $\dIBL^\PMC(\CycC(\bar{V}))$, which are isomorphic to $\dIBL(\CycC_{\mathrm{red}}(V))$ and $\dIBL^\PMC(\CycC_{\mathrm{red}}(V))$, respectively.
\qedhere
\end{RemarkList}
\end{Remark}

In the following list, we sum up our main reasons for considering units, augmentations and reduced Maurer-Cartan elements. Suppose that we are in the situation of Definition~\ref{Def:StrictlyReduced}, then:
\begin{itemize}
 \item  Proposition~\ref{Prop:Reduced} implies the splitting
\begin{equation}\label{Eq:IBLSPlit}
 \HIBL^\PMC(\CycC)[1] = \HIBL^\PMC(\CycC_{\mathrm{red}})[1] \oplus \langle \Susp \NOne^{2q-1 *} \mid q\in \N \rangle.
\end{equation}
Here $\NOne^{i*} \in \DBCyc V$ is the componentwise pullback $\varepsilon^*$ of $\NOne^{i*}\in \DBCyc(\R)$. To get this, we used
\begin{equation*} 
\HIBL^\PMC_*(\CycC_{\mathrm{red}}) =  r(\RedCycCoH^*(\mathcal{A}_\PMC))[3-n],
\end{equation*}
which can be seen by redoing the proof of Proposition~\ref{Prop:CyclicHom} with reduced cochains.
\item The subalgebra $\dIBL^\PMC(\CycC_{\mathrm{red}})\subset \dIBL^\PMC(\CycC)$ induces the subalgebra 
$$\IBL(\HIBL^\PMC(\CycC_{\mathrm{red}}))\subset \IBL(\HIBL^\PMC(\CycC)), $$
and any higher operation $\OPQ_{1lg}^\PMC$ which induces a map $\hat{\Ext}_1 \HIBL(\CycC) \rightarrow \hat{\Ext}_l \HIBL(\CycC)$ induces a map $\hat{\Ext}_1\HIBL(\CycC_{\mathrm{red}}) \rightarrow \hat{\Ext}_l\HIBL(\CycC_{\mathrm{red}})$ as well.
\item If $V$ is non-negatively graded, connected and simply-connected, then we have $\hat{\Ext}_k \CycC_{\mathrm{red}} \simeq \Ext_k \RedCycC$ for all $k\in \N_0$ by Proposition~\ref{Prop:SimplCon}, and hence $\dIBL^\PMC(\RedCycC)$ is completion-free. 
\end{itemize}

\begin{Proposition}[Operations on units]\label{Prop:Ones}
Suppose that $(V,\Pair,m_1,\NOne,\varepsilon)$ is a finite-dimensional augmented cyclic cochain complex of degree $2-n$ such that $n\ge 1$, and let~$\PMC$ be a strictly reduced Maurer-Cartan element for $\dIBL(\CycC(V))$. The following relations are the only relations containing $\NOne^{i*}$ which may be non-zero on the homology $\HIBL^\PMC(\CycC)$: 

For all $\Psi \in \RedCycC(V)$ and $l\ge 1$, $g\ge 0$, we have
\begin{align*}
\OPQ_{210}(\Susp\NOne^* \otimes \Psi) &= (-1)^{(n-2)\Abs{\Psi}} \OPQ_{210}(\Psi \otimes \Susp\NOne^*)  = (-1)^{n-2}\Psi \circ \iota_\NVol\quad\text{and} \\[\jot]
\OPQ_{1lg}^\PMC(\Susp\NOne^*) & = - \PMC_{lg} \circ \iota_\NVol,
\end{align*}
where $\iota_\NVol$ is defined as follows:
\begin{itemize}
\item The element $\NVol \in V[1]$ is the unique vector such that $\Pair(\NOne,\NVol) = 1$ and $\NVol \perp \bar{V}[1]$ with respect to $\Pair$. Note that $\Abs{\NVol} = n-1$ and that such $\NVol$ always exists due to non-degeneracy.
\item We start by defining $\iota_\NVol : \BCyc V \rightarrow \BCyc V$ by
$$ \iota_\NVol(v_1\dots v_k) := \sum_{i=1}^k (-1)^{\Abs{\NVol}(\Abs{v_1} + \dotsb + \Abs{v_{i-1}})} v_1 \dots v_{i-1} \NVol v_i \dots v_k $$
for all generating words $v_1 \dots v_k \in \BCyc V$. Next, for all $k\ge 1$, we define $\iota_\NVol : (\BCyc V)^{\otimes k} \rightarrow (\BCyc V)^{\otimes k}$ by
$$ \iota_\NVol(w_1 \otimes \dotsb \otimes w_k) :=  \begin{multlined}[t](-1)^{\Abs{\NVol}k}\sum_{j=1}^k (-1)^{\Abs{\NVol}(\Abs{w_1} + \dotsb + \Abs{w_{j-1}})} w_1 \otimes \dotsb \otimes w_{j-1} \\ \otimes  \iota_\NVol(w_j) \otimes w_{j+1} \otimes \dotsb \otimes w_k \end{multlined}$$
for all generating words $w_1$, $\dotsc$, $w_k \in \BCyc V$. Finally, we take the degree shift $\iota_\NVol: (\BCyc V[3-n])^{\otimes k} \rightarrow (\BCyc V[3-n])^{\otimes k}$ according to the degree shift convention \eqref{Eq:DegreeShiftConv}. 
\end{itemize}
\end{Proposition}
\begin{proof}
Pick a basis $(e_0, \dotsc, e_m)$ of $V[1]$ such that $e_0 = \NOne$ and $\bar{V}[1] = \langle e_1, \dotsc, e_m \rangle$. If $(e^0,\dotsc,e^m)$ is the dual basis, then we have $\NVol = e^0$. We will often use the following relation:
\begin{equation}\label{Eq:NiceFormula}
 \sum_{j=0}^m T^{1j} e_j = \sum_{j=0}^m (-1)^{\Abs{\NOne}} \Pair(\NVol,e^j) e_j =  - \NVol.
\end{equation}
We consider only those generating words $w = v_1 \dotsc v_k$ of $\BCyc V$ with either $v_i\in \bar{V}$ for each $i$ (shortly $w\in \BCyc \bar{V}$) or $v_i = \NOne$ for each $i$ with $k$ odd (i.e., $w=\NOne^{2j-1}$ for some $j$). Let $w_1$, $\dotsc$, $w_l$ with $w_j = v_{j 1} \dots v_{j k_j}$ denote such generating words. Clearly, if $\Phi \in \hat{\Ext}_l \CycC(V)$ is a $\OPQ_{110}^\PMC$-closed element which vanishes on all $w_1 \otimes \dotsb \otimes w_l$, then~\eqref{Eq:IBLSPlit} implies that $[\Phi] = 0$ in $\hat{\Ext}_l\HIBL(\CycC)$.

For $\Psi\in \RedCycC(V)$ and $q\ge 1$ odd, we compute using \eqref{Eq:NiceFormula} the following:
\allowdisplaybreaks
\begin{align*}
\OPQ_{210}(\Susp^2 \NOne^{q*}\otimes \psi)(\Susp w)  &= \sum \varepsilon( w\mapsto w^1 w^2)(-1)^{(n-1)\Abs{w^1}}T^{1j}\NOne^{q*}(\NOne w^1) \psi(e_j w^2) \\[\jot]
&= -\sum \varepsilon( w\mapsto w^1 w^2)(-1)^{(n-1)\Abs{w^1}}\NOne^{q*}(\NOne w^1) \psi(\NVol w^2)\\[\jot]
&=:(*). 
\end{align*}
Now, in order to get $(*)\neq 0$, we need either $q=1$ and $w\in \BCyc \bar{V}$, in which case
\allowdisplaybreaks
\begin{align*}
 (*)&=-\sum \varepsilon(w\mapsto \underbrace{w^1}_{=\emptyset} w^2)\psi(\NVol w^2) \\ &= -\sum_{j=1}^k (-1)^{\Abs{v}(\Abs{v_1} + \dotsb + \Abs{v_{j-1}})} \psi(v_1 \dotsc v_{j-1} \NVol v_{j+1}\dotsc v_k) \\[\jot]
 &=  - (\psi \circ \iota_\NVol)(w) = (-1)^{n-2} (\Psi \circ \iota_\NVol)(\W),
\end{align*}
or $q>1$ odd and $w = \NOne^{q-1}$, in which case
\allowdisplaybreaks
\begin{align*}
 (*) &= \sum \varepsilon(w\mapsto w^1 \underbrace{w^2}_{=\emptyset}) \NOne^{q*}(\NOne^q) \psi(\NVol) \\
 &= \psi(\NVol) \sum_{j=1}^{q-1} (-1)^j \\
 & = 0.
\end{align*}

Next, because $n\ge 1$, we get $T^{11} = 0$, and hence 
$$ \OPQ_{120}(\NOne^{q*}) = 0\quad\text{for all }q\in \N$$
on the chain level. Therefore, we have $\OPQ_{1lg}^\PMC = \OPQ_{210}\circ_1 \PMC_{lg}$ for all $l\ge 1$, $g\ge 0$, and  using Proposition~\ref{Prop:Formulafortwisted} and \eqref{Eq:NiceFormula}, we obtain 
\allowdisplaybreaks
\begin{align*}
&[(\OPQ_{210}\circ_1 \PMC_{lg})(\NOne^{q*})](\W_1 \otimes \dotsb \otimes \W_l) \\[\jot]
&\quad= \begin{multlined}[t]-\smash{\sum_{j=1}^l} \sum \varepsilon' \varepsilon(w_j \mapsto w_j^1 w_j^2) \NOne^{q*}(\NOne w_j^1) \PMC_{lg}(\W_1 \otimes \dotsb \otimes \W_{j-1} \otimes (\Susp \NVol w_j^2) \\[\jot] \otimes \W_{j+1}\otimes \dotsb \otimes \W_l)\end{multlined}\\
&\quad=:(**).
\end{align*}
In order to get $(**)\neq 0$, we need either $q=1$ and $w_j \in \BCyc \bar{V}$ for all $j$, in which case
\allowdisplaybreaks
\begin{align*}
(**) &= \begin{multlined}[t]-\smash{\sum_{j=1}^l} \smash{\sum_{i=1}^{k_j}} (-1)^{\Abs{\NVol}(\Abs{\W_1}+ \dotsb + \Abs{\W_{j-1}} + \Abs{\Susp})}(-1)^{\Abs{\NVol}(\Abs{v_1} + \dotsb + \Abs{v_{j-1}})}\PMC_{lg}(\W_1 \otimes \dotsb \\[\jot] \otimes \W_{j-1} \otimes (\Susp v_1 \dots v_{i-1} \NVol v_{i} \dots v_{k_j}) \otimes \W_{j+1}\otimes \dotsb\otimes \W_l)\end{multlined} \\[\jot]
 &= -(\PMC_{lg} \circ \iota_\NVol)(\W_1 \otimes \dotsb \otimes \W_l),
\end{align*}
or $q>1$ odd and $w_j = \NOne^{q-1}$ for some $j$, in which case
\allowdisplaybreaks
\begin{align*}
(**) &=\begin{multlined}[t]-\smash{\sum_{\substack{1\le j \le l \\ w_j = \NOne^{q-1}}}} \varepsilon' \Bigl(\sum \varepsilon(w_j \mapsto w_j^1 \underbrace{w_{j}^2}_{=\emptyset} \NOne^{q*}(\NOne w_j^1)) \Bigr) \PMC_{lg}(\W_1\otimes \dotsb \\ \otimes\W_{j-1}\otimes (\Susp\NVol)\otimes \W_{j+1} \otimes \dotsb \otimes \W_l )\end{multlined}\\[\jot]
&=\begin{multlined}[t]-\smash{\sum_{\substack{1\le j \le l \\ w_j = \NOne^{q-1}}}} \varepsilon' \smash{\underbrace{\Bigl( \sum_{i=1}^{q-1} (-1)^i \Bigr)}_{=0}} \PMC_{lg}(\W_1 \otimes \dotsb \otimes \W_{j-1} \otimes (\Susp \NVol) \otimes \W_{j+1} \otimes \\[\jot] \dotsb \otimes \W_l)\end{multlined}\\[\jot]
&=0.
\end{align*}

The only relation left to check is
$$ \OPQ_{210}(\Susp \NOne^{q_1 *},\Susp \NOne^{q_1 *}) = 0 \quad\text{for all } q_1, q_2 \in \N.$$
However, this is easy to see, and the proof is done. 
\end{proof}

\clearpage
\section{Twisted \texorpdfstring{$\IBLInfty$-structure}{IBL-infinity-structure} and string topology
}

%

In Section~\ref{Sec:Manifold1}, we consider the cyclic dga's $\DR(M)$, $\H_{\mathrm{dR}}(M)$ and $\Harm(M)$ for a closed oriented $n$-manifold $M$ (Proposition~\ref{Prop:DGAs}) and apply the theory from Section~\ref{Sec:Alg3} to the last two, which are finite-dimensional.

In Section~\ref{Section:Proof1}, we define the Green kernel $\GKer$ (Definition~\ref{Def:GreenKernel}). It is a primitive to the Schwartz kernel $\HKer$ of the harmonic projection~$\pi_\Harm$ (see Proposition~\ref{Lemma:HKer}) outside the diagonal and extends smoothly to the spherical blow-up of the diagonal. These ideas come from an early version of~\cite{Cieliebak2018}. We consider conditions (G1)--(G5) on a linear operator $\GOp$ and its Schwartz kernel $\GKer$ (see p.~\pageref{ConditionsG}) and show that $\GKer$ satisfying all these conditions always exists (Proposition~\ref{Prop:ExistenceG}). We also mention the standard Green kernel $\GKerStd$ (see \eqref{Eq:GStdStd}), which might be a canonical Green kernel satisfying (G1)--(G5). 

In Section~\ref{Section:Proof2}, we review ribbon graphs, labelings, compatibility of the order and orientation of internal edges, and the edge and vertex order from~\cite{Cieliebak2015} (Definitions~\ref{Def:Ribbon}, \ref{Def:Labeling}, \ref{Def:CompatLabeling} and~\ref{Def:EdgeVertex}). We then define $\PMC$ as a signed sum of integrals of products of Green kernels and harmonic forms which are associated to labeled trivalent ribbon graphs (Definition~\ref{Def:PushforwardMCdeRham}). We do not show that these integrals converge and that $\PMC$ satisfies the Maurer-Cartan equation, but we do show all other properties of a Maurer-Cartan element (Lemma~\ref{Lem:MCCond} and Proposition~\ref{Prop:FormalPushforwardProp}). We define the $Y$-graph, trees, circular graphs, vertices of types $A$, $B$, $C$ and their contributions $A_{\alpha_1, \alpha_2}$, $B_\alpha$, $C$, respectively (Definitions~\ref{Def:Graphs} and~\ref{Def:Contributions}). 

In Section~\ref{Sec:Vanishing}, we observe that vanishing of some special vertices in the graphs implies $\PMC_{lg} = \MC_{lg}$. For example, if all graphs, except for the $Y$-graph, with $\NOne$ at an external vertex vanish, which holds if $\GKer$ satisfies (G4) and (G5) (Proposition~\ref{Prop:COne}), then all higher operations $\OPQ_{1lg}^\PMC$ vanish on the chain level in dimensions $n>3$ (Proposition~\ref{Prop:PMCEqualsMC}). Next, if all graphs with an $A$-vertex vanish, then $\PMC_{10}=\MC_{10}$, and hence $\OPQ_{110}^\PMC = \OPQ_{110}^\MC$ (Proposition~\ref{Prop:Avertexvanish}). We show that $\PMC = \MC$ for simply-connected geometrically formal manifolds with $n\neq 2$ (Proposition~\ref{Prop:GeomForm}). Using the results of \cite{Cieliebak2018}, we argue that the chain complexes of $\OPQ_{110}^\MC$ and~$\OPQ_{110}^\PMC$ are quasi-isomorphic provided~$M$ is simply-connected and formal (Proposition~\ref{Prop:Formal}).


In Section~\ref{Sec:StringTopology}, we recall basic facts about the Chas-Sullivan operations $\StringOp_2$ and $\StringCoOp_2$ on the $\Sph{1}$-equivariant homology of the free loop space and formulate a version of the string topology conjecture for simply-connected manifolds (Conjecture~\ref{Conj:StringTopology}). \Correct[caption={DONE Mistake in reduced notation}]{Correct the notation here, if it is not already corrected. Anyway, define something like RedNTwistHIBL and replace the construction.}

\subsection{Canonical \texorpdfstring{$\dIBL$-structures on $\CycC(\H_{\mathrm{dR}}(M))$}{dIBL-structures on C(HdR(M))}} \label{Sec:Manifold1}
Let $M$ be an oriented closed Riemannian manifold of dimension $n$. We consider the following graded vector spaces:
$$\begin{aligned}
\DR^*(M) &\;\,\dots && \text{smooth de Rham forms}, \\
\Harm^*(M) &\;\,\dots &&\text{harmonic forms}, \\
\H_{\mathrm{dR}}^*(M) &\;\,\dots && \text{de Rham cohomology}.
\end{aligned}$$
Since $M$ is fixed, we often write just $\DR$, $\Harm$ and~$\HDR$. We consider the Hodge decomposition $\DR=\Harm\oplus\Im\Dd\oplus\Im\CoDd$, where~$\Dd$ is the de Rham differential and~$\CoDd$ the codifferential. We call the corresponding projection
\begin{equation}\label{Eq:HarmProj}
\pi_\Harm: \DR^*(M) \longrightarrow \Harm^*(M) 
\end{equation}
the \emph{harmonic projection} and the induced isomorphism $\pi_\Harm: \H_{\mathrm{dR}} \rightarrow \Harm$ mapping a cohomology class into its unique harmonic representative the \emph{Hodge isomorphism}.

\begin{Notation}[Updated notation for bar complexes] \label{Def:DeRham}
We use Notation~\ref{Def:Notation} for $V=\DR{}$, $\Harm{}$, $\H_{\mathrm{dR}}{}$ and $A=n-3$ with the following changes:
$$\tilde{v}\sim \eta \in V,\quad v \sim \alpha\in V[1],\quad w\sim \omega \in \BCyc V, \quad \W\sim \Omega\in \BCyc V[n-3]. $$
We use the formal symbols $\Susp$ and $\SuspU$ with $\Abs{\Susp} = n - 3$ and $\Abs{\SuspU} = -1$, so that $\alpha = \SuspU \eta$ and $\Omega = \Susp \omega$.
\end{Notation}

\begin{Proposition}[De Rham cyclic dga's]\label{Prop:DGAs}
Let $M$ be an oriented closed Riemannian manifold of dimension $n$. The quadruple $(\DR(M), \Pair, m_1, m_2)$ with the operations from~\eqref{Eq:DeRhamDGA} is a cyclic dga of degree $2-n$.
For the operations before the degree shift, we have
\begin{align*}
\tilde{m}_1(\eta_1) & = \Dd \eta_1, \\[\jot]
\tilde{m}_2(\eta_1, \eta_2) &= \eta_1 \wedge \eta_2, \\ 
\tilde{\Pair}(\eta_1, \eta_2) &=  \int_M \eta_1 \wedge \eta_2 =: (\eta_1,\eta_2),
\end{align*}
where $\Dd$ is the de Rham differential, $\wedge$ the wedge product and $\tilde{\mathcal{P}}$ the \emph{intersection pairing}. 
The operations restrict to $\H_{\mathrm{dR}}(M)$ and make $(\H_{\mathrm{dR}}(M),\Pair,m_1\equiv 0,m_2)$ into a cyclic dga. If we define $\mu_1 \equiv 0$ and
\begin{equation}\label{Eq:HarmProd}
\mu_2(\alpha_1, \alpha_2) := \pi_{\Harm}(m_2(\alpha_1, \alpha_2))\quad \text{for all }\alpha_1, \alpha_2 \in \Harm(M)[1],
\end{equation}
then $(\Harm(M), \Pair, \mu_1, \mu_2)$ is a cyclic dga as well, and $\pi_\Harm: \H_{\mathrm{dR}} \rightarrow \Harm$ is an isomorphism of cyclic dga's. All three dga's $\DR$, $\H_{\mathrm{dR}}$ and $\Harm$ are strictly unital and strictly augmented with the unit $\NOne:= \SuspU 1 \in \DR[1]$, where $1$ is the constant one.
\end{Proposition}

\begin{proof}
The relations \eqref{Eq:CycDGA} follow from the classical properties of $\Dd$ and $\wedge$ and from the Stokes' theorem for oriented closed manifolds. The Poincar\'e duality asserts that $(\cdot,\cdot)$ is non-degenerate on $\H_{\mathrm{dR}}{}$ and~$\Harm{}$, and thus they are cyclic dga's as well. The fact that $\pi_\Harm : \H_{\mathrm{dR}}{}\rightarrow \Harm{}$ is an isomorphism of vector spaces follows from the Hodge theory. As for compatibility with the product, given $\eta_1$, $\eta_2 \in \Harm{}$, then $\eta_1 \wedge \eta_2$ is closed, and since $\Ker \Dd = \Harm \oplus \Im \Dd$, we see that $\pi_{\Harm}(\eta_1 \wedge \eta_2) = \eta_1 \wedge \eta_2 + \Dd \eta$ for some $\eta\in \DR$ is a harmonic representative of the cohomology class $[\eta_1 \wedge \eta_2] = [\eta_1] \wedge [\eta_2]$. Unitality is obvious, and the construction of an augmentation map clear. Note that a strict augmentation for $\DR(M)$ is the evaluation at a point, for instance. 
\end{proof}

The facts (A) and (C) from the Introduction apply to the cyclic dga's $\Harm{}$ and~$\H_{\mathrm{dR}}{}$ (not to $\DR{}$ because it is infinite-dimensional!), and we get the canonical $\dIBL$-algebras $\dIBL(\CycC(\Harm))$ and $\dIBL(\CycC(\H_{\mathrm{dR}}))$ of bidegrees $(n-3,2)$ with the canonical Maurer-Cartan element $\MC = (\MC_{10})$. The Hodge isomorphism induces an isomorphism of these $\dIBL$-algebras, and hence we can use $\Harm$ and~$\HDR$ interchangeably.
We have $\OPQ_{110} \equiv 0$, and hence $\dIBL(\CycC(\Harm))$ is, in fact, an $\IBL$-algebra. However, we will denote it by $\dIBL$ and call it a $\dIBL$ algebra as a reminder of the canonical $\dIBL$-structure. The canonical Maurer-Cartan element~$\MC$ satisfies
\begin{equation}\label{Eq:CanonMC}
\MC_{10}(\Susp \alpha_1 \alpha_2 \alpha_3) = (-1)^{n-2 + \eta_2} \int_M \eta_1 \wedge \eta_2 \wedge \eta_3 \quad \text{for all } \alpha_1, \alpha_2, \alpha_3 \in \Harm[1].
\end{equation}
We get the canonical twisted $\dIBL$-algebra $\dIBL^\MC(\CycC(\Harm))$ from~\eqref{Eq:CanonMCTwist} with, in general, non-trivial boundary operator $\OPQ_{110}^\MC$ whose homology is the cyclic homology of $\HDR$ up to degree shifts.

\subsection{Green kernel \texorpdfstring{$\GKer$}{G}} \label{Section:Proof1}

We will use fiberwise integration and spherical blow-ups, which we now recall.

\begin{Definition}[Fiberwise integration] \label{Def:FibInt}
Let $\Pr: E \rightarrow B$ be a smooth oriented fiber bundle with an oriented fiber $F$ over an oriented manifold~$B$ with $\Bdd B = \emptyset$. We orient $E$ as $F\times B$. Let 
$\DR_{\mathrm{c}} (B)$ denote the space of forms with compact support and $\DR_{\mathrm{cv}}(E)$ the space of forms with compact vertical support. For any $\kappa\in \DR_{\mathrm{cv}}(E)$, let $\FInt{F} \kappa \in \DR(B)$ be the unique smooth form such that
\begin{equation*} \label{Eq:FiberInt}
\int_E \kappa \wedge \Pr^*\eta = \int_B \Bigl(\FInt{F}\kappa \Bigr) \wedge \eta\quad\text{for all }\eta\in \DR_{\mathrm{c}} (B).
\end{equation*}
\end{Definition}
 
\begin{Definition}[Spherical blow-up] Let $X$ be a smooth $n$-dimensional manifold and $Y\subset X$ a smooth $k$-dimensional submanifold. The \emph{blow-up} of $X$ at $Y$ is as a set defined by
$$ \Bl_Y X := X\backslash Y \sqcup P^+N Y, $$
where $P^+N Y$ is the real oriented projectivization of the normal bundle $NY$ of~$Y$ in $X$. This means that $P^+NY$ is the quotient of $\{ v\in NY\,
\vert\, v\neq 0\}$ by the relation $v\sim a v$ for all $a\in (0,\infty)$. The \emph{blow-down map} is defined by
\begin{align*}
 \pi : \Bl_Y X &\longrightarrow X \\
     p\in X\backslash Y &\longmapsto p, \\
     [v]_p\in P^+N Y &\longmapsto p.
\end{align*}
\end{Definition}

In the following, we will equip the blow-up with the structure of a smooth manifold with boundary such that its interior becomes diffeomorphic to $X\backslash Y$ via the blow-down map and the boundary becomes $P^+ NY$. Consider an adapted chart $(U,\psi)$ for $Y$ in $X$ with $\psi(U)=\R^{n}$ and $\psi(U\cap Y)=\{(0,y)\mid y\in \R^k \}$. It induces the bijection\Correct[caption={DONE Blow up charts}]{It is not a bijection if $\psi(U\cap Y) \neq \emptyset$! We need it centered at $(0,0)$! No we need actually only the intersection with $Y$!}
\begin{align*}
 \tilde{\psi}: \Bl_{U\cap Y}U &\longrightarrow [0,\infty) \times \Sph{n-k-1} \times \R^k \\
  p\in U\backslash Y &\longmapsto \Bigl (\Abs{\pi_1 \psi(p)}, \frac{\pi_1\psi(p)}{\Abs{\pi_1\psi(p)}},\pi_2 \psi(p) \Bigr), \\
  [v]\in P^+N_p Y & \longmapsto \Bigl(0,\frac{\pi_1 \Diff{\psi}(v)}{\Abs{\pi_1 \Diff{\psi} (v)}},\pi_2\psi(p)\Bigr),
\end{align*}
where $\pi_1$ and $\pi_2$ are the canonical projections to the factors of $\R^{n-k} \times \R^k$. Notice that we have the canonical inclusion $\Bl_{U\cap Y} U \subset \Bl_Y X$. It can be checked that for any two overlapping adapted charts $(U_1,\psi_1)$ and $(U_2,\psi_2)$, the transition function $\tilde{\psi}_1\circ \tilde{\psi}_2^{-1}$ is a diffeomorphism of manifolds with boundary. Therefore, we can use the charts $(\Bl_{U\cap Y}U, \tilde{\psi})$ to define a smooth atlas on~$\Bl_Y X$. If $X$ is oriented, we orient $\Bl_Y X$ so that $\pi$ restricts to an orientation preserving diffeomorphism of the interior.

An important fact is that if $X$ is compact, then \emph{$\Bl_Y X$ is compact}.

We are interested in the case when $X = M\times M$ for an oriented closed manifold $M$ and $Y = \Delta := \{(m,m) \mid m\in M\}$ is the \emph{diagonal}. Given a chart $\varphi: U \rightarrow \R^n$ on~$M$, the following is a smooth chart on $\Bl_\Diag(M\times M)$:
\begin{equation} \label{Eq:BlowUpChart}
\begin{aligned}
\tilde{\varphi}: \Bl_{\Diag}(U\times U) & \longrightarrow [0,\infty) \times \Sph{n-1} \times \R^n \\
(x,y)\in (U\times U)\backslash \Diag & \longmapsto  (r,w,u):= \begin{multlined}[t] \Bigl(\frac{1}{2}\Abs{\varphi(x)-\varphi(y)}, \frac{\varphi(x)-\varphi(y)}{\Abs{\varphi(x)-\varphi(y)}},\\\frac{1}{2}(\varphi(x)+\varphi(y))\Bigr),\end{multlined} \\
[(v,-v)]_{(x,x)} &\longmapsto \Bigl(0,\frac{\Dd \varphi_x(v)}{\Abs{\Dd \varphi_x(v)}},\varphi(x)\Bigr).
\end{aligned}
\end{equation}
The inverse relations for $r>0$ read
\begin{equation*} \label{Eq:BlowupRelations}
\varphi(x) = u+w r\quad\text{and}\quad\varphi(y)=u-w r.
\end{equation*}
We will denote by $M_i$ the $i$-th factor of $M\times M$; i.e., we will write $M \times M = M_1 \times M_2$. We denote the corresponding projection by $\Pr_i$. We define $\widetilde{\Pr}_i := \Pr_i\circ\pi$, where $\pi: \Bl_\Diag(M\times M) \rightarrow M \times M$ is the blow-down map. We also identify $(M\times M)\backslash \Diag$ with the interior of $\Bl_\Diag(M\times M)$ via $\pi$.

The map $\widetilde{\Pr}_2: \Bl_\Diag(M\times M) \rightarrow M_2$ is an oriented fiber bundle with fiber $\Bl_{*}(M_1)$, which is the blow-up of $M_1$ at a point (we shall assume that $M$ is connected). The \emph{fiberwise integration along~$\widetilde{\Pr}_2$} will be denoted by~$\FInt{\Bl_* M_1}$. 

\begin{Def}[Green kernel] \label{Def:GreenKernel}
Let $M$ be an oriented closed $n$-dimensional Riemannian manifold. Consider the harmonic projection $\pi_{\Harm}$ from \eqref{Eq:HarmProj}, and let $\iota_{\Harm}: \Harm(M) \xhookrightarrow{} \DR(M)$ be the inclusion. 
A smooth $(n-1)$-form $\GKer$ on $(M\times M)\backslash \Diag$ is called a \emph{Green kernel} if the following conditions are satisfied:
\begin{PlainList}
\item The form $\GKer$ admits a smooth extension to $\Bl_\Diag(M\times M)$. More precisely, the pullback $(\Restr{\pi}{\mathrm{int}})^* \GKer$ along the blow-down map restricted to the interior is a restriction of a smooth form on $\Bl_\Diag(M\times M)$. We denote this form by $\GKer$ again by uniqueness.
\item The operator $\GOp: \DR^*(M) \rightarrow \DR^{*-1}(M)$ defined by
\begin{equation} \label{Eq:SchwKer}
\GOp(\eta) := \FInt{\Bl_* M_1} \GKer \wedge \widetilde{\Pr}_1^* \eta \quad \text{for all }\eta\in \DR(M)
\end{equation}
satisfies
\begin{equation} \label{Eq:CochainHomotopy}
\Dd\circ\GOp + \GOp\circ\,\Dd = \iota_{\Harm}\circ \pi_{\Harm} - \Id.
\end{equation}
Any homogenous linear operator $\GOp: \DR^*(M) \rightarrow \DR^{*-1}(M)$ satisfying~\eqref{Eq:CochainHomotopy} will be called a \emph{Green operator}.
\item For the twist map $\tau: M \times M \rightarrow M \times M$ defined by $(x,y)\mapsto (y,x)$, the following symmetry property holds:
\begin{equation}\label{Eq:SymProp}
\tau^* \GKer = (-1)^n \GKer. 
\end{equation}
\end{PlainList}
\end{Def}

\begin{Remark}[On Green kernel]\phantomsection\label{Rem:GKer}
\begin{RemarkList}
\item Given a homogenous linear operator $\GOp: \DR^*(M) \rightarrow \DR^{*-1}(M)$, if there is a $\GKer \in \DR^{n-1}(\Bl_{\Diag}(M\times M))$ such that \eqref{Eq:SchwKer} holds, then it is unique.
\item Because $\tau: M\times M \rightarrow M\times M$ preserves $\Diag$, it extends to a diffeomorphism $\tilde{\tau}$ of $\Bl_\Diag(M\times M)$. The condition \eqref{Eq:SymProp} is then equivalent to $\tilde{\tau}^*\tilde{\GKer} = (-1)^n \tilde{\GKer}$ for the extension $\tilde{\GKer}$ of $\GKer$ to $\Bl_{\Diag}(M\times M)$. We denote both extensions by $\tau$ and $\GKer$, respectively.

\item Using the intersection pairing $(\cdot,\cdot)$, we have 
$$ \begin{aligned}(\GOp(\eta_1),\eta_2) &= \int_{\Bl_{\Diag}(M\times M)} \GKer \wedge \widetilde{\Pr}_1^*\eta_1 \wedge \widetilde{\Pr}_2^*\eta_2 \\ 
&= (-1)^n \int_{\Bl_{\Diag}(M\times M)} \tau^*\GKer \wedge \widetilde{\Pr}_2^*\eta_1 \wedge \widetilde{\Pr}_1^*\eta_2 \end{aligned}$$
and 
$$ \begin{aligned}(\eta_1, \GOp(\eta_2)) &= (-1)^{\eta_1(\eta_2 - 1)} (\GOp(\eta_2),\eta_1) \\ &= (-1)^{\eta_1} \int_{\Bl_\Diag(M\times M)} \GKer \wedge \widetilde{\Pr}_2^* \eta_1 \wedge \widetilde{\Pr}_1^* \eta_2 \end{aligned}$$
for all $\eta_1$, $\eta_2\in \DR(M)$. This implies the following:
\begin{equation}\label{Eq:GSA}
\tau^* \GKer = (-1)^n \GKer \quad \Longleftrightarrow \quad \begin{multlined}[t](\GOp(\eta_1),\eta_2) = (-1)^{\eta_1}(\eta_1, \GOp(\eta_2)) \\ \forall \eta_1, \eta_2 \in\DR(M). \end{multlined}
\end{equation}

\item Because $\Bl_\Diag(M\times M)$ is compact, $\GKer\in \DR(\Bl_\Diag(M\times M))$ induces an $L^1$-integrable form on $M\times M$.

\item In the literature, the term ``Green operator'' often denotes a generalized inverse of an elliptic pseudo-differential operator, e.g., of the Laplacian $\Delta$. This is not what we mean here.\qedhere
\end{RemarkList}
\end{Remark}

We will now prove three propositions which will allow us to rewrite~\eqref{Eq:CochainHomotopy} equivalently as a differential equation for $\GKer$ on $M\times M \backslash \Diag$.

\begin{Proposition}[Identities for fiberwise integration]\label{Prop:StokesForm}
In the situation of Definition~\ref{Def:FibInt},
assume that $F$ has a boundary $\Bdd F$. We orient $\Bdd F$ using $T_p F = N(p)\oplus T_p \Bdd F$ for $p\in \Bdd F$, where $N$ is an outward pointing normal vector field.
The following formulas hold for all $\kappa\in \DR_{\mathrm{cv}}(E)$ and $\eta\in \DR_{\mathrm{c}}(B)$:
\begin{itemize}
\item The \emph{projection formula}
$$\FInt{F}(\kappa\wedge \pi^* \eta) = \Bigl(\FInt{F}\kappa\Bigr)\wedge \eta, $$
\item \emph{Stokes' formula}
$$(-1)^F \Dd \FInt{F}\kappa  = \FInt{F} \Dd \kappa -  \FInt{\Bdd F} \kappa, $$
where $F$ in the exponent denotes the dimension of $F$.
\end{itemize}
\end{Proposition} 
\begin{proof} 
The projection formula is proven by a straightforward calculation from the definition.

As for Stokes' formula, we get the oriented fiber bundle $\Bdd E \rightarrow B$ with fiber~$\Bdd F$ by restricting an oriented trivialization of $E$. There are two orientations on $\Bdd E$ --- as the total space of $\Bdd E \rightarrow B$ and as the boundary of $E$. They agree due to our orientation convention. Using standard Stokes' theorem, we get
\allowdisplaybreaks
\begin{align*}
 (-1)^F \int_B \Dd \Bigl(\FInt{F}\kappa\Bigr) \wedge \eta & = (-1)^{\kappa + 1} \int_B \Bigl(\FInt{F}\kappa\Bigr)\wedge \Dd \eta \\ &=  (-1)^{\kappa + 1} \int_{E} \kappa \wedge \Dd \pi^*\eta \\
  & = \int_E \bigl(\Dd \kappa \wedge \pi^*\eta - \Dd(\kappa \wedge\pi^*\eta)\bigr) \\ 
  &= \int_B \Bigl(\FInt{F} \Dd \kappa\Bigr) \wedge \eta - \int_{\Bdd E} \kappa \wedge \pi^*\eta \\ 
  &= \int_B \Bigl(\FInt{F} \Dd \kappa - \FInt{\Bdd F}\kappa \Bigr) \wedge \eta.
\end{align*}
This proves the proposition.
\end{proof}

In what comes next, we will view the canonical projection $\Pr_2 : M_1 \times M_2 \rightarrow M_2$ as an oriented fiber bundle such that the orientation of the total space agrees with the product orientation. The fiberwise integration for this bundle will be denoted by $\FInt{M_1}$. 

\begin{Proposition}[Schwartz kernel of the harmonic projection] \label{Lemma:HKer}
Let $M$ be an oriented closed $n$-dimensional Riemannian manifold. Let $\Le_1$,~$\dotsc$, $\Le_m$ be a homogenous basis of $\Harm(M)$ which is orthonormal with respect to the $L^2$-inner product 
$$ (\eta_1, \eta_2)_{L^2} := \int_M \eta_1 \wedge \Star\eta_2\quad\text{for }\eta_1, \eta_2 \in \DR(M), $$
where $\Star$ denotes the Hodge star. The smooth form $\HKer \in \DR^n(M\times M)$ defined by
\begin{equation}\label{Eq:HKerHKerHKer}
\HKer := \sum_{i=1}^m (-1)^{n \Le_i}\Pr_1^*(\Star \Le_i) \wedge \Pr_2^*(\Le_i)
\end{equation}
satisfies the following properties:
\begin{ClaimList}
\item For all $\eta\in \DR(M)$, we have
\begin{equation*}
 \HPr(\eta) = \FInt{M_1} \HKer\wedge \Pr_1^*\eta.
\end{equation*} 
\item The form $\HKer$ is closed and Poincar\'e dual to $\Diag \subset M\times M$.
\item The following symmetry condition is satisfied:
\begin{equation} \label{Eq:SymHKer}
\tau^* \HKer = (-1)^n \HKer.
\end{equation}
\end{ClaimList}
\end{Proposition}
\begin{proof} 
\begin{ProofList}
\item For the purpose of the proof, we denote $\Harm(\eta) := \FInt{M_1} \HKer \wedge \Pr_1^* \eta$. For every $k=1$,~$\dotsc$, $m$, we use the projection formula to compute
\allowdisplaybreaks
\begin{align*}
\Harm(\Le_k) &= \sum_{i=1}^m (-1)^{\Le_i n+\Le_i \Le_k}\FInt{M_1} \Pr_1^*(\Star \Le_i \wedge \Le_k)\wedge \Pr_2^*(\Le_i) \\ &=\sum_{i=1}^m (-1)^{\Le_i(n+\Le_k) + \Le_k(n+\Le_i)}\Bigl(\int_M \Le_k \wedge \Star \Le_i\Bigr) \Le_i \\
 & = \Le_k.
\end{align*}
It is easy to see that $\Harm(\eta) \in \Harm(M)$ for all $\eta\in \DR(M)$. Therefore, $\Harm$ is a projection to $\Harm(M)$. Relations $\Harm(\Dd \eta) = \Harm(\CoDd \eta) = 0$ for all $\eta\in \DR(M)$ follow from the second line of the computation above with $\Le_k$ replaced by $\Dd\eta$ and $\CoDd\eta$ using that $\Im \CoDd \oplus \Im \Dd$ is $L^2$-orthogonal to $\Harm(M)$. We see that $\Harm = \pi_{\Harm}$. 

\item Using $\Dd \circ \Harm = \Harm \circ \Dd = 0$ and Stokes' theorem, we get 
$$  \FInt{M_1} \Dd \HKer \wedge \Pr_1^* \eta = (-1)^n \Dd \Harm(\eta) - \Harm(\Dd \eta) = 0\quad \text{for all }\eta\in \DR(M). $$
It follows that $\Dd \HKer = 0$. Using the K\"unneth formula, we can write a given $\kappa\in \DR(M\times M)$ with $\Dd \kappa = 0$ as $\kappa = \Pr_1^* \eta_1 \wedge \Pr_2^* \eta_2 + \Dd \eta$ for some $\eta_1$, $\eta_2 \in \Harm(M)$ and $\eta\in \DR(M)$. Then 
$$ \begin{aligned} \int_{M\times M} \HKer \wedge \kappa & = \int_{M\times M} \HKer \wedge \Pr_1^* \eta_1 \wedge \Pr_2^* \eta_2 \\ &= \int_M \Harm(\eta_1)\wedge \eta_2 \\ &= \int_M \eta_1 \wedge \eta_2 = \int_{\Diag} \kappa. \end{aligned} $$
This shows that $\HKer$ is Poincar\'e dual to $\Diag$.
\item It follows from the Hodge decomposition that
\begin{equation}\label{Eq:HKerHKer}
(\pi_\Harm(\eta_1), \eta_2) = (\pi_{\Harm}(\eta_1), \pi_{\Harm}(\eta_2)) = (\eta_1,\pi_\Harm(\eta_2))\quad\text{for all }\eta_1, \eta_2 \in \DR(M).
\end{equation}
As in (iii) of Remark \ref{Rem:GKer}, one shows that this is equivalent to \eqref{Eq:SymHKer}.\qedhere
\end{ProofList}
\end{proof}

\begin{Proposition}[Differential condition] \label{Prop:GKer}
Let $M$ be an oriented closed $n$-dimensional Riemannian manifold. For $\GKer \in \DR^{n-1} (\Bl_\Delta(M\times M))$,  the following claims are equivalent:
\begin{PlainList} 
\item The operator $\GOp: \DR^*(M) \rightarrow \DR^{*-1}(M)$ defined by $\GOp(\eta) := \FInt{\Bl_* M_1} \GKer \wedge \widetilde{\Pr}_1^*\eta$ for $\eta\in \DR(M)$ is a Green operator.
\item It holds \begin{equation} \label{Eq:GKer}
 \Dd \GKer = (-1)^{n} \HKer\quad\text{on }(M\times M)\backslash \Delta.
\end{equation} 
\end{PlainList}
\end{Proposition}
\begin{proof} 
Before we begin, note that~\eqref{Eq:GKer} is equivalent to the equation $\Dd \tilde{\GKer} = (-1)^n \pi^* \HKer$ on $\Bl_\Diag(M\times M)$ for the extension $\tilde{\GKer}$ of $\GKer$; we denote~$\tilde{\GKer}$ by~$\GKer$ and $\pi^*\HKer$ by $\HKer$ by uniqueness.

We will first prove $2) \Longrightarrow 1)$.
Using Stokes' formula, we get for every $\eta\in \DR(M)$ the following:
\begin{equation*}
\begin{aligned}
  \Dd \GOp (\eta) &= \Dd \FInt{\Bl_* M_1} \GKer\wedge \widetilde{\Pr}_1^* \eta \\ 
  &= (-1)^n \Bigl( \FInt{\Bl_* M_1} \Dd(\GKer\wedge\widetilde{\Pr}_1^* \eta) - \FInt{\Bdd \Bl_* M_1} \GKer \wedge \widetilde{\Pr}_1^* \eta \Bigr) \\ 
  &= \HPr(\eta) - \GOp(\Dd\eta) +  \FInt{\Bdd \Bl_* M_1} (-1)^{n+1}\GKer\wedge \widetilde{\Pr}_1^*\eta.
\end{aligned}
\end{equation*}
Since $\widetilde{\Pr}_1 = \widetilde{\Pr}_2$ on $\Bdd \Bl_\Diag (M\times M)$, we get with the help of the projection formula the following:
\begin{equation*}
 \FInt{\Bdd \Bl_* M_1} \GKer\wedge \widetilde{\Pr}_1^*\eta  = \FInt{\Bdd \Bl_* M_1} \GKer\wedge \widetilde{\Pr}_2^*\eta = \Bigl(\FInt{\Bdd \Bl_* M_1} \GKer \Bigr)\wedge\eta.
\end{equation*}
We will show that the $0$-form $\FInt{\Bdd \Bl_* M_1} \GKer$ is constant $(-1)^n$. Stokes' formula implies
\begin{equation*}
 \FInt{\Bdd \Bl_* M_1} \GKer = \FInt{\Bl_* M_1} \Dd \GKer = (-1)^n \FInt{\Bl_* M_1} \HKer. 
\end{equation*}
Using that $\HKer$ is Poincar\'e dual to $\Diag$, we get for every $\eta\in \DR^n(M)$ the following:
\begin{equation*}
\begin{aligned}
 \int_M \Bigl(\FInt{\Bl_* M_1} \HKer \Bigr) \wedge \eta &= \int_{\Bl_\Diag(M\times M)} \HKer\wedge \widetilde{\Pr}_2^*\eta 
 \\ &= \int_{M\times M} H\wedge \Pr_2^*\eta  \\ &= \int_\Diag \Pr_2^*\eta  = \int_M 1\wedge \eta.
\end{aligned}
\end{equation*}
The implication follows.

We will now prove $1) \Longrightarrow 2)$. Assume that~\eqref{Eq:CochainHomotopy} holds and that~$\GKer$ extends smoothly to the blow-up. Denote
$$K:= (-1)^n \Dd \GKer - \HKer\quad\text{and}\quad L:= -1 +  \int^{\Bdd \Bl_*(M_1)} (-1)^n \GKer. $$
Notice that $L$ is a function on $M$. From the previous computations, we deduce that 
$$ \int^{\Bl_*(M_1)} K\wedge \widetilde{\Pr}_1^*\eta = L\eta\quad\text{for all }\eta\in \DR(M), $$
and hence
$$ \int_{\Bl_{\Diag}(M\times M)} K \wedge \widetilde{\Pr}_1^*(\eta_1) \wedge \widetilde{\Pr}_2^*(\eta_2)  = \int_M L \eta_1 \wedge \eta_2 \quad\text{for all }\eta_1, \eta_2 \in \DR(M). $$
If $K(x,y) \neq 0$ for some $(x,y)\in (M\times M)\backslash\Delta$, we can choose $\eta_1$, $\eta_2$ with disjoint supports such that the left-hand side is non-zero. This is a contradiction. Consequently, we have $K\equiv 0$.
\end{proof}

In general, the \emph{Schwartz kernel} of a linear operator $\GOp: \DR(M) \rightarrow \DR(M)$ is a distributional form $\GKer$ on $M\times M$ which satisfies\footnote{We may consider such class of $\GOp$'s, e.g., pseudo-differential operators, such that $\GKer$ exists and is unique (c.f., the well-known Schwartz kernel theorem).} 
\begin{equation*} 
\GOp(\eta)(x) = \int_{y\in M_1} \GKer(y,x)\eta(y)\quad\text{for all }\eta\in \DR(M)\text{ and }x\in M_2.
\end{equation*}
We consider the following conditions on $\GOp$ and $\GKer$:\label{ConditionsG}\Correct[caption={Typo},disable]{There is ''of'' missing downstairs.} 
\begin{center}
\begin{minipage}{.9\textwidth}
\begin{enumerate}[label=(G\arabic*)]
\item The Schwartz kernel $\GKer$ of $\GOp$ is a restriction of a smooth form on $\Bl_\Diag(M\times M)$.
\item $\Dd\circ \GOp + \GOp\circ \Dd = \iota_{\Harm} \circ \pi_\Harm - \Id$.
\item $(\GOp(\eta_1),\eta_2) = (-1)^{\eta_1}(\eta_1,\GOp(\eta_2))$ for all $\eta_1$, $\eta_2\in \DR(M)$.
\item $\GOp \circ \pi_\Harm = \pi_\Harm \circ \GOp = 0$.
\item $\GOp \circ \GOp = 0$.
\end{enumerate}
\end{minipage}
\end{center}
Clearly, (G1)--(G3) are equivalent to $\GKer$ being a Green kernel from Definition~\ref{Def:GreenKernel}. Conditions (G4) and (G5) play a crucial role in the vanishing results for the formal pushforward Maurer-Cartan element $\PMC$ in Section~\ref{Sec:Vanishing} --- the more conditions are satisfied, the more vanishing we get.

The following lemma will be used in the proof of the upcoming proposition.
\begin{Lemma}\label{Lem:Smoothing}
Let $\GOp_1$, $\GOp_2$ be two linear operators $\DR(M) \rightarrow \DR(M)$ with Schwartz kernels $\GKer_1$, $\GKer_2 \in \DR(\Bl_\Diag(M\times M))$. Then $\GOp:= \GOp_1 \circ \GOp_2$ is a smoothing operator, i.e., its Schwartz kernel $\GKer$ is a smooth form on $M\times M$. 
\end{Lemma}
\begin{proof}
It holds $\GKer(x_1,x_2) = \pm \int_x \GKer_2(x_1,x) \GKer_1(x,x_2)$. The lemma follows from properties of convolution. See \cite{MyPhD} for details.
\end{proof}

A version of the following proposition can be found in \cite{Mnev2009}.
\begin{Proposition}[Existence of special Green operator] \label{Prop:ExistenceG}
Every oriented closed Riemannian manifold $M$ admits an operator~$\GOp: \DR(M) \rightarrow \DR(M)$ which satisfies (G1)--(G5).
\end{Proposition}

\begin{proof}[Proof of Proposition \ref{Prop:ExistenceG}]
Because $\HKer$ is Poincar\'e dual to $\Diag$, we have for any closed $\kappa \in \DR_{\mathrm{c}}((M\times M)\backslash \Delta)$ the following:
$$ \int_{(M\times M)\backslash \Diag} \HKer \wedge \kappa = \int_{M\times M} \HKer \wedge \kappa = \int_{\Diag} \kappa = 0. $$
Poincar\'e duality for non-compact oriented manifolds (see \cite{BottTu1982}) implies that $\HKer$ is exact on $(M\times M)\backslash \Diag$. Because a manifold with boundary is homotopy equivalent to its interior,  the restriction of the blow-down map induces an isomorphism $\pi^* : \HDR((M\times M)\backslash \Diag) \rightarrow \HDR(\Bl_{\Diag}(M\times M))$. It follows that $(-1)^n \pi^* \HKer$ admits a primitive $\GKer \in \DR(\Bl_{\Diag}(M\times M))$. According to Proposition~\ref{Prop:GKer}, the corresponding $\GOp$ satisfies (G1) and (G2).

If we define
$$ \tilde{\GKer} := \frac{1}{2}(\GKer + (-1)^n \tau^* \GKer)\in \DR^{n-1}(\Bl_{\Diag}(M\times M)), $$
then $\tilde{\GKer}$ satisfies $\tau^* \tilde{\GKer} = (-1)^n \tilde{\GKer}$ and is still a primitive to $(-1)^n \pi^* \HKer$. Proposition~\ref{Prop:GKer} and \eqref{Eq:GSA} imply that the corresponding $\GOp$ satisfies (G1)--(G3).

Given $\GOp$ satisfying (G1)--(G3), we will now show that we can arrange (G4). Let us define
$$ \tilde{\GOp}:= (\Id - \pi_{\Harm}) \circ \GOp \circ (\Id - \pi_{\Harm}). $$
Then $\tilde{\GOp}$ is a Green operator because
$$ \Dd\circ\tilde{\GOp} + \tilde{\GOp} \circ \Dd = (\Id-\pi_{\Harm})\circ (\Dd\circ \GOp + \GOp \circ \Dd) \circ (\Id-\pi_{\Harm}) = \Id-\pi_{\Harm}. $$
Using \eqref{Eq:HKerHKer} and \eqref{Eq:GSA}, we see that $\tilde{\GOp}$ satisfies (G3). Using the intersection pairing and Proposition~\ref{Lemma:HKer}, we can write 
$$ \pi_{\Harm}(\eta) = \sum_{i=1}^m (-1)^{(n+\eta)\Le_i} (\Star \Le_i, \eta) \Le_i \quad\text{for all }\eta\in\DR(M),$$
and hence we have for all $\eta_1$, $\eta_2\in \DR(M)$ the following:
$$ \begin{aligned}
&\bigl(\GOp(\pi_{\Harm}(\eta_1)),\eta_2\bigr) = \sum_{i=1}^m (-1)^{(n+\eta_1)\Le_i} (\Star \Le_i, \eta_1) \bigl(\GOp(\Le_i), \eta_2\bigr) \\ 
& \qquad = \sum_{i=1}^m (-1)^{(n+1)\Le_i}  \int_{M\times M}\Pr_1^*(\Star \Le_i)\wedge \Pr_2^*(\GOp(\Le_i))\wedge \Pr_1^*(\eta_1)\wedge \Pr_2^*(\eta_2).
\end{aligned} $$
It follows that the Schwartz kernel of $\GOp \circ \pi_{\Harm}$ is the smooth form
$$ \Kern_{\GOp\circ \pi_{\Harm}} := \sum_{i=1}^m (-1)^{(n+1)\Le_i}\Pr_1^*(\Star \Le_i)\wedge \Pr_2^*(\GOp(\Le_i)). $$
Moreover, if we replace $\GOp$ with 
$\pi_{\Harm}\circ \GOp$, we get the smooth Schwartz kernel $\Kern_{\pi_{\Harm}\circ\GOp\circ \pi_{\Harm}}$ of $(\pi_{\Harm}\circ \GOp) \circ \pi_{\Harm}$. In the same way, but now using in addition \eqref{Eq:GSA}, we can write
\allowdisplaybreaks
\begin{align*}
&\bigl(\pi_{\Harm}(\GOp(\eta_1)),\eta_2 \bigr) = (-1)^{\eta_1}\bigl( \eta_1, \GOp(\pi_{\Harm}(\eta_2))\bigr) = (-1)^{\eta_1 \eta_2} \bigl(\GOp(\pi_{\Harm}(\eta_2)),\eta_1\bigr) \\
 &\qquad = \sum_{i=1}^m (-1)^{\eta_1 \eta_2 + (n+1)\Le_i} \int_{M\times M}\Pr_1^*(\Star \Le_i)\wedge \Pr_2^*(\GOp(\Le_i))\wedge \Pr_1^*(\eta_2)\wedge \Pr_2^*(\eta_1) \\
 &\qquad = \sum_{i=1}^m (-1)^{(n+1) \Le_i + n} \int_{M \times M} \Pr_2^*(\Star \Le_i) \wedge \Pr_1^*(\GOp(\Le_i))\wedge \Pr_1^*(\eta_1)\wedge\Pr_2^*(\eta_2),
\end{align*}
where in the last equality we pulled back the integral along the twist map. It follows that the Schwartz kernel of $\pi_{\Harm}\circ \GOp$ is the smooth form
$$ \Kern_{\pi_{\Harm}\circ\GOp} := \sum_{i=1}^m (-1)^{\Le_i}\Pr_1^*(\GOp(\Le_i))\wedge \Pr_2^*(\Star \Le_i). $$
The Schwartz kernel of $\tilde{\GOp} = \GOp - \pi_{\Harm}\circ \GOp - \GOp \circ \pi_{\Harm} + \pi_{\Harm}\circ\GOp\circ\pi_{\Harm}$ is then
$$ \tilde{\GKer} = \GKer - \pi^* \Kern_{\GOp \circ \pi_{\Harm}} - \pi^* \Kern_{\pi_{\Harm}\circ\GOp} + \pi^* \Kern_{\pi_{\Harm}\circ\GOp\circ\pi_{\Harm}}, $$
which is a smooth form on $\Bl_{\Diag}(M\times M)$. Therefore, $\tilde{\GKer}$ satisfies (G1)--(G4).

Given $\GOp$ satisfying (G1)--(G4), we will show that we can arrange (G5). The trick from \cite{Mnev2009} is to define
$$ \tilde{\GOp} = \GOp \Dd \GOp. $$
Applying (G1) and (G2) repeatedly, we compute
\begin{equation}\label{Eq:ExprGreen}
\begin{aligned}
\Dd \GOp \GOp \GOp \Dd &= \Dd \GOp \GOp - \Dd \GOp \GOp \Dd \GOp = \Dd \GOp \GOp - \Dd \GOp \GOp + \Dd \GOp \Dd \GOp \GOp \\ &= \Dd \GOp \GOp - \Dd \Dd \GOp \GOp \GOp = \Dd \GOp \GOp = \GOp - \GOp \Dd \GOp, 
\end{aligned}
\end{equation}
and hence
$$ \tilde{\GOp} = \GOp - \Dd \GOp \GOp \GOp \Dd. $$
Clearly, $\tilde{\GOp}$ satisfies (G1) and (G2). As for (G3), we compute
$$ (\eta_1, \tilde{\GOp} \eta_2) = (-1)^{\eta_1}(\GOp \eta_1, \Dd \GOp \eta_2) = (\Dd \GOp \eta_1, \GOp \eta_2) = (-1)^{\eta_1}(\tilde{\GOp}\eta_1,\eta_2). $$
As for (G5), we have
\begin{equation}\label{Eq:GSquared}
\begin{aligned}
\tilde{\GOp} \tilde{\GOp} &= \GOp \Dd \GOp (\GOp\Dd) \GOp = \GOp \Dd \GOp \GOp - \GOp \Dd (\GOp \Dd) \GOp \GOp \\
&= \GOp \Dd \GOp \GOp - \GOp \Dd \GOp \GOp + \GOp \Dd \Dd \GOp \GOp = 0.
\end{aligned}
\end{equation}
In order to show (G4), we have to compute the Schwartz kernel of $\Dd\GOp\GOp\GOp\Dd$. By Lemma \ref{Lem:Smoothing}, the Schwartz kernel $\TKer$ of $\TOp:= \GOp \GOp \GOp$ is a smooth form on $M \times M$. Therefore, Stokes' formula without the boundary term applies, and we get 
$$ (\Dd \TOp \Dd)\eta = \Dd \int^{M_1} \TKer \wedge \Dd \pi_1^*(\eta) = \int^{M_1} \Dd \TKer \wedge \Dd \pi_1^*(\eta) = (-1)^{\TKer} \int^{M_1} \Dd_1 \Dd \TKer \wedge \pi_1^*(\eta). $$
Here $\Dd_1: \DR(M\times M) \rightarrow \DR(M\times M)$ is the operator defined in local coordinates by
$$ \Dd_1\bigl(f(x,y)\Diff{x}^I\Diff{y}^J\bigr) = \sum_{i=1}^n \frac{\partial f}{\partial x^i}(x,y)\Diff{x^i}\Diff{x}^I \Diff{y}^J. $$
It follows that the Schwartz kernel $\tilde{\GKer}$ of $\tilde{\GOp}$ satisfies 
$$ \tilde{\GKer} = \GKer + (-1)^{n} \Dd_1 \Dd \TKer $$
and is a smooth $(n-1)$-form on $\Bl_\Diag(M\times M)$. Conditions (G1)--(G5) are satisfied. \qedhere
\end{proof}

\begin{Remark}[Property (G5) in dimensions $1$ and $2$]
In dimension $1$, every operator of degree $-1$ satisfies (G5) from degree reasons. In dimension $2$, every operator satisfying (G1) and (G2) satisfies (G5) as well, which follows from \eqref{Eq:ExprGreen} and \eqref{Eq:GSquared}.
\end{Remark}

\begin{Remark}[The standard Green kernel]
Consider the Hodge-de Rham Laplacian $\Delta = \Dd \circ \CoDd + \CoDd \circ \Dd : \DR(M) \rightarrow \DR(M)$ and its ``Green operator'' $\GOp_\Delta$ of degree~$0$ (see (v) of Remark \ref{Rem:GKer} for the collision of terminology) which was defined in~\cite[Definition 6.9]{Warner1983}~by
$$\GOp_{\Delta} := (\Restr{\Delta}{\Harm(M)^{\perp}})^{-1} \circ \pi_{\Harm(M)^\perp}, $$
where $\perp$ denotes the $L^2$-orthogonal complement. We introduce the \emph{standard Green operator} by
\begin{equation}\label{Eq:GStdStd}
\GOp_{\mathrm{std}} := -\CoDd\GOp_{\Delta}.
\end{equation}
Using the properties of $\GOp_{\Delta}$, $\Dd$ and $\CoDd$, one can show that $\GOpStd$ satisfies (G2)--(G5) (this will be shown in \cite{MyPhD}).

As for (G1), the author was able to show it for flat manifolds (:= locally isometric to $\R^n$) by transforming the following formula inspired by \cite{Harris2004} to blow-up coordinates and explicitly computing the integral and limit:
$$ \GKerStd = - \lim_{t\to 0} \int_t^\infty \frac{1}{2}\CoDd K_\tau \Diff{\tau}, $$
where $\KKer_t(x,y) = \sum_i (-1)^{n e_i} e^{-\lambda_i t} (\Star e_i)(x) \wedge e_i(y)$ is the heat kernel of $\Delta$ and~$e_i$ the $L^2$-orthonormal eigenbasis of $\Delta$ with eigenvalues $\lambda_i$ (the signs come from our convention for fiberwise integration, c.f., \eqref{Eq:HKerHKerHKer}).
%
\end{Remark}

\subsection{Formal pushforward Maurer-Cartan element \texorpdfstring{$\PMC$}{n}} \label{Section:Proof2} 

We first recall ribbon graphs and their labelings based on~\cite{Cieliebak2015}. 

\begin{Definition}[Ribbon graph]\label{Def:Ribbon}
A \emph{graph} $\Gamma$ is a quadruple $(V,H,\mathcal{V},\mathcal{E})$, where~$V$ is a finite set of vertices, $H$ a finite set of half-edges, $\mathcal{V}: H \rightarrow V$ the ``vertex map'' and $\mathcal{E}: H \rightarrow H$ with $\mathcal{E} \circ \mathcal{E} = \Id$ and without fixed points the ``edge map''. The preimage $\mathcal{E}^{-1}(h_1) = \{h_1, h_2\}$ for some $h_1$, $h_2\in H$ is called an \emph{edge}; the set of edges is denoted by $E$. We assume that the graphs are \emph{connected}, i.e., that for any $\Vert_1$, $\Vert_2 \in V$ there exists a path in $E$ connecting $\Vert_1$ to $\Vert_2$.

A \emph{ribbon graph} is a graph $\Gamma$ which is equipped with a free transitive action $\Z_{d(\Vert)}\ \rotatebox[origin=c]{-90}{$\circlearrowright$}\  \mathcal{V}^{-1}(\Vert)$ for every $\Vert\in V$, where $$d(\Vert) := \Abs{\mathcal{V}^{-1}(\Vert)} $$
is the \emph{valency of $\Vert$}. We denote by $\Succ: H \rightarrow H$ the bijection induced by $1\in \Z_{d(\Vert)}$ for every $\Vert\in V$.

For a ribbon graph $\Gamma$, consider the set of sequences $(h_n)_{n\in \Z} \subset H$ such that the following conditions holds:
$$ \forall n\in \Z: \quad  h_{n+1}= \begin{cases} 
\mathcal{E}(h_n) & n\text{ even,}\\                                
\Succ(h_n) & n\text{ odd.}
\end{cases}$$
Two such sequences $(h_n)_{n\in \Z}$ and $(h'_n)_{n\in \Z}$ are equivalent if and only if there exist $n_0$, $n'_0 \in \Z$ both even or both odd such that $h_{n_0} = h'_{n_0}$. An equivalence class $[(h_n)_{n\in \Z}]$ is called a \emph{boundary (or a boundary component) of $\Gamma$.} The set of boundaries of $\Gamma$ is denoted by $\Bdd \Gamma$.

An \emph{IE ribbon graph} is a ribbon graph $\Gamma$ together with the decomposition $V = V_{\mathrm{int}} \sqcup V_{\mathrm{ext}}$ into \emph{internal} and \emph{external vertices} $V_{\mathrm{int}}$ and $V_{\mathrm{ext}}$ such that $d(\Vert) = 1$ for all $\Vert\in V_{\mathrm{ext}}$, respectively. This decomposition induces the decomposition $E = E_{\mathrm{int}} \sqcup E_{\mathrm{ext}}$, where an edge $\Edge$ is internal if it connects two internal vertices and is external otherwise. We allow only graphs with \emph{at least one internal vertex}. We often identify an external vertex with its unique adjacent half-edge or the unique adjacent external edge; we call either of these an \emph{external leg}. For any $\Boundary \in \Bdd \Gamma$, we define the \emph{valency of $\Boundary$} by
$$ s(\Boundary) := \Abs{\mathcal{V}(\Boundary) \cap V_{\mathrm{ext}}}, $$
where $\mathcal{V}(\Boundary) = \{\mathcal{V}(h_n) \mid n\in \Z\}$. We also have the free transitive $\Z_{s(\Boundary)}$-action on $\mathcal{V}(\Boundary)\cap V_{\mathrm{ext}}$ mapping $\Vert\in \mathcal{V}(\Boundary) \cap V_{\mathrm{ext}}$ to the next external vertex in the sequence $(\mathcal{V}(h_n))_{n\in\Z}$. We will denote this action by $\mathcal{N}$ again.

We say that an IE ribbon graph $\Gamma$ is \emph{reduced} if $s(\Boundary) \ge 1$ for all $\Boundary\in \Bdd \Gamma$.

The following letters will be used to denote the numerical invariants of a graph:
$$\begin{aligned}
k &\;\,\dots && \text{the number of internal vertices}, \\
s &\;\,\dots && \makebox[\widthof{the number of}]{---\ditto---} \text{ external vertices.}\\
l &\;\,\dots && \makebox[\widthof{the number of}]{---\ditto---}\text{ boundary components,} \\
e &\;\,\dots && \makebox[\widthof{the number of}]{---\ditto---} \text{ internal edges.}
\end{aligned}$$
Moreover, we define the \emph{genus $g\in \N_0$} so that the following \emph{Euler formula} holds:
\begin{equation} \label{Eq:EulerFormula}
 k - e + l = 2 - 2g.
\end{equation}\par
We denote by $\RG_{klg}$ the set of isomorphism classes of connected IE ribbon graphs with fixed $k$, $l$, $g$. We let $\RRG_{klg} \subset \RG_{klg}$ be the subset of reduced graphs. For $m\in \N_0$, we denote by $\RG_{klg}^{(m)} \subset \RG_{klg}$ the set of isomorphism classes of connected IE ribbon graphs with all internal vertices \emph{m-valent}, i.e., with 
$$ d(\Vert) = m\quad\text{for all }\Vert\in V_{\mathrm{int}}. $$
The notation $\Gamma \in \RG_{klg}$ means that $\Gamma$ is a representative of an equivalence class $[\Gamma]\in \RG_{klg}$.
\end{Definition}

\begin{Remark}[On ribbon graphs]\phantomsection
\begin{RemarkList}
\item An $m$-valent ribbon graph with $m\ge 2$ has a unique decomposition $V = V_{\mathrm{int}}\sqcup V_{\mathrm{ext}}$, and hence we can omit writing IE.
\item In this text, we will use only reduced ribbon graphs. Non-reduced ribbon graphs may play a role in the extension of the theory of $\dIBL^\PMC(\CycC(\Harm))$ to non-reduced cyclic cochains  or in the weak $\IBLInfty$-theory (see Remarks \ref{Rem:Weak} and~\ref{Rem:NWG}).\qedhere
\end{RemarkList}
\end{Remark}

\begin{Def}[Labeling] \label{Def:Labeling}
A \emph{labeling} of an IE ribbon graph $\Gamma$ is the triple $L = (L_1,L_2,L_3)$, where $L_i$ have the following meanings: 
\begin{itemize}
 \item The symbol $L_1$ represents an ordering of internal vertices ($=:L_1^v$), and of boundary components ($=:L_1^b$). Given $L_1$, we write $V_{\mathrm{ext}} = \{\Vert_1, \dotsc, \Vert_k\}$, $\Bdd \Gamma = \{\Boundary_1, \dotsc, \Boundary_l\}$ and denote
$$ d_i := d(\Vert_i)\quad\text{and}\quad s_j:= s(\Boundary_j). $$
 \item The symbol $L_2$ represents an ordering and orientation of internal edges. Given $L_2$, we write $E_{\mathrm{int}} = \{\Edge_1, \dotsc, \Edge_e\}$ and $\Edge_i =\{h_{i,1}, h_{i,2}\}$ for $h_{i,1}$, $h_{i,2}\in H$.
 \item The symbol $L_3$ represents an ordering of half-edges at every internal vertex ($=:L_3^v$) and of external vertices at every boundary component ($=:L_3^b$), both compatible with the \emph{ribbon structure} (:=\,the $\Z_m$-actions). Given $L_3$, we write $\mathcal{V}^{-1}(\Vert) = \{h_{\Vert,1}, \dotsc,h_{\Vert,d(\Vert)} \}$ and $\mathcal{V}(\Boundary) \cap V_{\mathrm{ext}}= \{\Vert_{\Boundary,1},\dotsc,\Vert_{\Boundary,s(\Boundary)}\}$ with $\Succ(h_{\Vert,i}) = h_{\Vert,i+1}$ and $\Succ(\Vert_{\Boundary,j}) = \Vert_{\Boundary,j+1}$ for all $i$, $j$, respectively. 
\end{itemize}
We sometimes call $L_i$ \emph{partial labelings} and $L$ a \emph{full labeling}. A ribbon graph $\Gamma$ together with a labeling $L$ is called a \emph{labeled ribbon graph}.
\end{Def}

Given a ribbon graph $\Gamma$, one can construct an oriented surface with boundary~$\Sigma_\Gamma$ --- the thickening of $\Gamma$ --- in the obvious way and a closed oriented surface~$\hat{\Sigma}_\Gamma$ by gluing oriented disks to the oriented boundaries of $\Sigma_\Gamma$. If partial labelings~$L_1$ and~$L_2$ are given, we obtain the following chain complex with oriented chain groups (vector spaces over $\R$):
\begin{equation} \label{Eq:OrientationComplex}
\begin{tikzcd}
C_2:= \langle \Boundary_1,\dotsc,\Boundary_l \rangle \arrow{r}{\Bdd_2} & C_1:= \langle \Edge_1,\dotsc, \Edge_e \rangle \arrow{r}{\Bdd_1} &  C_0:= \langle \Vert_k,\dotsc, \Vert_1 \rangle.
\end{tikzcd}
\end{equation}
Here $\Boundary_i$ stands for the oriented disc glued to the $i$-th boundary component of~$\Sigma_\Gamma$ and now being mapped into $\hat{\Sigma}_\Gamma$, $\Edge_i$ stands for the $1$-simplex in $\hat{\Sigma}_\Gamma$ corresponding to the $i$-th internal edge, $\Vert_i$ stands for the $0$-simplex in $\hat{\Sigma}_\Gamma$ corresponding to the $i$-th internal vertex, and the boundary map $\Bdd$ is the ``geometric'' boundary operator. The homology of this chain complex is isomorphic to the singular homology $\H(\hat{\Sigma}):=\H(\hat{\Sigma}_\Gamma;\R)$.

The orientation of $C_i$ (:=\,the order of generators in \eqref{Eq:OrientationComplex}) induces naturally an orientation of $\H(\hat{\Sigma}_\Gamma)$. The construction from \cite[Appendix A]{Cieliebak2015} is as follows. We pick complements $H_i$ of $\Im(\Bdd_{i+1})$ in $\Ker(\Bdd_{i})$ and complements $V_i$ of $\Ker(\Bdd_i)$ in $C_i$ and write 
$$\begin{tikzcd}
C_2 = V_2 \oplus H_2 \arrow{r}{\Bdd_2} & C_1 = V_1 \oplus H_1 \oplus \Im(\Bdd_2) \arrow{r}{\Bdd_1} & C_0 = \Im(\Bdd_1) \oplus H_0.
\end{tikzcd}$$
We orient $V_i$ arbitrarily and transfer the orientation to $\Im(\Bdd_{i})$ via  $\Bdd_{i} : V_i \overset{\simeq}{\rightarrow} \Im(\Bdd_{i})$. Then, assuming the direct sum orientation, orienting $H_i$ is equivalent to orienting $C_i$, and we obtain the orientation of $\H_i(\hat{\Sigma}_\Gamma)$ via the canonical projection $\pi: H_i \overset{\simeq}{\rightarrow} \H_i(\hat{\Sigma}_\Gamma) = \Ker(\Bdd_i)/\Im(\Bdd_{i+1})$. This construction does not depend on the choices of complements and orientations of $V_i$.

\begin{Definition}[Compatibility of $L_1$ and $L_2$]\label{Def:CompatLabeling}
Given a ribbon graph $\Gamma$ with partial labelings $L_1$ and $L_2$, we say that \emph{$L_2$ is compatible with $L_1$} if the orientation on $\H(\widehat{\Sigma}_\Gamma)$ induced by \eqref{Eq:OrientationComplex} agrees with the canonical orientation $$\H(\widehat{\Sigma}_\Gamma) =  \langle \Vert_1 + \dotsb + \Vert_k \rangle  \oplus \H_1(\widehat{\Sigma}_\Gamma) \oplus \langle \Boundary_1 + \dotsb + \Boundary_l\rangle, $$
where $\H_1(\widehat{\Sigma}_\Gamma)$ is oriented using the canonical symplectic intersection form.
\end{Definition}

Given a labeled IE ribbon graph $\Gamma$, the set of half-edges adjacent to internal vertices $\mathcal{V}^{-1}(V_{\mathrm{int}})$ can be ordered in two ways corresponding to writing 
$$ 2e + (s_1+ \dotsb +s_l) = d_1+ \dotsb +d_k. $$
This leads to the following definition.
\begin{Definition}[Edge order and vertex order]\label{Def:EdgeVertex}
For a labeled IE ribbon graph~$\Gamma$, we define the following two orders on the set of half-edges $H$:
\begin{itemize}
\item \emph{Edge order:} The first $2e$ half-edges $h_{i,j}^{\mathrm{e}}$ are the ones from internal edges; they are ordered according to $L_2$. They are followed by blocks of $s_1$,~$\dotsc$, $s_l$ half-edges $h_{i,j}^{\Boundary}$ which come from the boundary components $i=1$,~$\dotsc$, $l$, respectively, and which are ordered according to $L_3^b$ inside the blocks. Schematically, we have
$$ (h_{1,1}^{\Edge} h_{1,2}^{\Edge})\dots(h_{e,1}^{\Edge} h_{e,2}^{\Edge})(h^{\Boundary}_{1,1}\dots h^{\Boundary}_{1,s_1})\dots(h^{\Boundary}_{l,1}\dots h^{\Boundary}_{l,s_l}). $$
\item \emph{Vertex order:} It consists of blocks of $d_1$,~$\dotsc$, $d_k$ half-edges $h_{i,j}^{\Vert}$ which come from internal vertices $1$,~$\dotsc$, $k$, and which are ordered according to $L_3^v$ inside the blocks. Schematically, we have
$$ (h_{1,1}^{\Vert}\dots h_{1,d_1}^{\Vert})\dots (h_{k,1}^{\Vert}\dots h_{k,d_k}^{\Vert}). $$
\end{itemize}
	
	We denote by $\sigma_L \in \Perm_{\Abs{H}}$ the \emph{permutation from the edge to the vertex order} which is constructed such that the $i$-th half-edge in the edge order is the same as the $\sigma_L(i)$-th half-edge in the vertex order.
\end{Definition}

From now on, we will consider only reduced trivalent ribbon graphs $\TRRG_{klg}$ with $k$, $l \ge 1$, $g\ge 0$. We will often use the equation
\begin{equation} \label{Eq:TrivalentFormula}
2e + s = 3k.
\end{equation}

\begin{Def}[Formal pushforward Maurer-Cartan element]
\label{Def:PushforwardMCdeRham}
Let $M$ be an oriented closed Riemannian manifold, and let $\GKer \in \DR^{n-1}(\Bl_\Diag(M\times M))$ be a Green kernel  from Definition~\ref{Def:GreenKernel}. The \emph{formal pushforward Maurer-Cartan element $\PMC$} is the collection of 
$$ \PMC_{lg}\in \hat{\Ext}_l \CycC(\Harm(M))\quad\text{for all }l\ge 1, g\ge 0 $$
defined on generating words $\omega_i = \alpha_{i1} \dots \alpha_{is_i}\in \BCyc \Harm(M)$, where $\alpha_{ij} = \SuspU \eta_{ij}$ with $\eta_{ij}\in \Harm(M)$ for $s_i\ge 1$ and $i=1, \ldots, l$, 
by the formula
\begin{equation}
\label{Eq:PushforwardMCdeRham}
\begin{aligned} 
& \PMC_{lg}(\Susp^{l} \omega_1 \otimes \dotsb \otimes \omega_l) \\ 
& \qquad := \frac{1}{l!}\sum_{[\Gamma]\in \TRRG_{klg}} \frac{1}{\Abs{\Aut(\Gamma)}} (-1)^{s(k,l) + P(\omega)} \sum_{L_1,\,L_3^b} (-1)^{\sigma_L} I(\sigma_L),
\end{aligned}
\end{equation}
which we explain as follows:
\begin{itemize}
\item 
The second sum is over all partial labelings $L_1$ and $L_3^b$ of a representative~$\Gamma$ of~$[\Gamma]$. In every summand, we complete $L_1$ and $L_3^b$  to a full labeling $L = (L_1, L_2, L_3)$ by picking an arbitrary $L_3^v$ and an arbitrary $L_2$ compatible with $L_1$.

\item Suppose that $\Gamma$ and $L_1$ are \emph{admissible} with respect to the input $\omega_1$, $\dotsc$, $\omega_l$; this means that $\Gamma$ has $l$ boundary components and that the $i$-th boundary component has valency $s_i$ for every $i=1$, $\dotsc$, $l$. In this case, denoting $\sigma = \sigma_L$, we define
\begin{equation} \label{Eq:ISigma}
I(\sigma_L) := \begin{multlined}[t]\int_{x_1,\dotsc,x_k} \GKer(x_{\xi(\sigma_1)},x_{\xi(\sigma_2)}) \dotsm \GKer(x_{\xi(\sigma_{2e-1})},x_{\xi(\sigma_{2e})}) \\ \eta_{11}(x_{\xi(\sigma_{2e+1})}) \dotsm \eta_{ls_{l}}(x_{\xi(\sigma_{2e+s})}),\end{multlined}
\end{equation}
where $\xi: \{1,\dotsc, 3k\} \rightarrow \{1,\dotsc,k\}$ is the function defined by 
$$ \xi(3j-2) = \xi(3j-1) = \xi(3j) := j $$ for all $j=1, \dotsc, k$, $s = s_1 + \dotsb + s_l$, $\eta_{}(x_i)$ denotes the pullback of $\eta$ along the canonical projection $\pi_i: M^{\times k} \rightarrow M$ to the $i$-th component $M_i$, $\GKer(x_i, x_j)$ denotes the pullback of $\GKer$ along $\pi_{i} \times \pi_j : M^{\times k} \rightarrow M_i \times M_j$, and $\int_{x_1,\dotsc,x_k}$ denotes the integral of an $nk$-form over $k$ copies of $M$.

If $\Gamma$ and $L_1$ are not admissible, then we set $I(\sigma_L):= 0$.
\item $s(k,l) := k + kl(n-1) + \frac{1}{2} k(k-1) n \mod 2$.
\item $P(\omega) :=  \sum_{i=1}^l \sum_{j=1}^{s_i} (s - s_1 - \dotsb - s_{i-1} - j) \eta_{ij} \mod 2$.
\end{itemize}
\end{Def}
\noindent 
In order to show that $\PMC_{lg}$ is well-defined and that the collection $(\PMC_{lg})$ satisfies Definition~\ref{Def:MaurerCartan} for $\dIBL(\CycC(\Harm(M)))$, there are several things to check:
\begin{PlainList}
 \item The integral $I(\sigma_L)$ converges.
 \item The sums are finite.
 \item The product $(-1)^{\sigma_L} I(\sigma_L)$ is independent of the choice of $L_3^v$ and $L_2$ compatible with $L_1$.
 \item The sum over labelings is independent of the chosen representative $\Gamma$ in an isomorphism class from $\TRRG_{klg}$.
  \item The map $\PMC_{lg}: \BCyc\Harm(M)[3-n]^{\otimes l} \rightarrow \R$ is graded symmetric on permutations of its inputs $\Susp\omega_i$.
 \item The map $\PMC_{lg}$ is graded symmetric on cyclic permutations of the components $\alpha_{ij}$ of each $\omega_{i}$.
 \item The degree condition 1) from Definition \ref{Def:MaurerCartan} holds with $d = n-3$.
 \item The filtration-degree condition 2) from Definition \ref{Def:MaurerCartan} holds with $\gamma = 2$.
 \item The Maurer-Cartan equation \eqref{Eq:MaurerCartanEquation} holds.
\end{PlainList}

Conditions 1) and 9) will be proven in~\cite{Cieliebak2018} using the theory of iterated blow-ups. In this text, \underline{we will take 1) and 9) for granted.} 

\begin{Lemma} \label{Lem:MCCond}
Assuming 1), the conditions 2) -- 8) hold.
\end{Lemma}
\begin{proof}

As for 2), the fixed input $\omega_1$, $\dotsc$, $\omega_l$ fixes the number $s$ of external vertices of $\Gamma$ by admissibility. Expressing $e$ from~\eqref{Eq:EulerFormula} and plugging it  in~\eqref{Eq:TrivalentFormula} gives
\begin{equation} \label{Eq:MCk}
k = s + 2 l + 4g - 4.
\end{equation}
We see that all parameters are fixed. Now, there is only finitely many elements with fixed $s$ in $\TRRG_{klg}$, and each of them has only finitely many labelings. Therefore, the sums are finite.

As for 3), we have to consider the orientation of the complex~\eqref{Eq:OrientationComplex}. Clearly, if two $L_2$'s are compatible with $L_1$, then they differ by an even number of the following operations: a~transposition of two internal edges or a change of the orientation of an internal edge. The former operation introduces no sign in $(-1)^{\sigma_L}$ but generates the sign $(-1)^{n-1}$ in $I(\sigma_L)$ from swapping the corresponding~$\GKer$'s. The latter operation induces the sign $-1$ in $(-1)^{\sigma_L}$ and the sign $(-1)^n$ in $I(\sigma_L)$ from the symmetry $\GKer(x,y) = (-1)^n \GKer(y,x)$. Because the overall signs in $(-1)^{\sigma_L} I(\sigma_L)$ are the same, an even number of these operations preserves $(-1)^{\sigma_L} I(\sigma_L)$. This implies the independence of an $L_2$ compatible with~$L_1$. A~change in $L_3^v$ produces no sign in $(-1)^{\sigma_L}$ because every internal vertex is trivalent and a cyclic permutation of an odd number of elements is even. The integral $I(\sigma_L)$ remains unchanged because the change in $\sigma_L$ is compensated by the composition with $\xi$. Independence of the choice of $L_3^v$ follows.

As for 4), every isomorphism of ribbon graphs $\Gamma \rightarrow \Gamma'$ induces the bijection $L\mapsto L'$ of compatible labelings such that $\sigma_L = \sigma_{L'}$ ($L'$ is the ``pushforward'' labeling). The independence of the choice of a representative of $[\Gamma]$ follows.

As for 5), let $\mu\in \Perm_l$ be a permutation of the inputs $\Susp \omega_1$, $\dotsc$, $\Susp \omega_l$. The set of graphs which admit an admissible labeling is the same for both $\PMC_{lg}(\Susp^l \omega_1 \otimes \dotsb \otimes \omega_l)$ and $\PMC_{lg}(\Susp^l \omega_{\sigma_{1}^{-1}} \otimes \dotsb \otimes \omega_{\sigma_l^{-1}})$; we will pick one such $\Gamma$ and study the admissible labelings $L$ and $L'$, respectively. We write $\eta_i = \eta_{i1}\dots \eta_{i s_i}$ and $\Omega_i = \Susp \omega_i$ for all $i$, $j$, and denote by $I'(\sigma_{L'})$ the integral in the definition of $\PMC_{lg}(\Susp^l \omega_{\mu_1^{-1}} \otimes \dotsb \otimes \omega_{\mu_l^{-1}})$. Let $\tilde{\mu}\in \Perm_{3k}$ be the permutation which acts as the identity on $1$, $\dotsc$, $2e$ and as the block permutation determined by $\mu$ on $2e + 1$,~$\dotsc$, $2e+s$ divided into $l$ blocks of lengths $s_1$, $\dotsc$, $s_l$. For any $\sigma\in \Perm_{3k}$, we have
$$ \begin{aligned}
I'(\sigma) & = \begin{multlined}[t]\int_{x_1,\dotsc,x_k}\!\! \GKer(x_{\xi(\sigma_1)},x_{\xi(\sigma_2)}) \dotsm \GKer(x_{\xi(\sigma_{2e-1})},x_{\xi(\sigma_{2e})}) \\ \eta_{\mu_1^{-1} 1}(x_{\xi(\sigma_{2e+1})}) \dots \eta_{\mu_l^{-1} s_{\mu_l^{-1}}}(x_{\xi(\sigma_{2e+s})})\end{multlined} \\
 & = \begin{multlined}[t]\varepsilon(\mu,\eta) \int_{x_1,\dotsc,x_k} \!\! \GKer(x_{\xi((\sigma\circ\tilde{\mu})_1)},x_{\xi((\sigma\circ\tilde{\mu})_2)}) \dotsm \GKer(x_{\xi((\sigma\circ \tilde{\mu})_{2e-1})},x_{\xi((\sigma\circ\tilde{\mu})_{2e})}) \\ \eta_{11}(x_{\xi((\sigma\circ \tilde{\mu})_{2e+1})})\dots  \eta_{ls_l}(x_{\xi((\sigma\circ \tilde{\mu})_{2e+s})})\end{multlined} \\
 & = \varepsilon(\mu,\eta) I(\sigma \circ \tilde{\mu}).
\end{aligned} $$
The precomposition with $\tilde{\mu}$ corresponds to a bijection $(L_1,L_3^b)\mapsto (L_1', {L_3^{b}}')$ of partial labelings for $\PMC_{lg}(\Susp^l \omega_1 \dots \omega_l)$ and $\PMC_{lg}(\Susp^l \omega_{\mu_1^{-1}} \otimes \dotsb \otimes \omega_{\mu_l^{-1}})$, respectively. However, if $L_2$ is compatible with $L_1$, then in order to get an $L_2'$ compatible with $L_1'$, the labeling~$L_2$ has to be altered by as many operations of switching two internal edges or changing the orientation of an internal edge as there are transpositions in $\mu$. We explained in the proof of 3) that this produces the sign $(-1)^{(n-1)\mu}$ in $(-1)^{\sigma_{L'}}I(\sigma_{L'})$. Therefore, after the choice of compatible $L_2$ and~$L_2'$, we have
$$ (-1)^{\sigma_{L'}}I'(\sigma_{L'}) = (-1)^{(n-1)\mu} (-1)^{\tilde{\mu}} \varepsilon(\mu, \eta) (-1)^{\sigma_L} I(\sigma_L). $$ 
If we view $\eta$ as $\eta_{11} \dots \eta_{l s_l}$, we can understand $(-1)^{P(\omega)}$ as the Koszul sign $\varepsilon(\SuspU, \eta)$. Similarly, we write $(-1)^{P(\mu(\omega))}=\varepsilon(\SuspU,\mu(\eta))$, where we first view $\eta$ as $\eta_1 \otimes \dotsb \otimes \eta_l$ to apply $\mu$ and then as the list of components $\eta_{ij}$ to compute the Koszul sign (this is a little ambiguity in our notation). If we denote by $\overline{\mu}$ the permutation of $1$, $\dotsc$, $s$ permuting the $l$ blocks of lengths $s_1$, $\dotsc$, $s_l$ according to $\mu$, then $\overline{\mu}$ has the same sign as $\tilde{\mu}$, and the decomposition of $\varepsilon(\SuspU,\mu(\eta))$ into the moves
$$ \begin{aligned}
\SuspU_{1} \dots \SuspU_{s} \eta_{\mu_1^{-1}1} \dots \eta_{\mu_l^{-1} s_{\mu_l^{-1}}} &\xrightarrow{(1)} \SuspU_{\overline{\mu}_{1}} \dots \SuspU_{\overline{\mu}_{s}} \eta_{11} \dots \eta_{l s_l} \xrightarrow{(2)} \SuspU_{\overline{\mu}_1} \eta_{11} \dots \SuspU_{\overline{\mu}_s} \eta_{l s_l} \\ &\xrightarrow{(3)} \SuspU_{1} \eta_{\mu_1^{-1}1} \dots \SuspU_{s} \eta_{\mu_l^{-1} s_l} \end{aligned} $$
shows that
$$ (-1)^{P(\mu(\omega))} = \underbrace{(-1)^{\tilde{\mu}}\varepsilon(\mu,\eta)}_{(1)} \underbrace{(-1)^{P(\omega)}}_{(2)}\underbrace{\varepsilon(\mu,\omega)}_{(3)}. $$
Using this, we write
$$ (-1)^{P(\mu(\omega))} (-1)^{\sigma_{L'}} I'(\sigma_{L'}) = \varepsilon(\mu,\omega) (-1)^{(n-1)\mu} (-1)^{P(\omega)}(-1)^{\sigma_L} I(\sigma_L), $$
and compute
$$ \begin{aligned}
& \PMC_{lg}(\Omega_{\mu_1^{-1}} \otimes \dotsb \otimes \Omega_{\mu_l^{-1}}) \\ & \qquad =\varepsilon(\mu(\Susp), \mu(\omega)) \PMC_{lg}(\Susp^l \omega_{\mu_1^{-1}} \otimes \dotsb \otimes \omega_{\mu_l^{-1}}) \\ 
& \qquad= \varepsilon(\mu(\Susp), \mu(\omega)) (-1)^{\Abs{s}\mu} \varepsilon(\mu,\omega) \PMC_{lg}(\Susp^l \omega_{1} \otimes \dotsb \otimes \omega_{l}) \\
& \qquad= \underbrace{\varepsilon(\mu(\Susp), \mu(\omega))}_{(1)} \underbrace{(-1)^{\Abs{s}\mu} \varepsilon(\mu,\omega)}_{(2)} \underbrace{\varepsilon(\Susp,\omega)}_{(3)} \PMC_{lg}(\Susp\omega_{1} \otimes \dotsb \otimes \Susp \omega_{l}) \\
& \qquad= \varepsilon(\mu,\Omega) \PMC_{lg}(\Omega_{1} \otimes \dotsb \otimes \Omega_{l}).
\end{aligned} $$
We used $\Abs{s} = n-1\text{ mod }2$, and the last equality follows from the decomposition of $\varepsilon(\mu,\Omega)$ into the moves
$$  \begin{multlined} \Susp_1 \omega_1 \dots \Susp_l \omega_l \xrightarrow{(3)} \Susp_1 \dots \Susp_l \omega_1 \dots \omega_l \xrightarrow{(2)} \Susp_{\mu_1^{-1}}\dots \Susp_{\mu_l^{-1}} \omega_{\mu_1^{-1}} \dots \omega_{\mu_l^{-1}} \\ \xrightarrow{(1)} \Susp_{\mu_1^{-1}} \omega_{\mu_1^{-1}} \dots \Susp_{\mu_l^{-1}} \omega_{\mu_l^{-1}}. \end{multlined}$$
This proves the symmetry of $\PMC_{lg}$.

As for 6), fix an $i=1$, $\dotsc$, $l$ and let $\mu\in \Perm_{s_i}$ be a cyclic permutation permuting the components of $\omega_i = \alpha_{i1}\dots \alpha_{is_i}$. Similarly to the previous case, we denote by~$\tilde{\mu}$ the corresponding permutation of $1$, $\dotsc$, $3k$ and get a bijection $(L_1,L_3^b) \mapsto (L_1'=L_1,{L_3^b}')$ of admissible labelings of a given graph $\Gamma$ for $\PMC_{lg}(\Susp^l \omega_1 \otimes \dotsb \otimes \alpha_{i1}\dots \alpha_{is_i} \otimes \dotsb \otimes \omega_l)$ and $\PMC_{lg}(\Susp^l \omega_1 \otimes \dotsb \otimes \alpha_{i\mu_1^{-1}} \dots \alpha_{i\mu_{s_i}^{-1}}\otimes \dotsb \otimes \omega_l)$, respectively. This time, there is no change in $L_1$, and thus we can take $L_2'=L_2$, producing no sign. Therefore, we have
$$ (-1)^{\sigma_{L'}}I'(\sigma_{L'}) = (-1)^{\tilde{\mu}}\varepsilon(\mu,\eta_i)(-1)^{\sigma_{L}}I(\sigma_{L}), $$
where $\varepsilon(\mu,\eta_i)$ comes from permuting the forms in $I'(\sigma_{L'})$. Further, we deduce
$$ (-1)^{P(\mu(\omega))} = (-1)^{\tilde{\mu}} \varepsilon(\mu,\eta_i) (-1)^{P(\omega)} \varepsilon(\mu,\omega_i), $$
and hence
$$ \begin{aligned}
&\PMC_{lg}(\Susp^l \omega_1 \otimes \dotsb \otimes \alpha_{i\mu_1^{-1}} \dots \alpha_{i\mu_{s_i}^{-1}}\otimes \dotsb \otimes \omega_l) \\
&\qquad = \varepsilon(\mu,\omega_i) \PMC_{lg}(\Susp^l \omega_1 \otimes \dotsb \otimes \alpha_{i1}\dots \alpha_{is_i}\otimes\dotsb\otimes \omega_l ). \end{aligned}$$
This shows the symmetry of $\PMC_{lg}$ on cyclic permutations of the components of~$\omega_i$.

As for 7), suppose that $\PMC_{lg}(\Susp^l \omega_1 \otimes \dotsb \otimes \omega_l) \neq 0$, and let $D$ denote the total form-degree of the input $\eta_{11}$, $\dotsc$, $\eta_{l s_l}\in \Harm(M)$; i.e., we define
\begin{equation*}
 D := \deg(\eta_{11}) + \dotsb + \deg(\eta_{1s_1}) + \dotsb + \deg(\eta_{l 1}) + \dotsb + \deg(\eta_{l s_l}).
\end{equation*}
Clearly, we must have
\begin{equation}  \label{Eq:TotDeg}
nk = (n-1) e  + D, 
\end{equation}
where the left-hand side is the dimension of $M^{\times k}$ and the right-hand side the form-degree of the integrand of $I(\sigma_L)$. If we plug in $e$ from~\eqref{Eq:EulerFormula} and $k$ from~\eqref{Eq:MCk}, we get
$$ \begin{aligned} D &= nk - (n-1) e \\ 
   &= nk - (n-1)(k+l+2g-2) \\
   & = k - (n-1)(l+2g-2) \\
   & = s + 2l +4g - 4 - (n-1)(l+2g-2) \\
   & = s - (n-3)(l+2g-2). \end{aligned}$$
It follows that
$$ \Abs{\PMC_{lg}} = \Abs{\Susp^l} + \Abs{\omega_1} +  \dotsb + \Abs{\omega_l} =  l(n-3) + D - s = - 2(n-3)(g-1). $$
This is exactly the degree from Definition~\ref{Def:MaurerCartan}. 

As for 8), if $\PMC_{lg}(\Susp^l \omega_1 \otimes \dotsb\otimes \omega_l)\neq 0$, then
$$ s = k - 2l -4g -4 \ge 1 + 2(2-2g-l) = 1 + 2 \chi_{0lg}, $$
and hence $\PMC_{lg}\in \mathcal{F}_{1+2\chi_{0lg}}\hat{\Ext}_l \CycC$ for the filtration induced from the dual of the filtration of $\BCyc \Harm$ by weights. Therefore, we get 
$$ \|\PMC_{lg}\| \ge 1 + 2\chi_{0lg} > 2\chi_{0lg} \quad\text{for all }l\ge 1, g\ge 0. $$
This finishes the proof.
\end{proof}


\begin{Definition}[Vertices of types A, B, C and some special graphs]\label{Def:Graphs}
Let $\Gamma \in \TRG_{klg}$ be a trivalent ribbon graph and $\Vert$ its internal vertex. We say that~$\Vert$ is of \emph{type $A$, $B$ or~$C$} if it is connected to precisely $1$, $2$ or~$3$ internal vertices, respectively (see Figure~\ref{Fig:Vertices}). The graph $\Gamma$ is called (see Figures~\ref{Fig:Types} and~\ref{Fig:YOk}):
\begin{itemize}
\item a \emph{tree} if $[\Gamma] \in \RG_{k10}$ for some $k\ge 1$;
\item \emph{circular} if $[\Gamma]\in \RG_{k20}$ for some $k\ge 1$;
\item the \emph{$Y$-graph} is the unique tree with $k=1$;
\item an \emph{$O_k$-graph} if $\Gamma$ is circular with $k$ internal vertices and no $A$-vertex.
\end{itemize}
We denote the $Y$-graph simply by $Y$.
\end{Definition}

{ \begingroup
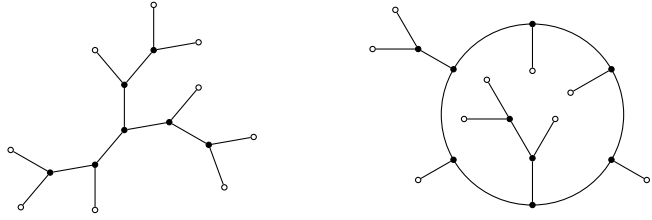
\begin{figure}[t]
\centering
\begin{subfigure}{0.4\textwidth}
\centering
\begin{tikzpicture}

\def\rad{1cm}
\def\angdist{140}
\def\posa{90}
\def\extlen{0.6cm}

\tikzset{point/.style = {draw, circle, fill=black, minimum size=2pt,inner sep=0pt}}

\coordinate[point] (C1) at (0,0) {};
\path (C1) node[point] (C11)  at +(\posa:\extlen) {};
\path (C1) node[point] (C12)  at +(\posa+\angdist:\extlen) {};
\path (C1) node[point] (C13)  at +(\posa+2*\angdist:\extlen) {};
\draw (C1) -- (C11);
\draw (C1) -- (C12);
\draw (C1) -- (C13);

\path (C11) node[point] (C111)  at +(-180+\posa+\angdist:\extlen) {};
\path (C11) node[point,style={fill=white}] (C112)  at +(-180+\posa-\angdist:\extlen) {};
\draw (C11) -- (C111);
\draw (C11) -- (C112);

\path (C12) node[point] (C121)  at +(-180+\posa+\angdist + \angdist:\extlen) {};
\path (C12) node[point,style={fill=white}] (C122)  at +(-180+\posa+\angdist - \angdist:\extlen) {};
\draw (C12) -- (C121);
\draw (C12) -- (C122);

\path (C13) node[point] (C131)  at +(-180+\posa+2*\angdist + \angdist:\extlen) {};
\path (C13) node[point,style={fill=white}] (C132)  at +(-180+\posa+2*\angdist - \angdist:\extlen) {};
\draw (C13) -- (C131);
\draw (C13) -- (C132);

\path (C121) node[point,style={fill=white}] (C1211)  at +(-180-180+\posa+\angdist + \angdist+\angdist:\extlen) {};
\path (C121) node[point,style={fill=white}] (C1212)  at +(-180-180+\posa+\angdist + \angdist-\angdist:\extlen) {};
\draw (C121) -- (C1211);
\draw (C121) -- (C1212);

\path (C111) node[point,style={fill=white}] (C1112)  at +(-180-180+\posa+\angdist+\angdist:\extlen) {};
\path (C111) node[point,style={fill=white}] (C1111)  at +(-180-180+\posa+\angdist-\angdist:\extlen) {};
\draw (C111) -- (C1111);
\draw (C111) -- (C1112);

\path (C131) node[point,style={fill=white}] (C1312)  at +(-180-180+\posa+2*\angdist + \angdist+\angdist:\extlen) {};
\path (C131) node[point,style={fill=white}] (C1311)  at +(-180-180+\posa+2*\angdist + \angdist-\angdist:\extlen) {};
\draw (C131) -- (C1311);
\draw (C131) -- (C1312);

\end{tikzpicture}
\end{subfigure}
\begin{subfigure}{0.4\textwidth}
\centering
\begin{tikzpicture}

\def\rad{1.2cm}
\def\angdist{60}
\def\posa{90}
\def\extlen{0.6cm}

\tikzset{point/.style = {draw, circle, fill=black, minimum size=2pt,inner sep=0pt}}

\coordinate (C1) at (0,0) {};

\path (C1) node[point] (C11)  at +(\posa:\rad) {};
\path (C1) node[point] (C12)  at +(\posa+\angdist:\rad) {};
\path (C1) node[point] (C13)  at +(\posa+2*\angdist:\rad) {};
\path (C1) node[point] (C14)  at +(\posa+3*\angdist:\rad) {};
\path (C1) node[point] (C15)  at +(\posa+4*\angdist:\rad) {};
\path (C1) node[point] (C16)  at +(\posa+5*\angdist:\rad) {};

\draw (C1) circle (\rad);

\path (C1) to coordinate[pos=1-\extlen/\rad,point,style={fill=white}] (C111) (C11);
\draw (C11) -- (C111);

\path (C1) to coordinate[pos=1+\extlen/\rad,point] (C121) (C12);
\path (C121) node[point,style={fill=white}] (C1211)  at +(\posa+\angdist+30:\extlen) {};
\path (C121) node[point,style={fill=white}] (C1212)  at +(\posa+\angdist-30:\extlen) {};

\draw (C12) -- (C121);
\draw (C121) -- (C1211);
\draw (C121) -- (C1212);

\path (C1) to coordinate[pos=1+\extlen/\rad,point,style={fill=white}] (C131) (C13);
\draw (C13) -- (C131);

\path (C1) to coordinate[pos=1-\extlen/\rad,point] (C141) (C14);
\path (C141) node[point] (C142)  at +(120:\extlen) {};
\path (C141) node[point,style={fill=white}] (C143)  at +(60:\extlen) {};
\path (C142) node[point,style={fill=white}] (C1421)  at +(180:\extlen) {};
\path (C142) node[point,style={fill=white}] (C1422)  at +(120:\extlen) {};
\draw (C14) -- (C141);
\draw (C141) -- (C142);
\draw (C141) -- (C143);
\draw (C142) -- (C1421);
\draw (C142) -- (C1422);

\path (C1) to coordinate[pos=1+\extlen/\rad,point,style={fill=white}] (C151) (C15);
\draw (C15) -- (C151);

\path (C1) to coordinate[pos=1-\extlen/\rad,point,style={fill=white}] (C161) (C16);
\draw (C16) -- (C161);

\end{tikzpicture}
\end{subfigure}
\caption{A tree and a circular graph. Internal vertices are denoted with a full dot and external vertices with an empty dot.}\label{Fig:Types}
\end{figure}
\endgroup }
{ \begingroup
\begin{figure}[t]
\centering
\begin{subfigure}{0.4\textwidth}
\centering
\begin{tikzpicture}

\def\rad{1cm}
\def\angdist{120}
\def\posa{90}
\def\extlen{1.3cm}

\tikzset{point/.style = {draw, circle, fill=black, minimum size=2pt,inner sep=0pt}}

\coordinate[point] (C1) at (0,0) {};
\path (C1) node[point,style={fill=white}] (C11)  at +(\posa:\extlen) {};
\path (C1) node[point,style={fill=white}] (C12)  at +(\posa+\angdist:\extlen) {};
\path (C1) node[point,style={fill=white}] (C13)  at +(\posa+2*\angdist:\extlen) {};
\draw (C1) -- (C11);
\draw (C1) -- (C12);
\draw (C1) -- (C13);

\end{tikzpicture}
\end{subfigure}
\begin{subfigure}{0.4\textwidth}
\centering
\begin{tikzpicture}

\def\rad{1.2cm}
\def\angdist{60}
\def\posa{90}
\def\extlen{0.6cm}

\tikzset{point/.style = {draw, circle, fill=black, minimum size=2pt,inner sep=0pt}}

\coordinate (C1) at (0,0) {};

\path (C1) node[point] (C11)  at +(\posa:\rad) {};
\path (C1) node[point] (C12)  at +(\posa+\angdist:\rad) {};
\path (C1) node[point] (C13)  at +(\posa+2*\angdist:\rad) {};
\path (C1) node[point] (C14)  at +(\posa+3*\angdist:\rad) {};
\path (C1) node[point] (C15)  at +(\posa+4*\angdist:\rad) {};
\path (C1) node[point] (C16)  at +(\posa+5*\angdist:\rad) {};

\draw (C1) circle (\rad);

\path (C1) to coordinate[pos=1-\extlen/\rad,point,style={fill=white}] (C111) (C11);
\draw (C11) -- (C111);

\path (C1) to coordinate[pos=1+\extlen/\rad,point,style={fill=white}] (C121) (C12);

\draw (C12) -- (C121);

\path (C1) to coordinate[pos=1-\extlen/\rad,point,style={fill=white}] (C131) (C13);
\draw (C13) -- (C131);

\path (C1) to coordinate[pos=1-\extlen/\rad,point,style={fill=white}] (C141) (C14);
\draw (C14) -- (C141);

\path (C1) to coordinate[pos=1+\extlen/\rad,point,style={fill=white}] (C151) (C15);
\draw (C15) -- (C151);

\path (C1) to coordinate[pos=1+\extlen/\rad,point,style={fill=white}] (C161) (C16);
\draw (C16) -- (C161);

\end{tikzpicture}
\end{subfigure}
\caption{The $Y$-graph and an $O_6$-graph.}\label{Fig:YOk}
\end{figure}
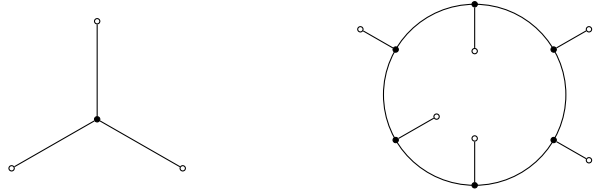
\endgroup }

\begin{Remark}[On $A$, $B$, $C$ vertices and special graphs]
We observe the following:
\begin{RemarkList}
\item A trivalent graph $\Gamma \neq Y$ has each internal vertex of type $A$, $B$ or $C$.
\item The term $\PMC_{10}$ is a sum over trees, and the term $\MC_{10}$ is the contribution of the $Y$-graph to $\PMC_{10}$ (see Proposition~\ref{Prop:FormalPushforwardProp} below). The term $\PMC_{20}$ is a sum over circular graphs.\qedhere
\end{RemarkList}
\end{Remark}

Wee will also denote by $A$, $B$, $C$ the numbers of internal vertices of the corresponding type. Under the change of variables
\begin{equation} \label{Eq:ChangeOfVariables}
\begin{aligned}
  s &= 2 A + B, \\
  e &= B + \frac{1}{2} A + \frac{3}{2} C, \\
  k &= A + B + C,
\end{aligned}
\end{equation}
the trivalent formula~\eqref{Eq:TrivalentFormula} becomes trivial and the Euler formula~\eqref{Eq:EulerFormula} becomes
\begin{equation} \label{Eq:GenusFormulaa}
 C - A = 2l - 4 + 4g.
\end{equation}

\begin{Proposition}[Formal pushforward Maurer-Cartan element]\label{Prop:FormalPushforwardProp}
The pushforward Maurer-Cartan element $\PMC = (\PMC_{lg})$ is a Maurer-Cartan element for $\dIBL(\Harm(M))$ which is compatible with $\MC$. In particular, the $\AInfty$-algebra $\Harm(M)_\PMC$ is homologically unital and augmented.
\end{Proposition}
\begin{proof}
The fact that $\PMC$ is a Maurer-Cartan element for $\dIBL(\Harm(M))$ follows from Lemma~\ref{Lem:MCCond} assuming 1) and 9) from \cite{Cieliebak2018}. \Correct[caption={DONE Redundant check of vanishing for $w\le 2$}]{Change it here because the definition of compatible changed!! In particular, the vanishing on words of length $1$ and $2$ follows from the filtration degree condition. Do not have to show it here.}

As for the compatibility with $\MC$,
the only graph contributing to $\PMC_{10}(\Susp \alpha_1 \alpha_2 \alpha_3)$  is the $Y$-graph with $k=1$. The group $\Aut(Y)$ consists of three rotations, and there is only one possible~$L_1$, no~$L_2$ and three~$L_3^b$. In Definition~\ref{Def:PushforwardMCdeRham}, we get $s(1,1) = n-2$, $(-1)^{\sigma_L} = 1$ because a cyclic permutation of an odd number of elements is even, and also $P(\alpha_1\alpha_2\alpha_3) =\eta_2$. Finally, we compute
\begin{align*}
 \PMC_{10}(\Susp \alpha_1 \alpha_2 \alpha_3) &= \frac{1}{3} (-1)^{n-2 + \eta_2}\sum_{L_3^b} \int_x \alpha_1(x_{\xi(\sigma_1)})\alpha_2(x_{\xi(\sigma_2)})\alpha_3(x_{\xi(\sigma_3)}) \\ 
 & = (-1)^{n-2 + \eta_2}\int_M \eta_1 \wedge \eta_2 \wedge \eta_3\\[\jot]
 & = \MC_{10}(\Susp \alpha_1 \alpha_2 \alpha_3).\qedhere
\end{align*}
\end{proof}
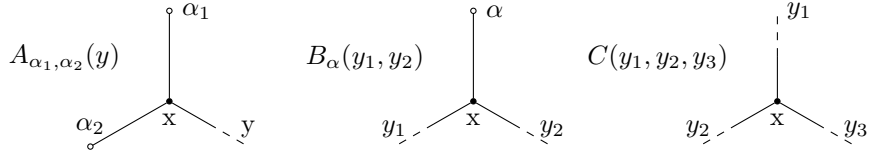
\begin{figure}[t]
\centering
\begin{tikzpicture}

\tikzset{point/.style = {draw, circle, fill=black, minimum size=2pt,inner sep=0pt}}

\def\dist{4}
\def\startangle{90}
\def\leglength{1.2}
\def\dashpart{0.5}

\node[point,label={[xshift=-0.5cm,yshift=0.25cm]125:$A_{\alpha_1,\alpha_2}(y)$},label={-90:x}] (A) at (0,0) {};
\node[point,label={[xshift=-0.5cm,yshift=0.25cm]125:$B_\alpha(y_1,y_2)$},label={-90:x}] (B) at ($(A)+(\dist,0)$) {};
\node[point,label={[xshift=-0.5cm,yshift=0.25cm]125:$C(y_1,y_2,y_3)$},label={-90:x}] (C) at ($(A)+(2*\dist,0)$) {};

\path (A) -- +(\startangle:\dashpart*\leglength) coordinate (A11) -- +(\startangle:\leglength) coordinate (A12);
\path (A) -- +(\startangle+120:\dashpart*\leglength) coordinate (A21) -- +(\startangle+120:\leglength) coordinate (A22);
\path (A) -- +(\startangle+240:\dashpart*\leglength) coordinate (A31) -- +(\startangle+240:\leglength) coordinate (A32);

\draw (A) -- (A12) node[point,style={fill=white},label={0:$\alpha_1$}] {};
\draw (A) -- (A22) node[point,style={fill=white},label={$\alpha_2$}] {};
\draw (A) -- (A31); \draw[dashed] (A31) -- (A32) node[at end,above]{y};

\path (B) -- +(\startangle:\dashpart*\leglength) coordinate (B11) -- +(\startangle:\leglength) coordinate (B12);
\path (B) -- +(\startangle+120:\dashpart*\leglength) coordinate (B21) -- +(\startangle+120:\leglength) coordinate (B22);
\path (B) -- +(\startangle+240:\dashpart*\leglength) coordinate (B31) -- +(\startangle+240:\leglength) coordinate (B32);

\draw (B) -- (B12) node[point,style={fill=white},label={0:$\alpha$}] {};
\draw (B) -- (B21); \draw[dashed] (B21) -- (B22) node[at end,above]{$y_1$};
\draw (B) -- (B31); \draw[dashed] (B31) -- (B32) node[at end,above]{$y_2$};

\path (C) -- +(\startangle:\dashpart*\leglength) coordinate (C11) -- +(\startangle:\leglength) coordinate (C12);
\path (C) -- +(\startangle+120:\dashpart*\leglength) coordinate (C21) -- +(\startangle+120:\leglength) coordinate (C22);
\path (C) -- +(\startangle+240:\dashpart*\leglength) coordinate (C31) -- +(\startangle+240:\leglength) coordinate (C32);

\draw (C) -- (C11); \draw[dashed] (C11) -- (C12) node[at end,above,right]{$y_1$};
\draw (C) -- (C21); \draw[dashed] (C21) -- (C22) node[at end,above]{$y_2$};
\draw (C) -- (C31); \draw[dashed] (C31) -- (C32) node[at end,above] {$y_3$};


\end{tikzpicture}
\caption{Trivalent vertices of types $A$, $B$ and $C$ with the corresponding forms $A_{\alpha_1,\alpha_2}$, $B_\alpha$ and $C$, respectively.}\label{Fig:Vertices}
\end{figure}
\begin{Definition}[Contributions of A, B, C vertices]\label{Def:Contributions}
Consider an internal vertex of type A, B or C as in Figure~\ref{Fig:Vertices}. We define the following smooth forms on $M$, $M^{\times 2}$ and $M^{\times 3}$, respectively:\footnote{The definitions can be made precise in local coordinates. Smoothness of $A_{\alpha_1, \alpha_2}$ is clear, smoothness of $B_\alpha$ follows from Lemma~\ref{Lem:Smoothing}, and smoothness of $C$ can be shown by a similar argument.}
$$\begin{aligned} 
 A_{\alpha_1, \alpha_2}(y)&:= \int_x \GKer(y,x)\eta_1(x)\eta_2(x),\\ 
 B_\alpha(y_1,y_2) &:= \int_x \GKer(y_1,x)\GKer(x,y_2)\eta(x), \\
 C(y_1,y_2,y_3) &:= \int_x \GKer(x,y_1)\GKer(x,y_2)\GKer(x,y_3).
\end{aligned}$$
\end{Definition}

\subsection{Results about vanishing of \texorpdfstring{$\PMC$}{n}}
\label{Sec:Vanishing}

In the situation of Definition~\ref{Def:PushforwardMCdeRham}, let $\Gamma \in \TRRG_{klg}$ be a reduced trivalent ribbon graph, $L=(L_1,L_2,L_3)$ its labeling, $x_i$ the integration variable associated to the $i$-th internal vertex, $\GKer(x_i,x_j)$ the Green kernel on the oriented internal edge between $x_i$ and~$x_j$, and $\alpha_{ij}\in \Harm(M)[1]$ the harmonic form on the $j$-th external vertex on the $i$-th boundary component. Recall that we denote by $\omega_i = \Susp \alpha_{i1}\dotsc\alpha_{is_i}$ the $i$-th input of $\PMC_{lg}$ and by $D$ the total form-degree of all inputs. 

By saying ``\emph{a graph vanishes}'' we mean that $I(\sigma_L) = 0$ in the given context.

\begin{Proposition}[Vanishing of graphs with $\NOne$] \label{Prop:PMCEqualsMC}
In the setting of Definition~\ref{Def:PushforwardMCdeRham}, suppose that the following condition is satisfied: 
\begin{description}
\item[($V_{\NOne}$)] Every graph $\Gamma \in \TRRG_{klg}$, $\Gamma \neq Y$ which has $\NOne = \SuspU 1\in \Harm(M)[1]$ at an external vertex vanishes. 
\end{description}
Then $\PMC$ is strictly reduced, and the following holds depending on the dimension $n$:
\begin{enumerate}[label=(\alph*)]
 \item For $n>3$: All graphs which are not trees or circular vanish. Therefore, $\PMC_{lg} = 0$ for all $(l,g)\neq (1,0)$, $(2,0)$, and it follows that all higher operations~$\OPQ_{1lg}^\PMC$ vanish on the chain level.
  \item For $n=3$: A tree vanishes unless all $\eta_{1}$,~$\dotsc$, $\eta_{s}$ are one-forms. Therefore, $\PMC_{10}(\Susp \alpha_1 \dots \alpha_s) \neq 0$ implies $\deg(\eta_i)=1$ for all $i$.
 \item For $n<3$: All trees except for $Y$ vanish. Therefore, we have $\PMC_{10} = \MC_{10}$, and consequently $\OPQ_{110}^\PMC = \OPQ_{110}^\MC$.
\end{enumerate}
Moreover, we have 
\begin{enumerate}[resume,label=(\alph*)]
 \item A circular graph vanishes unless all $\eta_{11}$, $\dotsc$, $\eta_{2s_2}$ are one-forms. Therefore, $\PMC_{20}(\Susp^2 \alpha_{11}\dots \alpha_{1s_1} \otimes \alpha_{21}\dots \alpha_{2s_2})\neq 0$ implies $\deg(\eta_{ij})=1$ for all $i$, $j$.
\end{enumerate}
In addition to $(V_{\NOne})$, suppose that $\HDR^1(M) = 0$. Then:
\begin{enumerate}[resume,label=(\alph*)]
 \item All circular graphs vanish. Therefore, we have $\PMC_{20} = 0$, and consequently $\OPQ_{120}^\PMC = \OPQ_{120}$.
\item For $n\le 6$: All trees except for $Y$ vanish. Therefore, we have $\PMC_{10} = \MC_{10}$, and consequently $\OPQ_{110}^\PMC = \OPQ_{110}^\MC$. 
\end{enumerate} 
\end{Proposition}

\begin{proof}
The proof is just combinatorics with $D$. Suppose that a trivalent ribbon graph $\Gamma\neq Y$ does not vanish on the input $\omega_1$, $\dotsc$, $\omega_l$. Because all external vertices of $\Gamma$ are adjacent to an $A$-vertex or a $B$-vertex, the assumption $(V_{\NOne})$ implies $D\ge s$, where $s$ is the total number of external vertices. A combination of~\eqref{Eq:TotDeg} and~\eqref{Eq:TrivalentFormula} yields
$$ nk - (n-1)e = D \ge s = 3k - 2e\quad\Equiv\quad(n-3)k \ge (n-3)e. $$
\begin{ProofList}[label=(\alph*)]
\item For $n>3$, we get $k \ge e$, which implies that $\Gamma$ is either a tree or a circular graph.
\item If $\Gamma$ is a tree, then $s = k + 2$ and $e = k-1$. From~\eqref{Eq:TotDeg} we get
\begin{equation} \label{Eq:TreeEq}
D = nk - (n-1)(k-1) = k+n-1.  
\end{equation}
Now $D$ is the sum of $s=k+2$ form-degrees $\deg(\eta_{ij})>0$, and hence~\eqref{Eq:TreeEq} for $n=3$ implies that $\deg(\eta_{ij}) = 1$ for all $i$, $j$.
\item For $n<3$, we get $e \ge k$, which implies that $\Gamma$ is not a tree.
\item If $\Gamma$ is a circular graph, then $e=k=s$, and we get using~\eqref{Eq:TotDeg} that
$$ D = nk - (n-1)k = k. $$
Here $D$ is the sum of $s=k$ form-degrees $\deg(\eta_{ij})>0$, and hence $\deg(\eta_{ij})=1$ for all $i$, $j$.
\end{ProofList}
We will now assume, in addition, that $\Harm^1(M) \simeq \HDR^1(M) = 0$.\Add[caption={DONE Add in addition}]{This is now assumed in addition to (1)!}
\begin{ProofList}[resume, label=(\alph*)]
\item We must have $D\ge 2 s$, which is in contradiction with $D = s$ for a circular graph. Therefore, $\PMC_{20} = 0$.
\item Finally, for a tree $\Gamma \neq Y$, we have
\begin{equation*}
 k+n-1 = D \ge 2 s = 2(k + 2)\quad\Equiv\quad  n-5 \ge k. \end{equation*}
This finishes the proof of the proposition.\qedhere
\end{ProofList}
\end{proof}


\begin{Proposition}[Green kernel with (G4) and (G5)]\label{Prop:COne}
In the setting of Definition~\ref{Def:PushforwardMCdeRham}, suppose that the Green kernel $\GKer$ satisfies (G4) and (G5). Then the condition ($V_{\NOne}$), and hence Proposition~\ref{Prop:PMCEqualsMC} holds.
\end{Proposition}
\begin{proof}
It is easy to see that $A_{\alpha_1,\alpha_2} = \GOp(\eta_1 \wedge \eta_2)$ for all $\alpha_1$, $ \alpha_2\in \Harm(M)[1]$, and that $-B_{\NOne}$ is the Schwartz kernel of $\GOp \circ \GOp$. Therefore, (G4) and (G5) imply $A_{\alpha_1,\NOne}=0$ and $B_{\NOne} = 0$, respectively.

As for the integral $I(\sigma_L)$,  one has to apply the Fubini theorem in order to integrate out single vertices $A_{\alpha_1, \NOne}$ and $B_{\NOne}$. This step relies on $L^1$-integrability of the integrand which follows from \cite{Cieliebak2018} (the integrand comes from a smooth form on a compact manifold with corners).
\end{proof}
\begin{Proposition}[Vanishing of $A$-vertices]\label{Prop:Avertexvanish}
In the setting of Definition~\ref{Def:PushforwardMCdeRham}, suppose that the following condition is satisfied:
\begin{description}
\item[($V_A$)] Every graph with an $A$-vertex vanishes.
\end{description}
Then we have $\PMC_{10} = \MC_{10}$, and the only contribution to $\PMC_{20}(\Susp^2 \alpha_{11}\dots \alpha_{1s_1} \otimes \alpha_{21}\dots \alpha_{2s_2})$ comes from $O_k$-graphs with $k = s_1 + s_2 = D$.
\end{Proposition}

\begin{proof}
The only trees and circular graphs which are not excluded by the assumption are the $Y$-graph and $O_k$-graphs, respectively (the external branches contract). The condition on form-degrees is obtained as in the proof of Proposition~\ref{Prop:PMCEqualsMC}.

To argue that $I(\sigma_L)=0$, we again need $L^1$-integrability as in the proof of Proposition \ref{Prop:COne}.
\end{proof}

\begin{Remark}[Integrability for trees]
Given a tree, we can start at a leaf and write $I(\sigma_L)$ as an iterative integral of contributions $A_{\alpha_1,\alpha_2}$ for $\alpha_1$, $\alpha_2 \in \DR(M)$. These are smooth forms, and hence integrability is guaranteed. Therefore, the result $\PMC_{10} = \MC_{10}$ is independent of the convergence results from~\cite{Cieliebak2018}.
\end{Remark}

\begin{Proposition}[$1$-connected geometrically formal manifolds] \label{Prop:GeomForm}
Let $M$ be a geometrically formal $n$-manifold and $\GKer$ a Green kernel satisfying (G4) and (G5) (it exists by Proposition~\ref{Prop:ExistenceG}). If $\HDR^1(M) = 0$, then the following holds:
\begin{description}
\item[$(n\neq 2)$]  All $Y \neq \Gamma \in \RRG_{klg}$ with $k$, $l\ge 1$, $g\ge 0$ vanish, and hence $\PMC = \MC$.
\item[$(n=2)$] All $Y\neq \Gamma \in \RRG_{kl0}$ with $k$, $l\ge 1$ vanish, and hence $\PMC_{l0} = \MC_{l0}$ for all~$l\ge 1$.
\end{description}
\end{Proposition}
\begin{proof}
Given $\eta_1$, $\eta_2 \in \Harm$, geometric formality implies $\eta_1 \wedge \eta_2 \in \Harm$, and hence $A_{\alpha_1,\alpha_2} = \GOp(\eta_1\wedge\eta_2) = 0$. We see that $(V_{\NOne})$ and $(V_{A})$ are satisfied, and hence the implications of Propositions~\ref{Prop:PMCEqualsMC} and~\ref{Prop:Avertexvanish} hold. The claim for $n>3$ follows.

As for $n=3$, Poincar\'e duality implies $\HDR^2(M;\R)=0$. Therefore, the total form-degree $D$ satisfies $D= n B$, where $B$ is the number of $B$-vertices. We see using \eqref{Eq:GenusFormulaa} that \eqref{Eq:TotDeg} is equivalent to
\begin{equation}\label{Eq:VerticesEq}
B+\frac{1}{2}(3-n) C = D = nB\quad\Equiv\quad (n-1)B = \frac{1}{2}(3-n) C.
\end{equation}
It follows that $B=0$, and hence all reduced graphs vanish.

As for $n=2$, we get from \eqref{Eq:VerticesEq} and \eqref{Eq:GenusFormulaa} that $B\ge l$ is equivalent to $g\ge 1$.
\qedhere
\end{proof}


\begin{Remark}[$\AInfty$-homotopy transfer]  \label{Rem:RemMu}
In~\cite{Cieliebak2018}, it will be shown that the $\AInfty$-algebra $\Harm(M)_\PMC = (\Harm(M),(\mu_k))$ induced by $\PMC_{10}$ agrees with the $\AInfty$-algebra obtained by the $\AInfty$-homotopy transfer
$$\begin{tikzcd}
\biggl(\ \begin{gathered}\DR(M) \\ m_1,\  m_2\end{gathered}\ \biggr)\arrow[rightsquigarrow]{r} & 
\biggl(\ \begin{gathered}
\Harm(M) \\
\mu_1\equiv 0,\ \mu_2 = \pi_\Harm m_2 (\iota_\Harm, \iota_\Harm),\ \mu_3,\ \dotsc
\end{gathered}\ \biggr)
\end{tikzcd}$$
 using the homotopy retract (see~\cite{Vallette2012})
$$\begin{tikzcd}[column sep=large]
(\DR(M),m_1) \arrow[loop left,"\GOp"] \arrow[shift left]{r}{\pi_\Harm}  & \arrow[shift left]{l}{\iota_\Harm} (\Harm(M),m_1 \equiv 0).
\end{tikzcd}$$
The operation $\mu_k$ of the transferred $\AInfty$-structure is computed as a sum over planar trees with a root and $k$ leaves decorated by $\iota_\Harm$ at the leaves, $\pi_\Harm$ at the root and~$\GOp$ at the internal edges (see \cite{Akaho2007}). The result of \cite{Cieliebak2018} is plausible because the part of $\PMC_{10}$ contributing to $\mu_k$ is a sum over trivalent ribbon trees with $k+1$ leaves.

In~\cite{Cieliebak2018}, they will also show that $\iota_1:= \iota_\Harm: \Harm \rightarrow \DR$ extends to an $\AInfty$-quasi-isomorphism $(\iota_k)_{k\ge 1}$ from $(\Harm,(\mu_k))$ to $(\DR,m_1,m_2)$. The induced chain map on the dual cyclic bar complexes is then the map $\HTP_{110}^\MC$ coming from the $\IBLInfty$-theory in the Introduction.
\end{Remark}


\begin{Proposition}[Twisted boundary operator for formal manifolds]\label{Prop:Formal}
In the setting of Definition~\ref{Def:PushforwardMCdeRham}, suppose that $M$ is formal in the sense of rational homotopy theory. Then there is a quasi-isomorphism
$$\begin{tikzcd}
\HHTP_{110}: (\CDBCyc \HDR(M)[3-n], \OPQ_{110}^\MC)\arrow{r}{} & (\CDBCyc \Harm(M)[3-n],\OPQ_{110}^\PMC). \end{tikzcd}$$
\end{Proposition}

\begin{proof}
Formality of $M$ is equivalent to the existence of a zig-zag of dga-quasi-isomorphisms (see \cite{Vallette2012}) 
$$\begin{tikzcd}[column sep=normal] (\H_{\mathrm{dR}}(M),m_1\equiv 0, m_2) \arrow[rightsquigarrow]{r} &\bullet\quad\dotsb\quad\bullet &\arrow[rightsquigarrow]{l} (\DR(M),m_1,m_2). \end{tikzcd}$$
Because a dga-quasi-isomorphism has a homotopy inverse in the category of $\AInfty$-algebras, we get a direct $\AInfty$-quasi-isomorphism 
$$\begin{tikzcd}
(g_k):\quad (\DR(M),m_1,m_2) \arrow[rightsquigarrow]{r} & (\H_{\mathrm{dR}}(M),m_1\equiv 0, m_2).
\end{tikzcd}$$
Precomposing with $(\iota_k)$ from Remark~\ref{Rem:RemMu}, we get the $\AInfty$-isomorphism 
$$\begin{tikzcd}
(h_k):\quad (\Harm(M),(\mu_k)) \arrow[rightsquigarrow]{r} & (\HDR(M),m_1\equiv 0,m_2). \end{tikzcd}$$
This induces the quasi-isomorphism $\HHTP_{110}$ of the corresponding cyclic cochain complexes (see \cite{MyPhD} for details).
\end{proof}

\begin{Remark}[On formality]\phantomsection
Geometrically formal manifolds include $\Sph{n}$, $\C P^n$ and Lie groups (see~\cite{Kotschick2000}). Any geometrically formal manifold is formal. Every simply-connected manifold of dimension at most $6$ is formal (see \cite{Miller1979}).
\end{Remark}
%
%

\subsection{Conjectured relation to string topology}\label{Sec:StringTopology}
\Add[inline,caption={DONE Simplify string top. and cyc. hom. comparison}]{Add somewhere that the map from $(\CDBCyc \HDR(M),b^*)$ to $(C(\StringSpace M),\Bdd)$ with no degree shifts does the job (after removing all degree shifts from definitions).}

Given a smooth connected oriented $n$-dimensional manifold $M$, we consider the equivariant homology of the free loop space $\Loop M := \{\gamma: \Sph{1} \rightarrow M \text{ continuous}\}$ with respect to the reparametrization action of $\Sph{1}$. It is defined as the singular homology of the Borel construction 
$$ \LoopBorel M := \EG\Sph{1} \times_{\Sph{1}} \Loop M := (\EG\Sph{1}\times \Loop M)/\Sph{1}, $$
where $\EG\Sph{1} = \Sph{\infty} \rightarrow \BG \Sph{1} = \CP^\infty$ is a model for the universal bundle for $\Sph{1}$, and we quotient out the diagonal action. We denote this homology by
$$ \StringH_*(\Loop M) := \H_*(\LoopBorel M). $$
The ``geometric versions'' of the homologies were defined in \cite{Sullivan1999} as the degree shifts
$$ \GeomH(\Loop M) := \H(\Loop M)[n]\quad\text{and}\quad \GeomStringH(\Loop M):= \StringH(\Loop M)[n]. $$

There is the \emph{loop product} $\LoopPr: \GeomH(\Loop M)^{\otimes 2} \rightarrow \GeomH(\Loop M)$ of degree $0$ which makes $\GeomH(\Loop M)$ into a graded commutative dga. There is also the \emph{loop coproduct} $\LoopCoPr: \ConstRedGeomH(\Loop M) \rightarrow \ConstRedGeomH(\Loop M)^{\otimes 2}$ of degree $1-2n$ which is graded cocommutative and coassociative and is a derivation of $\LoopPr$. The geometric construction of~$\LoopPr$ and~$\LoopCoPr$ on transverse smooth chains in $\Loop M$ was described in~\cite{Sullivan1999} and~\cite{Basu2011}, respectively. Here, the symbol $\ConstRedGeomH(\Loop M)$ stands for the degree shifted relative homology
$$  \ConstRedGeomH(\Loop M):=\H(\Loop M, M)[n] $$
with respect to constant loops $M \hookrightarrow \Loop M$. The geometric construction of~$\LoopCoPr$ does not work on the whole $\GeomH(\Loop M)$ because of the phenomenon of ``vanishing of small loops'' depicted in \cite[Figure 4, p.\,13]{Cieliebak2007}.
\Add[caption={DONE Reduced loop product}]{Is it really true? What about the example of torus.}

The projection $\EG\Sph{1} \times \Loop M \rightarrow \LoopBorel M$ is an $\Sph{1}$-principal bundle and thus induces a Gysin sequence. This sequence written using the geometric versions reads
\begin{equation}\label{Eq:Gysin}
\begin{tikzcd}
\dots\arrow{r}& \GeomH_i \arrow{r}{\Erase} & \GeomStringH_i \arrow{r}{\cap c} & \GeomStringH_{i-2} \arrow{r}{\Mark} & \GeomH_{i-1} \arrow{r} & \dots,
\end{tikzcd}
\end{equation}
where the map $\Mark$ adds a marked point in each string in a family in all possible positions, the map $\Erase$ erases the marked point of each string in a family, $c\in \StringH_{2}(\Loop M)$ is the Euler class of the circle bundle and $\cap$ the cap product.

The \emph{string bracket} $\tilde{\StringOp}_2: \GeomStringH(\Loop M)^{\otimes 2}\rightarrow \GeomStringH(\Loop M)$ and the \emph{string cobracket} $\tilde{\StringCoOp}_2: \ConstRedGeomStringH(\Loop M) \rightarrow \ConstRedGeomStringH(\Loop M)^{\otimes 2}$ are defined by
$$ \tilde{\StringOp}_2 := \Erase \circ \LoopPr \circ \Mark^{\otimes 2}\quad\text{and}\quad\tilde{\StringCoOp}_2 := \Erase^{\otimes 2} \circ \nu \circ \Mark. $$
Here, the symbol $\ConstRedGeomStringH(\Loop M)$ stands for the degree shifted relative $\Sph{1}$-equivariant homology
$$ \ConstRedGeomStringH(\Loop M) := \underbrace{\StringH(\EG \Sph{1} \times_{\Sph{1}} \Loop M, \EG \Sph{1} \times_{\Sph{1}} M)}_{\displaystyle =: \ConstRedStringH(\Loop M)}[n]. $$
Because $\Abs{\Mark} = 1$ and $\Abs{\Erase} = 0$, we have for all $\xi \in \ConstRedGeomStringH(\Loop M)$ and $\xi_1$, $\xi_2 \in \GeomStringH$ the relations\Correct[caption={DONE Typo}]{$\xi$ in the exponent should be $\Abs{\xi_1}$}
\begin{equation}\label{Eq:StringOpCoOp}
\begin{aligned}
\tilde{\StringOp}_2(\xi_1,\xi_2) &= (-1)^{\Abs{\xi_1}} \Erase(\Mark(\xi_1)\LoopPr\Mark(\xi_2)), \\
\tilde{\StringCoOp}_2(\xi) & = \sum \Erase(\nu^{1}) \otimes \Erase(\nu^2),
\end{aligned}
\end{equation}
where we write $\nu(\Mark(\xi)) = \sum \nu^1 \otimes \nu^2$. The operations $\tilde{\StringOp}_2$ and $\tilde{\StringCoOp}_2$ have degrees~$2$ and $2-2n$ with respect to the grading on $\GeomStringH(\Loop M)$, respectively. In fact, we will consider $\tilde{\StringOp}_2$ and $\tilde{\StringCoOp}_2$ given by \eqref{Eq:StringOpCoOp} as operations on the even degree shift $\StringH(\Loop M)[2-n] = \GeomStringH(\Loop M)[2-2n]$, which have degrees $2(2-n)$ and $0$, respectively. The symbols $\StringOp_2$ and $\StringCoOp_2$ will denote their degree shifts to $\StringH(\Loop M)[3-n]$, which have degrees of an $\IBL$-algebra from Definition \ref{Def:IBLInfty}.
%

In work in progress \cite{Cieliebak2018b}, they consider the map 
$$I_{\lambda,*}: \CycH_{-* - 1}(\DR^*(M)) \longrightarrow \StringCoH^*(\Loop M; \R)$$ defined on the chain level as a cyclic version of Chen's iterated integrals. Recall that $\CycH_{-*-1}(\DR^*) = \H_*(\CDBCyc \DR^*, \Hd^*)$,
where $\Hd: \B \DR = \bigoplus_{k\ge 1} \DR[1]^{\otimes k} \rightarrow \B \DR $ is the Hochschild differential of the de Rham dga $(\DR^*,m_1,m_2)$, and the grading on $\CycH_*(\DR^*)$ was chosen such that $\CycH_*(\DR^*) \simeq \ClasCycH_*(\DR^*)$ for the classical cyclic homology of a dga. They prove in \cite{Cieliebak2018b} that if $M$ is simply-connected, then the map~$I_{\lambda,*}$ induces an isomorphism $\RedCycH_{-*-1}(\DR^*(M))) \simeq \RedStringCoH^*(\Loop M)$, where 
$$ \RedStringCoH^*(\Loop M) := \StringCoH^*(\EG \Sph{1} \times_{\Sph{1}} \Loop M, \EG \Sph{1} \times_{\Sph{1}} \{x_0\}) $$
is the \emph{reduced $\Sph{1}$-equivariant cohomology} with respect to a base point $x_0 \in M$ (the constant loop at $x_0$). Dualizing their map, we obtain the isomorphism 
\begin{equation}\label{Eq:StringIsom}
\RedCycCoH^{-*-1}(\DR^*(M))\simeq \RedStringH_*(\Loop M; \R).
\end{equation}


Suppose from now on that $M$ is closed. Pick a Riemannian metric and a Green kernel $\GKer \in \DR^{n-1}(\Bl_\Diag(M\times M))$. We will assume that $\GKer$ satisfes (G1)--(G5) from Section~\ref{Section:Proof1} so that the formal pushforward Maurer-Cartan element~$\PMC$ is strictly reduced, and hence the twisted reduced $\IBLInfty$-algebra $\dIBL^\PMC\bigl(\RedCycC(\Harm)\bigr)$ and the induced $\IBL$-algebra $\IBL(\HIBL^{\PMC,\mathrm{red}}(\CycC(\Harm)))$ are well-defined.
Recall that $\HIBL^\PMC_*(\CycC(\Harm)) = \CycH_{n-3-*}(\Harm_\PMC)$, where $\Harm_\PMC$ is the $\AInfty$-algebra on $\Harm$ twisted by $\PMC_{10}$. From~\cite{Cieliebak2018}, we have
\begin{equation}\label{Eq:IBLIsom}
\CycCoH^*(\Harm^*(M)_\PMC) \simeq \CycCoH^*(\DR^*(M)).
\end{equation}

A combination of \eqref{Eq:StringIsom} and \eqref{Eq:IBLIsom} gives the following version of the string topology conjecture from the Introduction.

\begin{Conjecture}[String topology conjecture for simply-connected manifold]\label{Conj:StringTopology}
Let $M$ be an oriented closed manifold of dimension $n$. There is a chain map 
$$(C_*^{\mathrm{sing}}(\Loop_{\Sph{1}}M; \R),\Bdd)\longrightarrow (\CDBCyc\Harm(M),\OPQ_{110}^\PMC), $$
where $C_*^{\mathrm{sing}}$ denotes the (smooth) singular chain complex and $\Bdd$ the standard boundary operator, which, if $M$ is simply-connected, satisfies the following:
\begin{itemize}
\item It induces an isomorphism $\RedStringH_*(\Loop M; \R)[2-n] \simeq \HIBL_*^{\PMC,\mathrm{red}}\bigl(\CycC(\Harm(M))\bigr)$.
\item It intertwines $\StringOp_2$ on $\StringH(\Loop M; \R)$ and $\OPQ_{210}$.
\item The pullback of $\OPQ_{120}^\PMC$ to $\RedStringH(\Loop M; \R)$ is compatible with $\StringCoOp_2$ on $\ConstRedStringH(\Loop M; \R)$ under the morphism induced by the inclusion $(\Loop M, x_0) \rightarrow (\Loop M,M)$.
\end{itemize}
\end{Conjecture}

\begin{Remark}[On string topology conjecture]\phantomsection
\begin{RemarkList}
\item The conjecture can be interpreted as follows. There is an $\IBL$-structure on $\RedStringH(\Loop M;\R)$ compatible with Chas-Sullivan operations, and the $\IBLInfty$-algebra $\dIBL^\PMC(\RedCycC(\Harm(M)))$ is its chain model.
\item The loop coproduct $\tau$ is geometrically defined only on $\ConstRedStringH(\Loop M)$; the conjecture thus provides an extension of $\StringCoOp_2$ to $\RedStringH(\Loop M)$. In~\cite{Basu2011}, it is shown that the geometric definition of $\tau$ can be extended to $\H(\Loop M)$ for manifolds with zero Euler characteristic, i.e., $\chi(M) = 0$. This extension depends on the choice of a non-vanishing vector field on $M$. By homotopy invariance (see (v) below), our extension of $\StringCoOp_2$ should not depend on the Green kernel $\GKer$.
\item The loop product $\LoopPr$ is geometrically defined on $\H(\Loop M)$; however, it does not always induce an associative product on $\RedH(\Loop M) = \H(\Loop M, x_0)$. Indeed, the examples of $\T^2$ (see \cite{Basu2011}) and $\Sph{3}$ (see \cite{Sullivan1999}) show that $\H(x_0;\R) \subset \H(\Loop M;\R)$ is not an ideal with respect to $\LoopPr$. By \cite{Tamanoi2010}, this does not happen when $\chi(M) \neq 0$, and hence, in this case, $\LoopPr$ restricts to $\H(\Loop M, x_0; \R)$.
\item The computation for $\Sph{n}$ with $n\ge 2$ and the computation for $\CP^n$ in Section~\ref{Section:Computation} support the conjecture. The computation for $\Sph{1}$ in Section~\ref{Section:HomSphere} provides a counterexample for non-simply-connected $M$. In \cite{MyPhD}, surfaces of genus~$g\ge 1$ will be considered.

\item  We expect that if $M_1$ and $M_2$ are homotopy equivalent, then the $\IBLInfty$-algebras $\dIBL^\PMC(\CycC(\HDR(M_1)))$ and $\dIBL^\PMC(\CycC(\HDR(M_2)))$ are $\IBLInfty$-homotopy equivalent.
\qedhere
\end{RemarkList}
\end{Remark}

\clearpage
\section{Explicit computations}

\label{Section:Computation}
In Section~\ref{Sec:GreenSphere}, we solve the differential equation for the Green kernel~$\GKer$ for~$\Sph{n}$ (Proposition~\ref{Prop:GKerSph}) using the Relative Poincar\'e Lemma (Lemma~\ref{Lem:ChainHtpy}). In the rest of the section, we will be showing that $\GKer$ satisfies all properties of the Green kernel (Proposition~\ref{Proposition:GreenKernel}); the most work is to show that $\GKer$ extends smoothly to the blow-up (Proposition~\ref{Prop:GKerBdd}). Another Green kernel for~$\Sph{1}$ can be obtained in an alternative simple way by writing $\Sph{1} = \R / \Z$, and there are nice geometric formulas for $\GKer$ for $\Sph{2}$ (Example~\ref{Example:Circle}).

In Section \ref{Section:MCSphere}, we use $\GKer$ from Section~\ref{Sec:GreenSphere} to compute the formal pushforward Maurer-Cartan element $\PMC$ for $\Sph{n}$ (Proposition~\ref{Proposition:MCSphere}). We first prove that the condition~$(V_{\NOne})$ from Proposition~\ref{Prop:PMCEqualsMC} is satisfied (Lemma~\ref{Lemma:ABVanishing}) and then perform combinatorics with degrees to show vanishing of some more integrals (Proposition~\ref{Prop:TotalVanishing}). In fact, all the integrals vanish for $\Sph{n}$ with $n\ge 3$, and the only non-vanishing integrals for $\Sph{1}$ are the $O_k$-graphs with even $k$. We compute these integrals explicitly together with all signs and combinatorial coefficients required to obtain $\PMC_{20}$ (Lemmas~\ref{Lemma:IntegralFor1}, \ref{Lemma:Independence}, \ref{Lemma:SignForMCOnCircle} and  \ref{Lemma:CombinatorialCoefficientForMCOnCircle}). There might be some non-vanishing integrals associated to reduced graphs for $\Sph{2}$ as well as some non-vanishing integrals associated to graphs without external vertices for $\Sph{3}$; however, the simplest examples vanish (Remarks~\ref{Rem:GraphsTwoSphere} and~\ref{Rem:GraphsThreeSphere}).  

In the remaining Sections~\ref{Section:HomSphere} and~\ref{Section:CPn1}, we compute $\IBL(\HIBL^\PMC(\CycC(\Harm(M))))$ and the higher operations~$\OPQ_{1lg}^\PMC$ on $\HIBL^\PMC$ for $M = \Sph{n}$, $\CP^n$. As soon as we argue that $\PMC_{10} = \MC_{10}$ due to geometric formality, the computation of $\HIBL^\MC(\CycC(\Harm(\Sph{n})))$ and $\HIBL^\MC(\CycC(\Harm(\CP^n)))$ is an easy exercise in cyclic homology. The operations for $\Sph{2m}$ and $\CP^n$ vanish for degree reasons (Remark~\ref{Rem:DegRes}). Therefore, the integrals from Section \ref{Section:MCSphere} help only in the case of $\Sph{2m-1}$. We compare our results to Chas-Sullivan string topology from~\cite{Basu2011} and confirm Conjecture~\ref{Conj:StringTopology} for~$\Sph{n}$ with~$n\ge 2$ and for~$\CP^n$.


\subsection{Computation of \texorpdfstring{$\GKer$}{G} for \texorpdfstring{$\Sph{n}$}{Sn}}
\label{Sec:GreenSphere}

The standard Riemannian volume form on the round sphere $\Sph{n}\subset \R^{n+1}$ is the restriction of the following closed form on $\R^{n+1}\backslash \{0\}$:  
$$ \Vol(x) := \frac{1}{|x|^{n+1}}\sum_{i=1}^{n+1} (-1)^{i+1} x^i \Diff x_1 \dotsm \widehat{\Diff x_i} \dotsm \Diff x_{n+1}. $$
Here $\widehat{\Diff x_i}$ means that $\Diff{x_i}$ is omitted. We denote the Riemannian volume of $\Sph{n}$ by
$$ V:= \int_{\Sph{n}} \Vol. $$
The $n$-form $\HKer$ from Proposition~\ref{Lemma:HKer} reads
\begin{equation*}
\HKer =  \frac{1}{V}\bigl(\Pr_1^*\Vol +  (-1)^{n} \Pr_2^*\Vol\bigr).
\end{equation*}
According to Proposition~\ref{Prop:GKer}, the equation which we want to solve reads
\begin{equation} \label{Eq:GreenKernel}
\Dd \GKer = \frac{1}{V}\bigl((-1)^{n}\Pr_1^*\Vol +  \Pr_2^*\Vol\bigr).
\end{equation}
We denote 
$$\tilde{G}:=V\GKer\quad\text{and}\quad \tilde{H}:=V\HKer.$$

The following lemma will be used to construct a solution to~\eqref{Eq:GreenKernel}. 
\begin{Lem}[Relative Poincar\'e Lemma]\label{Lem:ChainHtpy}
Let $M$ be a smooth oriented manifold and $\psi: [0,1]\times M \rightarrow M$ a smooth map. Consider the operator $T: \DR^*(M) \rightarrow \DR^{*-1}(M)$ defined by
$$ T(\eta) := \FInt{[0,1]} \psi^*\eta\quad\text{for all }\eta\in \DR(M), $$
where we integrate along the fiber of the oriented fiber bundle $\Pr_2: [0,1]\times M \rightarrow M$. Then we have
$$ \Dd\circ T + T\circ \Dd = \psi_1^* - \psi_0^*. $$
\end{Lem}
\begin{proof}
Stokes' formula from Proposition~\ref{Prop:StokesForm} gives
$$ \Dd \FInt{[0,1]}\psi^*\eta = -\Bigl( \FInt{[0,1]} \Dd \psi^* \eta - \FInt{\Bdd [0,1]} \psi^* \eta \Bigr) = - \FInt{[0,1]} \psi^* \Dd \eta + \psi^*_1 \eta - \psi^*_0 \eta$$
for all $\eta \in \DR(M)$.
\end{proof}

\begin{Proposition}[Solution to \eqref{Eq:GreenKernel}] \label{Prop:GKerSph}
For all $(x,y)\in (\Sph{n}\times \Sph{n}) \backslash \Diag$, let
\begin{equation} \label{Eq:GreenKernelMC1}
\GKer (x,y) := (-1)^{n} \sum_{k=0}^{n-1} g_k(x,y)\omega_k(x,y),
\end{equation}
where
\begin{equation}\label{Eq:FunctionsGk1}
 g_k(x,y) := \int_{0}^1 \frac{t^k(t-1)^{n-1-k}}{(2t(t-1)(1+x\cdot y) + 1)^{\frac{n+1}{2}}} \Diff{t}
\end{equation}
and 
\begin{equation} \label{Eq:FormOmega1}
\omega_k(x,y) := \begin{multlined}[t]\frac{1}{k!} \frac{1}{(n-1-k)!} \sum_{\sigma \in \Perm_{n+1}} (-1)^\sigma x^{\sigma_1} y^{\sigma_2} \Diff{x^{\sigma_3}} \dotsm \Diff{x^{\sigma_{2+k}}} \\ \Diff{y^{\sigma_{3+k}}} \dotsm \Diff{y^{\sigma_{n+1}}}.\end{multlined}
\end{equation}
The form~\eqref{Eq:GreenKernelMC1} is a smoooth solution to~\eqref{Eq:GreenKernel} on $(\Sph{n}\times \Sph{n}) \backslash \Diag$.
\end{Proposition}
%
\begin{proof} 
Define the set
$$ N := (\R^{n+1}_{\neq 0}\times\R^{n+1}_{\neq 0})\backslash \{(x,a x) \mid x\in \R^{n+1}, a>0\}. $$
It is an open thickening of $(\Sph{n}\times \Sph{n})\backslash \Diag$ in $\R^{n+1} \times \R^{n+1}\backslash \Diag $. Consider the smooth deformation retraction
\begin{align*}
 \psi : [0,1] \times N & \longrightarrow N \\
 (t,x,y) &\longmapsto \psi_t(x,y) := (x,(1-t)y-tx)
\end{align*}
with 
$$ \psi_0(x,y)=(x,y) \quad \text{and}\quad \psi_1(x,y)= (x,-x)\quad\text{for all }(x,y)\in N. $$
The retraction is depicted in Figure~\ref{Fig:Retraction}.
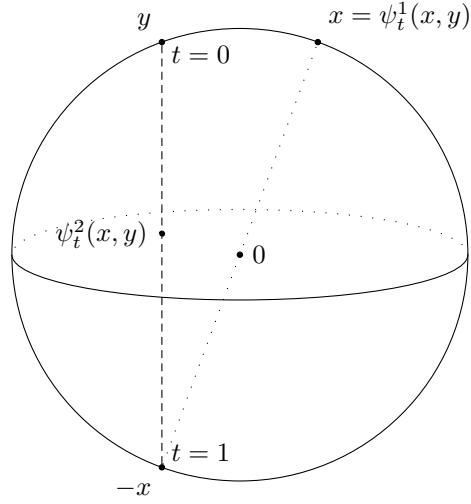
\begin{figure}
\centering
\begin{tikzpicture}
  \tikzset{point/.style = {draw, circle, fill=black, minimum size=2pt,inner sep=0pt}}
  \tikzset{->-/.style={decoration={markings,mark=at position #1 with {\arrow{>}}},postaction={decorate}}}
  \tikzset{-<-/.style={decoration={markings,mark=at position #1 with {\arrow{<}}},postaction={decorate}}}

  \def\rad{3cm}
  \def\posx{70}
  \def\posy{110}
  
  \node [point,label={0:0}] (C) at (0,0) {};
  \draw (C) circle (\rad);
  \draw (-\rad,0) arc (180:360:3 and 0.6);
  \draw[loosely dotted] (\rad,0) arc (0:180:3 and 0.6);

  \path (C) node[point,label={\posx:$x=\psi^{1}_t(x,y)$}] (x)  at +(\posx:\rad) {};
  \path (C) node[point,label={\posx+180:$-x$}] (mx)  at +(\posx+180:\rad) {};
  \path (C) node[point,label={\posy:$y$}] (y)  at +(\posy:\rad) {};
  
  \draw[loosely dotted] (mx) -- (x);
  \draw[densely dashed] (y) node [label={[label distance=-6pt,yshift=1pt,xshift=1pt]-50:$t=0$}] {} -- node[point,pos=0.45,label={[left,label distance=-3pt]0:$\psi_t^{2}(x,y)$}] {} (mx) node [label={[label distance=-6pt,yshift=11pt,xshift=1pt]-50:$t=1$}] {};
\end{tikzpicture}
\caption{Retraction $\psi_t = (\psi_t^1, \psi_t^2)$. A point of $\Sph{n}\times \Sph{n}$ is visualized as a pair of points on $\Sph{n}$.}\label{Fig:Retraction}
\end{figure}
Denote by $A: \R^{n+1} \rightarrow \R^{n+1}$, $x\mapsto -x$ the antipodal map. It is easy to see that 
$$ A^* \Vol = (-1)^{n+1} \Vol, $$
and hence
\begin{equation*}
\psi_1^* \tilde{\HKer} = \psi_1^* \Pr_1^*\Vol + (-1)^n \psi_1^*\Pr_2^* \Vol  = \Pr_1^*\Vol + (-1)^n \Pr_1^*A^* \Vol = 0. 
\end{equation*}
Define 
\begin{equation}  \label{Eq:GExpr}
\GKer := (-1)^{n+1} \FInt{[0,1]} \psi^*\HKer.
\end{equation}
Let $T: \DR^*(N)\rightarrow \DR^{*-1}(N)$ be the cochain homotopy from Lemma~\ref{Lem:ChainHtpy} associated to $\psi$. Because $\Dd \HKer=0$, we get
$$ \Dd \GKer = (-1)^{n+1} \Dd T(\HKer) = (-1)^{n+1}(\Dd T + T \Dd)\HKer = (-1)^{n+1}(\psi_1^*-\psi_0^*)\HKer = (-1)^{n}\HKer. $$
For every $i=1$,~$\dotsc$, $n+1$, we have
$$ \psi^*(\Diff{x^i}) = \Diff{x^i}\quad\text{and}\quad\psi^*(\Diff{y^i}) = (1-t)\Diff{y^i} - t\Diff{x^i} - (y^i+x^i)\Diff{t}. $$
We compute
\begin{equation*}
\begin{aligned}
 & (-1)^{n+1} \FInt{[0,1]} \psi^* \tilde{\HKer} = -\FInt{[0,1]} \psi^* \Pr_2^* \Vol \\
 & \quad = \FInt{[0,1]} \sum_{i=1}^{n+1} (-1)^{i}\frac{((1-t)y^i - t x^i)}{{\Abs{(1-t)y-tx}^{n+1}}} \psi^*(\Diff{y^1} \dotsm \widehat{\Diff{y^i}} \dotsm \Diff{y^{n+1}}) \\
 & \quad = \sum_{1\le i<j \le n+1} (-1)^{i+j}(x^i y^j - y^i x^j) \FInt{[0,1]}  \frac{\Diff{t} \psi^*(\Diff{y^1} \dotsm \widehat{\Diff{y^i}} \dotsm \widehat{\Diff{y^j}} \dotsm \Diff{y^{n+1}})}{\Abs{(1-t)y-tx}^{n+1}} \\ 
 & \quad = \begin{multlined}[t] (-1)^{n}\sum_{k=0}^{n-1} \Bigl(\int_0^1 \frac{t^k(t-1)^{n-1-k}}{\Abs{(1-t)y-tx}^{n+1}} \Diff{t}\Bigr) \sum_{1\le i<j\le n+1} (-1)^{i+j+1} (x^i y^j - y^i x^j) \\
 \sum_{\mathclap{\substack{\sigma: \{1,\dotsc,n-1\}\rightarrow \{1,\dotsc,\hat{i},\dotsc,\hat{j},\dotsc,n+1\} \\ \sigma_1<\dots < \sigma_k \\ \sigma_{k+1}<\dots < \sigma_{n-1}}}} (-1)^\sigma \Diff{x^{\sigma_1}} \dotsm \Diff{x^{\sigma_k}}\Diff{y^{\sigma_{k+1}}} \dotsm \Diff{y^{\sigma_{n-1}}}.\end{multlined}
\end{aligned}
\end{equation*}
The formulas~\eqref{Eq:FunctionsGk1} and~\eqref{Eq:FormOmega1} are obtained from this by writing 
$$ \Abs{(1-t)y - tx}^2 = 2t(t-1)(1+x \cdot y) + 1 $$
in the denominator of the integrand and by simple combinatorics in the form part, respectively. Smoothness of~$\GKer$ on $(\Sph{n}\times \Sph{n})\backslash \Diag$ follows from the expression~\eqref{Eq:GExpr}.
\end{proof}
Note that $g_k$ are smooth functions on $(\Sph{n}\times \Sph{n})\backslash \Diag$.
\begin{Example}[Green kernel for $\Sph{1}$ and $\Sph{2}$]\phantomsection\label{Example:Circle}
\begin{ExampleList}
\item Let 
$$\alpha: (\Sph{1}\times\Sph{1})\backslash \Diag \rightarrow (0,2\pi)$$
be the smooth function assigning to a pair $(x,y)\in  (\Sph{1}\times\Sph{1})\backslash \Diag$ the counterclockwise angle from $x$ to $y$. Let $\alpha_1$, $\alpha_2 \in [0,2\pi)$ be such that $x=\cos(\alpha_1)\StdBasis_1 + \sin(\alpha_1)\StdBasis_2$ and $y=\cos(\alpha_2)\StdBasis_1+\sin(\alpha_2)\StdBasis_2$ for the standard Euclidean basis $\StdBasis_1$, $\StdBasis_2$ of $\R^2$. It is easy to see that
$$ \alpha(x,y) = \begin{cases} \alpha_2 - \alpha_1 & \text{if }\alpha_1<\alpha_2, \\
\alpha_2-\alpha_1+2\pi & \text{if }\alpha_1>\alpha_2. \end{cases} $$
Therefore, we get
$$ \Diff{\alpha} = \Diff{\alpha_2} -\Diff{\alpha_1} = -2\pi H \quad\text{on } (\Sph{1}\times\Sph{1})\backslash \Diag. $$
On the other hand, we can compute $\GKer$ from~\eqref{Eq:GreenKernelMC1} as follows. Using the substitution $u=2t-1$, we get for all $x$, $y\in \Sph{1}$ with $x\neq \pm y$ the following:
\begin{equation*}
\begin{aligned}
 g_{0}(x,y) &= \int_0^1 \frac{\Diff{t}}{2t(t-1)(1+x\cdot y) +1} =  \frac{1}{1-x\cdot y}\int_{-1}^1 \frac{\Diff{u}}{\frac{1+x\cdot y}{1- x\cdot y}u^2 + 1} \\ &= \frac{2}{\sqrt{1-(x\cdot y)^2}}\arctan\Bigl(\sqrt{\frac{1+x\cdot y}{1-x\cdot y}}\Bigr) \\ &= \frac{\pi - \arccos(x\cdot y)}{\sqrt{1-(x\cdot y)^2}}  = \frac{\pi - \arccos(x\cdot y)}{ \Abs{x^1 y^2 - x^2 y^1}} = \frac{\pi - \alpha(x,y)}{x^1 y^2 - x^2 y^1}.
\end{aligned}
\end{equation*}
The third from last equality can be obtained by trigonometric considerations and the second from last equality by an algebraic manipulation with the denominator. We will explain the last equality. Consider the matrix 
$$ R=\begin{pmatrix}
0 & -1 \\ 1 & 0
\end{pmatrix} $$
representing the counterclockwise rotation by $\frac{\pi}{2}$. The function $\arccos: (-1,1) \rightarrow (0,\pi)$ satisfies
$$ \arccos(x\cdot y) = \begin{cases}
                        \alpha(x,y) & \text{if }y\cdot Rx > 0, \\
                        2\pi - \alpha(x,y) & \text{if }y\cdot Rx<0. 
                       \end{cases}$$
The last equality becomes clear when we notice that $x^1 y^2 - x^2 y^1 = y\cdot Rx$. 

Finally, we have $\omega_{0}(x,y) = x^1 y^2 - x^2 y^1$, and hence
$$ 2\pi G(x,y) =  - g_{0}(x,y) \omega_{0}(x,y) = \alpha(x,y) - \pi = \pi - \alpha(y,x). $$
\item For $n=2$, we get the formulas
\allowdisplaybreaks
\begin{align*}
g_0(x,y) & = - g_1(x,y) = \frac{1}{x\cdot y - 1}\quad\text{and} \\
\omega_0(x,y) &=(x^2 y^3 - x^3 y^2) \Diff{y^1} +(x^3 y^1 - x^1 y^3) \Diff{y^2} + (x^1 y^2 - x^2 y^1) \Diff{y^3} \\ &=\sum_{i=1}^3 (x\times y)^i \Diff{y^i}.
\end{align*}
The formula for $\omega_1(x,y)$ is obtained from the formula for $\omega_0(x,y)$ by replacing~$\Diff{y}$ with $\Diff{x}$.\qedhere
\end{ExampleList}
\end{Example}

Consider the diagonal action of the orthogonal group $O(n+1)$ on $\R^{n+1}\times\R^{n+1}$ by matrix multiplication.
\begin{Proposition}[Symmetries of $\GKer$]\label{Prop:SymmetryOfG}
Consider $\GKer$ from Proposition~\ref{Prop:GKerSph}. For all $R\in O(n+1)$, we have
$$ R^* \GKer = (-1)^R \GKer, $$
where $(-1)^R = \det(R)$. Moreover, if $\tau$ denotes the twist map, then
$$\tau^*\GKer = (-1)^{n}\GKer. $$ 
\end{Proposition}
\begin{proof} 
We will use the thickening $N$, the antipodal map $A$ and the expression~\eqref{Eq:GExpr} for~$\GKer$ from the proof of Proposition~\ref{Prop:GKerSph}.

It is easy to check that both $\tau$ and $R$ preserve $N$. Let~$\tilde{\tau}$ and $\tilde{R}$ be the isomorphisms of the fiber bundle $\Pr_2: [0,1]\times N \rightarrow N$ given by 
$$ \tilde{\tau}(t,x,y) := (1-t,y,x)\quad \text{and}\quad \tilde{R}(t,x,y) := (t,Rx,Ry) $$
for all $(t,x,y)\in [0,1]\times N$. Then $\tilde{\tau}$ covers $\tau$ and $\tilde{R}$ covers $R$. A simple computation directly from Definition~\ref{Def:FibInt} shows that the fiberwise integration commutes with the pullback along a bundle morphism if the bundle map and the base map are both either orientation preserving or reversing. In our case, we have 
$$ (-1)^{\tau + \tilde{\tau}} = -1\quad \text{and}\quad (-1)^{R+\tilde{R}} = 1. $$
Using this and the equation
$$ \Pr_2 \circ \psi \circ \tilde{\tau} = A\circ \Pr_2 \circ \psi, $$
we get firstly
\allowdisplaybreaks
\begin{align*}
\tau^* \FInt{[0,1]} \psi^*\tilde{\HKer} &=  - \FInt{[0,1]} \tilde{\tau}^* \psi^*\Pr_2^*\Vol \\ &= - \FInt{[0,1]} \psi^*\Pr_2^*A^* \Vol \\ &= (-1)^n \FInt{[0,1]} \psi^*\Pr_2^*\Vol \\ & = (-1)^n \FInt{[0,1]} \psi^*\tilde{\HKer}
\end{align*}
and secondly
$$ R^* \FInt{[0,1]} \psi^* \HKer = \FInt{[0,1]} \tilde{R}^* \psi^* \HKer = \FInt{[0,1]} \psi^* R^*\HKer = (-1)^{n+1} \FInt{[0,1]} \psi^* \HKer. $$
This proves the proposition.
\end{proof}
Both diffeomorphisms $R$ and $\tau$ preserve $\Delta$, and hence they extend to diffeomorphisms of $\Bl_\Diag(\Sph{n}\times\Sph{n})$. If also $\GKer$ extends, then the statement of Proposition~\ref{Prop:SymmetryOfG} holds for $\GKer$ on $\Bl_\Diag(\Sph{n}\times\Sph{n})$.

In the rest of the section, we will be proving that $\GKer$ extends smoothly to $\Bl_{\Diag}(\Sph{n}\times \Sph{n})$. This is a local problem at the boundary, where we introduce the following radial coordinates. Define the set
$$ X := \{(r,\eta,x)\in [0,\infty)\times \Sph{n}\times\Sph{n} \mid \eta\cdot x = 0\}, $$
and let $\kappa: X \longrightarrow \Bl_\Diag(\Sph{n}\times \Sph{n})$ be the map defined by
\begin{align*}
\kappa(r,\eta,x) &:= \begin{cases} 
\Bigl(x,\dfrac{x+r\eta}{\Abs{x+r\eta}}\Bigr)\in (\Sph{n}\times \Sph{n})\backslash \Diag & \text{for }r>0, \\[2ex]
[(-\eta,\eta)]\in P^+ N_{(x,x)}\Diag & \text{for }r=0.
\end{cases}
\end{align*}
For the upcoming computations, it is convention to define the map $\gamma: \R \rightarrow (-1,1)$ by
$$ \gamma(r) := \frac{r}{\sqrt{1+r^2}+1} \quad\text{for all }r\in \R. $$
It is a diffeomorphism with inverse $r = \frac{2 \gamma}{1-\gamma^2}$.

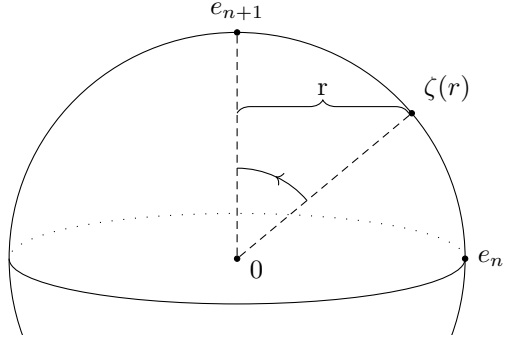
\begin{figure}[t]\centering
\begin{tikzpicture}
\clip (-3.5,-1) rectangle (3.6,3.5); 
  \tikzset{point/.style = {draw, circle, fill=black, minimum size=2pt,inner sep=0pt}}
  \tikzset{->-/.style={decoration={markings,mark=at position #1 with {\arrow{>}}},postaction={decorate}}}
  \tikzset{-<-/.style={decoration={markings,mark=at position #1 with {\arrow{<}}},postaction={decorate}}}
  \def\rad{3cm}
  \def\pos{40}

  \node [point,label={[yshift=-4pt]0:0}] (C) at (0,0) {};
  \draw (C) circle (\rad);
  \draw (-\rad,0) arc (180:360:3 and 0.6);
  \draw[loosely dotted] (\rad,0) arc (0:180:3 and 0.6);

  \path (C) node[point,label={0:$e_n$}] (en)  at +(0:\rad) {};
  \path (C) node[point,label={90:$e_{n+1}$}] (en1)  at +(90:\rad) {};
  \path (C) node[point,label={\pos:$\zeta(r)$}] (x) at +(\pos:\rad) {};
  
  \draw[decoration={markings, mark=at position 0.5 with {\arrow{>}}},postaction={decorate}] ([shift=(\pos:0.4*\rad)]C) arc (\pos:90:0.4*\rad);

  \draw[densely dashed] (C) -- (x);
  \draw[densely dashed] (C) -- (en1);
  
  \draw let \p{A}=(x) in [decorate,decoration={brace,amplitude=5pt}] (0,\y{A}) -- (x) node [midway,yshift=9pt]{r};
\end{tikzpicture}
\caption{The curve $\zeta:= \kappa \circ \zeta'$ is given by $\zeta(r)=\bigl(e_{n+1},\frac{e_{n+1}+r e_n}{\Abs{e_{n+1}+r e_n}}\bigr)$ for $r>0$.}\label{Fig:CurveOnSphere}
\end{figure}

\begin{Lem}[Parametrization of the collar neighborhood] \label{Lem:NewBlowupParam}
The subset $X\subset \R\times \R^{n+1}\times\R^{n+1}$ is a submanifold with boundary, and the map $\kappa: X \longrightarrow \Bl_\Diag(\Sph{n}\times \Sph{n})$ is an embedding onto a neighborhood of $\Bdd \Bl_\Diag(\Sph{n}\times \Sph{n})$.
\end{Lem}
\begin{proof}
The set $X$ is a Cartesian product of $[0,\infty)$ and a regular level set; therefore, it is a submanifold with boundary. The inclusion $\Sph{n}\times \Sph{n} \subset \R^{n+1}\times \R^{n+1}$ induces an embedding of manifolds with boundary $\Bl_\Diag(\Sph{n}\times\Sph{n})\subset \Bl_\Diag(\R^{n+1}\times \R^{n+1})$.  Consider the global chart $\tilde{\Id}: \Bl_\Diag(\R^{n+1}\times \R^{n+1}) \rightarrow [0,\infty) \times \Sph{n} \times \R^{n+1}$ from~\eqref{Eq:BlowUpChart} induced by the identity. We have
$$ \begin{aligned}Y &:= \tilde{\Id}(\Bl_\Diag(\Sph{n}\times \Sph{n})) \\ &= \{(\tilde{r},w,u)\in [0,\infty) \times \Sph{n} \times \R^{n+1} \mid \Abs{u}^2+\tilde{r}^2=1,\ w\cdot u = 0\}, \end{aligned}$$ 
where we denote $r$ on $Y$ by $\tilde{r}$ in order to distinguish it from $r$ on $X$. It suffices to prove the claim for the map $\mu:= \tilde{\Id}\circ\kappa: X \rightarrow Y$. For $(r,\eta,x)\in X$, we compute
$$ \mu(r,\eta,x) = \biggl( \frac{\gamma}{\sqrt{1+\gamma^2}}, \frac{1}{\sqrt{1+\gamma^2}}(\gamma x - \eta), \frac{1}{1+\gamma^2}(x+\gamma \eta) \biggr). $$
This formula defines a smooth map of $\R\times \R^{n+1}\times\R^{n+1}$.
It is a local diffeomorphism because its Jacobian is non-vanishing:
$$ |\Jac{\mu}| = \frac{\partial \tilde{r}}{\partial r} \Bigl(\frac{\partial w}{\partial \eta}\frac{\partial u}{\partial x}  - \frac{\partial w}{\partial x}\frac{\partial u}{\partial \eta}\Bigr)^{n+1} = (-1)^{n+1}(1+\gamma^2)^{-\frac{n+4}{2}} \frac{\partial \gamma}{\partial r}. $$
Moreover, the map $\mu$ is injective, maps $X$ into $Y$ and $\Bdd X$ onto $\Bdd Y$. The claim follows.
\end{proof}
Consider the action of $O(n+1)$ on $X$ defined by
$$ R\cdot(r,\eta,x) := (r,R\eta,Rx)\quad\text{for all }(r,\eta,x)\in X\text{ and }R\in O(n+1). $$
Via $\kappa$, this agrees with the diagonal action of $O(n+1)$ on $\Bl_\Diag(\Sph{n}\times\Sph{n})$. Denote
$$ \GKer':= \kappa^* \GKer \in \Omega^{n-1}(\Int(X)). $$
From Proposition~\ref{Prop:SymmetryOfG} we get
\begin{equation} \label{Eq:SymmetryOfGPrime}
R^* \GKer' = (-1)^R \GKer'\quad \text{for all }R\in O(n+1).
\end{equation}
Consider the smooth curve (see Figure~\ref{Fig:CurveOnSphere})
\begin{equation*}\label{Eq:CurveZetaDef}
 \begin{aligned} \zeta': [0,\infty) &\longrightarrow  X \\
                    r &\longmapsto (r,e_n,e_{n+1}). \end{aligned}
\end{equation*}
We have the following lemma.
\begin{Lem}[Smooth extension along the curve] \label{Lem:ExtAlongCurve}
The form $\GKer'$ extends smoothly to $X$ if and only if the map $\GKer'\circ \zeta' : (0,\infty)\rightarrow \Lambda^{n-1}T^* X$ extends smoothly to the interval $[0,\infty)$. 
\end{Lem}
\begin{proof}
As for the non-trivial implication, let $(0,\eta_0,x_0) \in X$ be a boundary point. Pick vectors $v_1$,~$\dotsc$, $v_{n-1}\in \R^{n+1}$ so that the vectors $v_1$,~$\dotsc$, $v_{n-1}$, $\eta_0$, $x_0$ are linearly independent, and define the set 
$$ U:=\{(r,\eta,x)\in X \mid v_1,\,\dotsc,\,v_{n-1},\,\eta,\,x \text{ are linearly independent}\}. $$
It is an open neighborhood of $(0,\eta_0,x_0)$ in $X$. Applying the Gram-Schmidt orthogonalization to $v_1$,~$\dotsc$, $v_{n-1}$, $\eta$, $x$, we find a smooth map $R: U \rightarrow O(n+1)$ such that 
$$ R(r,\eta,x)\cdot (r,\eta,x) = (r,e_n,e_{n+1}) \quad \text{for all }(r,\eta,x)\in U. $$
The equation~\eqref{Eq:SymmetryOfGPrime} implies
$$ \GKer'(r,\eta,x) =(-1)^R R(r,\eta,x)^*\bigl(\GKer'(r,e_n,e_{n+1})\bigr)\quad\text{for all }(r,\eta,x)\in \Int(U), $$
where $R(r,\eta,x)^*: \Lambda^* T^* X \rightarrow \Lambda^* T^* X$ is the smooth cotangential map which is induced by the diffeomorphism $R(r,\eta,x): X\rightarrow X$, and which maps the fiber over $z\in X$ to the fiber over $R(r,\eta,x)^{-1} z$. By the assumption, all maps in the composition are smooth in their arguments. The lemma follows.
\end{proof}

\begin{Lem}[Local expression at the boundary] \label{Lem:FormulaAlongCurve}
On the interval $(0,\infty)$, we have
\begin{equation*}\label{Eq:FormulaAlongCurve}
\tilde{\GKer}'\circ\zeta' = (-1)^{n+1}(1+\gamma^2)^{-\frac{n-1}{2}} \sum_{k=0}^{n-1} \gamma^{n-k} (h_{k}\circ\gamma)(\nu_{k}\circ\zeta'), 
\end{equation*}
where the functions $h_{k}: (0,1)\rightarrow \R$ are defined by
\begin{equation*}
  h_{k}(\gamma):= \int_{-1}^1 \frac{(u+\gamma^2)^{k}(u-1)^{n-1-k}}{(u^2+\gamma^2)^{\frac{n+1}{2}}} \Diff{u} \quad \text{for all }\gamma\in (0,1)
\end{equation*}
and the forms $\nu_{k}\in \Omega(X)$ are defined by
\begin{align*}
 \nu_{k}(r,x,\eta) &:= \frac{1}{k!(n-1-k)!}\sum_{\sigma\in \Perm_{n-1}} (-1)^\sigma \Diff{x^{\sigma_1}}\dotsm
\Diff{x^{\sigma_k}} \Diff{\eta^{\sigma_{k+1}}} \dotsm \Diff{\eta^{\sigma_{n-1}}}.
\end{align*}
\end{Lem}

\begin{proof}
We start with the following formula from the proof of Proposition~\ref{Prop:GKerSph}:
$$ \tilde{\GKer} =  \sum_{1\le i<j \le n+1} (-1)^{i+j}(x^i y^j - y^i x^j) \FInt{[0,1]}  \frac{\Diff{t} \psi^*(\Diff{y^1} \dotsm \widehat{\Diff{y^i}} \dotsm \widehat{\Diff{y^j}} \dotsm \Diff{y^{n+1}})}{\Abs{(1-t)y-tx}^{n+1}}. $$
We restrict to the points $(x,y)=\kappa(r,e_n,e_{n+1})$ with $r>0$. There, we have
\begin{align*}
&x^1= \dotsb =x^n=0,\ x^{n+1}=1, \\
&y^1 = \dotsb = y^{n-1} =0,\ y^n= \frac{2\gamma}{1+\gamma^2},\ y^{n+1} = \frac{1-\gamma^2}{1+\gamma^2}.
\end{align*}
Under the substitution $u = 2t-1$, we get
$$\Abs{(1-t)y-t x}^2 = \frac{4 t (t-1)}{1+\gamma^2}+1=\frac{u^2+\gamma^2}{1+\gamma^2}. $$
We make the following preliminary computations:
$$ \begin{aligned}
x^i y^j - y^i x^j & = 0\quad\text{for }1\le i \le n-1\text{ and }i<j\le n+1, \\
x^n y^{n+1} - y^n x^{n+1} & = -\frac{2\gamma}{1+\gamma^2}, \\
\kappa^*(\Diff{y^i}) & = \frac{1}{1+\gamma^2}\bigl((1-\gamma^2)\Diff{x^i}+ 2\gamma \Diff{\eta^i}\bigr)\quad\text{for } 1\le i \le n-1. \end{aligned} $$
We plug these in the formula for $\tilde{G}$ and get
%
\begin{align*}
 &\tilde{\GKer}'(\zeta'(r)) = 2\gamma(1+\gamma^2)^{\frac{n-1}{2}} \FInt{[0,1]}\Diff{t}\frac{\prod_{i=1}^{n-1} \bigl((1-t)\kappa^*(\Diff{y^i}) - t\Diff{x^i}\bigr)}{(u^2+\gamma^2)^{\frac{n+1}{2}}}  \\ 
&\quad=(-1)^{n+1}\gamma(1+\gamma^2)^{-\frac{n-1}{2}}\FInt{[-1,1]} \Diff{u} \frac{\prod_{i=1}^{n-1}\bigl((u+\gamma^2)\Diff{x^i} + \gamma(u-1)\Diff{\eta^i}\bigr)}{(u^2+\gamma^2)^{\frac{n+1}{2}}}  \\
&\quad =(-1)^{n+1}(1+\gamma^2)^{-\frac{n-1}{2}}\sum_{k=0}^{n-1} \gamma^{n-k}\Bigl( \int_{-1}^1 \frac{(u+\gamma^2)^{k}(u-1)^{n-1-k}}{(u^2+\gamma^2)^{\frac{n+1}{2}}} \Diff{u}\Bigr) \nu_{k}.
\end{align*}
The lemma follows.
\end{proof}
%
%
\begin{Lem}[Integrals depending on parameter] \label{Lem:GeneralIntegralExtension}
Let $n\in \N$, and let $l=0$,~$1$,~$\dotsc$, $n-1$. The function $F_{n,l}: (0,\infty)\rightarrow \R$ defined by
\begin{equation}\label{Eq:GeneralIntegral}
F_{n,l}(t) := \int_{-1}^1 \frac{t^{n-l} u^l}{(u^2+t^2)^{\frac{n+1}{2}}} \Diff{u}\quad\text{for all }t\in(0,\infty)
\end{equation}
extends smoothly to $\R$.
\end{Lem} 
\begin{proof}
We have
$$ F_{1,0}(t) = 2 \arctan\Bigl(\frac{1}{t}\Bigr)=\pi - 2 \arctan(t) \quad\text{for all }t\in (0,\infty). $$
The right-hand side is a smooth function on $\R$.

For $n\ge 2$, we deduce the recursive formula
$$ F_{n,0}(t) = \frac{1}{n-1}\Bigl((n-2)F_{n-2,0}(t)+\frac{2 t^{n-2}}{(1+t^2)^{\frac{n-1}{2}}}\Bigr).$$
If $l$ is odd, then $F_{n,l}\equiv 0$ for all $n$ because the integrand of \eqref{Eq:GeneralIntegral} is odd as a function of $u$.

For $n\ge 3$ and even $2\le l \le n-1$, we deduce yet another recursive formula
$$  F_{n,l}(t) = \frac{1}{n-l}\Bigl((l-1) F_{n,l-2}(t) -  \frac{2 t^{n-l}}{(1+t^2)^{\frac{n-1}{2}}} \Bigr).
$$
The claim for all $F_{n,l}$ follows by induction.
\end{proof}

\begin{Proposition}[Smooth extension to the boundary]\label{Prop:GKerBdd}
The form $\GKer$ from~\eqref{Eq:GreenKernelMC1} extends smoothly to $\Bl_\Diag(\Sph{n}\times\Sph{n})$.
\end{Proposition}
\begin{proof}
According to Lemmas~\ref{Lem:NewBlowupParam} and~\ref{Lem:ExtAlongCurve}, it suffices to show that the curve $\GKer'\circ \zeta': (0,\infty)\rightarrow \Lambda^{n-1}T^* X$ extends smoothly to $[0,\infty)$. Lemma~\ref{Lem:FormulaAlongCurve} gives an expression for $\GKer'\circ \zeta'$ as a linear combination of smooth forms $\nu_{k}\in \Omega^{n-1}(X)$ with coefficients $\gamma^{n-k}(h_{k}\circ \gamma)$ for $k=0$,~$\dotsc$, $n-1$ multiplied by the overall coefficient $(-1)^n (1+\gamma^2)^{-\frac{n-1}{2}}$. We expand
\begin{equation*}
\gamma^{n-k}(h_{k}\circ\gamma) = \sum_{a=0}^k\sum_{b=0}^{n-1-k}(-1)^{n-1-k-b} \binom{k}{a} \binom{n-1-k}{b} \int_{-1}^1 \frac{\gamma^{n+k-2a}u^{a+b}}{(u^2+\gamma^2)^{\frac{n+1}{2}}} \Diff{u}
\end{equation*}
and notice that we can write
\begin{equation*}
 \int_{-1}^1 \frac{\gamma^{n+k-2a}u^{a+b}}{(u^2+\gamma^2)^{\frac{n+1}{2}}} \Diff{u} = \gamma^{k-a+b}(F_{n,a+b}\circ \gamma)
\end{equation*}
for the function $F_{n,l}$ from~\eqref{Eq:GeneralIntegral} with $l:=a+b$. Because $0\le l \le n-1$, Lemma~\ref{Lem:GeneralIntegralExtension} asserts that $F_{n,l}$ extends smoothly to $[0,\infty)$. Because $k-a+b\ge 0$, the entire coefficient at $\nu_{k}$ extends smoothly to $[0,\infty)$ for every $k=0$,~$\dotsc$, $n-1$. The lemma follows.
\end{proof}


We summarize our results in the following proposition:

\begin{Proposition}[Green kernel for $\Sph{n}$]\label{Proposition:GreenKernel}
The form $\GKer$ from~\eqref{Eq:GreenKernelMC1} defines a Green kernel for $\Sph{n}$ satisfying Definition~\ref{Def:GreenKernel}. Moreover, we have the symmetries
\begin{align*}
R^* \GKer &= (-1)^R \GKer\quad \text{for all }R\in O(n+1)\text{ and} \\
\tau^* \GKer & = (-1)^n \GKer.
\end{align*}
\end{Proposition}
\begin{proof}
The proposition is a summary of Propositions~\ref{Prop:GKerSph},~\ref{Prop:SymmetryOfG} and~\ref{Prop:GKerBdd}.
\end{proof}

\begin{Remark}[Better notation due to R. Bryant, see  \cite{MO291535}] \label{Remark:Bryant}
Pick an oriented basis $e_1$,~$ \dotsc$, $e_{n+1}$ of $\R^{n+1}$ as generators of the exterior algebra $\Lambda^*(\R^{n+1})$, and view $x$, $y$, $\Diff{x}$, $\Diff{y}$ as $\Lambda^*(\R^{n+1})$-valued forms on~$\R^{n+1}$. For example, we view $x$ as the map $x\in \R^{n+1} \mapsto \sum_{i=1}^{n+1} x^i e_i \in \Lambda^1(\R^{n+1})$ and $\Diff{x}$ as the map $x\in \R^{n+1} \mapsto \sum_{i=1}^{n+1} (\Diff{x}_i)_x e_i \in \Lambda^1(\R^{n+1})$. There is a natural wedge product on the space of $\Lambda^*(\R^{n+1})$-valued forms. If $\omega$ is a top-form, we denote by $[\omega]$ the coefficient of $\omega$ at $e_1 \wedge \dotsm \wedge e_{n+1}$. Then it holds
$$ \omega_k(x,y) = \frac{1}{k!}\frac{1}{(n-1-k)!}[x\wedge y \wedge (\Diff{x})^{k} \wedge (\Diff{y})^{n-1-k}]. $$

Note that if we view $e_i$ as odd variables, then $[\cdot]$ corresponds to the odd integration~$\int \mathrm{D}e(\cdot)$. It would be interesting to know whether this notation simplifies some proofs, especially if Lemma~\ref{Lemma:ABVanishing} can be deduced from abstract algebraic facts or rules valid for odd integration.
\end{Remark}

\subsection{Computation of \texorpdfstring{$\PMC$}{n} for \texorpdfstring{$\Sph{n}$}{Sn}}
\label{Section:MCSphere}
We recall from Definition~\ref{Def:PushforwardMCdeRham} that the formal pushforward Maurer-Cartan element~$\PMC$ is computed as a sum over trivalent ribbon graphs decorated with the Green kernel~$\GKer$ at internal edges, integration variables $x_i$ at internal vertices and, in the case of $\Sph{n}$, with $\NOne$ or $\NVol$ at external vertices.
\begin{figure}[t]
\centering
\begin{tikzpicture}

\tikzset{point/.style = {draw, circle, fill=black, minimum size=2pt,inner sep=0pt}}

\def\dist{4}
\def\startangle{90}
\def\leglength{1.2}

\node[point,label={125:$x_1$}] (A) at (0,0) {};

\path[draw] (A) -- +(\startangle:\leglength) node[at end,right]{$\NVol$};
\path (A) -- +(\startangle:\leglength) node[point,style={fill=white}] {};
\path[draw] (A) -- +(\startangle+120:\leglength) node[at end, above, yshift=2pt]{$\NOne$};
\path (A) -- +(\startangle+120:\leglength) node[point,style={fill=white}] {};
\path[draw] (A) -- +(\startangle+240:\leglength) node[at end, above, yshift=2pt]{$\NOne$};
\path (A) -- +(\startangle+240:\leglength) node[point,style={fill=white}] {};

\end{tikzpicture}
\caption{The $Y$-graph for $\Sph{n}$.}\label{Fig:YGraph}
\end{figure}
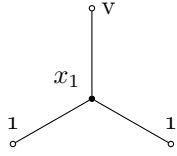

The canonical Maurer-Cartan element $\MC$ is the contribution of the Y-graph (see Figure~\ref{Fig:YGraph}), and it is easy to see that
\begin{equation*}
 \MC_{10}(\Susp \NVol \NOne \NOne ) = (-1)^n \MC_{10}(\Susp\NOne \NVol \NOne) = \MC_{10}(\Susp \NOne \NOne \NVol) = (-1)^{n-2}.
\end{equation*}
Throughout this section, we will be in the setting of Definition~\ref{Def:PushforwardMCdeRham}. In particular, $\Gamma \in \TRG_{klg}$ is a ribbon graph, $L$ its compatible labeling admissible with respect to an input $\omega_1$, $\dotsc$, $\omega_l$ and $I(\sigma_L)$ the corresponding integral.
\begin{Lemma}[Condition $(V_{\NOne})$ holds] \label{Lemma:ABVanishing}
Consider $\Sph{n}$ with the Green kernel $\GKer$ from~\eqref{Eq:GreenKernelMC1}. Then every graph $\Gamma \neq Y$ with~$\NOne$ at an external vertex vanishes.
\end{Lemma}
\begin{proof}
The only contribution of an $A$-vertex which does not vanish for degree reasons is
$$ A_{\NVol,\NOne}(y) = \int_x \GKer(x,y) \Vol(x). $$
From the symmetry of $\GKer$ and $\Vol$ under the action of $O(n+1)$, we get
$$ R^* A_{\NVol, \NOne} = (-1)^R A_{\NVol, \NOne}\quad\text{for all }R\in O(n+1). $$
Therefore, it suffices to check that $A_{\NVol, \NOne}(\StdBasis_1) = 0$, where $\StdBasis_1$, $\dotsc$, $\StdBasis_{n+1}$ denotes the standard basis of $\R^{n+1}$. Evaluation of~\eqref{Eq:FormOmega1} at $(x,\StdBasis_1)$ gives
$$ \omega_0(x,\StdBasis_1) = \frac{1}{(n-1)!} \sum_{\substack{\sigma\in\Perm_{n+1}\\ \sigma_2 = 1}} (-1)^\sigma x^{\sigma_1} \Diff{y^{\sigma_3}} \dotsm \Diff{y^{\sigma_{n+1}}}. $$
Therefore, we get
$$ \begin{aligned}
A_{\NVol, \NOne}(\StdBasis_1) &= (-1)^n \int_x g_0(x\cdot \StdBasis_1) \omega_0(x,\StdBasis_1) \Vol(x)  \\ &= \sum_{j=2}^{n+1} (-1)^{n+j+1} \Bigl(\int_x g_0(x^1) x^j \Vol(x)\Bigr) \Diff{y^2} \dotsm  \widehat{\Diff{y^j}} \dotsm \Diff{y^{n+1}},
\end{aligned}$$
where we view $g_0$ as a function of $x\cdot y$. For every $j=2$, $\dotsc$, $n+1$, consider the orientation reversing diffeomorphism 
$$ \begin{aligned} 
I_j: \Sph{n} &\longrightarrow \Sph{n} \\
   (x^1, \dotsc, x^{n+1}) &\longmapsto (x^1, \dotsc, -x^j, \dotsc, x^{n+1}). 
\end{aligned} $$   
Then we have
$$ \int_x g_0(x^1) x^j \Vol(x)  = - \int_x I_j^*\bigl( g_0(x^1) x^j \Vol(x) \bigr) = - \int_x g_0(x^1)(-x^j)(-\Vol(x)), $$
and it follows that $A_{\NVol,\NOne}(\StdBasis_1)=0$.

Let us now consider the contribution of a $B$-vertex with $\NOne$:
$$ B_{\NOne}(y,z) = \int_{x} \GKer(y,x) \GKer(x,z) = (-1)^n \int_x \GKer(y,x) \GKer(x,z). $$
For $n=1$, the degree of $\GKer(y,x)\GKer(x,z)$ is $0$, and hence $B_{\NOne}= 0$ trivially. Suppose that $n\ge 2$. As in the case of $A_{\NVol,\NOne}$, we get that 
$$ R^* B_{\NOne} = (-1)^R B_{\NOne} \quad\text{for all }R\in O(n+1). $$
Therefore, it suffices to check that $B_{\NOne}(\StdBasis_1, c_1 \StdBasis_1 + c_2 \StdBasis_2) = 0$ for all $(c_1,c_2)\in \Sph{1}$. We have
$$  B_{\NOne}(\StdBasis_1, c_1 \StdBasis_1 + c_2 \StdBasis_2) = \begin{multlined}[t] (-1)^n \sum_{a=1}^{n-1} \int_x g_a(x^1) g_{n-a}(c_1 x^1 + c_2 x^2)\omega_a(x,\StdBasis_1) \\ \omega_{n-a}(x,c_1 \StdBasis_1 + c_2 \StdBasis_2).\end{multlined}$$
We will show that for every $a=1$,~$\dotsc$, $n-1$ we can write
\begin{equation} \label{Eq:Mu}
\mu_a(x) := \omega_a(x,\StdBasis_1) \omega_{n-a}(x,c_1 \StdBasis_1 + c_1 \StdBasis_2) = \Bigl(\sum_{i=3}^{n+1} \pm  x^i \Vol(x)\Bigr) \eta_a(y,z)
\end{equation}
for some form $\eta_a(y,z)$. Then, using the same argument as for $A_{\NVol,\NOne}$, we will have
$$\begin{aligned}
& \int_x  g_a(x^1) g_{n-a}(c_1 x^1 + c_2 x^2) x^i \Vol(x) \\
&\qquad = - \int_x I^*_i \bigl(g_a(x^1) g_{n-a}(c_1 x^1 + c_2 x^2) x^i \Vol(x)\bigr) \\
&\qquad = - \int_x g_a(x^1) g_{n-a}(c_1 x^1 + c_2 x^2) (-x^i) (-\Vol(x))
\end{aligned}$$
for all $3 \le i \le n+ 1$, and hence $B_{\NOne}(\StdBasis_1, c_1 \StdBasis_1 + c_2 \StdBasis_2) = 0$.

In order to show \eqref{Eq:Mu}, we have to study the product of $\omega_i$'s. From~\eqref{Eq:FormOmega1} we get
\begin{equation}\label{Eq:GenProd}
\begin{aligned}
&\omega_a(x,y)\omega_{n-a}(x,z)\\
&\quad =\begin{multlined}[t] \frac{1}{a! (n-1-a)! (n-a)! (a-1)!} \sum_{\substack{\sigma,\,\mu\in \Perm_{n+1}}} (-1)^{\sigma + \mu} x^{\sigma_1} x^{\mu_1} y^{\sigma_2} z^{\mu_2}  \\ \Diff{x}^{\sigma_3} \dotsm \Diff{x}^{\sigma_{2+a}} \Diff{x}^{\mu_3}\dotsm\Diff{x}^{\mu_{2+n-a}} 
\Diff{y}^{\sigma_{3+a}}\dotsm \Diff{y}^{\sigma_{n+1}}\\\Diff{z}^{\mu_{3+n-a}}\dotsm \Diff{z}^{\mu_{n+1}}.
\end{multlined}
\end{aligned}
\end{equation}
In order to simplify this expression, we decompose $\sigma\in \Perm_{n+1}$ as
\begin{equation} \label{Eq:Decomp}
 \sigma = \sigma^5 \circ \sigma^4 \circ \sigma^3 \circ \sigma^2 \circ \sigma^1,
 \end{equation}
where $\sigma^1$, $\dotsc$, $\sigma^5\in \Perm_{n+1}$ are permutations defined as follows:
\begin{itemize}
 \item The permutation $\sigma^1$ is a shuffle permutation $\sigma^1 \in \Perm_{2+a,n-a-1}$ such that its first block denoted by $\sigma^1(1) = (\sigma^1_1,\dotsc,\sigma^1_{2+a})$ is equal to the ordered set $\{\sigma_1, \dotsc, \sigma_{2+a}\}$. The second block $\sigma^1(2)$ is then the ordered set $\{\sigma_{3+a},\dotsc,\sigma_{n+1}\}$, which will be denoted by $J_\sigma$.
 \item The permutation $\sigma^2$ acts on the block $\sigma^1(1)$ by moving $\sigma_2$ in front. We denote the new block $\sigma^1(1)\backslash\{\sigma_2\}$ by $I_\sigma$, so that we can write $\sigma^2 : \sigma^1(1) \mapsto (\sigma_2, I_\sigma)$. 
 \item The permutation $\sigma^3$ acts on the block $I_\sigma$ by moving $\sigma_1$ in front. Together with the previous step we get $\sigma^1(1)\mapsto (\sigma_2, \sigma_1,I_\sigma\backslash\{\sigma_1\})$.
 \item The permutation $\sigma^4$ is a transposition of $\sigma_1$ and $\sigma_2$.
 \item The permutation $\sigma^5$ is determined by the pair $(\sigma^{51},\sigma^{52}) \in \Perm_{a} \times \Perm_{n-1-a}$ of permutations $\sigma^{51}$ and $\sigma^{52}$ acting on blocks $I_\sigma\backslash\{\sigma_1\}$ and $J_\sigma$ to get $(\sigma_3,\dotsc,\sigma_{2+a})$ and $(\sigma_{3+a},\dotsc,\sigma_{n+1})$, respectively.
\end{itemize}

We define the decomposition $\mu^1$,~$\dotsc$, $\mu^5$ for $\mu\in \Perm_{n+1}$ from \eqref{Eq:GenProd} analogously with $a$ replaced by $n-a$. Using~\eqref{Eq:Decomp}, the product~\eqref{Eq:GenProd} can be written as
\begin{align*}
  &\begin{multlined}\frac{1}{a! (n-1-a)! (n-a)! (a-1)!} \sum_{\substack{\sigma^1,  \dotsc,\, \sigma^5 \\ \mu^1, \dotsc,\, \mu^5}} (-1)^{\sigma^1 + \dotsb + \sigma^5 + \mu^1 + \dotsb + \mu^5} x^{\sigma_1} x^{\mu_1} y^{\sigma_2} z^{\mu_2} \\ \Diff{x}^{\sigma^{51}(I_\sigma\backslash \{\sigma_1\})} \Diff{x}^{\mu^{51}(I_\mu\backslash\{\mu_1\})} \Diff{y}^{\sigma^{52}(J_\sigma)} \Diff{z}^{\mu^{52}(J_\mu)} \end{multlined} \\[5pt]
 &\qquad =\begin{multlined}[t]- \sum_{\sigma^1,\, \mu^1} (-1)^{\sigma^1 + \mu^1} \Bigl(\sum_{\sigma^2,\, \mu^2} (-1)^{\sigma^2 + \mu^2} \sum_{\sigma^3,\, \mu^3} (-1)^{\sigma^3 + \mu^3} x^{\sigma_1} x^{\mu_1} y^{\sigma_2} z^{\mu_2} \\ \Diff{x}^{I_\sigma \backslash \{\sigma_1\}} \Diff{x}^{I_\mu\backslash \{\mu_1\}}\Bigr) \Diff{y}^{J_\sigma} \Diff{z}^{J_\mu},\end{multlined} \end{align*}
where $-1$ comes from $(-1)^{\sigma^4}$ and $\sigma^5$ is compensated by permutations of forms. For fixed $\sigma^1$ and $\mu^1$, consider the coefficient at $\Diff{y}^{J_\sigma}\Diff{z}^{J_\mu}$ in the brackets. If we evaluate it at $y=\StdBasis_1$, $z=c_1 \StdBasis_1 + c_2 \StdBasis_2$, we get
\begin{align*}
&  c_1 \overbrace{\sum_{\substack{\sigma^3,\, \mu^3 \\ \sigma_2 = 1 \\ \mu_2 = 1}} (-1)^{\sigma^3 + \mu^3} x^{\sigma_1} x^{\mu_1}  \Diff{x}^{I_\sigma\backslash\{\sigma_1\}} \Diff{x}^{I_\mu \backslash\{\mu_1\}}}^{=:\displaystyle\mathrm{I}} \\ 
& {} + (-1)^{\mu^2} c_2  \overbrace{\sum_{\substack{\sigma^3,\, \mu^3 \\ \sigma_2 = 1 \\ \mu_2 = 2}} (-1)^{\sigma^3 + \mu^3} x^{\sigma_1} x^{\mu_1}  \Diff{x}^{I_\sigma\backslash\{\sigma_1\}} \Diff{x}^{I_\mu \backslash\{\mu_1\}}}^{=:\displaystyle\mathrm{II}},
 \end{align*}
where $(-1)^{\mu^2} = -1$ if and only if $1\in I_{\mu}$.

More generally, for multiindices $I_1$, $I_2 \subset \{1,\dotsc, n+1\}$ of lengths $a+1$ and $n-a+1$,  respectively, consider the sum
\begin{equation} \label{Eq:Sum}
S(I_1,I_2) := \sum_{\substack{i_1 \in I_1 \\ i_2 \in I_2}} (-1)^{(i_1,I_1) + (i_2,I_2)} x^{i_1} x^{i_2} \Diff{x}^{I_1\backslash\{i_1\}} \Diff{x}^{I_2\backslash\{i_2\}},
\end{equation}
where $(i_j, I_j)$ is the number of transpositions required to move $i_j$ in front of $I_j$. The following implication holds: 
\begin{equation*}
 S(I_1, I_2)\neq 0\; \Implies \; 1 \le \Abs{I_1 \cap I_2} \le 2.
\end{equation*}
We distinguish the two cases left: 
\begin{description}[font=\normalfont\itshape]
\item[Case $I_1 \cap I_2 =\{i,j\}$ with $i < j$: ] We get
\allowdisplaybreaks
\begin{align*} 
S(I_1, I_2) &= \begin{multlined}[t] (-1)^{(i,I_1) + (j,I_2)} x^{i} x^j \Diff{x}^{I_1 \backslash\{i\}} \Diff{x}^{I_2 \backslash \{j\}} \\ {}+ (-1)^{(j,I_1) + (i,I_2)} x^{j} x^i \Diff{x}^{I_1 \backslash\{j\}} \Diff{x}^{I_2 \backslash \{i\}} \end{multlined} \\
&=\begin{multlined}[t] (-1)^{(i,I_1) + (j,I_2) + (j,I_1)+1 + (i,I_2)} x^i x^j \Diff{x}^j \Diff{x}^{I_1\backslash\{i,j\}}\\\Diff{x}^i \Diff{x}^{I_2\backslash\{i,j\}} + (-1)^{(j,I_1) + (i,I_2)+(i,I_1)+ (j,I_2)+1} x^i x^j \\ \Diff{x}^i \Diff{x}^{I_1\backslash\{i,j\}}\Diff{x}^j \Diff{x}^{I_2\backslash\{i,j\}} \end{multlined} \\
&= \pm(-1 +  1) x^i x^j \Diff{x}^i \Diff{x}^{I_1\backslash\{i,j\}} \Diff{x}^j\Diff{x}^{I_2\backslash\{i,j\}},
\end{align*}
where in the last step we switched $\Diff{x}^i \leftrightarrow \Diff{x}^j$ in the first summand. Therefore, it holds $S(I_1, I_2) = 0$.
\item[Case $I_1 \cap I_2 = \{i\}$:] We must have $I_1 \cup I_2 = \{1,\dotsc,n+1\}$. A non-zero summand in~\eqref{Eq:Sum} has either $i_1 = i$ and $i_2\in I_2$, in which case $$I_1\backslash \{i_1\} \cup I_2\backslash\{i_2\}= \{1, \dotsc, \widehat{i_2}, \dotsc, n+1\}, $$ or $i_2 = i$ and $i_1\in I_1$ with $i_1 \neq i$, in which case $$I_1\backslash \{i_1\} \cup I_2\backslash\{i_2\}= \{1, \dotsc, \widehat{i_1}, \dotsc, n+1\}.$$ Indices $i_2$ from the first case and $i_1$ from the second case constitute $\{1,\dotsc,n+1\}$. Therefore, for some signs $\pm$, we can write
\begin{equation*} \label{Eq:AlternatingSum}
S(I_1,I_2) = x^i \sum_{j=1}^{n+1}  \pm  x^j \Diff{x^1} \dotsm \widehat{\Diff{x^j}} \dotsm \Diff{x^{n+1}}.
\end{equation*}
We will prove that the signs alternate, and hence $S(I_1,I_2)= \pm x^i \Vol(x)$. Suppose that $j$, $j+1 \in I_{1}$ for some $j\in \{1,\ldots,n\}$. The two summands in~\eqref{Eq:Sum} with $(i_1, i_2)=(j, i)$ and $(i_1, i_2)= (j+1,i)$, respectively, give
\allowdisplaybreaks
\begin{align*}
& \begin{multlined}[t](-1)^{(j,I_1) + (i,I_2)} x^{j} x^{i} \Diff{x}^{I_1\backslash\{j\}} \Diff{x}^{I_2\backslash\{i\}} + (-1)^{(j+1,I_1) + (i,I_2)} x^{j+1} x^{i} \\ \Diff{x}^{I_1\backslash\{j+1\}}\Diff{x}^{I_2\backslash\{i\}} \end{multlined}
\\ &\quad= \begin{multlined}[t] (-1)^{(i,I_2)} x^{j} x^{i} \Diff{x}^{j+1}\Diff{x}^{I_1\backslash\{j, j+1\}} \Diff{x}^{I_2\backslash\{i\}}  \\ {}+ (-1)^{1 + (i,I_2)} x^{j+1} x^{i} \Diff{x}^j \Diff{x}^{I_1\backslash\{j, j+1\}} \Diff{x}^{I_2\backslash\{i\}} \end{multlined} \\
&\quad = (-1)^{(i,I_2)}x^i (x^j \Diff{x}^{j+1} - x^{j+1} \Diff{x}^j) \Diff{x}^{I_1\backslash\{j, j+1\}} \Diff{x}^{I_2\backslash\{i\}}.
\end{align*}
The signs clearly alternate. A symmetric argument holds when $j$, $j+1\in I_2$. Now assume that $j \in I_1$ and $j+1\in I_2$. The two summands in~\eqref{Eq:Sum} which have $(i_1, i_2)=(j,i)$ and $(i_1, i_2) = (i,j+1)$, respectively, give
\allowdisplaybreaks
\begin{align*}
& \begin{multlined}[t] (-1)^{(j,I_1)+(i,I_2)} x^j x^i \Diff{x}^{I_1\backslash\{j\}}\Diff{x}^{I_2\backslash\{i\}} + (-1)^{(i,I_1) + (j+1,I_2)} x^i x^{j+1} \\ \Diff{x}^{I_1\backslash\{i\}}\Diff{x}^{I_2\backslash\{j+1\}} \end{multlined}\\
&\quad=\begin{multlined}[t]
(-1)^{(j,I_1)+(i,I_1\backslash\{j\}) + (a + 1)} x^j x^i \Diff{x^{I_1\backslash\{i, j\}}}\Diff{x}^{I_2} \\ {}+ (-1)^{(i,I_1) + (j+1,I_2) + (j,I_1\backslash\{i\})} x^i x^{j+1} \Diff{x}^j \Diff{x}^{I_1\backslash\{i, j\}} \Diff{x}^{I_2\backslash\{j+1\}}
\end{multlined}\\&\quad=
\begin{multlined}[t]
(-1)^{(j,I_1) + (i,I_1\backslash\{j\})+(j+1,I_2)} x^i x^j \Diff{x}^{j+1} \Diff{x}^{I_1\backslash\{i,j\}} \Diff{x}^{I_2\backslash\{j+1\}} \\ {}+ (-1)^{1 + (j,I_1) + (i,I_1\backslash\{j\})+(j+1,I_2)} x^i x^{j+1} \Diff{x}^j \Diff{x}^{I_1\backslash\{i,j\}} \Diff{x}^{I_2\backslash\{j+1\}} 
\end{multlined}\\
&\quad= \begin{multlined}[t] (-1)^{(j,I_1) + (i,I_1\backslash\{j\})+(j+1,I_2)} x^i (x^j \Diff{x}^{j+1} - x^{j+1} \Diff{x}^j) \\ \Diff{x}^{I_1\backslash\{i,j\}} \Diff{x}^{I_2\backslash\{j+1\}}. \end{multlined}
\end{align*}
The signs alternate again. A symmetric argument holds for $j\in I_2$ and $j+1\in I_1$.
\end{description}	 

 Back to the original problem, we have $\mathrm{I} = S(I_\sigma, I_\mu)$ with $I_\sigma, I_\mu \subset \{2,\dotsc, n+1\}$. It follows that the first case applies, and hence $\mathrm{I} = 0$.  We have $\mathrm{II} = S(I_\sigma, I_\mu)$ with $I_\sigma \subset \{2,\dotsc, n+1\}$ and $I_\mu \subset \{1, \widehat{2}, \dotsc, n+1\}$. It follows that either the first case or the second case with $i\ge 3$ applies. This proves~\eqref{Eq:Mu}. Consequently, we get $B_{\NOne} = 0$ also for $n\ge 2$.
 
The last paragraph of the proof of Proposition~\ref{Prop:COne} finishes the proof.
\end{proof}

We summarize the consequences in the following proposition. The main argument is the same as in the proof of Proposition~\eqref{Prop:GeomForm}.

\begin{Proposition}[Vanishing of graphs for $\Sph{n}$] \label{Prop:TotalVanishing}
Consider $\Sph{n}$ with the Green kernel~\eqref{Eq:GreenKernelMC1}. Only the following trivalent ribbon graphs $\Gamma \neq Y$ do not necessarily vanish:
\begin{description}[font=\normalfont\itshape]
 \item[($n=1$):] The $O_{k}$-graph with $k\in 2\N$ internal vertices of type $B$ with $\NVol$ at the external vertex (see Figure~\ref{Fig:Gamma0}).
 \item[($n=2$):] It must hold $A=0$, $C=2B$ and all $B$ vertices must have $\NVol$ at the external vertex. Moreover, if $\Gamma$ is reduced, it must have $g\ge 1$.
 \item[($n=3$):] There is no external vertex and $4 \mid C$ holds.
 \item[($n>3$):] All graphs vanish.
\end{description}
\end{Proposition}
\begin{proof}
Lemma~\ref{Lemma:ABVanishing} implies that $A=0$ and that the total form-degree $D$ satisfies $D= n B$. Therefore, we get from \eqref{Eq:VerticesEq} the following: for $n>3$ there is neither a $B$-vertex nor a $C$-vertex; for $n=3$, there is no $B$-vertex; for $n=2$, we have $C=2B$; and for $n=1$, there is no $C$-vertex.

Consider the pullback of $I(\sigma_L)$ along the (multi)diagonal action of an $R\in O(n+1)$ with $\Det(R)=-1$ on $(\Sph{n})^{\times k}$. We get schematically
$$ \int_{(\Sph{n})^{\times k}} \GKer^{e}\Vol^{s} = (-1)^{k +e + s} \int_{(\Sph{n})^{\times k}} G^{e}\Vol^{s}. $$
Therefore, $k+ e + s$ has to be even. If we plug-in from~\eqref{Eq:ChangeOfVariables}, we get
$$ k+ e + s =  \begin{cases}
3 B & \text{for }n=1, \\
8 B & \text{for }n=2, \\
\frac{5}{2} C & \text{for }n=3.
\end{cases} $$

A non-vanishing reduced graph must have $B\ge l$. For $n=2$, so that $C=2B$, the formula~\eqref{Eq:GenusFormulaa} gives $g\ge 1$.
\end{proof}

\begin{Remark}[Graphs for $\Sph{2}$]\phantomsection\label{Rem:GraphsTwoSphere}

{ \begingroup
\begin{figure}
\centering
\begin{subfigure}{0.45\textwidth}
\centering
\begin{tikzpicture}
 \tikzset{point/.style = {draw, circle, fill=black, minimum size=2pt,inner sep=0pt}}
 \tikzset{->-/.style={decoration={markings,mark=at position #1 with {\arrow{>}}},postaction={decorate}}}
 
 \def\lenI{3cm}
 \def\lenII{1cm}
 \def\lenext{0.5cm}
 
 \coordinate (x1) at (0,0) {};
 \coordinate (x2) at (\lenI,0);
 \coordinate (x3) at (0.5*\lenI,-1*\lenII);

 \coordinate (p4) at (0.5*\lenI,\lenII);
 \coordinate (p5) at (0.6*\lenI,0.4*\lenII);
 \coordinate (p6) at (0.8*\lenI,1.3*\lenII);

 \draw (x1) to[out=90,in=180] (p4);
 \draw (p4) to[out=0,in=90] (x2);
 \draw (x1) to[out=-90,in=180] (x3);
 \draw (x3) to[out=0,in=-90] (x2);
 \draw (x1) to[out=0,in=-120] (p5);
 \draw[dashed] (p5) to[out=60,in=180] (p6);
 \draw (p6) to[out=0,in=0] (x2);
 
 \node[point,label={180:$x_1$}] at (x1) {};
 \node[point,label={180:$x_2$}] at (x2) {};
 \node[point,label={-90:$x_3$}] at (x3) {};
 
 \draw (x3) -- +(90:\lenext) node[point,,style={fill=white},label={180:$\NVol$}] {};
\end{tikzpicture}
\caption{$P_1$}
\end{subfigure}
\begin{subfigure}{0.45\textwidth}
\centering
\begin{tikzpicture}
 \tikzset{point/.style = {draw, circle, fill=black, minimum size=2pt,inner sep=0pt}}
 \tikzset{->-/.style={decoration={markings,mark=at position #1 with {\arrow{>}}},postaction={decorate}}}
 
 \def\rad{1.5cm}
 \def\ang{60}
 \def\dashsep{0.5cm}
 \def\lenext{0.5cm}
 
 \coordinate (C) at (0,0);
 \draw (C) circle (\rad);

 \path (C) node[point,label={0:$x_5$}] (x5)  at +(0*\ang:\rad){};
 \path (C) node[point,label={\ang:$x_3$}] (x3)  at +(1*\ang:\rad){};
 \path (C) node[point,label={2*\ang:$x_4$}] (x4)  at +(2*\ang:\rad){};
 \path (C) node[point,label={0:$x_6$}] (x6)  at +(3*\ang:\rad){};
 \path (C) node[point,label={4*\ang:$x_1$}] (x1)  at +(4*\ang:\rad){};
 \path (C) node[point,label={5*\ang:$x_2$}] (x2)  at +(5*\ang:\rad){};
 
 \draw (x5) -- +(180:\lenext) node[point,style={fill=white},label={90:$\NVol$}] {};
 \draw (x6) -- +(180:\lenext) node[point,style={fill=white},label={90:$\NVol$}] {};
 \draw (x1) -- (x3);
 
 \path (C) coordinate (p1)  at +(2*\ang:\dashsep){};
 \path (C) coordinate (p2)  at +(-1*\ang:\dashsep){};
 
 \draw (x2) -- (p2);
 \draw (x4) -- (p1);
 \draw[dashed] (p1) -- (p2);
\end{tikzpicture}
\caption{$P_2$}
\end{subfigure}
\caption{Graphs $P_1$ with $(l,g)=(1,1)$ and $P_2$ with $(l,g)=(2,1)$ for $n=2$.}\label{Fig:P1P2}
\end{figure}
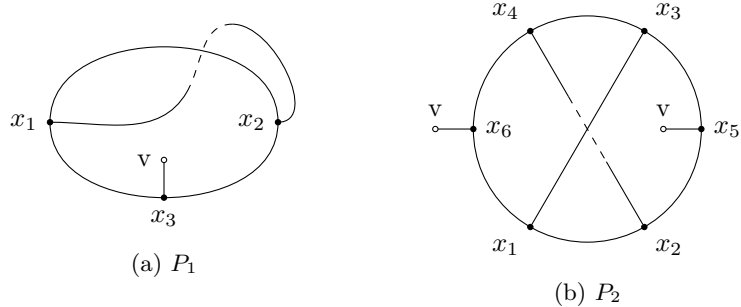
\endgroup }
The simplest possibly non-vanishing graph for $\Sph{2}$ has $A= 0$, $B=1$, $C=2$. If it is reduced, we must have $l = g = 1$, and hence it will contribute to $\PMC_{11}$. Up to an isomorphism, there is only one such graph, which we denote by $P_1$ (see Figure~\ref{Fig:P1P2}). However, we see that the pair of internal vertices $x_1$ and $x_2$ is connected by two edges, which implies that $P_1 =0$. Indeed, $\GKer(x,y)$ has odd degree, and hence we have\footnote{We recall from Section~\ref{Section:Proof2} that the notation $\GKer(x_i,x_j)$ means $(\pi_{i} \times \pi_j)^* \GKer$ and not just the evaluation at $(x_i,x_j)$.}
$$ \GKer(x,y)\GKer(y,x) = \GKer(x,y)^2 = 0 $$
by the symmetry on the pullback along the twist map. It follows that $\PMC_{11} = 0$.

The second simplest possibly non-vanishing reduced graph is the graph~$P_2$ from Figure~\ref{Fig:P1P2}. Let 
\begin{equation*}
\eta(x_1,x_2,x_3,x_4,x_5) :=\begin{multlined}[t]\GKer(x_1,x_2)\GKer(x_1,x_3)\GKer(x_4,x_2)\GKer(x_4,x_3)\GKer(x_3,x_5)\\ \GKer(x_2,x_5)\Vol(x_5) \end{multlined}
\end{equation*}
denote the form in the integrand coming from the part of the graph on the right-hand side of the vertical axis going through $x_1$, $x_4$. If $\tau_{1,4}$ denotes the exchange of $x_1$ and $x_4$, then clearly $\tau_{1,4}^* \eta = \eta$ because the graph is symmetric with respect to the horizontal axis going through $x_5$, $x_6$. Using this, we compute
\begin{align*}
& \int_{x_1,x_2,x_3,x_4,x_5,x_6} \NVol(x_6)\GKer(x_1,x_6)\GKer(x_4,x_6) \eta(x_1,x_2,x_3,x_4,x_5) \\
& = \int_{\tau_{1,4}(x_1,x_2,x_3,x_4,x_5,x_6)} \tau_{1,4}^*\bigl(\NVol(x_6)\GKer(x_1,x_6)\GKer(x_4,x_6) \eta(x_1,x_2,x_3,x_4,x_5)\bigr) \\
& = \int_{x_4,x_2,x_3,x_1,x_5,x_6} \NVol(x_6)\GKer(x_4,x_6)\GKer(x_1,x_6) \eta(x_4,x_2,x_3,x_1,x_5)\\
& = 
 -\int_{x_1,x_2,x_3,x_4,x_5,x_6} \NVol(x_6)\GKer(x_1,x_6)\GKer(x_4,x_6)  \eta(x_1,x_2,x_3,x_4,x_5),
\end{align*}
where the minus sign comes from switching the first two $\GKer$'s. We see that $P_2$ vanishes. The other variants with $x_5$ moved on the edge $x_3$, $x_4$ and~$x_2$, $x_4$ vanish by a similar argument using the compositions $\tau_{1,3}\circ\tau_{5,6}$ and $\tau_{1,2}\circ\tau_{5,6}$, respectively. We conclude that $\PMC_{21} = 0$, and hence $\OPQ_{121}^\PMC = 0$. 

We sum up some general observations about the integrals for $\Sph{2}$:

\begin{itemize}
\item We have $B_{\NOne} \neq 0$ and $C \neq 0$ for the corresponding forms.
\item We have the multiplication formula (c.f., Example~\ref{Example:Circle})
$$ \omega_1(x,y) \omega_1(x,z) = x\cdot(y\times z) \Vol(x). $$
\item The number $(-1)^{\sigma_L} I(\sigma_L)$ does not depend on the choice of $L_1$ provided a compatible $L_2$ is chosen.
\item It holds $\sum_{L_3^b} (-1)^{\sigma_L} I(\sigma_L) = 0$ whenever there is a boundary component with even number of $\NVol$'s.
\item If there is a $B$-vertex $x$ such that the underlying graph (after forgetting the ribbon structure) is symmetric on the reflection along an axis going through $x$, then $I(\sigma_L) = 0$. \qedhere
\end{itemize}

\Correct[caption={DONE Nonsense about lower bound}]{$g\ge 1$ is equivalent to $B \ge l$ and moreover we must have $C=2B$. What else was I thinking here?}
\end{Remark}

\begin{Remark}[Graphs for $\Sph{3}$]\label{Rem:GraphsThreeSphere}
{ \begingroup
\begin{figure}
\centering
\begin{subfigure}{0.45\textwidth}
\centering
\begin{tikzpicture}
 \tikzset{point/.style = {draw, circle, fill=black, minimum size=2pt,inner sep=0pt}}
 \tikzset{->-/.style={decoration={markings,mark=at position #1 with {\arrow{>}}},postaction={decorate}}}
 \def\rad{1cm}
 \def\len{3cm}
 
 \node (C1) at (0,0) {};
 \node (C2) at (\len,0) {};
 \draw (C1) circle (\rad);
 \draw (C2) circle (\rad);
 
 \path (C1) node[point,label={180:$x_1$}] (x1)  at +(0:\rad) {};
 \path (C2) node[point,label={0:$x_2$}] (x2)  at +(180:\rad) {};
 
 \draw (x1) -- (x2);
  
\end{tikzpicture}
\caption{$K_1$}
\end{subfigure}
\begin{subfigure}{0.45\textwidth}
\centering
\begin{tikzpicture}
\tikzset{point/.style = {draw, circle, fill=black, minimum size=2pt,inner sep=0pt}}
\node[point,label={180:$x_1$}] (x1) at (0,0) {};
\node[point,label={0:$x_2$}] (x2) at (3,0) {};
\draw[bend left=70] (x1) to (x2);
\draw[bend right=70] (x1) to (x2);
\draw (x1) -- (x2);
\end{tikzpicture}
\caption{$K_2$}
\end{subfigure}
\begin{subfigure}{0.45\textwidth}
\centering
\begin{tikzpicture}
 \tikzset{point/.style = {draw, circle, fill=black, minimum size=2pt,inner sep=0pt}}
 \tikzset{->-/.style={decoration={markings,mark=at position #1 with {\arrow{>}}},postaction={decorate}}}
 
 \def\lenI{2.5cm}
 \def\lenII{1cm}
 \def\lenext{0.6cm}
 
 \coordinate[point,label={180:$x_1$}] (x1) at (0,0) {};
 \coordinate[point,label={0:$x_2$}] (x2) at (\lenI,0);
 \path (x1) coordinate[point,label={0:$x_3$}] (x3) at +(60:\lenI);
 
 \draw (x1) -- (x2);
 \draw (x2) -- (x3);
 \draw (x3) -- (x1);

 \draw (x3) -- +(90:\lenext) node[point,style={fill=white},label={180:$\NVol$}] {};
 \draw (x1) -- +(35:\lenext) node[point,style={fill=white},label={35:$\NOne$}] {};
 \draw (x2) -- +(145:\lenext) node[point,style={fill=white},label={145:$\NOne$}] {};

\end{tikzpicture}
\caption{Tadpole}
\end{subfigure}
\caption{Graphs $K_1$ and $K_2$ from the Chern-Simons theory and the tadpole graph with $(l,g)=(2,0)$ for $n=3$.}\label{Fig:K1K2}
\end{figure}
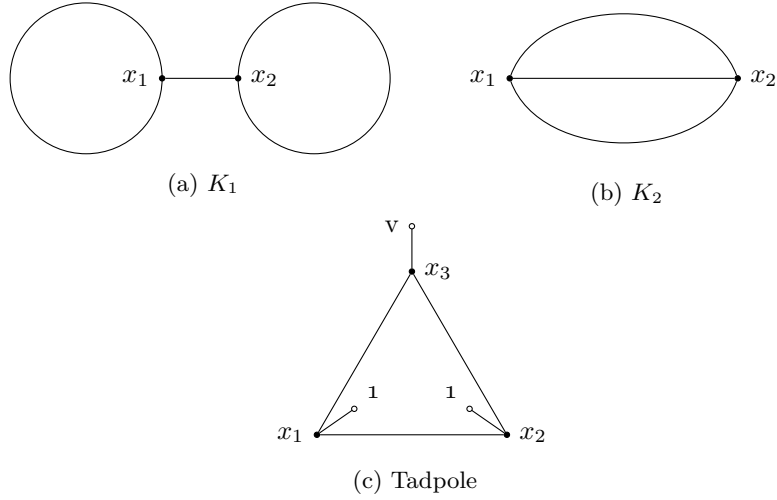
\endgroup }

For $\Sph{3}$, we consider the non-reduced graphs $K_1$ and $K_2$ and the tadpole graph from Figure~\ref{Fig:K1K2}. The graphs $K_1$ and $K_2$ appear in the definition of the Chern-Simons topological invariant in~\cite{Kohno2002} (with a gauge group). The corresponding integrals from our theory vanish ``algebraically'', i.e., at the level of wedge products of $\omega_i$. Indeed, every summand in $K_1$ contains 
$$ \omega_a(x_1,x_1) = 0\quad \text{for some }a=0, 1, 2, $$
and, for degree reasons, the form part of $K_2$ can contain   only
$$ \omega_1(x_1,x_2)^3=0\quad\text{or}\quad\omega_0(x_1,x_2)\omega_{1}(x_1,x_2)\omega_{2}(x_1,x_2)=0.$$ 
The tadpole graph contains only
\begin{equation*}
 \omega_2(x_1,x_3)\omega_1(x_1,x_2)\omega_2(x_2,x_3) = 0.\qedhere \end{equation*}
\qedhere
\end{Remark}

Equations in Remarks~\ref{Rem:GraphsTwoSphere} and~\ref{Rem:GraphsThreeSphere} were checked by the computer. The program for Wolfram Mathematica~10.4 will be made available at \cite{sourcecode}.

We will now compute $\PMC_{20}$ for $\Sph{1}$, which according to Proposition~\ref{Prop:TotalVanishing} consists only of contributions from the $O_k$-graphs with $k$ even. 
By analogy with the finite dimensional case (see~Appendix~\ref{Section:Appendix}), we expect that the number $(-1)^{\sigma_L}I(\sigma_L)$ does not depend on $L$. All inputs are namely the same and the degrees even, i.e., $|m_2^+| = -2$, $|\SuspU^2\GKer| = -2$ and $|\NVol| = 0$. 

We fix $s_1$, $s_2\ge 1$ such that $k=s_1+s_2$ is even and make the ansatz
\begin{equation*}
n_{20}(\Susp \NVol^{s_1}\otimes  \Susp \NVol^{s_2}) := \varepsilon(s_1,s_2) C(s_1,s_2) I(k),
\end{equation*}
where $I(k)$ is the integral
\begin{equation} \label{Eq:Ik}
\frac{1}{V^k}\int_{x_1, \dotsc, x_{k}} G(x_1,x_2) \dotsm G(x_{k-1},x_{k})G(x_{k},x_1)\Vol(x_1) \dotsm \Vol(x_{k}), \end{equation}
$\varepsilon(s_1,s_2)$ a sign and $C(s_1,s_2)$ a combinatorial coefficient to be determined.

We fix a circle in the plane with $k$ points (=\,internal vertices) and denote by $O(s_1,s_2)$ the set of ribbon graphs constructed by attaching external legs from which $s_1$ points in the interior and $s_2$ in the exterior, or the other way round, so that $O(s_1,s_2) = O(s_2, s_1)$ (see Figure~\ref{Fig:Gamma0}). Recall that the ribbon structure is induced from the counterclockwise orientation of the plane. It is easy to see that all graphs in $O(s_1,s_2)$ admit a labeling which is admissible with respect to $\Susp \NVol^{s_1} \otimes \Susp \NVol^{s_2}$, and that $O(s_1,s_2)$ contains a representative of every such $O_k$-graph. 

\begin{Lemma}[Integral for the $O_k$-graph for $\Sph{1}$] \label{Lemma:IntegralFor1}
For every even $k\ge 2$, the integral $I(k)$ is equal to
 \begin{equation}\label{Eq:TheFormulaForIk}
 (-1)^{k+1} \frac{1}{2^k}\sum_{i=2, 4, \dotsc ,k} \frac{i}{(i+1)!} \sum_{\substack{i_1+ \dotsb +i_r = k-i \\ i_1, \dotsc, i_r \in 2\N,\, r\in \N}} (-1)^r \frac{1}{(i_1+1)! \dotsm (i_r+1)!}.
 \end{equation}
\end{Lemma}

\begin{proof}
Denote $\bar{\GKer}(x,y) := -2\pi \GKer$. For all $k$, $l \ge 1$, we consider the more general integral
$$ I(k,l) := \int_{x_1, \dotsc, x_k} \bar{\GKer}(x_1,x_2) \dotsm \bar{\GKer}(x_{k-1},x_k) \bar{\GKer}(x_k,x_1)^l\Vol(x_1) \dotsm \Vol(x_k). $$
Taking the pullback along $(x_1, x_2, \dotsc, x_{k-1}, x_k) \mapsto (x_k, x_{k-1}, \dotsc, x_2, x_1)$ and using the antisymmetry of $\bar{\GKer}(x,y)$, we get $I(k,l) = 0$ whenever $k+l$ is even. We will compute $I(k,1)$ for $k\in 2\N$ from a recursive relation which arises from successive integration.

For the recursion step, we need to evaluate the integral 
$$\int_{y} \bar{\GKer}(x,y)\bar{\GKer}(y,z)^l \Vol(y)$$
for fixed $(x,z)\in (\Sph{1}\times \Sph{1})\backslash\Diag$. Pick the chart $g: \Sph{1}\backslash\{z\} \rightarrow (-\pi,\pi)$ defined by 
$$ g(y)= \bar{\GKer}(y,z) =  \pi - \alpha(y,z) \quad\text{for } y\in \Sph{1}\backslash\{z\}, $$
where the angle $\alpha$ was defined in Example~\ref{Example:Circle}. It holds $\Diff{g}(y) = \Vol(y)$ and
$$ \bar{\GKer}(x,y) = \begin{cases} \bar{\GKer}(x,z) - g(y) - \pi & \text{for }-\pi< g(y)< \bar{\GKer}(x,z), \\
\bar{\GKer}(x,z) - g(y) + \pi & \text{for }\bar{\GKer}(x,z)<g(y)<\pi.
 \end{cases}$$
We compute
\allowdisplaybreaks
\begin{align*}
\int_{y} \bar{\GKer}(x,y)\bar{\GKer}(y,z)^l \Vol(y) &= \begin{multlined}[t] \int_{-\pi}^{\pi} (\bar{\GKer}(x,z) - g)g^l \Diff{g} - \pi \int_{-\pi}^{\bar{\GKer}(x,z)} g^l \Diff{g} \\ {}+ \pi \int_{\bar{\GKer}(x,z)}^\pi g^l \Diff{g} \end{multlined} \\
 &=  \frac{2\pi}{l+1} \begin{cases}
  \pi^l \bar{\GKer}(x,z) - \bar{\GKer}(x,z)^{l+1} & \text{for }l\text{ even},\\[2ex]
  \dfrac{\pi^{l+1}}{l+2} - \bar{\GKer}(x,z)^{l+1} & \text{for }l \text{ odd}.
 \end{cases}
\end{align*}

From now on, $\int$ will stand for the Riemannian integral, i.e., $\int f := \int f\Vol$ for a function $f$. We compute
\allowdisplaybreaks
\begin{align*}
I(2,l) &= \int_{x_1, x_2} \bar{\GKer}(x_1,x_2)\bar{\GKer}(x_2,x_1)^{l} = - \int_{y z} \bar{\GKer}(y,z)^{l+1} =- 2\pi \int_{-\pi}^\pi g^{l+1} \Diff{g} \\ &= \begin{cases} 0 & \text{for }l \text{ even}, \\[2ex] 
 - \dfrac{4 \pi^{l+3}}{l+2} &\text{for }l\text{ odd}. \end{cases}
\end{align*}
For $k\ge 4$ even and $l$ odd, we compute 
\allowdisplaybreaks
\begin{align*}
 I(k,l) &= \begin{multlined}[t]\frac{2\pi}{l+1}\int_{x_1, \dotsc, x_{k-1}}\bar{\GKer}(x_1,x_2)\dotsm \bar{\GKer}(x_{k-2},x_{k-1}) \\ \Bigl(\frac{\pi^{l+1}}{l+2} - \bar{\GKer}(x_{k-1},x_1)^{l+1}\Bigr) \end{multlined} \\ 
&\begin{multlined}
=\frac{-4\pi^2}{(l+1)(l+2)} \int_{x_1, \dotsc, x_{k-2}}\ \bar{\GKer}(x_1,x_2) \dotsm \bar{\GKer}(x_{k-3},x_{k-2})\\ \Bigl(\pi^{l+1} \bar{\GKer}(x_{k-2},x_1)
 -\bar{\GKer}(x_{k-2},x_1)^{l+2}\Bigr)
\end{multlined} \\
&= \frac{4\pi^2}{(l+1)(l+2)}\bigl(-\pi^{l+1} I(k-2,1)+I(k-2,l+2)\bigr).
\end{align*}
%
For the second equality, we used $\int_{x_1} \bar{\GKer}(x_1,x_2) = 0$ to show that the term multiplied by $\frac{\pi^{l+1}}{l+2}$ vanishes. It follows that
\allowdisplaybreaks
\begin{align*}
I(k,1) &= \frac{(2\pi)^{k-2}}{(k-1)!} I(2,k-1) - \sum_{l=2, 4, \dotsc, k-2} \frac{(2\pi^2)^{k-l}}{(k-l+1)!} I(l,1) \\ 
&=-\frac{k(2\pi^2)^k}{(k+1)!}-\sum_{l=2, 4, \dotsc, k-2} \frac{(2\pi^2)^{k-l}}{(k-l+1)!} I(l,1)\qquad\text{for all }k=2,\,4,\,\dotsc
\end{align*}
This is a recursive equation of the form $a_k = c_k + \sum_{l=1}^{k-1} d_{k-l} a_l$. Its solution is $a_k = \sum_{i=1}^k c_i D_{k-i}$ with $D_0:= 1$ and $D_i = \sum d_{i_1} \dotsm d_{i_r}$, where we sum over all $r=1$,~$\dotsc$, $i$ and $i_1$,~$\dotsc$, $i_r\in \N$ such that $i_1+ \dotsb + i_r = i$. Therefore, we get 
$$ I(k,1) = -(2\pi^2)^k \sum_{i=2, 4, \dotsc, k} \frac{i}{(i+1)!} \sum_{\substack{i_1 + \dotsb + i_r = k-i \\ i_1, \dotsc, i_r \in 2\N, r\in \N}} (-1)^r \frac{1}{(i_1+1)! \dotsm (i_r+1)!}.$$
The result has to be multiplied by $(-1)^k(2\pi)^{-2k}$ in order to get $I(k)$. 
\end{proof} 


\begin{Lemma}[Independence of labeling]\label{Lemma:Independence}
The summand $(-1)^{\sigma_L} I(\sigma_L)$ in the definition of $\PMC_{20}(\Susp\NVol^{s_1}\otimes \Susp\NVol^{s_2})$ for $\Sph{1}$ is independent of the choice of $\Gamma\in O(s_1,s_2)$ and its labeling $L$ which is compatible and admissible with respect to the input.
\end{Lemma}
\begin{proof}
Pick $\Gamma\in O(s_1,s_2)$ and its admissible labeling $L$. Let $L'$ be an other admissible labeling of $\Gamma$. We distinguish the following situations:
\begin{itemize}
 \item Suppose that $L$ and $L'$ differ by a permutation $\mu$ in $L_3^b$. A similar argument as in the proof of Lemma~\ref{Lem:MCCond} shows that $(-1)^{\sigma_{L'}} = (-1)^\mu (-1)^{\sigma_L}$ and $I(\sigma_L') = (-1)^\mu I(\sigma_L)$, where the sign in the integral comes from the permutation of $\Vol$'s, which have form-degree $1$. Hence $(-1)^{\sigma_{L'}}I(\sigma_{L'}) = (-1)^{\sigma_{L}}I(\sigma_{L})$.
\item Suppose that the boundaries are permuted, i.e., that~$L$ and~$L'$ differ in~$L_1^b$. Notice that~$s_1=s_2$ because otherwise one of~$L$ or~$L'$ would not be admissible. The sign from changing~$L_1^b$ cancels as in the previous case. 
\item Suppose that $L$ and $L'$ differ in $L_2$. It was explained in the proof of Lemma~\ref{Lem:MCCond} that a single change of $L_2$ induces the sign $(-1)^{n-1} = 1$ in $(-1)^{\sigma_L} I(\sigma_L)$.  
\item A cyclic permutation in $L_3^v$ induces a sign neither in $(-1)^{\sigma_L}$ nor in $I(\sigma_L)$.
 \item A permutation $\mu$ in $L_1^v$ induces $(-1)^\mu$ in $(-1)^{\sigma_L}$ and a change in $I(\sigma_L)$, which can be realized by taking the pullback along $\mu: (x_1,\ldots,x_k)\mapsto (x_{\mu_1},\ldots,x_{\mu_k})$. However, the sign of the Jacobian is $(-1)^\mu$, which cancels the sign from $(-1)^{\sigma_L}$.
\end{itemize}

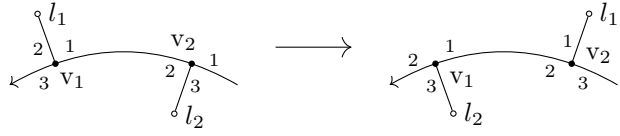
\begin{figure}[t]
\centering
\begin{tikzpicture}

\def\rad{10cm}
\def\angdist{30}
\def\posa{75}
\def\extlen{0.7cm}
\def\delta{5}
\def\startpos{0.2}
\def\endpos{0.8}
\def\labeldelta{0.003}
\def\labelrad{0.15}

\tikzset{point/.style = {draw, circle, fill=black, minimum size=2pt, inner sep=0pt}}
\tikzset{->-/.style={decoration={markings,mark=at position #1 with {\arrow{>}}},postaction={decorate}}}
\tikzset{-<-/.style={decoration={markings,mark=at position #1 with {\arrow{<}}},postaction={decorate}}}

\draw [<-] (0,0) to[bend left] coordinate [pos=\startpos-\labeldelta] coordinate [pos=\startpos-0.01, below] (v11d) coordinate [pos=\startpos, below] (v11) coordinate [pos=\startpos+\labeldelta] coordinate [pos=\endpos-0.01, below] (v12d) coordinate [pos=\endpos, below] (v12) (3,0);

\node[point] at (v11) {};
\node[point] at (v12) {};

\path (v11) node at +(50:2*\labelrad){\scriptsize{1}};
\path (v11) node at +(140:2*\labelrad){\scriptsize{2}};
\path (v11) node at +(240:2*\labelrad){\scriptsize{3}};

\path (v12) node at +(10:2*\labelrad){\scriptsize{1}};
\path (v12) node at +(200:2*\labelrad){\scriptsize{2}};
\path (v12) node at +(280:2*\labelrad){\scriptsize{3}};

\draw [<-] (5,0) to[bend left] coordinate [pos=\startpos-0.01, below] (v21d) coordinate [pos=\startpos, below] (v21) coordinate [pos=\endpos-0.01, below] (v22d) coordinate [pos=\endpos, below] (v22) (8,0);
\draw [-{>[scale=1.5]}] (3.5,0.5) -- (4.5,0.5);

\node[point] at (v21) {};
\node[point] at (v22) {};

\path (v11d) -- (v11) -- ([turn]90:\extlen) coordinate[point,style={fill=white}] (v11l);
\path (v12d) -- (v12) -- ([turn]-90:\extlen) coordinate[point,style={fill=white}] (v12l);

\draw (v11) node [inner sep=0pt,label={[yshift=1pt,xshift=-2pt]-80:$\Vert_1$}] {}  node [inner sep=0pt] {} -- (v11l) node [label={[xshift=-3pt]0:$l_1$}] {};
\draw (v12) node [inner sep=0pt,label={[xshift=-3pt,yshift=1pt]90:$\Vert_2$}] {} node [inner sep=0pt] {} node [inner sep=0pt] {} node [inner sep=0pt] {} -- (v12l)  node [label={[xshift=-3pt,yshift=-1pt]0:$l_2$}] {};

\path (v21d) -- (v21) -- ([turn]-90:\extlen) coordinate[point,style={fill=white}] (v21l);
\path (v22d) -- (v22) -- ([turn]90:\extlen) coordinate[point,style={fill=white}] (v22l);

\path (v21) node at +(50:2*\labelrad){\scriptsize{1}};
\path (v21) node at +(170:2*\labelrad){\scriptsize{2}};
\path (v21) node at +(260:2*\labelrad){\scriptsize{3}};

\path (v22) node at +(100:2*\labelrad){\scriptsize{1}};
\path (v22) node at +(200:2*\labelrad){\scriptsize{2}};
\path (v22) node at +(300:2*\labelrad){\scriptsize{3}};

\draw (v21) node [inner sep=0pt,label={[yshift=-6pt,xshift=1pt]0:$\Vert_1$}] {} node [inner sep=0pt] {}  node [inner sep=0pt] {}  node [inner sep=0pt] {} -- (v21l) node [label={[xshift=-3pt]0:$l_2$}] {};

\draw (v22) node [inner sep=0pt,label={[yshift=3pt,xshift=1pt]0:$\Vert_2$}] {} node [inner sep=0pt] {}  node [inner sep=0pt] {}  node [inner sep=0pt] {} -- (v22l) node [label={[xshift=-3pt]0:$l_1$}] {};;

\end{tikzpicture}
\caption{Swapping adjacent legs.}\label{Fig:TwoLegs}
\end{figure}

Next, we prove the independence of $\Gamma\in O(s_1,s_2)$. Let $L$ be an admissible and compatible labeling of $\Gamma$. Pick two adjacent internal vertices with external legs pointing to different regions, i.e., one to the interior of the circle and the other to the exterior. Suppose that the vertices are labeled by $\Vert_1$ and $\Vert_2$ and the legs by $l_1$ and $l_2$, respectively. Let $\Gamma'\in O(s_1,s_2)$ be the graph with the two legs turned inside out (see Figure~\ref{Fig:TwoLegs}). We can construct an admissible and compatible labeling $L'$ of $\Gamma'$ by making the following changes to $L$: The new leg at $\Vert_1$ will be labeled by $l_2$ and the new leg at $\Vert_2$ by $l_1$.
The cyclic orderings at $\Vert_1$ and $\Vert_2$, respectively, have to be modified by a transposition in order to get compatibility with the new ribbon structure. All other labelings can be copied from $L$.
In total, we get
$$ (-1)^{\sigma_L - \sigma_{L'}} = -1. $$
This sign is compensated by swapping the one-forms in $I(\sigma_L)$:
$$ \Vol (x_{\Vert_1}) \dots \Vol(x_{\Vert_2}) \;\longleftrightarrow\;\Vol(x_{\Vert_2}) \dots \Vol(x_{\Vert_1}). $$
The independence of $\Gamma\in O(s_1,s_2)$ follows from the fact that we can span the entire $O(s_1,s_2)$ by repeating the swap-of-legs operation.
\end{proof}

\begin{Lemma}[Sign]\label{Lemma:SignForMCOnCircle}
We have
$$\varepsilon(s_1,s_2) = (-1)^{s_1+1}. $$
\end{Lemma}
\begin{proof} 

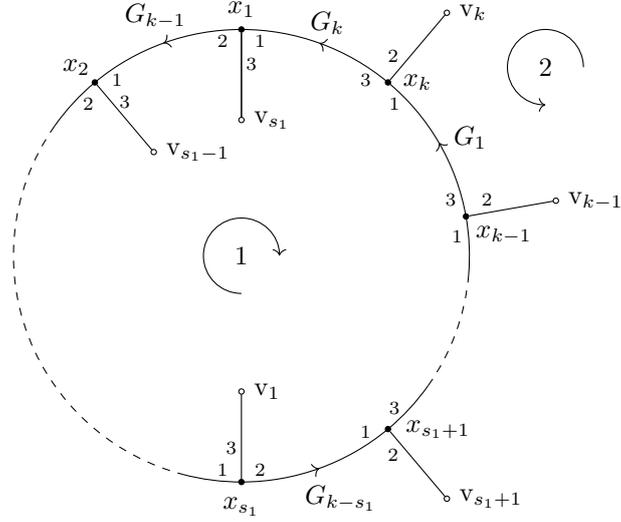
\begin{figure}
\centering
\begin{tikzpicture}

\def\rad{3cm}
\def\angdist{40}
\def\posa{90}
\def\posb{270}
\def\delta{15}
\def\orientrad{0.5cm}
\def\orientang{270}
\def\extlen{1.2cm}
\def\halflabeldelta{5}
\def\halflabeldeltarad{0.15cm}
\def\verthalflabeldeltarad{0.45cm}
\def\verthalflabeldelta{3}

\tikzset{point/.style = {draw, circle, fill=black, minimum size=2pt,inner sep=0pt}}
\tikzset{->-/.style={decoration={markings,mark=at position #1 with {\arrow{>}}},postaction={decorate}}}

\node (C) at (2,1) {1};
\node (D) at (6,3.5) {2};

\draw[->-=0.5] ([shift=(\posa-2*\angdist:\rad)]C) arc (\posa-2*\angdist:\posa-\angdist:\rad) node[midway,right=2pt,yshift=2pt]{$\GKer_{1}$};
\draw[->-=0.5] ([shift=(\posa-\angdist:\rad)]C) arc (\posa-\angdist:\posa:\rad) node[midway,above,xshift=3pt]{$\GKer_k$};
\draw[->-=0.5] ([shift=(\posa:\rad)]C) arc (\posa:\posa+\angdist:\rad) node[midway,above=2pt,xshift=-3pt]{$\GKer_{k-1}$};
\draw[->-=0.5] ([shift=(\posb:\rad)]C) arc (\posb:\posb+\angdist:\rad) node[midway,below=3pt,xshift=8pt]{$\GKer_{k-s_1}$};

\draw([shift=(\posb-\delta:\rad)]C) arc (\posb-\delta:\posb:\rad);
\draw([shift=(\posb+\angdist:\rad)]C) arc (\posb+\angdist:\posb+\angdist+\delta:\rad);
\draw([shift=(\posa+\angdist:\rad)]C) arc (\posa+\angdist:\posa+\angdist+\delta:\rad);
\draw([shift=(\posa-2*\angdist-\delta:\rad)]C) arc (\posa-2*\angdist-\delta:\posa-2*\angdist:\rad);

\draw[->] ([shift=(\orientang:\orientrad)]C) arc (\orientang:0:\orientrad);
\draw[->] ([shift=(0:\orientrad)]D) arc (0:\orientang:\orientrad);

\draw[dashed]([shift=(\posa+\angdist+\delta:\rad)]C) arc (\posa+\angdist+\delta:\posb-\delta:\rad);
\draw[dashed]([shift=(\posb+\angdist+\delta:\rad)]C) arc (\posb+\angdist+\delta:360+\posa-2*\angdist-\delta:\rad);

\path (C) node[point,label={90:$x_1$}] (x1)  at +(\posa:\rad){};
\path (C) node[point,label={[label distance=-3pt]160:$x_2$}] (x2)  at +(\posa+\angdist:\rad){};
\path (C) node[point,label={[label distance=]0:$x_k$}] (xk)  at +(\posa-\angdist:\rad){};
\path (C) node[point,label={[label distance=-1pt]-15:$x_{k-1}$}] (xkk) at +(\posa-2*\angdist:\rad){};

\path (C) node[point,label={[label distance=]-90:$x_{s_1}$}] (xa) at +(\posb:\rad){};
\path (C) node[point,label={[label distance=2pt]360:$x_{s_1+1}$}] (xaa) at +(\posb+\angdist:\rad){};

\path (C) -- (x1) -- ([turn]180:\extlen) coordinate (vx1) {};
\path (C) -- (x2) -- ([turn]180:\extlen) coordinate (vx2) {};
\path (C) -- (xk) -- ([turn]0:\extlen) coordinate (vxk) {};
\path (C) -- (xkk) -- ([turn]0:\extlen) coordinate (vxkk) {};
\path (C) -- (xa) -- ([turn]180:\extlen) coordinate (vxa) {};
\path (C) -- (xaa) -- ([turn]0:\extlen) coordinate (vxaa) {};

\path[draw] (x1)--(vx1);

\draw (x1) -- (vx1) node[point,style={fill=white},label={0:$\NVol_{s_1}$}] {};
\draw (x2) -- (vx2) node[point,style={fill=white},label={0:$\NVol_{s_1-1}$}] {};
\draw (xk) -- (vxk) node[point,style={fill=white},label={0:$\NVol_{k}$}] {};
\draw (xkk) -- (vxkk) node[point,style={fill=white},label={0:$\NVol_{k-1}$}] {};
\draw (xa) -- (vxa) node[point,style={fill=white},label={0:$\NVol_{1}$}] {};
\draw (xaa) -- (vxaa) node[point,style={fill=white},label={0:$\NVol_{s_1+1}$}] {};


\path (C) +(\posa+\halflabeldelta:\rad-\halflabeldeltarad) node {{\scriptsize 2}};
\path (C) +(\posa-\halflabeldelta:\rad-\halflabeldeltarad) node {{\scriptsize 1}};
\path (C) +(\posa+\angdist+\halflabeldelta:\rad-\halflabeldeltarad) node {{\scriptsize 2}};
\path (C) +(\posa+\angdist-\halflabeldelta:\rad-\halflabeldeltarad) node {{\scriptsize 1}};
\path (C) +(\posa-\angdist+\halflabeldelta:\rad-\halflabeldeltarad) node {{\scriptsize 3}};
\path (C) +(\posa-\angdist-\halflabeldelta:\rad-\halflabeldeltarad) node {{\scriptsize 1}};
\path (C) +(\posa-2*\angdist+\halflabeldelta:\rad-\halflabeldeltarad) node {{\scriptsize 3}};
\path (C) +(\posa-2*\angdist-\halflabeldelta:\rad-\halflabeldeltarad) node {{\scriptsize 1}};
\path (C) +(\posb+\halflabeldelta:\rad-\halflabeldeltarad) node {{\scriptsize 2}};
\path (C) +(\posb-\halflabeldelta:\rad-\halflabeldeltarad) node {{\scriptsize 1}};
\path (C) +(\posb+\angdist+\halflabeldelta:\rad-\halflabeldeltarad) node {{\scriptsize 3}};
\path (C) +(\posb+\angdist-\halflabeldelta:\rad-\halflabeldeltarad) node {{\scriptsize 1}};

\path (C) +(\posa-\verthalflabeldelta:\rad-\verthalflabeldeltarad) node {{\scriptsize 3}};
\path (C) +(\posa+\angdist-\verthalflabeldelta:\rad-\verthalflabeldeltarad) node {{\scriptsize 3}};
\path (C) +(\posa-\angdist+\verthalflabeldelta:\rad+.7*\verthalflabeldeltarad) node {{\scriptsize 2}};
\path (C) +(\posa-2*\angdist+\verthalflabeldelta:\rad+.7*\verthalflabeldeltarad) node {{\scriptsize 2}};

\path (C) +(\posb+\angdist-\verthalflabeldelta:\rad+.7*\verthalflabeldeltarad) node {{\scriptsize 2}};
\path (C) +(\posb-\verthalflabeldelta:\rad-\verthalflabeldeltarad) node {{\scriptsize 3}};
\end{tikzpicture}
\caption{The graph $\Gamma^*$ with the labeling $L^*$. It can be checked that~$L_1$ and~$L_2$ are compatible.}\label{Fig:Gamma0}
\end{figure}

By Lemma~\ref{Lemma:Independence}, in order to compute $(-1)^{\sigma_L}I(\sigma_L)$, we can pick $\Gamma^*\in O(s_1,s_2)$ and its admissible and compatible labeling $L^*$ from Figure~\ref{Fig:Gamma0}. We abbreviate $\sigma_0 = \sigma_{L^*}$. The corresponding integral \eqref{Eq:ISigma} reads
$$ I(\sigma_0) = \begin{multlined}[t] \frac{1}{V^k}\int_{x_1, \dotsc, x_k} G(x_{k-1},x_k) \dotsm G(x_{1},x_2)G(x_k,x_1) \Vol(x_{s_1}) \dotsm \Vol(x_1)\\ \Vol(x_{s_1+1}) \dotsm \Vol(x_k). \end{multlined} $$
It differs from $I(k)$ from~\eqref{Eq:Ik} in the order of $\GKer$'s and $\Vol$'s. A reordering produces the sign
$$(-1)^{\frac{1}{2}s_1(s_1-1)}.$$ 
We will compute $(-1)^{\sigma_0}$ by ordering half-edges from the edge order back to the vertex order while looking at Figure~\ref{Fig:Gamma0}. The steps are as follows: 
\begin{itemize}
 \item Transpose half-edges at internal vertices so that the first half-edge goes inside the vertex and the third outside with respect to the counterclockwise orientation. This gives $(-1)^a$.
 \item Permute external legs so that $\NVol_i$ is at $x_i$ for all $i=1$, $\dotsc$, $k$. This gives 
 $$(-1)^{\frac{1}{2}s_1(s_1-1)}. $$
 \item Permute internal edges so that $\GKer_i$ starts at the third half-edge of $x_i$ and ends at the first half-edge of $x_{i+1}$ for all $i=1$, $\dotsc$, $k-1$. This does not produce any sign as swapping of two $\GKer$'s requires two transpositions.
 \item  At this point, we have the permutation
 $$\begingroup\setlength\arraycolsep{4pt}      \begin{pmatrix}
 1 & 2 &\dots & 2(e-1)-1 & 2(e-1) & 2e-1 & 2e & 2e + 1 & \dots & 3k \\
 3 & 4 &\dots & 3k-3 & 3k-2 &  3k &  1 & 2 & \dots & 3k-1
 \end{pmatrix}.\endgroup$$
We interpret the last line as $\GKer_1\dots \GKer_k \NVol_1\dots \NVol_k$ and permute it to the sequence $\NVol_1 \GKer_1 \NVol_2 \GKer_2\dots \NVol_k \GKer_k$, which does not produce any sign. We end up with 
$$ \sigma_0' = \begin{pmatrix}
 1 & 2 & 3 & \dots & 3k-1 & 3k \\
 2 & 3 & 4 & \dots & 3k & 1
\end{pmatrix}. $$
It is now easy to see that
$$ (-1)^{\sigma_0'} = (-1)^{3k-1}. $$
\end{itemize}
In total, we get
$$ (-1)^{\sigma_0} = (-1)^{s_1 + \frac{1}{2}s_1(s_1-1) + k + 1}. $$
As for the other signs in Definition~\ref{Def:PushforwardMCdeRham}, we have $s(k,l) = k + \frac{1}{2}k(k-1)$ and $P(\NVol^k) = \frac{1}{2}k(k-1)$. There is no sign from $\Susp^2 \NVol^{s_1}\otimes \NVol^{s_2} = \Susp \NVol^{s_1} \otimes \Susp \NVol^{s_2}$ since $\Abs{\Susp} = -2$. Multiplying everything together, we get $\varepsilon(s_1,s_2)$.
\end{proof}
\begin{figure}
\centering
\begin{tikzpicture}

\def\rad{1cm}
\def\angdist{60}
\def\posa{90}
\def\extlen{0.5cm}
\def\delta{20}

\tikzset{point/.style = {draw, circle, fill=black, minimum size=2pt,inner sep=0pt}}
\tikzset{->-/.style={decoration={markings,mark=at position #1 with {\arrow{>}}},postaction={decorate}}}
\tikzset{-<-/.style={decoration={markings,mark=at position #1 with {\arrow{<}}},postaction={decorate}}}

\coordinate (C1) at (0,0) {};
\coordinate (C2) at (4*\rad,0) {};
\coordinate (C3) at (8*\rad,0) {};

\path (C1) node[point,label={270:1}] (C11)  at +(\posa:\rad) {};
\path (C1) node[point,label={[label distance=-3pt]30:2}] (C12)  at +(\posa-\angdist:\rad) {};
\path (C1) node[point,label={[label distance=-3pt]290:k}] (C1k)  at +(\posa+\angdist:\rad) {};

\draw ([shift=(\posa:\rad)]C1) arc (\posa:\posa+\angdist:\rad);
\draw[-<-=0.5] ([shift=(\posa-\angdist:\rad)]C1) arc (\posa-\angdist:\posa:\rad);

\draw ([shift=(\posa+\angdist:\rad)]C1) arc (\posa+\angdist:\posa+\angdist+\delta:\rad);
\draw ([shift=(\posa-\angdist-\delta:\rad)]C1) arc (\posa-\angdist-\delta:\posa-\angdist:\rad);

\draw[dashed]([shift=(\posa+\angdist+\delta:\rad)]C1) arc (\posa+\angdist+\delta:360+\posa-\angdist-\delta:\rad);


\path (C2) node[point,label={[label distance=-1pt]90:$\bar{1}$}] (C21)  at +(\posa:\rad) {};
\path (C2) node[point,label={[label distance=-3pt]30:$\bar{k}$}] (C2k)  at +(\posa-\angdist:\rad) {};
\path (C2) node[point,label={[label distance=-3pt]290:$\bar{2}$}] (C22)  at +(\posa+\angdist:\rad) {};

\draw[->-=0.5]  ([shift=(\posa:\rad)]C2) arc (\posa:\posa+\angdist:\rad);
\draw ([shift=(\posa-\angdist:\rad)]C2) arc (\posa-\angdist:\posa:\rad);

\draw ([shift=(\posa+\angdist:\rad)]C2) arc (\posa+\angdist:\posa+\angdist+\delta:\rad);
\draw ([shift=(\posa-\angdist-\delta:\rad)]C2) arc (\posa-\angdist-\delta:\posa-\angdist:\rad);

\draw[dashed]([shift=(\posa+\angdist+\delta:\rad)]C2) arc (\posa+\angdist+\delta:360+\posa-\angdist-\delta:\rad);


\path (C3) node[point,label={[label distance=0pt]90:$\bar{k}$}] (C3k)  at +(\posa:\rad) {};
\path (C3) node[point,label={[label distance=-3pt]120:$\bar{1}$}] (C31)  at +(\posa+\angdist:\rad) {};
\path (C3) node[point,label={[label distance=-3pt]60:$\bar{2}$}] (C32)  at +(\posa+2*\angdist:\rad) {};

\draw[->-=0.5] ([shift=(\posa+\angdist:\rad)]C3) arc (\posa+\angdist:\posa+2*\angdist:\rad);
\draw ([shift=(\posa:\rad)]C3) arc (\posa:\posa+\angdist:\rad);

\draw ([shift=(\posa+2*\angdist:\rad)]C3) arc (\posa+2*\angdist:\posa+2*\angdist+\delta:\rad);
\draw ([shift=(\posa-\delta:\rad)]C3) arc (\posa-\delta:\posa:\rad);

\draw[dashed]([shift=(\posa+2*\angdist+\delta:\rad)]C3) arc (\posa+2*\angdist+\delta:360+\posa-\delta:\rad);


\draw[-{>[scale=1.0]}] ($(C1)+(1.5*\rad,0)$) -- ($(C1)+(2.5*\rad,0)$) node[midway,above=2pt]{Inv};
\draw[-{>[scale=1.0]}] ($(C2)+(1.5*\rad,0)$) -- ($(C2)+(2.5*\rad,0)$) node[midway,above]{(-1)};


\path (C1) -- (C11) -- ([turn]0:\extlen) coordinate (C11x){};
\path[draw] (C11)--(C11x);
\path (C1) -- (C12) -- ([turn]180:\extlen) coordinate (C12x){};
\path[draw] (C12)--(C12x);
\path (C1) -- (C1k) -- ([turn]0:\extlen) coordinate (C1kx){};
\path[draw] (C1k)--(C1kx);

\path (C2) -- (C21) -- ([turn]180:\extlen) coordinate (C21x){};
\path[draw] (C21)--(C21x);
\path (C2) -- (C22) -- ([turn]0:\extlen) coordinate (C22x){};
\path[draw] (C22)--(C22x);
\path (C2) -- (C2k) -- ([turn]180:\extlen) coordinate (C2kx){};
\path[draw] (C2k)--(C2kx);

\path (C3) -- (C31) -- ([turn]180:\extlen) coordinate (C31x){};
\path[draw] (C31)--(C31x);
\path (C3) -- (C32) -- ([turn]0:\extlen) coordinate (C32x){};
\path[draw] (C32)--(C32x);
\path (C3) -- (C3k) -- ([turn]180:\extlen) coordinate (C3kx){};
\path[draw] (C3k)--(C3kx);

\node[point,style={fill=white}] at (C11x) {};
\node[point,style={fill=white}] at (C12x) {};
\node[point,style={fill=white}] at (C1kx) {};

\node[point,style={fill=white}] at (C21x) {};
\node[point,style={fill=white}] at (C22x) {};
\node[point,style={fill=white}] at (C2kx) {};

\node[point,style={fill=white}] at (C31x) {};
\node[point,style={fill=white}] at (C32x) {};
\node[point,style={fill=white}] at (C3kx) {};
\end{tikzpicture}
\caption{The mirror isomorphism $M: 1 \dots k \mapsto \bar{k} \dots\bar{1}$ is a composition of the inversion and the counterclockwise rotation by one place.}\label{Fig:Mirror}
\end{figure}
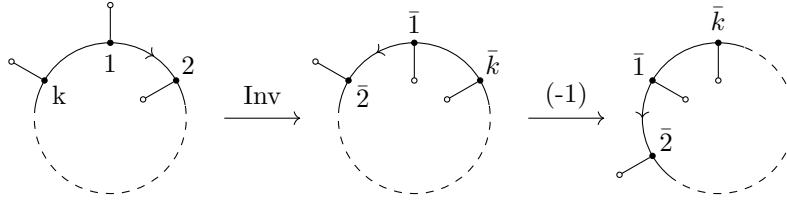
\begin{Lemma}[Combinatorial coefficient]
\label{Lemma:CombinatorialCoefficientForMCOnCircle}
It holds $$C(s_1,s_2) = \frac{1}{2} a k! \binom{k-1}{s_1}.$$
\end{Lemma}
\begin{proof}
Every isomorphism of ribbon graphs $\Gamma$ and $\Gamma'$ from $O(s_1,s_2)$ is a composition of the clockwise rotation $(r)$ for $r\in \Z$ and the mirror operation $M$ defined as follows: If $1$,~$\dotsc$, $k$ label internal vertices in the clockwise direction starting from the north-pole, then the result of $M$ is $\bar{k}$,~$\dotsc$, $\bar{1}$, where $\bar{i}$ means that the external leg is reversed (see Figure~\ref{Fig:Mirror}). These operations satisfy
$$ (r+k) = (r),\; (r)(-r)=\Id,\; M^2 = \Id,\; (r)M = M(-r), $$
and hence generate a group $\mathrm{G}$ which is isomorphic to the dihedral group $\Z_k \rtimes \Z_2$. The orbit space $O(s_1,s_2)/\mathrm{G}$ is in $1:1$ correspondence with isomorphism classes of admissible $O_k$-graphs and $\Aut(\Gamma)$ is in $1:1$ correspondence with $\Stab{\Gamma}$. From the orbit-stabilizer formula, we get
\allowdisplaybreaks
\begin{align*}
\sum_{\substack{[\Gamma] \text{ admiss. }\\O_k-\text{graph}}} \frac{1}{\Abs{\Aut(\Gamma)}} &= \sum_{[\Gamma]\in O(s_1,s_2)/\mathrm{G}} \frac{1}{\Abs{\Stab{\Gamma}}} = \sum_{\Gamma\in O(s_1,s_2)} \frac{1}{\Abs{\Orb{\Gamma}}\Abs{\Stab{\Gamma}}} \\ &= \frac{\Abs{O(s_1,s_2)}}{\Abs{\mathrm{G}}} = \frac{1}{2k}{\binom{k}{s_1}}\times\begin{cases} 1 & \text{for }s_1=s_2, \\ 2 & \text{for }s_1\neq s_2.\end{cases}
\end{align*}
The two cases are compensated in the sum over labelings: For $s_1=s_2$, both labelings $L_1^b$ are admissible, and hence we get the factor $2$.

Next, we multiply by $k! s_1(k-s_1)$, which is the number of $L_1^v$ and $L_3^b$. There is also the factor $\frac{1}{l!} = \frac{1}{2}$. Multiplying everything together, we get $C(s_1,s_2)$.
\end{proof}

Before we summarize the results of our computations (see Proposition~\ref{Proposition:MCSphere} below), we show directly that $\PMC$ is a Maurer-Cartan element.

\begin{Lemma}[Maurer-Cartan equation for $\Sph{n}$] \label{Lem:MCEquation}
Consider $\Sph{n}$ with the Green kernel from~\eqref{Eq:GreenKernelMC1}. The collection $(\PMC_{lg})$ satisfies the Maurer-Cartan equation~\eqref{Eq:MaurerCartanEquation} for $\dIBL(\CycC(\Harm(\Sph{n})))$.
\end{Lemma}

\begin{proof}
%
We will show that for every $l\ge 1$, $g\ge 0$ all summands in the relation corresponding to $(l,g)$ vanish. The summands for $(l,g)= (1,0)$ are $\OPQ_{110}(\PMC_{10})$ and $\frac{1}{2}\OPQ_{210}(\PMC_{10},\PMC_{10})$, and the summand for $(l,g)=(2,0)$ is $\OPQ_{120}(\PMC_{10})$.
The first term vanishes trivially as $\OPQ_{110}=0$, while the other two terms vanish by~\cite[Proposition 12.5]{Cieliebak2015} because $\PMC_{10} = \MC_{10}$
is the canonical Maurer-Cartan element. For $(l,g)\neq (1,0)$, we have the following four situations:
\begin{description}[font=\normalfont\itshape]
\item[$\OPQ_{210}\circ_2 \PMC_{lg}$, $l\ge 2$:] 
Let $\Psi = \Psi_1 \cdots \Psi_l\in \Ext_l \CycC$ be a summand of $\PMC_{lg}$. 
 From Proposition~\ref{Prop:TotalVanishing} it follows that the summands can be chosen such that $\Psi_1$,~$\dotsc$, $\Psi_l \in \DBCycRed\Harm(\Sph{n})[3-n]$, i.e., such that $\Psi_i$ evaluates to $0$ whenever $\NOne$ is a part of its argument. From Definition~\ref{Def:CircS}, we compute
$$\OPQ_{210}\circ_2 (\Psi_1\cdots\Psi_l) = \sum_{\sigma\in \Perm_{2,l-2}} \varepsilon(\sigma,\Psi)\OPQ_{210}(\Psi_{\sigma_1^{-1}}\cdot\Psi_{\sigma_2^{-1}})\cdot\Psi_{\sigma_3^{-1}}\cdots \Psi_{\sigma_l^{-l}}. $$
We clearly have $\OPQ_{210}(\Psi_{\sigma_1^{-1}}\cdot\Psi_{\sigma_2^{-1}})=0$ because $\OPQ_{210}$ feeds $\NOne$ into one of its inputs. It follows that $\OPQ_{210}\circ_2 \PMC_{lg} = 0$.

\item[$\OPQ_{210}\circ_{1,1}(\PMC_{l_1 g_1} \odot \PMC_{l_2 g_2})$, $(l_i,g_i) \neq (1,0)$:] A similar argument as above.
\item[$\OPQ_{120}\circ_1 \PMC_{lg}$, $(l,g)\neq(1,0)$:] A similar argument as above using that $\OPQ_{120}$ also feeds~$\NOne$ into its input. 
\item[$\OPQ_{210}\circ_{1,1} (\PMC_{10}\odot \PMC_{lg})$, $(l,g)\neq (1,0)$:] 
As in the case of $\OPQ_{210}\circ_2 \PMC_{lg}$, let $\Psi_1$,~$\dotsc$, $\Psi_l \in \DBCycRed\Harm(\Sph{n})[3-n]$. Recall that we write $\Omega_i = \Susp \omega_i \in \BCyc \Harm(\Sph{n})[3-n]$ and $\Omega = \Omega_1\otimes \dotsb \otimes \Omega_l$. From Definition~\ref{Def:CircS}, we compute
\allowdisplaybreaks
\begin{align*}
&[\OPQ_{210} \circ_{1,1}(\PMC_{10}\odot \Psi)](\Omega_1 \otimes \dotsb \otimes \Omega_l) 
\\ &\quad = \begin{multlined}[t] \Bigl[ \sum_{i=1}^l (-1)^{\Abs{\Psi_i}(\Abs{\Psi_1} + \dotsb + \Abs{\Psi_{i-1}})} \OPQ_{210}(\PMC_{10}\Psi_i)\Psi_1\cdots \widehat{\Psi_i}\cdots \Psi_l\Bigr]\\ (\Omega_1 \otimes \dotsb \otimes \Omega_l) \end{multlined} \\ 
 &\quad = \begin{multlined}[t] \vphantom{\sum_{i}}\smash{\sum_{\substack{\mu \in \Perm_l \\ i=1, \dotsc, l}}} \frac{1}{l!} (-1)^{\Abs{\Psi_i}(\Abs{\Psi_1} + \dotsb + \Abs{\Psi_{i-1}})} \varepsilon(\mu,\Omega) (\OPQ_{210}(\PMC_{10}\cdot\Psi_i)\otimes \Psi_1\otimes \dotsb \\ \widehat{\Psi_i}\dotsb\otimes \Psi_l)(\Omega_{\mu_1^{-1}} \otimes \dotsb \otimes \Omega_{\mu_l^{-1}}). \end{multlined}
\end{align*}
For every $i=1$, $\dotsc$, $l$, we have
\allowdisplaybreaks
 \begin{align*} 
 & \OPQ_{210}(\PMC_{10}\cdot \Psi_i)(\Omega) = \OPQ_{210}(\PMC_{10}\otimes \Psi_i)(\Omega) \\
 & \qquad = \begin{multlined}[t]- \sum \varepsilon(\omega\mapsto \omega^1 \omega^2)[(-1)^{(n-1)\Abs{\omega^1}} \PMC_{10}(\Susp \NOne \omega^1) \Psi_i(\Susp \NVol \omega^2) \\ {}+ (-1)^{\Abs{\omega^1}} \PMC_{10}(\Susp \NVol \omega^1) \Psi_i(\Susp \NOne \omega^2)] \end{multlined} \\ 
 & \qquad = - \sum \varepsilon(\omega\mapsto \omega^1 \omega^2)(-1)^{(n-1)\Abs{\omega^1}} \PMC_{10}(\Susp \NOne \omega^1) \Psi_i(\Susp \NVol \omega^2).
\end{align*}
This can be non-zero only if $\omega = \NOne \NVol^{s-1}$ for some $s\ge 2$ (up to a cyclic permutation). For this input, we get
\allowdisplaybreaks
\begin{align*} &\OPQ_{210}(\PMC_{10}\cdot \Psi_i)(\Susp \NOne \NVol^{s-1}) \\ &\quad = \begin{multlined}[t] -[\varepsilon(\NOne \NVol^{s-1} \mapsto \NOne\NVol^{s-1})\PMC_{10}(\Susp \NOne\NOne\NVol)\Psi_i(\Susp\NVol^{s-1}) \\ {}+ \varepsilon(\NOne \NVol^{s-1} \mapsto \NVol\NOne\NVol^{s-2})\PMC_{10}(\Susp \NOne\NVol\NOne)  \Psi_i(\Susp\NVol^{s-1})] \end{multlined} \\
 &\quad= (-1)^{n-3}[1+(-1)^{ns+s-1}] \Psi_i(\Susp\NVol^{s-1}).
\end{align*}
The prefactor in brackets is $0$ for $n$ odd or $s$ even, whereas $\NVol^{s-1} = 0$ for $n$ even and $s$ odd. Therefore,  we have $\OPQ_{210}\circ_{1,1} (\PMC_{10}\odot \PMC_{lg}) = 0$. \qedhere
\end{description}
\end{proof}


\begin{Proposition}[Formal pushforward Maurer-Cartan element for $\Sph{n}$] \label{Proposition:MCSphere}
Consider the round sphere $\Sph{n}$ with the Green kernel~\eqref{Eq:GreenKernelMC1}. The formal pushforward Maurer-Cartan element $\PMC$ is a strictly reduced Maurer-Cartan element for $\dIBL(\CycC(\Harm(\Sph{n})))$ which satisfies
$$ \PMC_{10} = \MC_{10}\quad\text{for all }n\in \N$$
plus the following properties depending on the dimension:
\begin{description}[font=\normalfont\itshape]
 \item[($n = 1$):] It holds $\PMC_{lg}=0$ for all $l\ge 1$, $g\ge 0$ such that $(l,g) \neq (1,0)$, $(2,0)$; the only non-trivial relation for $\PMC_{20}$ is
  \begin{equation}\label{Eq:MC20}
  \PMC_{20}(\Susp \NVol^{s_1}\otimes  \Susp \NVol^{s_2}) = (-1)^{s_1+1} \frac{1}{2} s_1 k! \binom{k-1}{s_1} I(k),
  \end{equation}
  where $s_1$, $s_2\ge 1$ are such that $k = s_1 + s_2$ is even, and $I(k)$ is given by~\eqref{Eq:TheFormulaForIk}.
 \item[($n=2$):] It holds $\PMC_{l0}=0$ for all $l\ge 2$. We also have $\PMC_{11}=0$.
 \item[($n\ge 3$):] It holds $\PMC_{lg} = 0$ for all $l\ge 1$, $g\ge 0$ such that $(l,g) \neq (1,0)$.
\end{description}
\end{Proposition}

Notice that $\PMC_{20}\not\in \Ext_2 \CycC(\Harm(\Sph{1}))$, i.e.~$\PMC_{20}$ is a long cochain, because it is non-zero in infinitely many weights.

\subsection{Twisted \texorpdfstring{$\IBLInfty$-structure for $\Sph{n}$}{IBL-infinity-structure for Sn}
}
\label{Section:HomSphere}
Let $e_0$, $e_1$ be the basis of $\Harm(\Sph{n})[1]$ defined by
\begin{equation*} \label{Eq:BasisOfHarm}
 e_0 := \NOne := \SuspU 1, \quad e_1:=\NVol := \frac{1}{V}\SuspU\Vol.
\end{equation*}
The degrees satisfy
$$ \Abs{\NOne} = -1, \quad \Abs{\NVol} = n-1. $$
The matrix of the pairing $\Pair$ with respect to the basis $e_0$, $e_1$ reads
$$ \Pair=\begin{pmatrix}
 0 & 1 \\
 (-1)^{n} & 0
\end{pmatrix}. $$
The dual basis $e^0$, $e^1$ to $e_0$, $e_1$ with respect to $\Pair$ is thus
$$ e^0= \NVol,\quad e^1 = (-1)^{n} \NOne. $$
It follows that the matrix $(T^{ij})$ from~\eqref{Eq:PropagatorT} satisfies
\begin{equation*} 
 (T^{ij}) = - \begin{pmatrix}
0 & 1 \\
1 & 0
\end{pmatrix}.
\end{equation*}

We clearly have
$$ \CRedDBCyc \Harm(\Sph{1}) = \Bigl\{ \sum_{k=1}^\infty c_k \NVol^{k*} \BigMid c_k \in \R \Bigr\}, $$
where $\NVol^{k*}$ is the dual to the cyclic word $\NVol^k = \NVol \dots \NVol$ of length $k$. Observe that the cyclic symmetry gives
\begin{equation*}
\NVol^i = (-1)^{(n-1)(i-1)} \NVol^i\quad\text{for all }i\ge 1.
\end{equation*}
Therefore, $\NVol^{i*} = 0$ holds if both $n$ and $i$ are even.

For $n\ge 2$, the vector space $\Harm(\Sph{n})$ is connected and simply-connected, and Proposition~\ref{Prop:SimplCon} implies that there are no long reduced cyclic cochains (i.e., we have only finite sums of $\NVol^{k*}$'s).

The product $\mu_2: \Harm[1]^{\otimes 2} \rightarrow \Harm[1]$ from \eqref{Eq:HarmProd} has the following matrix with respect to the basis $\NOne$, $\NVol$:
\begin{equation*} 
 \mu_2 = \begin{pmatrix} \NOne &  \NVol \\ (-1)^n \NVol & 0 \end{pmatrix}.
\end{equation*}
Because $\mu_2(\NVol, \NVol) = 0$, we get
\begin{equation*}
\HIBL^\MC(\RedCycC(\Harm(\Sph{n})))[1] = \begin{cases}
                        \langle \Susp \NVol^{i*} \mid i \ge 1 \rangle & \text{for }n\ge 3\text{ odd}, \\
                        \langle  \Susp \NVol^{2i-1*} \mid i\ge 1 \rangle & \text{for }n\text{ even},\\
 \bigl\{ \Susp\sum_{k=1}^\infty c_k \NVol^{k*} \mid c_k\in \R \bigr\} & \text{for }n=1. 
\end{cases}
\end{equation*}
Because we are in the strictly unital and strictly augmented case, we obtain
\begin{equation} \label{Eq:HIBLSn}
\HIBL^\MC(\CycC)[1] = \begin{cases}
\langle \Susp \NVol^{i*}, \Susp \NOne^{2j-1*} \mid i, j \ge 1\rangle & \text{for }n\ge 3 \text{ odd}, \\
\langle \Susp \NVol^{2 i-1*}, \Susp \NOne^{2j-1*} \mid i, j \ge 1\rangle &\text{for }n\text{ even}, \\
 \bigl\langle \Susp\sum_{k=1}^\infty c_k \NVol^{k*}, \Susp \NOne^{2j-1*}\mid c_k\in \R, j \ge 1\bigr\rangle & \text{for }n=1. 
\end{cases}
\end{equation}
The canonical $\IBL$-operations can be written as
\begin{align*}
\OPQ_{210}(\Susp^2 \psi_1 \otimes \psi_2)(\Susp \omega) &= \begin{multlined}[t]-\sum \varepsilon(\omega\mapsto \omega^1 \omega^2)[(-1)^{(n-1)\Abs{\omega^1}} \psi_1(e_0 \omega^1) \\ \psi_2(e_1 \omega^2) + (-1)^{\Abs{\omega_1}}\psi_1(e_1 \omega^1) \psi_2(e_0 \omega^2)], \end{multlined} \\
\OPQ_{120}(\Susp \psi)(\Susp^2 \omega_1\otimes \omega_2) & = \begin{multlined}[t] - \frac{1}{2} \sum \varepsilon(\omega_1\mapsto \omega_{1}^{1}) \varepsilon(\omega_2\mapsto \omega_{2}^{1}) [(-1)^{(n-1)\Abs{\omega_{1}^{1}}} \\ \psi(e_0 \omega_{1}^{1} e_1 \omega_{2}^{1})  + (-1)^{\Abs{\omega_{1}^{1}}}\psi(e_1 \omega_{1}^{1} e_0 \omega_{2}^{1})] \end{multlined}
\end{align*}
for all $\psi$, $\psi_1$, $\psi_2 \in \CDBCyc\Harm$ and generating words $\omega$, $\omega_1$, $\omega_2\in \BCyc\Harm$. For all $k$, $k_1$, $k_2 \ge 1$, we have
$$ \OPQ_{210}((\Susp \NVol^{k_1*}) \cdot (\Susp \NVol^{k_2*})) = 0\quad\text{and}\quad \OPQ_{120}(\Susp \NVol^{k*}) = 0$$
because both $\OPQ_{210}$ and $\OPQ_{120}$ feed $\NOne$ into their inputs. For the \emph{canonically twisted reduced $\IBL$-algebra}, this implies the following:
$$ \IBL\bigl(\HIBL^\MC(\RedCycC)\bigr) = \bigl(\HIBL^\MC(\RedCycC), \OPQ_{210} \equiv 0, \OPQ_{120} \equiv 0 \bigr)\quad \text{for all }n\in \N.  $$
By Proposition~\ref{Prop:Ones}, the only possibly non-zero relation  of $\IBL(\HIBL^\MC(\CycC))$ is   
$$\begin{aligned}
& \OPQ_{210}(\Susp \NOne^* \otimes \Susp \NVol^{k*}) \\[\jot]
&\qquad = (-1)^{n-2} \Susp (\NVol^{k*} \circ \iota_\NVol) \\ 
&\qquad = (-1)^{n-2}\bigl(\sum_{i=1}^{k-1} (-1)^{i \Abs{\NVol}}\bigr)\Susp\NVol^{k-1 *}
= \begin{cases}
   -(k-1) \Susp \NVol^{k-1*} & \text{for }n\text{ odd},\\
    0 & \text{for }n\text{ even}.
  \end{cases}\end{aligned}$$
The reason for $0$ for even $n$ is that either $k$ is odd, in which case $\sum_{i=1}^{k-1} (-1)^i = 0$, or $k$ is even, in which case $\NVol^{k*} = 0$. Therefore, for the \emph{canonically twisted $\IBL$-algebra}, we have
\begin{equation*}
\IBL\bigl(\HIBL^\MC(\CycC)\bigr) = \bigl(\HIBL^\MC(\CycC), \OPQ_{210}, \OPQ_{120} \equiv 0 \bigr)\quad \text{for all }n\in \N,
\end{equation*}
where $\HIBL^\MC(\CycC)$ is given by \eqref{Eq:HIBLSn} and $\OPQ_{210}$ satisfies the following:
\begin{description}[font=\normalfont\itshape]
\item[($n$ even):] $\OPQ_{210} \equiv 0$.
\item[($n\ge 3$ odd):] The non-trivial relations are
$$ \OPQ_{210}(\Susp \NOne^* \otimes \Susp \NVol^{k*}) = \OPQ_{210}(\Susp \NVol^{k*} \otimes \Susp \NOne^*) = -(k-1) \NVol^{k-1*}\quad\text{for }k\ge 2.  $$
\item[($n=1$):]  The non-trivial relations are
$$ \OPQ_{210}\Bigl(\Susp \NOne^* \otimes \Susp\sum_{k=1}^\infty c_k \NVol^{k*}\Bigr) = - \Susp \sum_{k=1}^\infty k c_{k+1} \NVol^{k*} \quad\text{for }c_k\in \R. $$
\end{description}
Recall that the twist by $\MC$ does not produce any higher operation $\OPQ_{1lg}^\MC$.

We will now consider $\dIBL^\PMC(\CycC(\Harm(\Sph{n})))$. Recall that $\OPQ_{110}^\PMC = \OPQ_{210}\circ_1 \PMC_{10}$, $\OPQ_{210}^\PMC = \OPQ_{210}$ and $\OPQ_{120}^\PMC = \OPQ_{120} + \OPQ_{210}\circ_1 \PMC_{20}$. By Proposition~\ref{Proposition:MCSphere}, we have $\PMC_{10} = \MC_{10}$ for all $n\in \N$ and $\PMC_{20} = 0$ for all $n\ge 2$. It follows that $\OPQ_{110}^\PMC = \OPQ_{110}^\MC$ for all $n\in \N$ and that the only non-trivial twist may occur in $\OPQ_{120}^\PMC$ for $\Sph{1}$. Using~\eqref{Eq:Twistn2}, we get for all $\psi\in \CDBCyc \Harm(\Sph{n})$ and generating words $\omega_1$, $\omega_2 \in \BCyc \Harm(\Sph{n})$ the following:
\begin{equation}
\begin{aligned}
& (\OPQ_{210}\circ_1 \PMC_{20})(\Susp \psi)(\Susp \omega_1 \otimes \Susp \omega_2) \label{Eq:CoprodTwist}\\
& \quad = \begin{multlined}[t] (-1)^{n-2}\Bigl[ \sum \varepsilon(\omega_1 \mapsto \omega_1^1 \omega_1^2)\psi(\NOne \omega_1^1)\PMC_{20}(\Susp \NVol \omega_1^2 \otimes \Susp \omega_2) \\ {}+ (-1)^{(n-3+\Abs{\omega_1})(n-3+\Abs{\omega_2})} \sum \varepsilon(\omega_2 \mapsto \omega_2^1 \omega_2^2) \psi(\NOne \omega_2^1)  \\ \PMC_{20}(\Susp \NVol \omega_2^2 \otimes \Susp \omega_1)\Bigr].  \end{multlined}\end{aligned}
\end{equation}

In this paragraph, we suppose that $n=1$ and compute $\OPQ_{120}^\PMC$. Clearly, $(\OPQ_{210}\circ_1\PMC_{20})(\Susp \NVol^{k*}) = 0$ for all $k\ge 1$ since~$\NOne$ is fed into $\NVol^{k*}$. A non-zero evaluation of $(\OPQ_{210}\circ_1\PMC_{20})(\Susp\NOne^{k*})$ for some $k\ge 1$ odd is possible only on $\Susp \NOne^{k-1}\NVol^{k_1}\otimes \Susp\NVol^{k_2}$ for $k_1$, $k_2\ge 0$ (up to a transposition of arguments and their cyclic permutation). If $k>1$, only the first summand of~\eqref{Eq:CoprodTwist} contributes, and we get
\begin{equation*}
\begin{aligned}
&(\OPQ_{210}\circ_1\PMC_{20})(\Susp\NOne^{k*})(\Susp \NOne^{k-1}\NVol^{k_1}\otimes \Susp\NVol^{k_2})  \\
& \qquad= \begin{multlined}[t] (-1)^{n-2} \sum \varepsilon(\NOne^{k-1}\NVol^{k_{1}} \mapsto \omega_1 \omega_2) \NOne^{k*}(\NOne \omega_1) \PMC_{20}(\Susp \NVol \omega_2\otimes \Susp \NVol^{k_2}) \end{multlined} \\ & \qquad = (-1)^{n-2} \NOne^{k*}(\NOne\NOne^{k-1}) \PMC_{20}(\Susp \NVol\NVol^{k_{1}}\otimes \Susp \NVol^{k_2})  \\ & \qquad = - \PMC_{20}(\Susp \NVol^{k_{1}+1}\otimes \Susp\NVol^{k_2}).
\end{aligned}
\end{equation*}
According to Proposition~\ref{Proposition:MCSphere}, this is non-zero if and only if $k_{1}+k_2$ is odd. It follows that 
$$ \OPQ_{120}^\PMC \neq \OPQ_{120}^\MC = \OPQ_{120}\quad\text{on the chain level for }\Sph{1}. $$
However, the chains $\Susp \NOne^{k-1}\NVol^{k_1}\otimes \Susp \NVol^{k_2}$ for $k>1$ do not survive to the homology (c.f., \eqref{Eq:HIBLSn}). The only possibility is thus $k=1$. In this case, both summands of~\eqref{Eq:CoprodTwist} contribute, and  using~\eqref{Eq:MC20}, we get for all $k_1$, $k_2 \ge 1$ the following:
\allowdisplaybreaks
\begin{align*}
&(\OPQ_{210}\circ_1\PMC_{20})(\Susp\NOne^*)(\Susp\NVol^{k_1} \otimes \Susp\NVol^{k_2}) \\ &\qquad = \begin{multlined}[t](-1)^{n-2}\Bigl[\sum \varepsilon(\NVol^{k_1} \mapsto \NVol^0 \NVol^{k_1}) \NOne^*(\NOne) \PMC_{20}(\Susp\NVol^{k_1+1}\otimes \Susp \NVol^{k_2}) \\ {}+ (-1)^{(n-3 + k_1(n-1))(n-3 + k_2(n-1))} \sum \varepsilon(\NVol^{k_2} \mapsto \NVol^0 \NVol^{k_2})  \NOne^*(\NOne)\\ \PMC_{20}(\Susp \NVol^{k_2 + 1} \otimes\Susp\NVol^{k_1})\Bigr] \end{multlined}
\\&\qquad = -  k_1 \PMC_{20}(\Susp\NVol^{k_1+1}\otimes\Susp\NVol^{k_2}) -  k_2 \PMC_{20}(\Susp\NVol^{k_2+1}\otimes\Susp\NVol^{k_1}) \\ 
&\qquad = \begin{multlined}[t] -\frac{1}{2}(k_1+k_2+1)!I(k_1+k_2+1)\Bigl[(-1)^{k_1} k_1 (k_1+1) \binom{k_1+k_2}{k_1+1} \\ {}+ (-1)^{k_2} k_2  (k_2+1) \binom{k_1+k_2}{k_2+1}\Bigr] \end{multlined} \\ 
&\qquad =  -\frac{1}{2}(k_1+k_2+1)! k_1 k_2 \binom{k_1+k_2}{k_1} \underbrace{I(k_1 + k_2 + 1) [(-1)^{k_1} + (-1)^{k_2}]}_{=:(*)}.
\end{align*}
Denoting $k:= k_1 + k_2 + 1$, we have that $(-1)^{k_1} + (-1)^{k_2} = 0$ for $k$ even and $I(k) = 0$ for $k$ odd. Therefore, $(*) = 0$ for any $k_1$, $k_2\ge 1$.
This implies that 
$$ \OPQ_{120}^\PMC = \OPQ_{120}^\MC = \OPQ_{120}\quad\text{on the homology for }\Sph{1}. $$
We conclude that the \emph{twisted $\IBL$-algebra} satisfies
$$ \IBL\bigl(\HIBL^\PMC(\CycC(\Harm(\Sph{n})))\bigr) = \IBL\bigl(\HIBL^\MC(\CycC(\Harm(\Sph{n})))\bigr) \quad\text{for all }n\in \N. $$

As for the \emph{higher twisted operations}, combining Proposition~\ref{Prop:dIBL} and Proposition~\ref{Proposition:MCSphere}, we see that for~$\Sph{n}$ with $n\in \N\backslash\{2\}$ all higher operations~$\OPQ_{1lg}^\PMC$ vanish already on the chain level. For $n=2$, we have that $\OPQ_{1l0}^\PMC = 0$ for all $l\ge 3$ and $\OPQ_{111}^\PMC = 0$ on the chain level. However, we did not prove that all higher operations vanish on the chain level. As for the operations induced on the homology, the graded vector space~$\HIBL^\PMC(\CycC(\Harm(\Sph{2})))$ is concentrated in even degrees and $\OPQ_{1lg}^\PMC$ are odd (see Definition~\ref{Def:IBLInfty}). Therefore, all higher operations vanish also on $\HIBL^\PMC(\CycC(\Harm(\Sph{2})))$.

The string topology $\StringH_*(\Sph{n})$ and the string operations $\StringOp_2$ and $\StringCoOp_2$ were computed in \cite{Basu2011} for all $n\in \N$.
We review their results and basic ideas below:

We will consider \emph{even spheres} first. The minimal model for the Borel construction $\LoopBorel \Sph{2m}$ for $m\in \N$ is denoted by $\Lambda^{\Sph{1}}(2,m)$ --- it is the free graded commutative dga (=:cdga) over~$\R$ generated by homogenous vectors $x_1$, $y_1$, $x_2$, $y_2$, $u$ of degrees
$$ \Abs{x_1} = 2m,\quad \Abs{y_1} = 2m - 1, \quad\Abs{x_2} = 4m-1,\quad \Abs{y_2} = 2(2m - 1),\quad\Abs{u} = 2, $$
whose differential $\Dd$ satisfies
$$ \Dd y_1 = 0, \quad \Dd x_1 = y_1 u, \quad \Dd y_2 = - 2 x_1 y_1, \quad\Dd x_2 = x_1^2 + y_2 u. $$
The minimal model for the loop space $\Loop \Sph{2m}$ is the dga $\Lambda(2,m)$ which is obtained from $\Lambda^{\Sph{1}}(2,m)$ by setting $u = 0$. A computation (see \cite[Theorem 3.6]{Basu2011}) gives the following for all $m\in \N$:
\begin{equation}\label{Eq:EvenSphereString}
\begin{aligned}
\H^*(\Loop \Sph{2m}; \R) &\simeq \H_*(\Lambda(2,m), \Dd) =\langle y_2^i x_1 - 2 i y_1 x_2 y_2^{i-1}, y_1 y_2^j, 1 \mid i, j\in \N_0 \rangle,  \\
\StringCoH^*(\Loop \Sph{2m}; \R) &\simeq \H_*(\Lambda^{\Sph{1}}(2,m),\Dd) = \langle y_1 y_2^i, u^j \mid i, j\in \N_0\rangle,
\end{aligned}
\end{equation}
where $y_2^0 :=u^0 := 1$ is the unit in $\Lambda^{\mathrlap{\Sph{1}}\hphantom{S}}(2,m)$ and $\langle \cdot \rangle$ denotes the linear span over~$\R$. Clearly, the cohomology groups are degree-wise finite-dimensional, and hence, using the universal coefficient theorem, they are isomorphic to the corresponding homology groups. We can thus identify $\H_*(\Loop \Sph{2m}; \R)$ and $\StringH_*(\Loop\Sph{2m}; \R)$ with the vector spaces on the right hand side of \eqref{Eq:EvenSphereString}. We have $\StringH_{2k}= \langle u^k \rangle$ for all $k\in \N_0$, and hence the multiplication with $u$ induces an isomorphism $\StringH_{2k} \simeq \StringH_{2k+2}$. This corresponds to the cap product with the Euler class in \eqref{Eq:Gysin}, and exactness of the sequence implies $\Mark(\StringH_{2k}) = \Erase(\StringH_{2k}) = 0$. Using this and degree considerations, we get $\StringOp_2=\StringCoOp_2 = 0$.

We will now consider \emph{odd spheres} with $n\ge 3$. The minimal model for $\LoopBorel \Sph{2m+1}$ 
for $m\in \N$ is denoted simply by $\Lambda(x,y,u)$ --- it is the free cdga on homogenous vectors $x$, $y$, $u$ of degrees
$$ \Abs{x} = 2m+1, \quad \Abs{y} = 2m,\quad \Abs{u} = 2, $$
such that
$$ \Dd x = y u, \quad \Dd y = \Dd u = 0. $$
We get immediately
$$\begin{aligned}
\H^*(\Loop \Sph{2m+1}; \R) & \simeq \langle x^i, y^j \mid i, j \in \N_0 \rangle,  \\
\StringCoH^*(\Loop \Sph{2m+1}; \R) &\simeq \langle y^i, u^j \mid i,j \in \N_0 \rangle, \end{aligned}$$
and we can again identify $\H_*$ and $\StringH_*$ with the vector spaces on the right hand side. Clearly, $\StringH_{2k-1} = 0$ for all $k\in \N$, and hence $\StringOp_2 = \StringCoOp_2 = 0$ for degree reasons (the operations are odd).

We will now consider \emph{the circle} $\Sph{1}$. For every $i\in \Z$, let $\alpha_i : \Sph{1} \rightarrow \Sph{1}$ and $\theta_i : \Sph{1} \rightarrow \Loop \Sph{1}$ be the maps defined by
$$ \alpha_i(z) := z^i\quad\text{and}\quad\theta_i(w) := w \alpha_i \quad \text{for all }w,z\in \Sph{1}\subset \C. $$
By examining the equivariant homology of connected components of $\Loop \Sph{1}$ containing~$\alpha_i$ separately as in \cite[Section 2.1.4]{Basu2011}, we get 
$$\begin{aligned}
\H_*(\Loop \Sph{1}; \R) &=\langle \alpha_i, \theta_j \mid i,j\in \Z\rangle, \\
\StringH_*(\Loop \Sph{1}; \R) &= \langle u^i, \theta_0 u^j, \alpha_k \mid i, j \in \N_0, k\in \Z\backslash\{0\} \rangle,
\end{aligned}$$
where $u$ corresponds to the Euler class and
$$ \Abs{u} = 2, \quad \Abs{\theta_i} =1, \quad \Abs{\alpha_i} = 0 $$
are the degrees in the singular chain complex. On \cite[p. 21]{Basu2011} they show that the string cobracket $\StringCoOp_2$ is $0$ and that all non-trivial relations for the string bracket $\StringOp_2: \StringH(\Loop \Sph{1})[2]^{\otimes 2}\rightarrow \StringH(\Loop \Sph{1})[2]$ are the following:
\begin{equation*}
\StringOp_2( \Susp \alpha_{k}, \Susp \alpha_{-k}) = k^2 \Susp\theta_0 \quad\forall k \in \N.
\end{equation*}

We will now compare the reduced $\IBL$-structures motivated by Conjecture~\ref{Conj:StringTopology}. The point-reduced versions $\RedStringH_*(\Loop \Sph{n})$ for $n\ge 2$ are obtained from $\StringH_*(\Loop \Sph{n})$ by deleting $u^i$. We have the following isomorphisms of graded vector spaces:
$$ \begin{aligned}
  \HIBL^{\PMC}_*(\RedCycC(\Harm(\Sph{n})))[1] &\longrightarrow  \RedStringH_*(\Loop \Sph{n})[3-n] && \\ 
       \Susp \NVol^i &\longmapsto \Susp y^i  &&\text{for }n> 1\text{ odd}, \\  
    \Susp \NVol^{2i+1} &\longmapsto \Susp y_1 y_2^i  &&\text{for }n\text{ even}.
  \end{aligned}$$
Because all operations are trivial, it induces the isomorphism
$$ \IBL\bigl(\HIBL^{\PMC}_*(\RedCycC(\Harm(\Sph{n})))\bigr) \simeq \IBL\bigl(\RedStringH_*(\Loop \Sph{n})[2-n]\bigr) \quad\text{for }n\ge 2. $$
For $n = 1$, the reduced homology is seemingly different.

\begin{Remark}[Triviality for degree reasons]\label{Rem:DegRes}\Modify[caption={DONE Too dense text}]{Add paragraphs here---too dense. Ans also add $\Sph{1}$!!}
%
Both $\StringH_*(\Loop \Sph{2m-1})[3-n]$ and $\HIBL^\PMC_*(\CycC(\Harm(\Sph{2m})))[1]$ are concentrated in even degrees, and hence any $\IBLInfty$-structure must be trivial for degree reasons. On the other hand, $\StringH_*(\Loop \Sph{2m})$ and $\HIBL_*^\PMC(\CycC (\Harm(\Sph{2m-1})))[1]$ have both even and odd degrees, and hence an additional argument is needed to prove vanishing of the $\IBL$-structure. This is not the case of the reduced homology, which is again concentrated in even degree.\qedhere

\Add[caption={DONE Non-trivial degree}]{Add here the computation of which relations on homology are not implied automatically by degree reasons.}
\end{Remark}


\subsection{Twisted \texorpdfstring{$\IBLInfty$-structure for $\CP^n$}{IBL-infinity structure for CPn}
}
\label{Section:CPn1}
Let $\Kaehler\in \DR^2(\CP^n)$ be the Fubini--Study K\"ahler form on $\CP^n$ (see \cite[Examples 3.1.9]{Huybrechts2004}). The powers of $\Kaehler$ are harmonic,\footnote{This follows by induction on the power of $\Kaehler$ using the fact that, on a general K\"ahler manifold $M$, the Lefschetz operator $\Lef: \DR(M)\rightarrow \DR(M)$ defined by $\Lef(\eta):= \eta \wedge \Kaehler$ for all $\eta\in \DR(M)$ commutes with the Hodge--de Rham Laplacian~$\Delta$ (see \cite[Chapter 3]{Huybrechts2004}). } and we get easily
$$ \Harm(\CP^n) = \langle 1,\Kaehler, \dotsc, \Kaehler^n \rangle. $$
We denote the Riemannian volume of $\CP^n$ by
$$ V:= \int_{\CP^n} \frac{1}{n!}\Kaehler^n. $$
Consider the basis $e_0$, $\dotsc$, $e_n$ of $\Harm(\CP^n)[1]$ defined for all $i=0$, $\dotsc$, $n$ by
$$ e_i:= \frac{\NK^i}{(n! V)^{\frac{i}{n}}}, \quad \text{where}\quad\NK^i:= \SuspU \Kaehler^i. $$
The matrix of the pairing $\Pair$ from~\eqref{Eq:DeRhamDGA} with respect to the basis $e_0$,~$\dotsc$,~$e_{n}$ reads:
$$ (\Pair^{ij}) = \begin{pmatrix}
0 & \dotsb & 1 \\
\vdots & {\displaystyle\, .^{{\displaystyle \, .^{\displaystyle\,.}}}} & \vdots \\
1 & \dotsb & 0
\end{pmatrix}. $$
The basis $e^0$, $\dotsc$, $e^n$ dual to $e_0$, $\dotsc$, $e_n$ with respect to $\Pair$ thus satisfies
$$ e^i = e_{n-i}\quad\text{for all }i=0,\dotsc,n. $$
Therefore, the following holds for the matrix $(T^{ij})$ from~\eqref{Eq:PropagatorT}:
\begin{equation*}
(T^{ij}) = - (\Pair^{ij}).
\end{equation*}
For all $1\le i, j, k \le n$, we have
$$ \mu_2(e_i, e_j) = e_{i+j}\quad \text{and}\quad
\MC_{10}(\Susp e_{i}e_j e_k) = \delta_{i+j+k,n}. $$
For $\psi$, $\psi_1$, $\psi_2 \in \CDBCyc \Harm$ and generating words $\omega$, $\omega_1$, $\omega_2 \in \BCyc\Harm$, we chave
\begin{equation*}
\begin{aligned}
\OPQ_{210}(\Susp^2 \psi_1 \otimes \psi_2)(\Susp \omega) &= -\sum_{i=0}^n \sum \varepsilon(\omega\mapsto \omega^1\omega^2)(-1)^{\Abs{\omega^1}} \psi_1(e_i \omega^1)\psi_2(e_{n-i}\omega^2), \\
\OPQ_{120}(\Susp \psi)(\Susp^2 \omega_1 \otimes \omega_2) & = - \sum_{i=0}^n \sum \varepsilon(\omega_1 \mapsto \omega_1^1)\varepsilon(\omega_2\mapsto \omega_2^1) (-1)^{\Abs{\omega_1}} \psi(e_i \omega_1^1 e_{n-i} \omega_2^1).
\end{aligned}
\end{equation*}

The cyclic homology of $\Harm(\CP^n)$ is that of the truncated polynomial algebra $$ A :=\R[x]/(x^{n+1})\quad\text{with }\Deg(x)=2. $$
The computation of $\ClasCycH_*(A)$ for $\Abs{x}=0$ over a field is the goal of \cite[Exercise 4.1.8.]{LodayCyclic} or \cite[Exercise 9.1.1]{Weibel1994}. The case of $\Abs{x} = d$ can be solved by taking suitable degree shifts in the proposed projective resolution which is used to compute $\H\H(A)$. Unfortunately, using a non-canonical projective resolution, we lose the concrete form of the cyclic cycles and obtain just the following result (the full computation will be provided in \cite{MyPhD}):

For all $i=1$, $\dotsc$, $n$ and $k\in \N_0$, there are cycles $\tilde{t}_{2k+1,i}\in  \tilde{D}_q(A)$ of weights $2k+1$ and degrees $d(i+(n+1)k)$ which form a basis of $\ClasCycH_*(A)$. We apply the degree shift $U: \tilde{D}_*(A) \rightarrow D_*(A)$ from Proposition~\ref{Prop:DGA} to get the generators
$$ t_{w,i} := U(\tilde{t}_{w,i}) \in D_*^\lambda( \Harm(\CP^n)) $$
of weights $w$ and degrees $2i+ (w-1)n -1$, so that
$$ H^\lambda(\Harm(\CP^n)) = \langle t_{w,i}, \NOne^{w} \mid w\in \N \text{ odd}, i=1,\dotsc, n\rangle. $$
By the universal coefficient theorem we have $H_\lambda^* = (H^\lambda_*)'$ with respect to the grading by the degree. Given $d\in \Z$, the equation $d= 2i + (w-1)n - 1$ has only finitely many solution $(w,i) \in \N \times \{1,\dotsc,n\}$, and hence we get
\begin{equation}\label{Eq:CPnHom}
\HIBL^\MC(\CycC(\Harm(\CP^n))) = \langle \Susp t_{w,i}^*, \Susp\NOne^{w*} \mid w\in \N \text{ odd}, i=1,\dotsc, n \rangle,
\end{equation}
where $t_{w,i}^*$ and $\NOne^{w*} \in \DBCyc \Harm$ are the duals to $t_{w,i}$ and $\NOne^{w}$, respectively (see Remark~\ref{Rem:UCT}). Notice that both $\Abs{\Susp t_{w,i}^*}$ and $\Abs{\Susp \NOne^{w*}}$ are even since $\Abs{\Susp} = 2n-3$.

Because $\CP^n$ is geometrically formal, Proposition~\ref{Prop:GeomForm} implies that $\PMC_{10} = \MC_{10}$. Because $\HIBL^\MC(\CycC)$ is concentrated in even degrees and because a general $\IBLInfty$-operation $\OPQ_{klg}$ is odd, all operations vanish on the homology. Therefore, for the \emph{twisted $\IBL$-algebras} we have
\begin{equation*}
\IBL(\HIBL^\PMC(\CycC)) = \IBL(\HIBL^\MC(\CycC)) = (\HIBL^\MC(\CycC), \OPQ_{210} \equiv 0, \OPQ_{120}\equiv 0),
\end{equation*}
where $\HIBL^\MC(\CycC)$ is given by \eqref{Eq:CPnHom}.

According to \cite[Section 3.1.2]{Basu2011}, the minimal model for the Borel construction $\LoopBorel \CP^n$ is the cdga $\Lambda^{\mathrlap{\Sph{1}}\hphantom{S}}(n+1,1)$, which is freely generated (over $\R$) by the homogenous vectors $x_1$, $x_2$, $y_1$, $y_2$, $u$ of degrees
$$ \Abs{x_1} = 2,\quad \Abs{x_2} = 2n+1,\quad \Abs{y_1} = 1,\quad \Abs{y_2} = 2n, \quad \Abs{u} = 2, $$
and whose differential $\Dd$ satisfies
$$ \Dd y_1 = 0,\quad \Dd x_1 = y_1 u,\quad \Dd y_2 = -(n+1) x_1^n y_1,\quad \Dd x_2 = x_1^{n+1} + y_2 u. $$
By \cite[Theorem 3.6]{Basu2011}, the string cohomology $\StringCoH^*(\Loop \CP^n; \R)\simeq \H_*(\Lambda^{\mathrlap{\Sph{1}}\hphantom{S}}(n+1,1),\Dd)$ satisfies for all $m\in \N_0$ the following:
\begin{align*}
\StringCoH^m(\Loop \CP^n; \R) = \begin{cases} 
\langle u^j \rangle & \text{if }m=2j, \\
\langle y_1 y_2^p x_1^q \mid 0\le q \le n-1, p\ge 0; q + n p = j\rangle & \text{if }m=2j+1.
\end{cases}
\end{align*}
The right-hand side can be identified with $\StringH_*(\Loop \CP^n; \R)$ by the universal coefficient theorem. According to \cite[Proposition 3.7]{Basu2011}, we have $\StringOp_2 = 0$ and $\StringCoOp_2 = 0$. We conclude that the map
$$\begin{aligned}
 \HIBL^\PMC_*(\RedCycC(\Harm(\CP^n)))[1] & \longrightarrow \RedStringH_*(\Loop \CP^n; \R)[3-n] \\
\Susp t^*_{2k+1,l} & \longmapsto \Susp y_1 y_2^k x_1^{l-1}\qquad\text{for }k\ge 0\text{ and }l=1,\dotsc, n
\end{aligned} $$
induces an isomorphism of $\IBL$-algebras
$$ \IBL(\HIBL^\PMC_*(\RedCycC(\Harm(\CP^n)))) \simeq \IBL(\RedStringH_*(\Loop \CP^n; \R)[3-n]). $$

%
%
\clearpage
\def\appendixname{Appendix}
\begin{appendices}

\section{Evaluation of labeled ribbon graphs}
\label{Section:Appendix}
\Add[caption={DONE No higher canonical operations}]{Notice that this fact and the proof also show that we do not have canonical higher operations. They are not well defined.}

In this appendix, we define the propagator $P$ and the graph pairing $\langle \cdot, \cdot \rangle^P_\Gamma$ (Definition~\ref{Def:EvalRibGraph}), which encapsulates the contribution of a ribbon graph~$\Gamma$ to the map $f_{klg}: (\DBCyc V)^{\otimes k} \rightarrow (\DBCyc V)^{\otimes l}$ defined as a sum of contributions of ribbon graphs (Proposition~\ref{Prop:GraphPairing}). Such maps were already defined in \cite[Section 11]{Cieliebak2015} using coordinates; here we use an invariant framework inspired by~\cite{Mnev2017}. As an example, we work out in details expressions for the canonical $\dIBL$-operations $\OPQ_{210}$ and~$\OPQ_{120}$ (Example~\ref{Ex:Canon}). We also explain the technicality of identifying symmetric maps with maps on symmetric powers (Remark~\ref{Rem:SymMaps}).

Next, we define the notion of an algebraic Schwartz kernel (Definition~\ref{Def:LinSchw}) and show that the matrix $(T^{ij})$ from Definition~\ref{Def:CanonicaldIBL} corresponds to the Schwartz kernel of the identity $\Id$ up to a sign. Assuming that the Green kernel $\GKer$ from Definition~\ref{Def:GreenKernel} is algebraic, we deduce the signs in Definition~\ref{Def:PushforwardMCdeRham} using the formula from \cite[Remark 12.10]{Cieliebak2015} for the genuine pushforward Maurer-Cartan element~$\PMC$ in the finite-dimensional case. Establishing the formal analogy between the de Rham case and the finite-dimensional case is our main application of the invariant framework. Finally, we sketch how to obtain signs for the Fr\'echet $\dIBL$-structure on $\DR(M)$ (Remark~\ref{Rem:Frechet}).

Throughout this appendix, we will use Notation~\ref{Def:Notation} without further remarks.
%
\Modify[inline,caption={DONE Modify definition of $P$}]{Modify the definition of a propagator. Define first graph pairing, and then say that such $P$ in the graph pairing is a propagator.}

\begin{Def}[Propagator \& graph pairing]\label{Def:EvalRibGraph}
Let $V$ be a graded vector space. The tensor $\PKer\in V[1]^{\otimes 2}$ is called a \emph{propagator} if it satisfies the following symmetry condition:
\begin{equation}\label{Eq:SymmetryCondition}
 \tau(\PKer) = (-1)^{\Abs{\PKer}} \PKer.
\end{equation}
The map $\tau$ is the twist map defined by $\tau(v_1 \otimes v_2) = (-1)^{\Abs{v_1}\Abs{v_2}} v_2 \otimes v_1$ for all $v_1$, $v_2\in V[1]$.

For a ribbon graph $\Gamma \in \RRG_{klg}$ and its labeling $L$, consider the permutation~$\sigma_L$ from Definition~\ref{Def:EdgeVertex}. It acts on tensor powers according to Definition~\ref{Def:Permutations} and thus defines the map
\begin{equation*} \sigma_L: (V[1]^{\otimes 2})^{\otimes e} \otimes V[1]^{\otimes s_1} \otimes \dotsb \otimes V[1]^{\otimes s_l} \longrightarrow V[1]^{\otimes d_1} \otimes \dotsb \otimes V[1]^{\otimes d_k},
\end{equation*}
where $d_i$ and $s_i$ are the valencies of internal vertices $1$, $\dotsc$, $k$ and boundary components $1$, $\dotsc$, $l$, respectively, and $e$ is the number of internal edges. We extend $\sigma_L$ by $0$ to other combinations of tensor powers. The \emph{graph pairing}
$$ \langle \cdot, \cdot \rangle_{\Gamma}^P\ :\ (\DBCyc V)^{\otimes k} \otimes (\BCyc V)^{\otimes l} \longrightarrow \R  $$
is defined for all $\psi_1$, $\dotsc$, $\psi_{k} \in \DBCyc V$ and generating words $w_i = v_{i1} \dots v_{i m_i}$ with $v_{ij} \in V[1]$ for $m_{i} \in \N$ and $i=1$, $\dotsc$, $l$ by the following formula:
\begin{align*}
&\langle \psi_1\otimes \dotsb \otimes \psi_{k}, w_1 \otimes \dotsb \otimes w_l \rangle_{\Gamma}^P \\
 &\quad := \begin{multlined}[t]\smash{\sum_{L_1,\,L_3^b}}\vphantom{\sum_{L}} (\psi_1 \otimes \dotsb \otimes \psi_k)\bigl(\sigma_L(P^{\otimes e}\otimes  (v_{11}\otimes \dotsb \otimes v_{1 m_1}) \otimes \dotsb \\ \otimes (v_{l1}\otimes \dotsb \otimes v_{l m_{l}})\bigr), \end{multlined}
\end{align*}
where we use the pairing from Definition~\ref{Def:Pairings} and in every summand an $L_2$ compatible with $L_1$ and an $L_{3}^v$ are chosen arbitrarily to get a full labeling $L$ of~$\Gamma$. The graph pairing extends to $\langle \cdot, \cdot\rangle_\Gamma^P\ :\ \RTen \DBCyc V \otimes \RTen \BCyc V \rightarrow \R$. 
\end{Def}

\begin{Proposition}\label{Prop:GraphPairing}
In the setting of Definition~\ref{Def:EvalRibGraph}, we denote $w = w_1 \otimes \dotsb \otimes w_l$ and $\psi = \psi_1 \otimes \dotsb\otimes \psi_k$ and have the following:
\begin{ClaimList}
\item The number $\psi(\sigma_L(P^{\otimes e}\otimes w))$ does not depend on the choice of $L_{3}^v$ and an $L_2$ compatible with $L_1$. Moreover, $\langle\cdot, \cdot \rangle_\Gamma^P$ does not depend on the representative of $[\Gamma]\in \RRG_{klg}$.
\item If $V$ is finite-dimensional, then for every $k$, $l \ge 1$, $g \ge 0$ there is a unique linear map
$$ f_{klg}: (\DBCyc V)^{\otimes k} \rightarrow (\DBCyc V)^{\otimes l} $$ such that
\begin{equation*}
\begin{aligned}
&f_{klg}(\psi_1\otimes \dotsb \otimes \psi_k)(w_1 \otimes \dotsb \otimes w_l) \\
 &\qquad\qquad =  \frac{1}{l!} \sum_{[\Gamma]\in \RRG_{klg}} \frac{1}{\Abs{\Aut(\Gamma)}} \langle \psi_1\otimes \dotsb \otimes \psi_k, w_1 \otimes \dotsb \otimes w_l\rangle^P_{\Gamma}. \end{aligned}
\end{equation*}
\item The following holds for the map $f_{klg}$ from b):
\begin{itemize}
\item It is homogenous of degree
\begin{equation} \label{Eq:DegreeForm} \Abs{f_{klg}} = -\Abs{P}(k+l-2+2g). \end{equation}
\item The filtration degree satisfies
\begin{equation} \label{Eq:FiltrDegreeForm} 
 \Norm{f_{klg}} \ge - 2 (k+l-2+2g).
\end{equation}
\item For all $\eta\in\Perm_l$ and $\mu\in\Perm_k$, we have
\begin{equation} \label{Eq:SymmetryForm} \eta \circ f_{klg} \circ \mu = (-1)^{\Abs{P}(\eta + \mu)}f_{klg}.
\end{equation}
\end{itemize}
\end{ClaimList}
\end{Proposition}

\begin{proof}
\begin{ProofList}
\item Let us denote by $\bar{i}$ and $ij$ the operations on $L_2$ given by $\mathrm{e}_i \mapsto -\mathrm{e}_i$ and $\mathrm{e}_i \leftrightarrow \mathrm{e}_j$, respectively. An even number of these operations does not change the orientation of~\eqref{Eq:OrientationComplex}. Their effect in $\sigma_L$ acting on $P^{\otimes e}\otimes w$ is
\begin{equation*}
\bar{i}: P_i\mapsto \tau(P_i) = (-1)^{\Abs{P}} P_i\quad\text{and}\quad ij: P_i \dots P_j\mapsto (-1)^{\Abs{P}} P_j \dots P_i.
\end{equation*}
Therefore, an even number of them does not change $\sigma_L(P^{\otimes e}\otimes w)$. This proves the independence of the choice of a compatible $L_2$. The independence of the choice of $L_{3}^v$ is clear since $\psi_i$ are cyclic symmetric.

An isomorphism of ribbon graphs $\eta: \Gamma \rightarrow \Gamma'$ induces the map of compatible labelings $L \mapsto L' = \eta_* L$ such that $\sigma_{L} = \sigma_{L'}$. The independence of the choice of a representative of $[\Gamma]$ follows.

\item Suppose that $\psi = \psi_1 \otimes \dotsb \otimes \psi_k$ with $\psi_i \in (\DBCyc V)_{r_i}^{c_i}$, where $r_i\in \N$ and $c_i \in \Z$ for $i=1$, $\dotsc$, $k$. A~general element of $(\DBCyc V)^{\otimes k}$ is then a finite linear combination of such $\psi$'s. 

First of all, let us argue that the sum $\sum_{\RRG_{klg}}$ is finite. The number of internal edges~$e$ is fixed from \eqref{Eq:EulerFormula}. Therefore, the number of contributing graphs $(V_{\mathrm{int}}, E_{\mathrm{int}})$ is finite. In order to bound the number of external vertices, we notice that $d_1 = r_1$, $\dotsc$, $d_k = r_k$ must hold for $\psi(\sigma_L(P^{\otimes e}\otimes w))$ to be non-zero. Therefore, the sum is finite.

We now have the linear functional 
$$ f_{klg}(\psi) := \frac{1}{l!}\sum_{[\Gamma]\in \RRG_{klg} } \frac{1}{\Abs{\Aut(\Gamma)}} \langle \psi \mid \cdot \rangle_\Gamma^{\PKer}: (\BCyc V)^{\otimes l} \rightarrow \R $$ 
and need to show that $f_{klg}(\psi)\in (\DBCyc V)^{\otimes l} \subset (\BCyc V)^{\otimes l*}$. Because $V$ is finite-dimensional, the weight-filtration of $\BCyc V$ satisfies (WG1) \& (WG2) (see \eqref{Eq:WGs} and Proposition~\ref{Prop:Compl}), and hence we have
$$ (\DBCyc V)^{\otimes l} = (\BCyc V)''^{\otimes l} = \bigl((\BCyc V)^{\otimes l}\bigr)'' $$
for the weight-graded duals. Therefore, it suffices to show that $f_{klg}(\psi)$ vanishes on all but finitely many degrees and weights of $(\BCyc V)^{\otimes k}$. However, the relation $f_{klg}(\psi)(w) \neq 0$ for a generating word $w\in (\BCyc V)^{\otimes k}$ implies
\begin{equation}\label{Eq:UUUU}
\begin{aligned}
\Abs{w} &= \Abs{\psi} - e \Abs{\PKer}\quad\text{and} \\
k(w) &= k(\psi) - 2 e,
\end{aligned}
\end{equation}
where $k$ denotes the weight, and hence $f_{klg}(\psi)\in (\DBCyc V)^{\otimes l}$ indeed holds.

\item The formulas~\eqref{Eq:DegreeForm} and~\eqref{Eq:FiltrDegreeForm} follow from~\eqref{Eq:UUUU} and~\eqref{Eq:EulerFormula}.

As for the symmetry \eqref{Eq:SymmetryForm}, suppose that $L$ and $L'$ are compatible labelings of the same graph $\Gamma$ such that $L_1'$ differs from $L_1$ by a permutation $\mu\in \Perm_{k}$ of internal vertices and a permutation $\eta\in \Perm_l$ of boundary components. Viewing $\mu$ and $\eta$ as block permutations in the vertex and edge order, respectively, we get
$$ \sigma_{L'}(P^{\otimes e}\otimes w)=(-1)^{\Abs{P}(\eta+\mu)}\mu(\sigma_L(P^{\otimes e}\otimes \eta( w))). $$
The sign comes from the difference of $L_2$ and $L_2'$ which compensates the change of the orientation of~\eqref{Eq:OrientationComplex} caused by $\mu$ and $\eta$.\qedhere
\end{ProofList}
\end{proof}

Given $\mu\in \Perm_k$ and $\psi = \psi_1 \otimes \dotsb \otimes \psi_k \in (\DBCyc V)^{\otimes k}$, it is easy to see that
$$ \varepsilon(\mu, \Psi) = \varepsilon(\mu(\Susp),\mu(\psi)) \varepsilon(\mu,\Susp)\varepsilon(\Susp,\psi)\varepsilon(\mu,\psi), $$
where $\Psi = (\Susp \psi_1) \otimes \dotsb \otimes (\Susp \psi_k) \in (\DBCyc V[A])^{\otimes k}$ and $\varepsilon(\mu,\Susp) = (-1)^{\Abs{s}\mu}$. If $A= - \Abs{\PKer}$, then we get from \eqref{Eq:SymmetryForm} that the degree shift $\HTP_{klg}: (\DBCyc V [A])^{\otimes k} \rightarrow (\DBCyc V[A])^{\otimes l}$ has the following symmetries:
\begin{equation}\label{Eq:SymMap}
\forall \mu \in \Perm_k, \eta\in \Perm_l: \quad \eta \circ \HTP_{klg} \circ \mu = \HTP_{klg}.
\end{equation}
Note that the degrees satisfy
\begin{equation}\label{Eq:DegDegShift}
\Abs{\HTP_{klg}} = \Abs{f_{klg}} + (k-l) A.
\end{equation}

\begin{Remark}[Symmetric maps versus maps on symmetric powers]\label{Rem:SymMaps}
In the situation above, we define~$\tilde{\HTP}_{klg}$ as the unique map such that the solid lines of the following diagram commute:    
$$\begin{tikzcd}
(\DBCyc V[A])^{\otimes k} \arrow{r}{\HTP_{klg}} \arrow[two heads]{d}{\pi} & (\DBCyc V[A])^{\otimes l} \arrow[two heads,swap]{d}{\pi}\\
\arrow[bend left,dotted]{u}{\iota}\Sym_k \DBCyc V[A] \arrow{r}{\tilde{\HTP}_{klg}} & \Sym_l \DBCyc V[A].\arrow[bend right,swap,dotted]{u}{\iota}
\end{tikzcd}$$
The symmetry condition \eqref{Eq:SymMap} provides the existence of $\tilde{\HTP}_{klg}$ and implies commutativity of the dotted diagram as well. Moreover, for all $\psi_1$, $\dotsc$, $\psi_k \in \DBCyc V$ and $w_1$, $\dotsc$, $w_l\in \BCyc V$, we have 
$$ \tilde{\HTP}_{klg}(\Susp^k \psi_1\cdots \psi_k)(\Susp^l w_1\cdots w_l) = \HTP_{klg}(\Susp^k \psi_1 \otimes \dotsb \otimes \psi_k)(\Susp^l w_1 \otimes \dotsb\otimes w_l), $$
where we use the pairing from Definition~\ref{Def:Pairings}. We denote $\tilde{\HTP}_{klg}$ again by $\HTP_{klg}$.
\end{Remark}

\begin{Definition}[Algebraic Schwartz kernel]\label{Def:LinSchw}
Let $V$ be a graded vector space and $\Pair: V\otimes V \rightarrow \R$ a non-degenerate pairing on $V$. We extend $\Pair$ to a non-degenerate pairing $\Pair: V^{\otimes k}\otimes V^{\otimes k}\rightarrow \R$ for $k\ge 1$ by setting
\begin{equation*} 
\Pair(v_{11} \otimes \dotsb \otimes v_{1k}, v_{21}\otimes \dotsb \otimes v_{2k}) := \varepsilon(v_1, v_2)\Pair(v_{11},v_{21}) \dots \Pair(v_{1k},v_{2k})
\end{equation*}
for all $v_{11}$, $\dotsc$, $v_{1k}$, $v_{21}$, $\dotsc$,  $v_{2k}\in V$, where $\varepsilon$ is the Koszul sign (see Definition~\ref{Def:Koszul}). For $k=0$, we let $\Pair: \R\otimes \R \rightarrow \R$ be the multiplication on $\R$.

For $k$, $l \ge 0$, we say that $\Kern_L\in V^{\otimes k + l}$ is the \emph{algebraic Schwartz kernel} of a linear operator $L: V^{\otimes k} \rightarrow V^{\otimes l}$ if the following is satisfied:
\begin{equation} \label{Eq:KernelEquation}
\forall w_1 \in V^{\otimes k},  w_2 \in V^{\otimes l}:\quad \Pair(L(w_1),w_2) = \Pair(\Kern_L, w_1\otimes w_2).
\end{equation}
We usually omit writing ``algebraic'' if it is clear from the context (i.e., if we do not consider any ``extensions'' of $V^{\otimes k}$). 
\end{Definition}

In the situation of Definition~\ref{Def:LinSchw}, let $(e_i)\subset V$ be a basis and $(e^i)$ its dual basis such that $\Pair(e_i,e^j)=\delta_{ij}$. We define the coordinates $K_L^{ij} \in \R$ and $L^{ij}\in \R$ by 
$$ \Kern_L = \sum_{i,j} K_L^{ij} e_i \otimes e_j\quad\text{and}\quad L^{ij} := \Pair(L (e^i),e^j).$$
From \eqref{Eq:KernelEquation} we have
\begin{equation} \label{Eq:KernelCoordinates}
\Kern_L^{ij} = (-1)^{(\Abs{L}+1)(\Abs{\Pair}+\Abs{e_i})} L^{ij}\quad \text{for all }i,j.
\end{equation} 

From now on, we will be in the situation of (A) and (B) in the Introduction; in particular, we put $V[1]$ in place of $V$ in Definition~\ref{Def:LinSchw}. Let $\Kern_{\Id} \in V[1]^{\otimes 2}$ be the Schwartz kernel of the identity $\Id: V[1]\rightarrow V[1]$ and $\Kern_{\GOp} \in V[1]^{\otimes 2}$ the Schwartz kernel of the cochain homotopy $\GOp: V[1] \rightarrow V[1]$. From~\eqref{Eq:KernelCoordinates}, we get
\begin{equation*} 
\Kern_{\GOp}^{ij} = \GOp^{ij} \quad \text{and} \quad {\Kern_{\Id}}^{ij} = (-1)^{\Abs{e_i} + \Abs{\Pair}} \Pair(e^i,e^j)\quad\text{for all }i,j. 
\end{equation*}
We see that the tensor $\TKer = \sum_{i,j} \TKer^{ij} e_i \otimes e_j$ from~\eqref{Eq:PropagatorT} can be expressed as
\begin{equation*}
\TKer = (-1)^{n-2} \Kern_{\Id}.
\end{equation*}
This is the invariant meaning of $\TKer$. Note that the degrees satisfy 
$$ \Abs{\TKer} = n-2\quad\text{and}\quad \Abs{\Kern_{\GOp}} = n - 3. $$

The assumption~\eqref{Eq:ConditionOnG} on $\GOp$ is equivalent to graded antisymmetry of the bilinear form $\GOp^+:= \Pair\circ (\GOp\otimes \Id): V[1]^{\otimes 2} \rightarrow \R$. This is further equivalent to
$$ \tau(\Kern_{\GOp}) = (-1)^{\Abs{\Kern_{\GOp}}}\Kern_{\GOp}. $$
Therefore, $\Kern_{\GOp}$ satisfies~\eqref{Eq:SymmetryCondition}, and hence it can be used as a propagator for the construction of $f_{klg}$ for every $k$, $l\ge 1$, $g\ge 0$. We have from~\eqref{Eq:SymmetryForm} that the degree shift $\HTP_{klg}: (\DBCyc V[3-n])^{\otimes k} \rightarrow (\DBCyc V[3-n])^{\otimes l}$ is symmetric. Moreover, using~\eqref{Eq:DegreeForm},~\eqref{Eq:FiltrDegreeForm} and \eqref{Eq:DegDegShift}, we obtain
$$\begin{aligned}
\Abs{\HTP_{klg}} &= - 2d(k+g-1), \\
\Norm{\HTP_{klg}}&\ge \gamma (2-2g-k-l),
\end{aligned}$$
where $(d,\gamma) = (n-3,2)$. These are the degree and filtration conditions on an $\IBLInfty$-morphism from~\cite[Definition~2.8 and (8.3)]{Cieliebak2015}. As a matter of fact, our $\HTP=(\HTP_{klg})_{k,l\ge 1, g\ge 0}$ is precisely the $\IBLInfty$-homotopy from \cite[Theorem 11.3]{Cieliebak2015}.

Graded antisymmetry of $\Pair$ is equivalent to 
$$ \tau(\TKer)= (-1)^{\Abs{\TKer}+1}\TKer. $$
Visibly, $\TKer$ does not satisfy~\eqref{Eq:SymmetryCondition}. Nevertheless, we can still use it to define $f_{210}$ and~$f_{120}$ since the corresponding graphs $\Gamma$ (see Figure~\ref{Fig:GammasGraphs}) have only one internal edge $\mathrm{e}$, and, for a given $L_1$, there is a unique compatible~$L_2$ determined by the orientation of $\mathrm{e}$ (see Example~\ref{Ex:Canon} for the compatibility condition). As for the symmetry of the resulting maps, a transposition of internal vertices or boundary components in~\eqref{Eq:OrientationComplex} can be compensated only by $\mathrm{e}\mapsto-\mathrm{e}$, which produces $(-1)^{\Abs{\TKer}+1}$ (c.f., the proof of Proposition~\ref{Prop:GraphPairing} (a)). Therefore, if we shift the degrees by $A= - \Abs{\TKer} +  1 = n- 3$, we obtain symmetric maps $\OPQ_{210}: (\DBCyc V[A])^{\otimes 2} \rightarrow \DBCyc V[A]$ and $\OPQ_{120}: \DBCyc V[A] \rightarrow (\DBCyc V[A])^{\otimes 2}$. We show in Example~\ref{Ex:Canon} below that these operations agree with those defined in Definition~\ref{Def:CanonicaldIBL}.

\begin{Example}[The canonical $\dIBL$-operations]\label{Ex:Canon}
We have
\begin{equation}\label{Eq:DefByGraphs}
\begin{aligned}
f_{210}(\psi_1 \otimes \psi_2)(w) & = \frac{1}{1!} \sum_{[\Gamma]\in \RRG_{210}} \frac{1}{\Abs{\Aut(\Gamma)}} \langle \psi_1 \otimes \psi_2 \mid w \rangle_\Gamma^P\quad\text{and} \\
f_{120}(\psi)(w_1 \otimes w_2) & = \frac{1}{2!} \sum_{[\Gamma]\in \RRG_{120}} \frac{1}{\Abs{\Aut(\Gamma)}} \langle \psi \mid w_1 \otimes w_2 \rangle_\Gamma^P.
\end{aligned}
\end{equation}
We parametrize $\RG_{210}$ by the ribbon graphs $\Gamma_{k_1, k_2}$ with $1\le k_1\le k_2$ and $\RG_{120}$ by the ribbon graphs $\Gamma^{s_1, s_2}$ with $0\le s_1 \le s_2$; these graphs are depicted in Figure~\ref{Fig:GammasGraphs}. We have 
$\RRG_{210} = \RG_{210}\backslash \{[\Gamma_{1,1}]\}$ and $\RRG_{120} = \RG_{120}\backslash \{[\Gamma^{0,0}], [\Gamma^{0,1}]\}$. We also have 
$$\Abs{\Aut(\Gamma_{k_1,k_2})} = \begin{cases} 1 & \text{if }k_1 \neq k_2, \\ 
              2 & \text{if }k_1 = k_2, \end{cases}$$
and likewise for $\Gamma^{s_1,s_2}$. We fix labelings~$L_{3}^v$ and parametrize $L_{3}^b$ by $c=1$, $\dotsc$, $k_1 + k_2 - 2$ for $\Gamma_{k_1,k_2}$ and by $c_1 = 1$, $\dotsc$, $s_1$ and $c_2 = 1$, $\dotsc$, $s_2$ for $\Gamma^{s_1,s_2}$ as it is indicated in Figure~\ref{Fig:GammasGraphs}.

There are two possible labelings $L_{1}^v$ for $\Gamma_{k_1, k_2}$ and two possible labelings $L_{1}^b$ for $\Gamma^{s_1, s_2}$; this is the only freedom in choosing a full labeling $L$ because $L_3$ is fixed and $L_2$ is just the orientation of the single internal edge, which is uniquely determined by $L_1$. For both $\Gamma_{k_1, k_2}$ and $\Gamma^{s_1, s_2}$, we will denote the two possible full labelings by $L^1$ and $L^2$. They can be depicted as follows:
\begin{figure}[t]
\centering
\begin{tikzpicture}
\tikzset{int/.style = {draw, circle, fill=black, minimum size=2pt,inner sep=0pt}}
\tikzset{ext/.style = {draw, circle, fill=white, minimum size=2pt,inner sep=0pt}}
\tikzset{->-/.style={decoration={markings,mark=at position #1 with {\arrow{>}}},postaction={decorate}}} 
\def\FirstAng{70}
\def\AngDif{50}
\def\AngDifTwo{70}
\def\ExtLen{1.4} 
\def\ArcAngDelta{15}
\def\ArcAngDeltaLab{30}
\coordinate[int] (V1) at (0,0);
\coordinate[int] (V2) at (3,0);
\coordinate[ext,label={$\scriptstyle c$}] (C1) at ($(V1) + (\FirstAng:\ExtLen)$);
\coordinate[ext,label={[below]$\scriptstyle c+k_1-2$}] (C3) at ($(V1) + (-\FirstAng:\ExtLen)$);
\coordinate[ext,label={[below]$\scriptstyle c+k_1-1$}] (C4) at ($(V2) + (180+\FirstAng:\ExtLen)$);
\coordinate[ext,label={[above]$\scriptstyle c + k_1 + k_2 - 3$}] (C6) at ($(V2) + (180-\FirstAng:\ExtLen)$);
\draw[->-={0.5},dotted] ($([shift=(\FirstAng+\ArcAngDelta:\ExtLen)]V1)$) arc (\FirstAng+\ArcAngDelta:360-\FirstAng-\ArcAngDeltaLab:\ExtLen);

\draw[->-={0.5},dotted] ($([shift=(180+\FirstAng+\ArcAngDeltaLab:\ExtLen)]V2)$) arc (180+\FirstAng+\ArcAngDeltaLab:180-\FirstAng+360-\ArcAngDeltaLab:\ExtLen);

\draw (V1) -- (V2) node[pos=0.13,above] {$\scriptstyle 1$} node[pos=0.87,above] {$\scriptstyle 1$};
\draw (V1) -- (C1) node[pos=0.5,left] {$\scriptstyle 2$};
\draw (V1) -- (C3) node[pos=0.4,left] {$\scriptstyle k_1$};
\draw (V2) -- (C4) node[pos=0.5,right] {$\scriptstyle 2$};
\draw (V2) -- (C6) node[pos=0.5,right] {$\scriptstyle k_2$};
\end{tikzpicture}
\quad
\begin{tikzpicture}
\tikzset{int/.style = {draw, circle, fill=black, minimum size=2pt,inner sep=0pt}}
\tikzset{ext/.style = {draw, circle, fill=white, minimum size=2pt,inner sep=0pt}}
\tikzset{->-/.style={decoration={markings,mark=at position #1 with {\arrow{>}}},postaction={decorate}}} 
\def\CircRad{1.4}
\def\CircPos{45}
\def\CIAng{0}
\def\CIIIAng{90}
\def\CIVAng{180}
\def\CVIAng{290}
\def\AngDif{40}
\def\AngDifTwo{70}
\def\ExtLen{1.4}
\def\ArcAngDeltaI{15}
\def\ArcAngDeltaII{50}
\coordinate[int] (C) at (0,0);
\coordinate (R) at ($(C) + (\CircPos:\CircRad)$);
\coordinate[ext,label={[above right]$\scriptstyle c_1$}] (C1) at ($(C) + (\CIAng:\ExtLen)$);
\coordinate[ext,label={[above right]$\scriptstyle c_1+s_1-1$}] (C3) at ($(C) + (\CIIIAng:\ExtLen)$);
\coordinate[ext,label={[left]$\scriptstyle c_2$}] (C4) at ($(C) + (\CIVAng:\ExtLen)$);
\coordinate[ext,label={[right]$\scriptstyle c_2 + s_2 -1$}] (C6) at ($(C) + (\CVIAng:\ExtLen)$);

\draw ($([shift=(0:\CircRad)]R)$) arc (0:360:\CircRad) node[pos=0.5,label={[left]$\scriptstyle s_1 + 2$}] {} node[pos=0.75,label={[below,yshift=-1mm]$\scriptstyle 1$}] {};

\draw[->-={0.5},dotted] ($([shift=(\CIAng+\ArcAngDeltaI:\ExtLen)]C)$) arc (\CIAng+\ArcAngDeltaI:\CIIIAng-\ArcAngDeltaI:\ExtLen);
\draw[->-={0.5},dotted] ($([shift=(\CIVAng+\ArcAngDeltaI:\ExtLen)]C)$) arc (\CIVAng+\ArcAngDeltaI:\CVIAng-\ArcAngDeltaI:\ExtLen);

\draw (C) -- (C1) node[pos=0.3,above] {$\scriptstyle 2$};
\draw (C) -- (C3) node[pos=0.5,right] {$\scriptstyle s_1+1$};
\draw (C) -- (C4) node[pos=0.5,above] {$\scriptstyle s_1+3$};
\draw (C) -- (C6) node[pos=0.5,below left] {$\scriptstyle s_1 + s_2 + 2$};
\end{tikzpicture}
\caption{Graphs $\Gamma_{k_1,k_2}$ and $\Gamma^{s_1, s_2}$ with fixed labelings $L_3$.}
\label{Fig:GammasGraphs}
\end{figure}
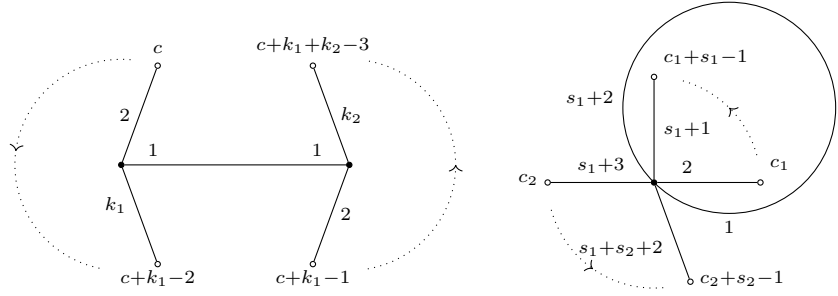
\begin{equation}\label{Eq:OrientLabel}
\begin{tabular}{c|m{1.5cm} m{1.5cm}}
 & \makebox[1.5cm]{$\Gamma_{k_1,k_2}$} & \makebox[1.5cm]{$\Gamma^{s_1,s_2}$} \\\hline
\rule{0pt}{4ex}$L^1$ &\centering{\begin{tikzpicture}
\tikzset{->-/.style={decoration={markings,mark=at position #1 with {\arrow{>}}},postaction={decorate}}} 
\node (V1) at (0,0) {1};
\node(V2) at (1,0) {2};
\draw[->-={0.5}] (V1) -- (V2);
\end{tikzpicture}}
& \begin{tikzpicture}
\tikzset{->-/.style={decoration={markings,mark=at position #1 with {\arrow{>}}},postaction={decorate}}} 
\node (V1) at (1,0) {1};
\node(V2) at (0,0) {2};
\draw[->-={0.5}] (V1) -- (V2);
\end{tikzpicture}\\[1ex]
$L^2$ & \begin{tikzpicture}
\tikzset{->-/.style={decoration={markings,mark=at position #1 with {\arrow{>}}},postaction={decorate}}} 
\def\Rad{0.4}
\node (B1) at (0,0) {1};
\node (B2) at (0.8,0) {2};
\draw[->-={0.5}] ($([shift=(0:\Rad)]B1)$) arc (0:360:\Rad);
\end{tikzpicture} & \begin{tikzpicture}
\tikzset{->-/.style={decoration={markings,mark=at position #1 with {\arrow{>}}},postaction={decorate}}} 
\def\Rad{0.4}
\node (B1) at (0,0) {2};
\node (B2) at (0.8,0) {1};
\draw[->-={0.5}] ($([shift=(360:\Rad)]B1)$) arc (360:0:\Rad);
\end{tikzpicture}
\end{tabular}
\end{equation}
Let us check that the indicated $L_1$ and $L_2$ are compatible. For the complexes $C_2 \rightarrow C_1 \rightarrow C_0$ from \eqref{Eq:OrientationComplex}, we have the following:
$$\begin{aligned}
\Gamma_{k_1,k_2}:\qquad &\langle \mathrm{b} \rangle \xrightarrow{\Bdd_2 = 0}  \langle \mathrm{e} \rangle \xrightarrow{\Bdd_1} \langle \mathrm{v}_2 - \mathrm{v}_1 \rangle \oplus \langle \mathrm{v}_1 + \mathrm{v}_2 \rangle,  \\
\Gamma^{s_1, s_2}:\qquad &\langle \mathrm{b}_1 - \mathrm{b}_2 \rangle \oplus \langle \mathrm{b}_1 + \mathrm{b}_2 \rangle  \xrightarrow{\Bdd_2} \langle \mathrm{e} \rangle \xrightarrow{\Bdd_1 = 0} \langle \mathrm{v} \rangle.  
\end{aligned}$$
As for $\Gamma_{k_1,k_2}$, the basis $\mathrm{v}_2 - \mathrm{v}_1$, $\mathrm{v}_1 + \mathrm{v}_2$ of $C_0$ is positively oriented with respect to the basis $\mathrm{v}_2$, $\mathrm{v}_1$. Therefore, $\mathrm{e}$ has to be oriented such that $\Bdd_1 \mathrm{e} = \mathrm{v}_2 - \mathrm{v}_1$; i.e., it is a path from $\mathrm{v}_1$ to $\mathrm{v}_2$. As for $\Gamma^{s_1,s_2}$, the basis $\mathrm{b}_1-\mathrm{b}_2$, $\mathrm{b}_1 + \mathrm{b}_2$ of $C_2$ is positively oriented with respect to $\mathrm{b}_1$, $\mathrm{b}_2$. Therefore, $\mathrm{e}$ has to be oriented such that $\mathrm{e} = \Bdd_2 (\mathrm{b}_1 - \mathrm{b}_2)$. Recall that we orient the boundary of a $2$-simplex by the ``outer normal first'' convention. We conclude that the labelings from~\eqref{Eq:OrientLabel} are indeed compatible.

As for $f_{210}$, the permutations $\sigma_1:=\sigma_{L^1}$ and $\sigma_2:= \sigma_{L^2}$ corresponding to the labelings $L^1$ and $L^2$, respectively, read
$$\begin{aligned}
\sigma_{1} &= \biggl(\begin{array}{cc|ccc}
 1 & 2     & \dots & c+2 & \dots \\
 1 & k_1+1 & \undermat{k_1+k_1-2}{\dots& c+2 & \dots}\dots & 2& \dots 
\end{array}\biggr)\quad\text{and}\\[4ex]
\sigma_{2} &= \biggl(\begin{array}{cc|ccc}
 1 & 2     & \dots & c+2 & \dots  \\
 1 & k_2+1 & \undermat{k_1+k_2-2}{\dots & k_2 + 2 & \dots}\dots & k_2 + 2 & \dots
\end{array}\biggr).\\[3ex]
\end{aligned}$$
The underbracket marks the block which represents a cyclic permutation of the remaining indices. We see that 
$$\begin{aligned}
\sigma_{1}&:  V^{\otimes 2} \otimes V^{\otimes s} \longrightarrow V^{\otimes k_1}\otimes V^{\otimes k_2},\quad e_i e_j w \longmapsto e_i w^1 e_j w^2,  \\
\sigma_{2}&: V^{\otimes 2}\otimes V^{\otimes s} \longrightarrow V^{\otimes k_2}\otimes V^{\otimes k_1},\quad e_i e_j w  \longmapsto e_i w^2 e_j w^1,
\end{aligned}$$
where $w^1 = w_c \dots w_{c+k_1-2}$, $w^2 = w_{c+k_1-1}\dots w_{c+k_1+k_2-3}$ and $s := k_1 + k_2 - 2$. Defining $\tilde{w}^1 := w^2$ and $\tilde{w}^2 := w^1$, The Koszul sign of $\sigma_2$ can be written as
$$ \varepsilon(w\mapsto w^1 w^2) (-1)^{\Abs{w^2}\Abs{e_j} + \Abs{w^1}\Abs{w^2}} = \varepsilon(w\mapsto \tilde{w}^1 \tilde{w}^2 ) (-1)^{\Abs{\tilde{w}^1} \Abs{e_j}}. $$
We use these facts to rewrite \eqref{Eq:DefByGraphs} as follows:
$$\begin{aligned}
& f_{210}(\psi_1 \otimes \psi_2)(w) = \\ &\quad \sum_{\substack{1\le k_1 < k_2}} \sum_{i,j} \TKer^{ij} \Bigl(\sum_{k(w^1) = k_1 - 1} \varepsilon(w\mapsto w^1 w^2)(-1)^{\Abs{w^1}\Abs{e_j}} \psi_1(e_i w^1) \psi_2(e_j w^2) \\ &\quad {}+ \sum_{k(w^1) = k_2 - 1} \varepsilon(w\mapsto w^1 w^2) (-1)^{\Abs{w^2}\Abs{e_j} + \Abs{w^1}\Abs{w^2}} \psi_1(e_i w^2) \psi_2(e_j w^1)\Bigr) \\ 
&\quad + \sum_{1< k_1 = k_2} \frac{1}{2} \Bigl(\sum_{k(w^1) = k_1 - 1} \varepsilon(w\mapsto w^1 w^2)(-1)^{\Abs{w^1}\Abs{e_j}} \psi_1(e_i w^1) \psi_2(e_j w^2) \\ 
&\quad {}+ \sum_{k(w^1) = k_2 - 1} \varepsilon(w\mapsto w^1 w^2) (-1)^{\Abs{w^2}\Abs{e_j} + \Abs{w^1}\Abs{w^2}} \psi_1(e_i w^2) \psi_2(e_j w^1)\Bigr) \\
&\quad =\sum_{\substack{k_1, k_2 \ge 1 \\ k_1 + k_2 > 2}} \sum_{\substack{ k(w^1) = k_1 - 1 \\ k(w^2) = k_2 - 1}}  \TKer^{ij}\varepsilon(w \mapsto w^1 w^2) (-1)^{\Abs{w^1}\Abs{e_j}} \psi_1(e_i w^1) \psi_2(e_j w^2).
\end{aligned}$$
This coincides with the formula from Definition~\ref{Def:CanonicaldIBL}.

As for $f_{120}$, the permutations $\sigma_1:=\sigma_{L^1}$ and $\sigma_2:= \sigma_{L^2}$ corresponding to the labelings $L^1$ and $L^2$, respectively, read
$$\begin{aligned}
\sigma_{1} &= \biggl(\begin{array}{cc|ccc|ccc}
1 & 2      & \dots & c_1+2 & \dots & \dots & c_2 + s_1 + 2 & \dots \\
 1 & s_1 + 2& \undermat{s_1}{\dots & c_1+2 & \dots}\dots & 2 & \dots &\undermat{s_2}{\dots & c_2 + s_1 + 2 & \dots}\dots & s_1 + 3 & \dots
\end{array}\biggr)\quad\text{and} \\[4ex]
\sigma_{2} &= \biggl(\begin{array}{cc|ccc|ccc}
 1       & 2 &  \dots  & c_2 + 2 & \dots & \dots & c_1 + s_2 + 2 & \dots  \\
 s_1 + 2 & 1 & \undermat{s_2}{\dots & c_2+2 & \dots} \dots  & s_1 + 3 & \dots & \undermat{s_1}{\dots & c_1 + s_2 + 2 & \dots} \dots &  2     & \dots
\end{array}\biggr), \\[3ex]
\end{aligned}$$
where the underbracketed blocks denote cyclic permutations of consecutive indices on the corresponding boundary component. We see that 
$$\begin{aligned}
\sigma_1 &: V^{\otimes 2}\otimes V^{\otimes s_1}\otimes V^{\otimes s_2} \longrightarrow V^{\otimes k}, \quad e_i e_j w_1 w_2 \longmapsto e_i w_1^1 e_j w_2^1, \\
\sigma_2 &:  V^{\otimes 2}\otimes V^{\otimes s_2} \otimes V^{\otimes s_1} \longrightarrow V^{\otimes k}, \quad  e_i e_j w_1 w_2 \longmapsto e_j w_2^1 e_i w_1^1,
\end{aligned}$$
where $w_i^1$ denotes a cyclic permutation and $k:= s_1 + s_2 +2$. The Koszul sign of~$\sigma_2$ can be written as
$$\begin{aligned}&(-1)^{\Abs{e_i}\Abs{e_j} + \Abs{w_1}\Abs{w_2} + \Abs{e_i}\Abs{w_2}} \varepsilon(w_1 \mapsto w_1^1)\varepsilon(w_2\mapsto w_2^1)\\
&\qquad= (-1)^{(\Abs{e_i} + \Abs{w_1})(\Abs{e_j} + \Abs{w_2}) + \Abs{w_1} \Abs{e_j}}\varepsilon(w_1 \mapsto w_1^1)\varepsilon(w_2\mapsto w_2^1).\end{aligned}$$
We use this fact and the cyclic symmetry of $\psi$ to rewrite \eqref{Eq:DefByGraphs} as follows:
$$\begin{aligned}
&f_{120}(\psi)(w_1 \otimes w_2)
\\&\; = \begin{aligned}[t]
& \begin{multlined}[t]\sum_{0\le s_1 < s_2} \Bigl(\delta\bigl(\substack{k(w_1) = s_1 \\ k(w_2) = s_2}\bigr)\sum \TKer^{ij} \varepsilon(w_1\mapsto w_1^1) \varepsilon(w_2\mapsto w_2^1)(-1)^{\Abs{w_1} \Abs{e_j}} \\
\psi(e_i w_1^1 e_j w_2^1) +\delta\bigl(\substack{k(w_1) = s_2 \\ k(w_2) = s_1}\bigr) \sum \TKer^{ij} \varepsilon(w_1\mapsto w_1^1)\varepsilon(w_2\mapsto w_2^1)\\ (-1)^{\Abs{e_i} \Abs{e_j} + \Abs{w_2} \Abs{w_1} + \Abs{e_i}\Abs{w_2}} \psi(e_j w_2^1 e_i w_1^1) \Bigr)
\end{multlined}\\
&\begin{multlined}{}+\sum_{0< s_1 = s_2} \delta\bigl(\substack{k(w_1) = k(w_1) = s_1 \\ k(w_2) = k(w_2) = s_2}\bigr) 
\frac{1}{2} \Bigl(\sum \TKer^{ij} \varepsilon(w_1\mapsto w_1^1) \varepsilon(w_2\mapsto w_2^1)\\ (-1)^{\Abs{w_1} \Abs{e_j}} \psi(e_i w_1^1 e_j w_2^1) + \sum \TKer^{ij} \varepsilon(w_1\mapsto w_1^1)\varepsilon(w_2\mapsto w_2^1) \\ (-1)^{\Abs{e_i} \Abs{e_j} + \Abs{w_2} \Abs{w_1} + \Abs{e_i}\Abs{w_2}} \psi(e_j w_2^1 e_i w_1^1) \Bigr) \end{multlined}
\end{aligned} \\
&\; = \begin{multlined}[t] \smash{\sum_{\substack{s_1, s_2 \ge 0 \\ s_1 + s_2 > 0}}} \delta\bigl(\substack{k(w_1) = s_1 \\ k(w_2) = s_2}\bigr) \sum \TKer^{ij} \varepsilon(w_1 \mapsto w_1^1)\varepsilon(w_2 \mapsto w_2^1)(-1)^{\Abs{w_1}\Abs{e_j}} \\ \psi(e_i w_1^1 e_j w_2^1).\end{multlined}
\end{aligned}$$
This coincides with the formula from Definition~\ref{Def:CanonicaldIBL}.
\end{Example}

We will now establish a formal analogy between the finite-dimensional and the de Rham case, which will explain the signs in Definition~\ref{Def:PushforwardMCdeRham}.

\textbf{The finite-dimensional case.} Consider the situation of (A) -- (D) in the Introduction. To recall briefly, we have a finite-dimensional cyclic dga $(V,\Pair,m_1, m_2)$ and a subcomplex $\Harm \subset V$ such that there is a projection $\pi: V[1] \rightarrow \Harm[1]$ chain homotopic to $\Id$ via a chain homotopy $\GOp: V[1] \rightarrow V[1]$. Using~$m_2$, one constructs the canonical Maurer-Cartan element $\MC$ for $\dIBL(\CycC(V))$. Using the algebraic Schwartz kernel $\Kern_{\GOp}$ of $\GOp$, one constructs the $\IBLInfty$-quasi-isomorphism $\HTP = (\HTP_{klg}): \dIBL(\CycC(V)) \rightarrow \dIBL(\CycC(\Harm))$. The Maurer-Cartan element $\MC$ is then pushed forward along $\HTP$ to obtain the Maurer-Cartan element $\PMC := \HTP_* \MC$  for $\dIBL(\CycC(\Harm))$ (see \cite[Lemma 9.5]{Cieliebak2015}). The formula for~$\PMC$ given in~\cite[Remark~12.10]{Cieliebak2015} reads
\begin{equation} \label{Eq:PushforwardMC}
\begin{aligned} &\PMC_{lg}(\Susp^l w_1 \otimes \dotsb \otimes  w_l) \\
&\qquad=\frac{1}{l!}\sum_{[\Gamma]\in \TRRG_{klg}} \frac{1}{\Abs{\Aut(\Gamma)}} (-1)^{k(n-2)}\langle (m_2^+)^{\otimes k}, w_1 \otimes \dotsb \otimes w_l \rangle_{\Gamma}^{\Kern_{\GOp}}.
\end{aligned}\end{equation}
Here the artificial sign $(-1)^{k(n-2)}$ is added because our sign conventions for $m_2^+$ differ (see Remark~\ref{Rem:mukplus}).

\textbf{The de Rham case.}
We are in the setting of Definition~\ref{Def:PushforwardMCdeRham}. To recall briefly, we have the cyclic dga $(\DR(M), \Pair, m_1, m_2)$, the subspace of harmonic forms $\Harm\subset \DR$, the harmonic projection $\pi_\Harm: \DR\rightarrow \Harm$ and a Green kernel $\GKer\in \DR(\Bl_\Diag(M\times M))$, which is the Schwartz kernel of a chain homotopy $\GOp: \DR \rightarrow \DR$ between $\pi_\Harm$ and~$\Id$. In analogy with the finite-dimensional case, the canonical Maurer-Cartan element~\eqref{Eq:CanonMC} for $\dIBL(\Harm)$ satisfies $\MC_{10} = (-1)^{n-2}m_2^+$ with $m_2^+ = \Pair(m_2 \otimes \Id)$. Because $\dim(\DR) = \infty$, Definition~\ref{Def:CanonicaldIBL} does not give the canonical $\dIBL$-structure on $\CycC(\DR)$, and hence we have neither $\HTP$ nor~$\PMC$ in the standard sense.

In order to deduce the formal analogy, we embed $\DR(M)^{\otimes 2}$ into $\DR(\Bl_{\Diag}(M\times M))$ using the external wedge product $(\eta_1,\eta_2)\mapsto \tilde{\pi}_1^*\eta_1 \wedge \tilde{\pi}_2^*\eta_2$ and suppose that the Green kernel $\GKer$ satisfies $\GKer \in \DR^{\otimes 2}$. This never happens, so what follows is just a formal computation.

\begin{Proposition}\label{Prop:FinDimAnalog}
In the de Rham case, suppose that $\GKer\in \DR(M)^{\otimes 2}$. Then \eqref{Eq:PushforwardMC} reduces to \eqref{Eq:PushforwardMCdeRham}.
\end{Proposition}
\begin{proof}
Consider the intersection pairing $\tilde{\Pair}$ and its degree shift $\Pair$ (see Proposition~\ref{Prop:DGAs}). According to Definition~\ref{Def:LinSchw}, they extend to pairings on $\DR(M)^{\otimes k}$ and $\DR(M)[1]^{\otimes k}$ for all $k\ge 1$, respectively. For all $\eta_{11}$, $\eta_{12}$, $\eta_{21}$, $\eta_{22}\in \DR(M)$, we have:
\begin{equation}\label{Eq:PairComp}
\begin{aligned}
&\Pair(\SuspU^2 \eta_{11}\otimes \eta_{12},\SuspU^2 \eta_{21}\otimes \eta_{22}) \\ 
&\qquad = (-1)^{\eta_{11} + \eta_{21}} \Pair(\SuspU\eta_{11}\otimes \SuspU\eta_{12}, \SuspU \eta_{21}\otimes \SuspU\eta_{22}) \\
&\qquad = (-1)^{\eta_{11} + \eta_{21} + (1+\eta_{12})(1+\eta_{21})} \Pair(\SuspU \eta_{11}, \SuspU \eta_{21}) \Pair(\SuspU \eta_{12},\SuspU \eta_{22})\\
&\qquad =(-1)^{1+\eta_{12}\eta_{21}} \tilde{P}(\eta_{11},\eta_{21})\tilde{\Pair}(\eta_{12},\eta_{22}) \\
&\qquad = - \tilde{\Pair}(\eta_{11}\otimes \eta_{12}, \eta_{21}\otimes \eta_{22}).
\end{aligned}
\end{equation}
One can also check that
$$ \tilde{\Pair}(\eta_{11} \otimes \eta_{12}, \eta_{21}\otimes\eta_{22}) = \int_{x,y} \eta_{11}(x)\eta_{12}(y)\eta_{21}(x)\eta_{22}(y). $$

For the Green operator $\GOp: \DR(M) \rightarrow \DR(M)$ and its Green kernel $\GKer \in \DR(M)^{\otimes 2}$, we have the following:
$$ \forall \eta_1, \eta_2\in \DR(M): \quad \tilde{\Pair}(\GOp(\eta_1),\eta_2) = \int_{x,y} \GKer(x,y)\eta_1(x)\eta_2(y) = \tilde{\Pair}(\GKer, \eta_1\otimes \eta_2). $$
From this and \eqref{Eq:PairComp}, we obtain
$$\begin{aligned}
\Pair(\GOp(\SuspU \eta_1),\eta_2) &= \Pair(\SuspU \GOp(\eta_1),\SuspU \eta_2) = (-1)^{1+\eta_1} \tilde{\Pair}(\GOp(\eta_1),\eta_2) \\ &= (-1)^{1+\eta_1} \tilde{\Pair}(\GKer,\eta_1\otimes \eta_2) = (-1)^{\eta_1}\Pair(\SuspU^2 \GKer, \SuspU^2 \eta_1 \otimes \eta_2)  \\ &= \Pair(\SuspU^2 \GKer, \SuspU\eta_1 \otimes \Susp\eta_2).
\end{aligned}$$
Therefore, the element $\SuspU^{2} \GKer \in V[1]^{\otimes 2}$ corresponds to the Schwartz kernel $\Kern_{\GOp}$ of $\GOp: V[1] \rightarrow V[1]$. We write this correspondence as
$$ \Kern_{\GOp}\in V[1]^{\otimes 2}\ \sim\ \SuspU^2 \GKer \in \Bl_\Diag(M\times M)[2]. $$

Let us check that $\SuspU^{2}\GKer$ satisfies~\eqref{Eq:SymmetryCondition}. First of all, if we embed $\DR(M)^{\otimes k}$ into $\DR(M^{\times k})$ using the external wedge product $\eta_1 \otimes \dotsb \otimes \eta_k \mapsto \pi_1^*\eta_1 \wedge \dotsm \wedge \pi_k^* \eta_k =: \eta_1(x_1)\wedge \dotsm \wedge \eta_k(x_k)$, then for all $\eta_1$, $\dotsc$, $\eta_k \in \DR(M)$ we have
$$\sigma(\eta_1 \otimes \dotsb \otimes \eta_k)(x_1,\dotsc,x_k) = \eta_1(x_{\sigma_1})\wedge \dotsc \wedge \eta_k(x_{\sigma_k}), $$
where the action on the left-hand side is given by~\eqref{Eq:Perm}. Now, the symmetry property \eqref{Eq:SymProp} implies
$$ \tau(\SuspU^{2}\GKer) = - \SuspU^{2}\tau^*(\GKer) = (-1)^{n+1} \SuspU^{2} \GKer = (-1)^{\Abs{\SuspU^2 \GKer}} \SuspU^2 \GKer. $$
Therefore, the symmetry condition \eqref{Eq:SymmetryCondition} is indeed satisfied.

Let $\Gamma\in \TRRG_{klg}$, and let $L$ be a labeling of $\Gamma$. We abbreviate $\sigma:= \sigma_L \in \Perm_{3k} $. Given $\eta_{ij}\in \DR(M)$ for $j=1$, $\dotsc$, $s_i$ and $i=1$, $\dotsc$, $l$, where $s_i$ is the valency of the $i$-th boundary component, we set $\eta_i = \eta_{i1}\otimes \dotsb \otimes \eta_{is_i}$, $\eta =\eta_1 \otimes \dotsb \otimes \eta_l$, $\alpha_{ij}=\SuspU \eta_{ij}$, $\omega_i = \alpha_{i1}\otimes \dotsb \otimes \alpha_{i s_i}$ and $\omega = \omega_1\otimes \dotsb \otimes \omega_l$. We denote $s := s_1 + \dotsb + s_l$, so that $3k = 2e + s$, where $e$ is the number of internal edges. We have
\begin{align*}(m_2^+)^{\otimes k} \bigl(\sigma((\SuspU^{ 2} \GKer)^{\otimes e} \otimes \omega)\bigr) &= \varepsilon(\SuspU,\eta)(m_2^+)^{\otimes k} \bigl(\sigma((\SuspU^{ 2} \GKer)^{\otimes e} \otimes \SuspU^{s} \eta)\bigr) \\
& = (-1)^{s e (n-1)}\varepsilon(\SuspU,\eta)(m_2^+)^{\otimes k} \bigl(\sigma(\SuspU^{2e+s}\GKer^{\otimes e} \otimes \eta)\bigr) \\
&= \underbrace{(-1)^{\sigma + s e (n-1)} \varepsilon(\SuspU,\eta) }_{=: \varepsilon_1}(m_2^+)^{\otimes k} \bigl(\SuspU^{2e+s} \sigma( \GKer^{\otimes e} \otimes \eta)\bigr),
\end{align*}
where $\varepsilon(\SuspU,\eta)$ is the Koszul sign to order $\SuspU^s \eta_{11}\dots\eta_{l s_l} \mapsto \SuspU\eta_{11} \dots \SuspU \eta_{l s_l}$ and $m_2^+: \DR(M)[1]^{\otimes 3} \rightarrow \R$ is given by $m_2^+ = \Pair(m_2 \otimes \Id)$. We denote $\kappa := \GKer^{\otimes e} \otimes \eta = \kappa_1 \otimes \dotsb \otimes \kappa_{3k}$, $\kappa_i \in \DR(M)[1]$ and compute
\begin{align*}
 & (m_2^+)^{\otimes k} (\SuspU^{3k} \sigma(\kappa)) \\
 &\quad \underset{\mathclap{\substack{\uparrow\rule{0pt}{1.7ex} \\ \ \ \ \Abs{m_2^+} = 3 - n}}}{=} \varepsilon(\sigma, \kappa) (m_2^+)^{\otimes k} (\SuspU^{3k} \kappa_{\sigma_1^{-1}} \otimes \dotsb \otimes \kappa_{\sigma_{3k}^{-1}})  \\
 & \quad = \begin{multlined}[t] (-1)^{\frac{1}{2}k(k-1)n}\varepsilon(\sigma,\kappa) (m_2^+)^{\otimes k}\bigl(\SuspU^3(\kappa_{\sigma_1^{-1}} \otimes \kappa_{\sigma_2^{-1}} \otimes \kappa_{\sigma_3^{-1}}) \otimes \dotsb \\ \otimes \SuspU^3(\kappa_{\sigma_{3k-2}^{-1}} \otimes \kappa_{\sigma_{3k-1}^{-1}} \otimes \kappa_{\sigma_{3k}^{-1}}) \bigr)\end{multlined}\\
 &\quad = \begin{multlined}[t]\overbrace{(-1)^{\frac{1}{2} k(k-1) n  + \kappa_{\sigma^{-1}_{2}} + \dotsb + \kappa_{\sigma^{-1}_{3k-1}}} \varepsilon(\sigma,\kappa)}^{=:\varepsilon_2} (m_2^+)^{\otimes k} \bigl((\SuspU \kappa_{\sigma^{-1}_1} \otimes \SuspU \kappa_{\sigma^{-1}_2} \\ \otimes \SuspU \kappa_{\sigma^{-1}_3}) \otimes \dotsb \otimes (\SuspU \kappa_{\sigma^{-1}_{3k-2}} \otimes \SuspU \kappa_{\sigma^{-1}_{3k-1}} \otimes \SuspU \kappa_{\sigma^{-1}_{3k}})\bigr). \end{multlined}
\end{align*}
Next, using the formula \eqref{Eq:ChernSimons} for $m_2^+$, we get
\begin{align*}
&(m_2^+)^{\otimes k}\bigl((\SuspU \kappa_{\sigma^{-1}_{1}} \otimes \SuspU \kappa_{\sigma^{-1}_{2}} \otimes \SuspU \kappa_{\sigma^{-1}_{3}}) \otimes \dotsb \otimes (\SuspU \kappa_{\sigma^{-1}_{3k-2}} \otimes \SuspU \kappa_{\sigma^{-1}_{3k-1}} \otimes \SuspU \kappa_{\sigma^{-1}_{3k}})\bigr) \\
&\quad= \begin{multlined}[t](-1)^{\kappa_{\sigma^{-1}_{2}} + \dotsb + \kappa_{\sigma^{-1}_{3k-1}}} \Bigl(\int_{x_1} \kappa_{\sigma^{-1}_{1}}(x_1)\kappa_{\sigma^{-1}_{2}}(x_1)\kappa_{\sigma^{-1}_{3}}(x_1) \Bigr) \dotsm \\ \Bigl(\int_{x_k} \kappa_{\sigma^{-1}_{3k-2}}(x_k)\kappa_{\sigma^{-1}_{3k-1}}(x_k) \kappa_{\sigma^{-1}_{3k}}(x_k) \Bigr)\end{multlined} \\
&\quad= \begin{multlined}[t](-1)^{\kappa_{\sigma^{-1}_{2}} + \dotsb + \kappa_{\sigma^{-1}_{3k-1}}} \int_{x_1,\dotsc,x_k} \kappa_{\sigma_{1}^{-1}}(x_1)\kappa_{\sigma_{2}^{-1}}(x_1)\kappa_{\sigma_{3}^{-1}}(x_1) \dotsm \\ \kappa_{\sigma_{3k-2}^{-1}}(x_k) \kappa_{\sigma_{3k-1}^{-1}}(x_k) \kappa_{\sigma_{3k}^{-1}}(x_k) \end{multlined}\\
&\quad = \begin{multlined}[t]\overbrace{(-1)^{\kappa_{\sigma^{-1}_{2}} + \dotsb + \kappa_{\sigma^{-1}_{3k-1}}} \varepsilon(\sigma,\kappa)}^{=:\varepsilon_3} \int_{x_1,\dotsc,x_k} \kappa_{1}(x_{\xi(\sigma_{1})})\kappa_{2}(x_{\xi(\sigma_{2})}) \kappa_{3}(x_{\xi(\sigma_{3})}) \dotsm \\ \kappa_{3k-2}(x_{\xi(\sigma_{3k-2})}) \kappa_{3k-1}(x_{\xi(\sigma_{3k-1})})\kappa_{3k}(x_{\xi(\sigma_{3k})}), \end{multlined}
\end{align*}
where $\xi(3j-2) = \xi(3j-1) = \xi(3j) = j$ for $j=1$, $\dotsc$, $k$ (see Definition~\ref{Def:PushforwardMCdeRham}). In total, we have 
\begin{align*} &(m_2^+)^{\otimes k} \bigl(\sigma((\SuspU^{ 2} \GKer)^{\otimes e} \otimes \omega)\bigr) \\ 
&\quad = \begin{multlined}[t]\varepsilon_1 \varepsilon_2 \varepsilon_3  \int_{x_1,\dotsc,x_k} \GKer(x_{\xi(\sigma_1)},x_{\xi(\sigma_2)}) \dotsm  \GKer(x_{\xi(\sigma_{2e-1})},x_{\xi(\sigma_{2e})}) \\ \alpha_{11}(x_{\xi(\sigma_{2e+1})}) \dotsm \alpha_{ls_{l}}(x_{\xi(\sigma_{2e+s})}),\end{multlined}
\end{align*}
where
$$ \varepsilon_1 \varepsilon_2 \varepsilon_3 = (-1)^{\sigma + s e (n-1)+ \frac{1}{2}k(k-1)n} \varepsilon(\SuspU,\eta). $$
Using~\eqref{Eq:EulerFormula},~\eqref{Eq:TrivalentFormula} and $\varepsilon(\SuspU,\eta) = (-1)^{P(\omega)}$, we get the total sign
$$ (-1)^{k(n-2)}\varepsilon_1 \varepsilon_2 \varepsilon_3  = (-1)^{s(k,l) + \sigma + P(\omega)},  $$
where $(-1)^{k(n-2)}$ is the artificial sign from~\eqref{Eq:PushforwardMC}. This proves the proposition.
\end{proof}

\begin{Remark}[Signs for the Fr\'echet $\dIBL$-structure on $\DR(M)$]\label{Rem:Frechet}
In \cite[Section~13]{Cieliebak2015}, they consider the weight-graded nuclear Fr\'echet space $\DBCyc \DR(M)_{\infty} \subset \DBCyc \DR(M)$ generated by $\varphi\in \DBCyc \DR(M)$ which have a smooth Schwartz kernel $k_\varphi\in \DR(M^{\times k})$; they showed that there is a canonical Fr\'echet $\dIBL$-structure on $\DBCyc \DR(M)_{\infty}[2-n]$. In order to deduce the signs, we can consider the subspace $\DBCyc \DR(M)_{\mathrm{alg}} \subset \DBCyc \DR(M)_\infty$ generated by $\varphi\in \DBCyc \DR(M)$ with an algebraic Schwartz kernel $\Kern_\varphi \in \DR(M)[1]^{\otimes k}$, rewrite \eqref{Eq:DefByGraphs} in terms of $\Kern_\varphi$ and extend the obtained formulas to $\DBCyc \DR(M)_\infty$. This may be done in \cite{MyPhD}.
\end{Remark}

\end{appendices}
\clearpage
\section*{References}\Correct[inline,caption={DONE Bibtex not handling primaryClass correctly}]{Rewrite the bib file so that the source maps for biblatex do not have to be used to correct the bib file. This is necessary to repair the references on arXive, where only bibtex and no biber is used. Bibtex namely ignores source maps and does not correct the entries.}
\normalem
\addcontentsline{toc}{section}{References}
\printbibliography[heading=none]
\end{document}